\documentclass[twocolumn]{aastex7}

\usepackage{graphicx}
\usepackage{float}
\usepackage{amsmath}
\usepackage{amssymb}
\usepackage{subfigure}
\usepackage{cases}
\usepackage{mathrsfs}
\usepackage[flushleft]{threeparttable}
\usepackage{pifont}
\usepackage{epstopdf}
\usepackage{xcolor}
\usepackage[all]{hypcap}
\usepackage{mwe,tikz}
\usepackage{bm}
\usepackage{tabularx}
\usepackage{multirow}
\usepackage{makecell}
\usepackage{longtable}
\usepackage{xspace}
\usepackage{rotating}
\usepackage{upgreek}
\usepackage{booktabs}
\usepackage{textcomp}
\usepackage{makecell}
\usepackage{enumitem}
\usepackage[utf8]{inputenc}
\usepackage{csquotes}
\usepackage{soul}
\usepackage{bbding}
\usepackage{array} 
\usepackage[T1]{fontenc}
\usepackage{float}
\usepackage{tikz}
\usepackage{appendix}
\usepackage{cleveref}
\usepackage[percent]{overpic} 
\usepackage{ulem} 
\usepackage{CJKutf8}
\usepackage{placeins} 
\usepackage{subcaption}

\newcommand{\chinesename}{\textnormal{(}张钊\textnormal{)}}

\newcommand{\e}{\text{e}}

\newcommand{\p}{\partial}

\newcommand{\Msun}{{\mathrm  M_{\odot}}}
\newcommand{\himpc}{h^{-1}{\rm Mpc}}
\newcommand{\Gthree}{\small GADGET3-OSAKA}
\newcommand{\Gfour}{\small GADGET4-OSAKA}

\newcommand{\fdiffmath}{\ensuremath{\lowercase{f}_{\mathrm{diff}}}}
\newcommand{\fhalo}{\ensuremath{\lowercase{f}_{\mathrm{Halos}}}}

\begin{document}
\begin{CJK}{UTF8}{gbsn}
\title{Probing the cosmic baryon distribution and the impact of AGN feedback with FRBs in CROCODILE simulation}
\correspondingauthor{Zhao Joseph Zhang}
\email{zzhang@astro-osaka.jp}

\author[0000-0001-6869-2996]{Zhao Joseph Zhang \chinesename}
\affiliation{Theoretical Astrophysics, Department of Earth and Space Science, The University of Osaka, 1-1 Machikaneyama, Toyonaka, Osaka 560-0043, Japan}
\email{zzhang@astro-osaka.jp}

\author[0000-0001-7457-8487]{Kentaro Nagamine}
\affiliation{Theoretical Astrophysics, Department of Earth and Space Science, The University of Osaka, 1-1 Machikaneyama, Toyonaka, Osaka 560-0043, Japan}
\affiliation{Theoretical Joint Research, Forefront Research Center, Graduate School of Science, The University of Osaka, Toyonaka, Osaka 560-0043, Japan}
\affiliation{Kavli IPMU (WPI), UTIAS, The University of Tokyo, Kashiwa, Chiba 277-8583, Japan}
\affiliation{Department of Physics \& Astronomy, University of Nevada, Las Vegas, 4505 S. Maryland Pkwy, Las Vegas, NV 89154-4002, USA}
\affiliation{Nevada Center for Astrophysics, University of Nevada, Las Vegas, 4505 S. Maryland Pkwy, Las Vegas, NV 89154-4002, USA}
\email{kn@astro-osaka.jp}

\author[0000-0002-5712-6865]{Yuri Oku}
\affiliation{Center for Cosmology and Computational Astrophysics, the Institute for Advanced Study in Physics,  Zhejiang University, China} 
\email{oku@zju.edu.cn}

\author[0000-0001-9299-5719]{Khee-Gan Lee}
\affiliation{Kavli IPMU (WPI), UTIAS, The University of Tokyo, Kashiwa, Chiba 277-8583, Japan}
\affiliation{Center for Data-Driven Discovery, Kavli IPMU (WPI), UTIAS, The University of Tokyo, Kashiwa, Chiba 277-8583, Japan}
\email{kglee@ipmu.jp}

\author[0000-0002-5045-6052]{Keita Fukushima}
\affiliation{Institute for Data Innovation in Science, Seoul National University, Seoul 08826, Republic of Korea}
\email{k-fukushima@astro-osaka.jp}

\author[0009-0007-4378-406X]{Kazuki Tomaru}
\affiliation{Theoretical Astrophysics, Department of Earth and Space Science, The University of Osaka, 1-1 Machikaneyama, Toyonaka, Osaka 560-0043, Japan}
\email{tomaru@astro-osaka.jp}

\author[0000-0002-9725-2524]{Bing Zhang}
\affiliation{Department of Physics \& Astronomy, University of Nevada, Las Vegas, 4505 S. Maryland Pkwy, Las Vegas, NV 89154-4002, USA}
\affiliation{Nevada Center for Astrophysics, University of Nevada, Las Vegas, 4505 S. Maryland Pkwy, Las Vegas, NV 89154-4002, USA}
\email{bing.zhang@unlv.edu}

\author[0009-0000-7415-8239]{Isabel Medlock}
\affiliation{Department of Astronomy, Yale University, New Haven, CT 06520, USA}
\email{isabel.medlock@yale.edu}

\author[0000-0002-6766-5942]{Daisuke Nagai}
\affiliation{Department of Astronomy, Yale University, New Haven, CT 06520, USA}
\affiliation{Department of Physics, Yale University, New Haven, CT 06520, USA}
\email{daisuke.nagai@yale.edu}

\begin{abstract}
We investigate the missing baryon problem using Fast Radio Bursts (FRBs) to trace cosmic baryons. Our CROCODILE simulations, performed with the {\small GADGET3/4-OSAKA} smoothed particle hydrodynamics code, include star formation, supernova (SN) and active galactic nuclei (AGN) feedback.
We generate light cones from LSS simulations to compute gas density profiles and dispersion measures (DMs) measurable by FRBs.
Our results show that AGN feedback reduces central gas densities in halos, reshaping the boundary between the circumgalactic medium (CGM) and intergalactic medium (IGM).
Zoom-in simulations reveal that AGN feedback significantly modulates the DM contributions from foreground halos along different sight lines.
Using the DM--redshift (DM--$z$) relation up to $z=1$, we constrain the diffuse baryon mass fraction at $z = 1$ to $f_{\mathrm{diff}} = 0.865^{+0.101}{-0.165}$ (fiducial) and $f{\mathrm{diff}} = 0.856^{+0.101}{-0.162}$ (NoBH), which includes contributions from both IGM ($f\text{IGM}$) and halos ($f_\text{Halos}$), serving as upper limits.
We further separate and quantify the redshift evolution of $f_{\mathrm{CGM}}$, $f_{\mathrm{IGM}}$, and $\langle f_{\mathrm{diff, obs}} \rangle$, using both phase-based and structure-based definitions.
Our study provides a framework for understanding baryon distribution across cosmic structures, FRB host galaxies, and the role of AGN in shaping foreground DM contributions.
\end{abstract}


\section{introduction}
Understanding the spatial distribution and evolution of baryons is crucial for studying cosmic structure formation. Observations \citep{DS2005ApJ...624..555D,DS2008ApJ...679..194D, Tilton2012ApJ...759..112T, Tripp2006ASPC..348..341T,Richter2004,Richter2006,Salucci1999MNRAS.309..923S,Reiprich2002ApJ...567..716R,Zwaan2003AJ....125.2842Z, Prochaska2011ApJ...740...91P} and simulations \citep[e.g.,][]{Shull2012, Cen2012ApJ...753...17C} suggest that a large fraction ($\sim$50--70\%) of baryons reside in the IGM, yet a significant fraction remains unaccounted for — the so-called “missing baryon problem”.
Hydrodynamical cosmological simulations \citep{CO1999ApJ...514....1C, Dave2001ApJ...552..473D,Yoshikawa2001ApJ...558..520Y} indicate that these baryons likely exist in the warm-hot IGM (WHIM, $10^5-10^7$\,K).  

Observations using X-ray \citep{Nicastro2005a,Nicastro2005b,Yao2012ApJ...746..166Y, Ren2014, Nicastro2018, Kovacs2019, Zhang2024}, Ly$\alpha$ \citep{Penton2004, DS2008ApJ...679..194D, Tejos2016, Kovacs2019, Hu2024}, and the SZ effect \citep{deGraaff2019, Singari2020, Erciyes2023, Tanimura2023, Hadzhiyska2025} have placed significant constraints on the WHIM. However, these methods are limited by observational uncertainties, necessitating alternative approaches to trace ionized baryons. Beyond direct observations, cosmological simulations \citep{Martizzi2019, Gheller2019, Medlock2021} have been extensively employed to study the baryon distribution in the WHIM, further highlighting its role in solving the missing baryon problem.

Fast Radio Bursts (FRBs), powerful transient radio pulses of extragalactic origin, were first discovered by \citet{Lorimer2007}. Subsequent observations by \citet{Thornton2013} confirmed their cosmological distances and further established FRBs as a new class of astrophysical transients. These bursts offer a promising avenue for mapping baryons through their DM, which traces the column density of electrons along the line of sight (LoS). Even before FRBs were discovered, \citet{Ioka2003} and \citet{Inoue2004} discussed using the DM from gamma-ray Bursts (GRBs) to study cosmic reionization and the IGM.

\cite{McQuinn2014} was the first to use simulations to demonstrate how FRBs can be utilized to probe cosmology and galaxy physics. This study laid the theoretical foundation for using FRBs as cosmic probes. Later, \cite{DZ2014} and \cite{Zheng2014} further explored the DM--$z$ relation in the context of FRBs, highlighting its implications for cosmology and baryon tracing. Subsequently, \cite{Macquart2020Nature} established an observational correlation between DM and FRB redshifts, now known as the Macquart relation, which links DM with FRB redshifts and provides a new method to quantify baryon content in the IGM. Building upon this, \cite{Li2019} proposed a cosmology-independent approach to estimate \( f_{\mathrm{IGM}} \)\footnote{Here, $f_{\text{IGM}}$ is inferred from the total observed DM including the contribution from the foreground halos. Therefore, to be more precise, it is better described as the diffuse gas fraction, $f_{\text{diff}}$, as we will discuss later in the paper. I.e., $\text{DM}_{\rm diff} = \text{DM}_{\rm IGM} + \text{DM}_{\rm Halos}.$ See Eq.~(\ref{DM_FRB}) and associated text for more details.} using FRB \(\mathrm{DM}\) and luminosity distances (\( d_L \)), demonstrating that as few as 50 FRBs can constrain \( f_{\mathrm{IGM}} \)\footnote{Same as above, this refers to $f_{\text{diff}}$ as defined later.} within 12\% uncertainty. Subsequently, \cite{ZJZ2021} applied this relation along with IllustrisTNG \citep{Weinberger2017, Pillepich2018MNRAS.473.4077P} simulation data to constrain the cosmological \(\mathrm{DM}\) of FRBs and analyze its redshift evolution. fiducialWith instruments like Parkes, ASKAP, Arecibo, CHIME, and DSA-110 detecting thousands of FRBs annually \citep{Lorimer2007, Spitler2014, Bannister2017, Ravi2023}, FRBs are emerging as a powerful probe of cosmic baryon distribution (e.g., \citealt{Bhandari2021, Shirasaki2022, Zhang2023RvMP...95c5005Z, Glowacki:2024, Wu2024ChPhL}).

Recent studies have refined estimates of the baryon fraction in the IGM ($f_{\rm IGM}$), leveraging well-localized FRB samples \citep{Yang2022ApJ...940L..29Y,Wang2023ApJ...944...50W}. 
However, disentangling DM contributions from the CGM of foreground galaxies remains challenging. Observational studies have highlighted the role of foreground galaxy halos in contributing to the total DM budget. \citet{Prochaska2019Sci...366..231P} and \citet{Li2019ApJ...884L..26L} demonstrated that intervening galaxy environments can significantly impact FRB dispersion, underscoring the necessity of accounting for these contributions in DM-based cosmological studies. Additionally, \cite{Connor&Ravi2022} and \cite{WuMQ2023} quantified the CGM's contribution to DM using CHIME/FRB data, while \citet{Lee2022,Lee2023} and \citet{Simha2023ApJ...954...71S} incorporated galaxy density reconstructions to refine baryon fraction estimates.

Large-scale structure (LSS) simulations have further elucidated baryon redistribution mechanisms. \cite{Sorini2022MNRAS.516..883S} and \cite{Khrykin2024MNRAS.529..537K} highlight the roles of SN and AGN feedback, with AGN-driven outflows being particularly effective at expelling baryons into the IGM. \cite{Zhu2021} and \cite{Mo2025} used RAMSES and IllustrisTNG-100 (TNG-100), respectively, to examine the contributions of various baryonic components to the DM and scattering of FRBs. \cite{Medlock2024} used CAMELS simulations \citep{CAMELS_presentation} to analyze DM profiles of dark matter halos, demonstrating the significance of CGM contributions.  
\cite{KHRYKIN2024} refined these estimates by separately quantifying CGM and IGM fractions, reporting a lower $f_{\rm IGM}$ of $\sim 60$\%, highlighting the need for precise modeling of CGM contributions when interpreting FRB observations. In contrast, \cite{Connor2024} found a significantly higher $f_{\rm IGM} \approx 0.80$, suggesting that the majority of baryons reside in the IGM, with only $\sim$11\% in galaxy halos (CGM). Using a large sample of localized FRBs from DSA-110, they constrained the cosmic baryon budget and found results that align closely with predictions from IllustrisTNG and SIMBA \citep{Dave2019} simulations, which also favor a baryon-rich IGM scenario. Their findings reinforce the effectiveness of FRBs in tracing diffuse baryons while implying a more limited role for the CGM. The discrepancy between these results underscores the importance of improving theoretical models of baryon redistribution and CGM-IGM interactions.

Recent work by \cite{Sorini2025} further explored the impact of baryons on the internal structure of dark matter haloes using TNG and MillenniumTNG simulations, demonstrating that baryonic feedback—including both stellar and AGN processes—can reduce total halo mass by up to 20\%. Moreover, baryonic effects, especially AGN feedback, can flatten the concentration–mass relation and reshape halo density profiles within the intermediate-mass range ($10^{11.3}$–$10^{13}\,M_\odot$). This restructuring has direct implications for interpreting FRB sight lines through massive systems and for calibrating CGM-related DM contributions.

Beyond the IGM, the host galaxy (HG) also contributes significantly to the DM of FRBs, with variations arising from different HG types and source locations within the host. The nature of FRB host galaxies provides crucial clues about their progenitors. \citet{Macquart2020Nature} first analyzed five ASKAP-localized FRBs and found that their host properties resemble those of core-collapse SNe, Type Ia SNe, and short GRBs \citep{Heintz2020, Bhandari2022}. This sample has since been expanded to over 90 localized FRBs \citep{Gordon2023a, Bhardwaj2024, Kocz2019, Law2024, Sharma2024, Connor2024, Shannon2024, Chime2025}, significantly improving our understanding of FRB host properties.

FRB host galaxies exhibit a wide range of stellar populations. Early studies \citep{Li2020ApJ...899L...6L} suggested that FRBs are predominantly found in intermediate-to-old stellar populations, consistent with both magnetars formed via extreme explosive events and those arising from standard stellar evolution. Recent large-sample analyses \citep{Sharma2024, Chime2025} further refined this classification, showing that most FRBs reside in Milky Way (MW)-like spirals, with a smaller fraction in dwarf and elliptical galaxies.

Observational studies have extensively explored the DM contribution from host galaxies. From FRB data, the typical DM$_{\text{HG}}$ is approximately 100~pc cm$^{-3}$, as shown in \citet{Li2019ApJ...884L..26L}. More directly, this value can also be derived from a simple linear fit to the DM$_{\text{E}}$--$z$ relation, which exhibits a y-intercept of around 100~pc cm$^{-3}$, as presented in \citet{Zhang2023RvMP...95c5005Z}.

From a numerical simulation perspective, \citet{Zhang2020} utilized the TNG-100 simulation, which has a 100~Mpc box size, to study the relationship between the DM contribution of FRB host galaxies and different FRB types, their physical origins (central engines), and their locations within their host galaxies.\citet{Mo2023} complemented this research by analyzing both Illustris and TNG simulations, and further distinguished between young and old progenitor populations, revealing significant differences in the resulting host DM distributions. More recently, \citet{Theis2024} used multiple simulation suites (TNG, SIMBA, ASTRID) to systematically analyze the DM contributions of FRB host galaxies and their dependence on galaxy formation models. Additionally, analytical models such as those in \citet{Reischke2024} provide a complementary approach to understanding host galaxy DM contributions and their impact on FRB sight lines.

The AGORA (Assembling Galaxies of Resolved Anatomy) project \citep{Kim2014ApJS..210...14K, Santi2021ApJ, Santi2024ApJ} plays a crucial role in benchmarking numerical models of galaxy formation, providing high-resolution simulations that enhance our understanding of FRB host galaxy DM contributions. We utilize one of their simulations for our analysis of the host galaxy's contribution to the total DM.

Additionally, the contribution of the MW's DM to FRB sight lines must be carefully accounted for. Empirical models such as NE2001 \citep{Cordes_Lazio2002} and YMW16 \citep{Yao2017} are widely used to estimate the Galactic DM contribution by modeling the distribution of free electrons in the disk and halo. \citet{Prochaska2019MNRAS.485..648P} further investigated the Galactic DM component, highlighting its significant impact on the total observed DM. Accurately characterizing these contributions is essential for isolating extragalactic DM components in FRB studies.

Motivated by these studies, we employ the SPH-based  {\small GADGET3/4-OSAKA} code \citep{Shimizu2019MNRAS.484.2632S,Nagamine2021,Oku2022,Romano2022a,Romano2022b,Fukushima2023MNRAS.525.3760F,Oku2024}, incorporating star formation and feedback models to simulate FRB sight lines. By combining zoom-in and LSS simulations, we disentangle the host galaxy and IGM contributions to DM. We further introduce a light cone generation technique using randomly rotated simulation boxes to mitigate boundary artifacts, enhancing our ability to trace DM-z evolution and constrain $f_{\rm IGM}$\footnote{See previous footnotes regarding \( f_{\mathrm{diff}} \).}.
\section{Dispersion measure (DM) of FRB} \label{sec:FRB_DM}

FRBs are brief but intense bursts of radio waves, emitting energies between $10^{38}$ and $10^{42}$ erg/s. 
Their DM, as first reported in the discovery of FRB~010724 by \citet{Lorimer2007}, results from frequency-dependent delays as radio waves propagate through plasma. This occurs due to interactions with free electrons, which alter the refractive index and slow lower-frequency waves.

The DM quantifies this effect as the time delay relative to a vacuum \citep{DZ2014, Zheng2014}:
\begin{equation}
\Delta t_{obs} = \int\frac{\nu_p^2}{\nu_{c}^2} \frac{dl}{c} =  \frac{e^2}{2\pi m_e c \nu_{c}^2} \mathrm{DM}, \label{eq:dt_obs}
\end{equation}
where $\nu_p = (n_{e,z}e^2/\pi m)^{1/2}$ is the plasma frequency, $\nu_c$ is comoving frequency, $c$ is the light speed $e$, $m_e$, $n_e$ are the charge, mass and number density of electrons, $t_{\rm obs}$ is the observed time. For a plasma at a redshift $z$, the rest-frame delay time is 
\begin{equation}
\Delta t_{z} = \frac{e^2}{2\pi m_e c \nu_z^2} \int_0^L n_{e,z} \, dl_z = \frac{e^2}{2\pi m_e c \nu_z^2} \mathrm{DM}_z, \label{eq:dt_z}
\end{equation}
where $L$ is the propagation distance of the EM wave through the plasma, $\nu_z$ is the rest-frame frequency at the source (or proper coordinate system), $\int_0^L n_{e,z} dl_z = \text{DM}_z$ is the rest-frame DM. In the observed frame, $\Delta t_{z} = \Delta t_{\rm obs}/(1+z)$ and $\nu_{z} = \nu_{c}(1+z)$. So equation \ref{eq:dt_z} can be adapted as 
\begin{equation}
\Delta t_{\rm obs} = \frac{e^2}{2\pi m_e c \nu_{c}^2} \int_0^L \frac{n_{e,z}}{1+z} \, dl_z.  \label{eq:Dt}
\end{equation}
Comparing this with Eq.\,(\ref{eq:dt_obs}), the measured DM at the observer frame is 
\begin{equation}
\text{DM} = \int_0^L \frac{n_{e,z}(l)}{1+z} \, dl_z. \label{eq:DM}
\end{equation}

When analyzing simulation results, we account for the DM in the comoving frame. Therefore, the emitting-frame length $dl_z$ is written as a cosmological distance incorporating the expansion of the Universe as follows:
\begin{equation}
dl_z = \frac{c \, dz}{(1 + z) H_0 E(z)}, \label{eq:dlz}
\end{equation}
where $H_0$ is the Hubble constant, $\Omega_m$ and $\Omega_{\Lambda}$ are the matter density and dark energy density parameters, and $E(z) = \sqrt{\Omega_m (1 + z)^3 + \Omega_{\Lambda}}$. Using the comoving electron density $n_{e, c} = n_{e, z}/(1+z)^3$, the equation \ref{eq:DM} becomes
\begin{equation}
\text{DM} = \frac{c}{H_0} \int_0^z \frac{n_{e,c} (1+z')}{E(z')} \, dz'. \label{eq:DM_cos}
\end{equation}

For a cosmological FRB, the total observed DM consists of contributions from multiple sources:
\begin{equation}
\text{DM}_{\rm FRB} = \frac{\text{DM}_{\rm HG}}{1+z} + \text{DM}_{\rm IGM} + \text{DM}_{\rm Halos} + \text{DM}_{\rm MW}. \label{DM_FRB}
\end{equation}
Here, $\text{DM}_{\rm HG}$ is the host galaxy's contribution, which includes the host-halo (HH) and source contributions, i.e., $\text{DM}_{\rm HG} = \text{DM}_{\rm source} + \text{DM}_{\rm HH}$. $\text{DM}_{\rm MW}$ arises from the MW, $\text{DM}_{\rm IGM}$ is for the IGM (excluding the intervening halos), and 
$\text{DM}_{\rm Halos}$ is for the gas in the intervening halos along the LoS (the foreground galaxies and their halos). 

In this paper, we also use the diffuse dispersion measure, $\text{DM}_{\rm diff}$\footnote{Note that \citet{Macquart2020Nature} referred to this quantity as DM$_{\rm cosmic}$, which is equivalent to our DM$_{\rm diff}$. However, our $f_{\rm diff}$ explicitly excludes stars and compact stellar remnants (e.g. black holes), whereas $f_{\rm cosmic}$ may include them, since DM only traces ionized components along the LoS, while $f_{\rm cosmic}$ may account for all baryons regardless of their ionization state. See Appendix~\ref{APP:f_diff_appendix} for a detailed discussion of $f_{\rm diff}$}, to represent the sum of $\text{DM}_{\rm IGM}$ and $\text{DM}_{\rm Halos}$, 
i.e., $\text{DM}_{\rm diff} = \text{DM}_{\rm IGM} + \text{DM}_{\rm Halos}.$
The total observed $\text{DM}_{\rm FRB}$ for FRBs at redshifts $z \lesssim 1$ is typically $\lesssim 10^3$ pc cm$^{-3}$ \citep{Lorimer2007,Thornton2013}.

Using our cosmological simulations, we can compute the total FRB DM contribution from the diffuse baryonic medium as follows: 
\begin{equation}
\text{DM}_{\text{diff}} = f_{\text{diff}} \frac{\rho_{c,0} \Omega_{b}}{m_p}\frac{c}{H_0}\int_0^z \frac{\chi_e(z') (1 + z')}{E(z')} \, dz'
\label{eq:dm_igm}
\end{equation}
where $\rho_{c, 0}$ is the critical mass density and $\Omega_b$ is the baryon density parameter at $z = 0$.  $f_{\rm diff} = f_{\rm IGM} + f_{\rm Halos}$ represents the baryon fraction in the diffuse medium (including both IGM and halos), $m_p$ is the proton mass, and $\chi_e(z)$ is the electron abundance from ionized hydrogen and helium, which is approximated by $\chi_e(z) \approx Y_H \,\chi_{e,\mathrm{H}}(z) + Y_{\rm He}\, \chi_{e,\mathrm{He}}(z)/2$, where $Y_{\rm H}$ and $Y_{\rm He}$ are the mass fractions of hydrogen and helium, respectively, assuming complete ionization and ignoring heavier elements.

This formulation aligns with previous works on the Macquart relationship and allows us to constrain $f_{\rm diff}$ using our simulation results, providing a robust framework to understand the baryonic distribution in the Universe.

\section{LSS Simulations and their Results} 
\label{sec:large-Scale_Structure }

\subsection{Simulation Setup}

Our results utilize data from the {\small GADGET3/4-OSAKA} simulation suite. The {\small GADGET-3} code, an evolved version of {\small GADGET-2} \citep{Springel2005}, serves as the foundation for our {\Gthree} simulations, incorporating enhancements for star formation and supernova feedback \citep{Shimizu2019MNRAS.484.2632S, Oku2022}. The {\Gfour} simulation extends this with additional feedback mechanisms, particularly within the CROCODILE dataset\footnote{\url{https://sites.google.com/view/crocodilesimulation/home}} \citep{Oku2024}, which implements updated SN and thermal AGN feedback models, providing a comprehensive range of runs with diverse feedback parameters.

We analyze multiple cosmological simulations across a range of box sizes and feedback implementations:
$25\,\himpc$ with AGN feedback (L25N512$_{\text{fiducial}}$),
$50\,\himpc$ with (L50N512$_{\text{fiducial}}$) and without (L50N512$_{\text{NoBH}}$) AGN feedback, 
$100\,\himpc$ with (L100N1024$_{\text{fiducial}}$) and without (L100N1024$_{\text{NoBH}}$) AGN feedback, 
and $500\,\himpc$ with stellar feedback but no AGN feedback (L500N1024$_{\text{NoBH}}$). AGN feedback is not included in the current 500 Mpc run due to data availability

Table~\ref{tab:simulation} summarizes the parameters, including box size, total particle count $N_p$, dark matter and gas-particle masses ($m_{\rm DM}$, $m_{\rm gas}$), and gravitational softening length $\epsilon_{\rm grav}$. Initial conditions were generated using {\small MUSIC2} \citep{HahnAbel2013}, with Planck 2018 cosmological parameters \citep{Planck2018}: $H_0 = 67.74$\,km\,s$^{-1}$\,Mpc$^{-1}$, $\Omega_m = 0.3099$, $\Omega_\Lambda = 0.6901$, and $\Omega_b = 0.04889$. Our treatment of supermassive black holes (SMBHs) follows the model used in the EAGLE simulation \citep{Schaye2015, Crain2015}. Black hole seeding is performed using the Friend-of-Friends (FoF) halo finder, where a gas particle with the minimum gravitational potential is converted into a black hole particle if the total FoF halo mass ($M_{\text{halo, FoF}}$) exceeds \(M_{\text{seeding, FoF}} = 10^{10} h^{-1} M_{\odot}\) and the stellar mass in the halo is greater than \(M_{\text{seeding, star}} = 10^8\,h^{-1}\,M_{\odot}\), provided the halo does not already contain a black hole. Black holes are repositioned within the HH based on the potential minimum and assigned the velocity of the FoF halo. The accretion onto black holes follows the Bondi-Hoyle prescription, and feedback is implemented via a thermal quasar mode, affecting the surrounding gas distribution. For more details on the SMBH and AGN feedback implementation in CROCODILE simulations, see \cite{Oku2024}.


Appendices \ref{App:LSS_overview}--\ref{App:boxsize_effect} provide details on these simulations, contextualizing the role of AGN feedback in our FRB-related results. This supplementary information supports our primary analyses while maintaining a focus on FRB implications in the main text.
The following sections present results from LSS simulations, followed by zoom-in cosmological hydrodynamic simulations in Section~\ref{sec:Host_Galaxy}.

\begin{table*}[ht]
\centering
\caption{Simulation Parameters for Different Simulations}
\resizebox{\textwidth}{!}{
\begin{tabular}{|c|c|c|c|c|c|c|}
\hline
\textbf{Simulation} & \textbf{Box Size} & \textbf{Particle Numbers} & \textbf{Dark Matter Mass} & \textbf{Gas Mass} & \textbf{$\epsilon_G^\dagger$} & \textbf{AGN Feedback} \\
 & ($h^{-1}\,\text{cMpc}$) & & ($h^{-1}\,M_\odot$) & ($h^{-1}\,M_\odot$) & ($h^{-1}\, \text{ckpc}$ / $h^{-1}\, \text{pkpc}$) & \\
\hline
L25N512\textsubscript{fiducial} & 25 & $2 \times 512^3$ & $8.43 \times 10^6$ & $1.58 \times 10^6$ & 1.63 / 0.25 & \checkmark \\
L50N512\textsubscript{fiducial} & 50 & $2 \times 512^3$ & $6.75 \times 10^7$ & $1.26 \times 10^7$ & 3.38 / 0.50 & \checkmark \\
L50N512\textsubscript{NoBH} & 50 & $2 \times 512^3$ & $6.75 \times 10^7$ & $1.26 \times 10^7$ & 3.38 / 0.50 & \ding{53} \\
L100N1024\textsubscript{fiducial} & 100 & $2 \times 1024^3$ & $6.75 \times 10^7$ & $1.26 \times 10^7$ & 3.38 / 0.50 & \checkmark \\
L100N1024\textsubscript{NoBH} & 100 & $2 \times 1024^3$ & $6.75 \times 10^7$ & $1.26 \times 10^7$ & 3.38 / 0.50 & \ding{53}\\
L500N1024\textsubscript{NoBH} & 500 & $2 \times 1024^3$ & $8.43 \times 10^7$ & $1.58 \times 10^7$ & 16 / 2.50 & \ding{53} \\
\hline
\end{tabular}
}

\vspace{5pt} 
\noindent \\
$^\dagger$ $\epsilon_G$ represents the gravitational softening length. The two values listed correspond to the softening comoving length (left, in $h^{-1}$ ckpc) and the softening max physical length (right, in $h^{-1}$ pkpc), which is the resolution limit of the simulation.

\label{tab:simulation}
\end{table*}

\begin{figure}[htbp]
    \centering
        \includegraphics[width=0.48\textwidth]{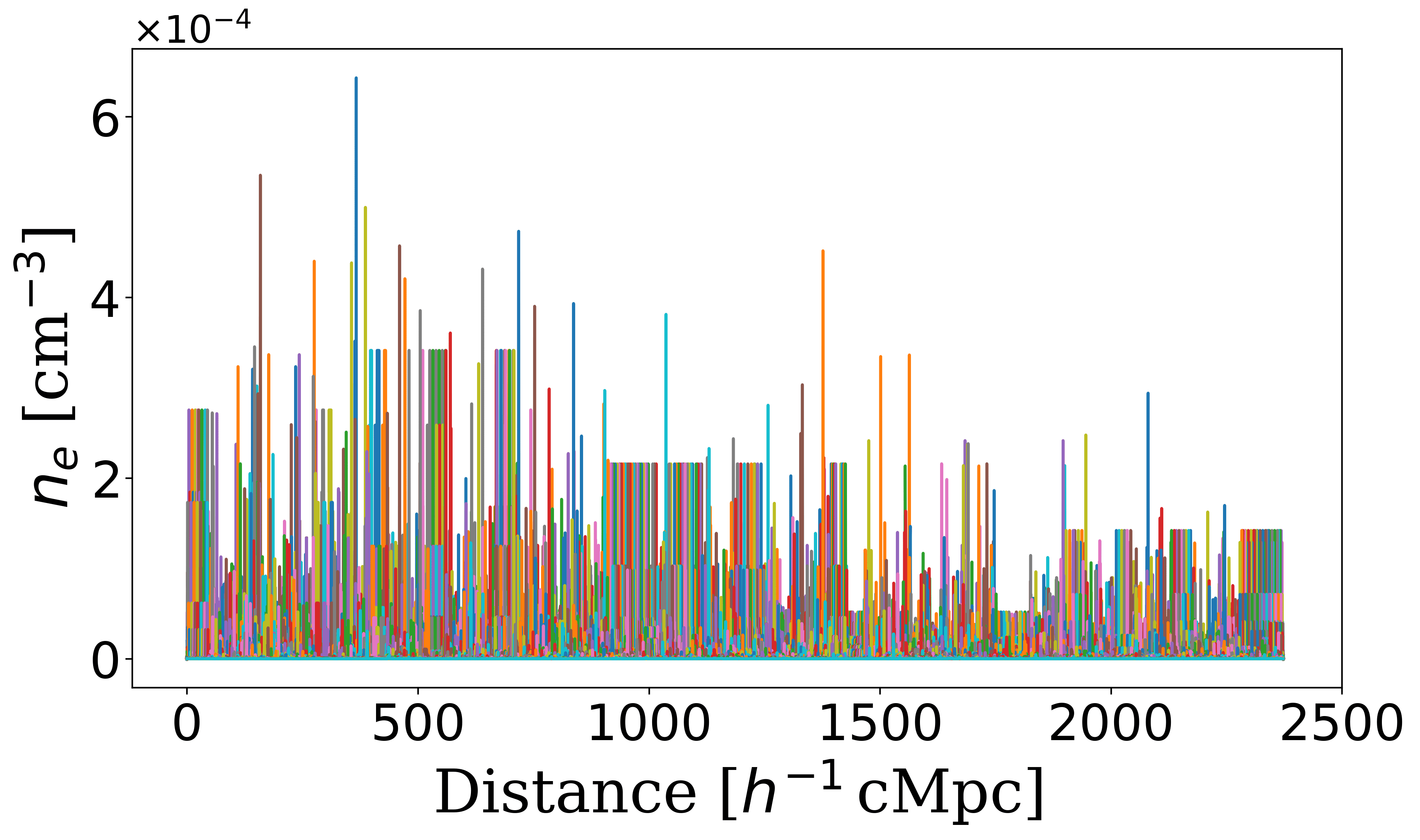}
    \caption{
    Electron density ($n_e$) distribution along 10,000 light cones in the fiducial model with AGN feedback. The integration spans 2500 Mpc in comoving space from $z = 0$ to $z \approx 1$. The NoBH model exhibits a similar distribution, with no visually significant differences, and is therefore omitted for clarity. Different colors represent different lines of sight.
    }

    \label{fig:ne_LC}
\end{figure}


\begin{figure*}[htbp]
    \centering
    \includegraphics[width=0.49\textwidth]{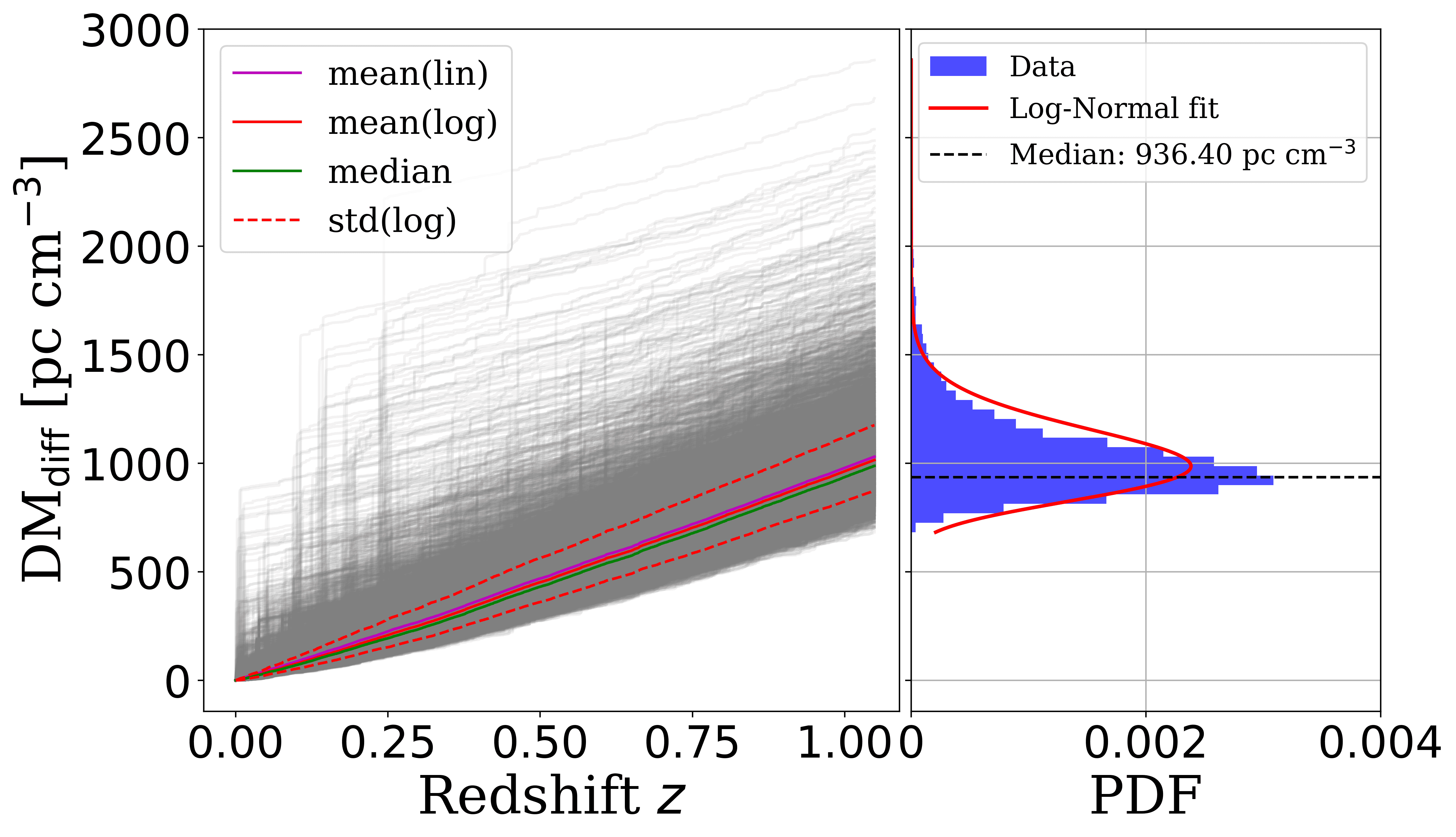}
    \includegraphics[width=0.49\textwidth]{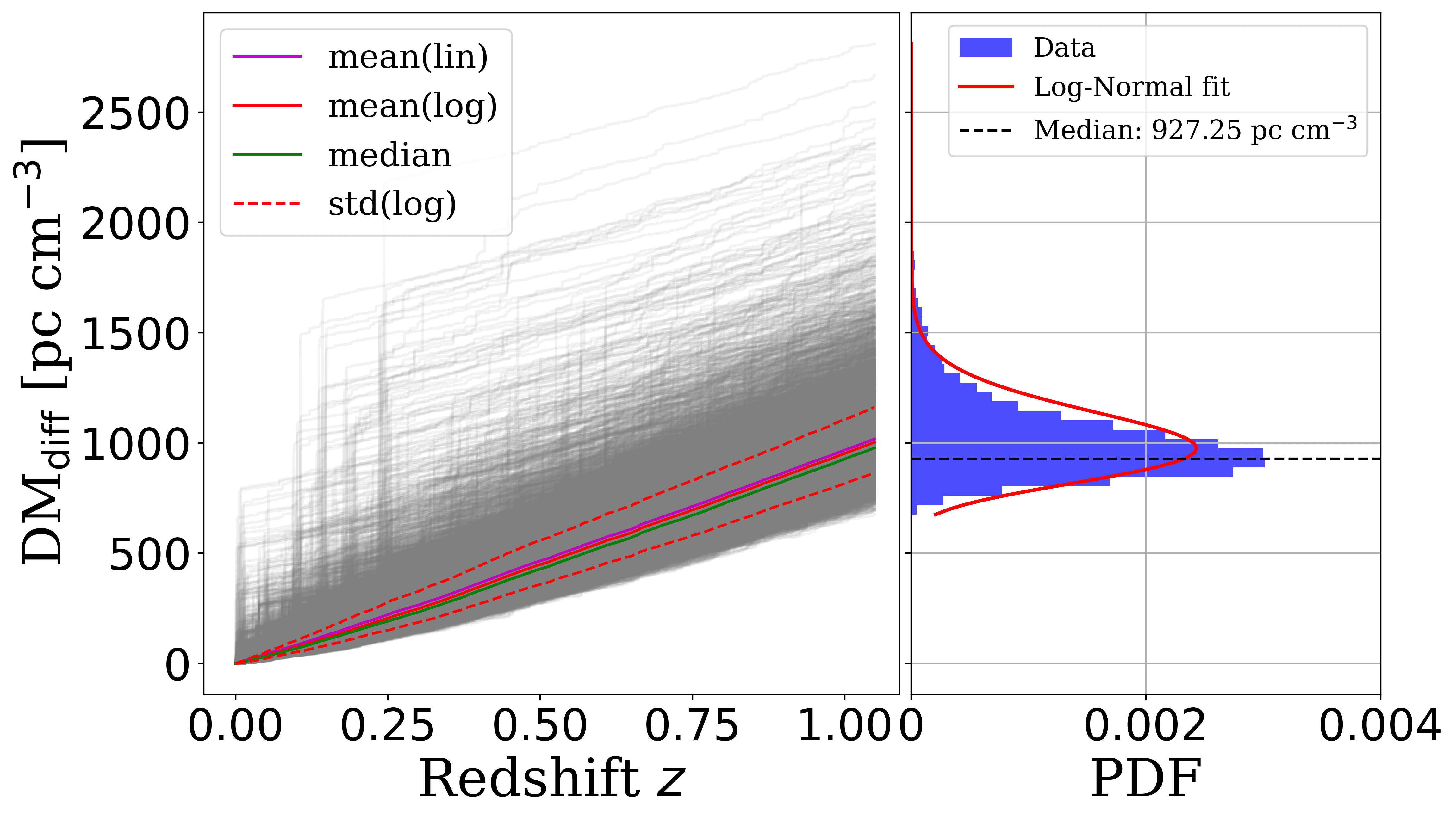}
    \caption{DM of all LoS are shown for the L100 fiducial model with AGN feedback (left panel), and the NoBH model (right panel). In each panel, the left subpanel shows the DM-$z$ relationship, with linear and logarithmic averages indicated, alongside the median and standard deviation of the DM values. The right sub-panel presents the histogram of DM values at $z$ = 1, with the best-fit log-normal distributions shown in red curves, and the dashed lines mark the median DM values of 936.40 $\mathrm{pc \, cm^{-3}}$ for the fiducial model and 927.25 $\mathrm{pc \, cm^{-3}}$ for the NoBH model.}
    \label{fig:combined_dm}
\end{figure*}

\begin{figure*}[htbp]
    \centering
     \includegraphics[width=0.49\textwidth]{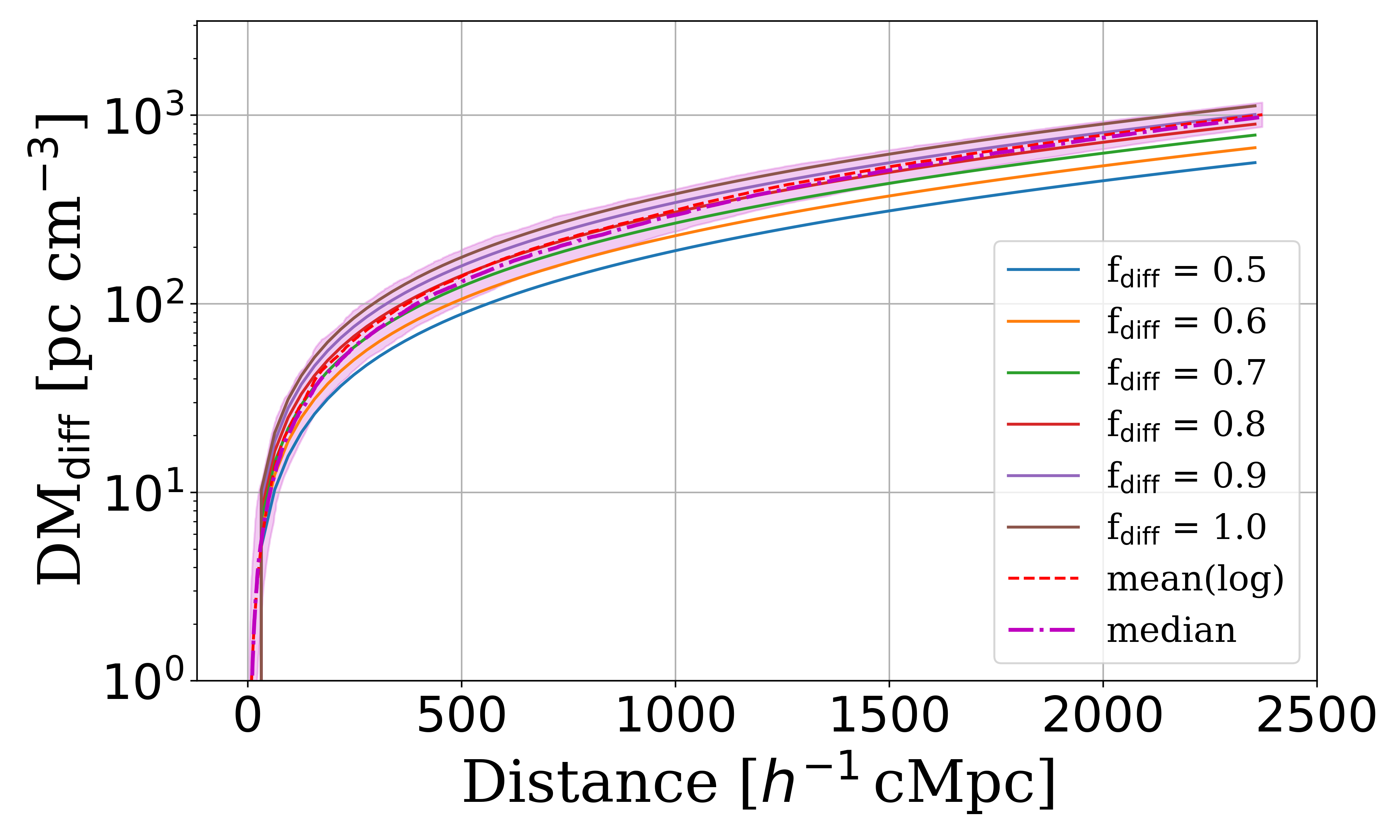}
    \includegraphics[width=0.49\textwidth]{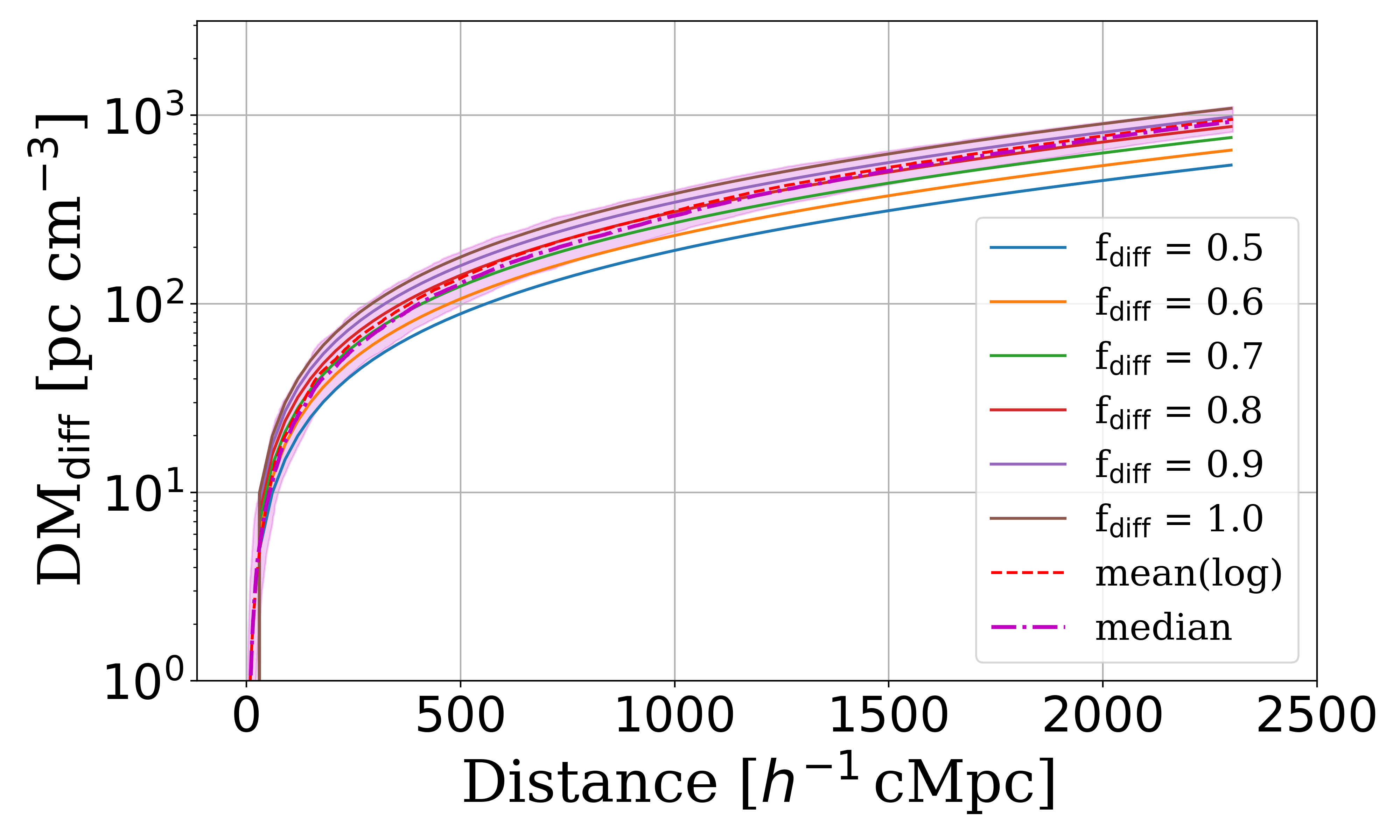}
    \caption{The dispersion measure up to certain distances, obtained from the light-cone data of CROCODILE simulation (log mean and median) and the model results calculated from Eq.~\ref{eq:DM_cos} for different values of $f_{\mathrm{diff}}$ (for both IGM and intervening halo gas).  The left panel shows results for the fiducial simulation, while the right panel corresponds to the NoBH simulation. The magenta-shaded region represents the 16th–84th percentile DM uncertainty range obtained from our simulations. The red dashed line and the purple dash-dotted line indicate the logarithmic mean and the median DM value from the simulation, respectively. The solid lines correspond to theoretical DM-$z$ relations derived from the Macquart relation for different values of $f_{\mathrm{diff}}$ ranging from 0.5 to 1.0.} \label{fig:DM_f_IGM} 
\end{figure*}

\subsection{Cosmological Dispersion Measure in LSS Simulations}
\label{subsec:DM_in_LSS}
To construct FRB sight lines, we randomly rotate LSS simulation boxes and seamlessly connect the LoS across redshifts. This method ensures comprehensive cosmic structure sampling beyond fixed parallel directions. 
The detailed method of light-cone generation is described in Appendix~\ref{App:LC}. We note that in our light-cone setup, FRB sight lines are initiated from a $100 \times 100$ uniformly spaced grid on the $z = 0$ plane. This ensures broad spatial sampling, however, without enforcing that each origin or endpoint resides within a halo. This modeling choice and its possible impact on DM estimation are discussed in Appendix~\ref{APP:halo_los}.

Figure~\ref{fig:ne_LC} shows the electron density distributions along 10000 sight lines in the L100 fiducial simulation (with AGN feedback). Each colored line represents the electron density along an individual sightline, reflecting the underlying baryonic distribution. The x-axis denotes the comoving distance. Although the NoBH model was also analyzed, its distribution is visually indistinguishable from the fiducial case, so it has been omitted for clarity.

By integrating electron density along sight lines from $z=0$, we obtain the cumulative DM as a function of redshift (Figure~\ref{fig:combined_dm}), spanning $ z = 0 $ to $z \approx 1$ ($\sim 2500$ Mpc comoving distance). 
The fiducial model (left panel) shows a slightly higher average DM due to AGN feedback redistributing gas in the IGM. However, this effect is modest since CROCODILE lacks strong jet feedback, making AGN-driven outflows weaker than those in simulations like SIMBA or IllustrisTNG. Since we still include foreground halos, the DM here represents $\mathrm{DM}_{\mathrm{diff}}$ rather than $\mathrm{DM}_{\mathrm{IGM}}$, reducing the influence of AGN feedback on the commonly used $f_{\mathrm{IGM}}$ parameter, which in fact represents $f_{\mathrm{diff}}$.

The subpanel in each figure presents a histogram of DM values, well described by a log-normal distribution:
\begin{equation}
P(x) = \frac{1}{\sqrt{2\pi}\sigma x} \exp\left(-\frac{(\ln x - \ln \mu)^2}{2\sigma^2}\right),
\label{eq:lognormal}
\end{equation}
where $\mu$ is the median value of DM and $\sigma$ is standard deviation of $\ln(\mathrm{DM})$. The mean and standard deviation of DM in this log-normal framework are given by:
\begin{equation}
    E(\text{DM}) = \mu e^{\frac{1}{2} \sigma^2}, \quad \mathrm{Std}(\text{DM}) = \mu e^{\frac{1}{2} \sigma^2} \sqrt{e^{\sigma^2} - 1}.
    \label{eq:lognorm_mean_std}
\end{equation}
The long tails in both models are due to intervening halos with large amounts of ionized plasma.

For the fiducial model, we find $\mu_{\mathrm{fiducial}} = 0.1727 \pm 0.0017$ and $\sigma_{\mathrm{fiducial}} = 963.34 \pm 1.91\,$pc$\,$cm$^{-3}$. The corresponding mean and sta
ndard deviation of DM, integrated up to $z=1$, are $E(\text{DM})_{\mathrm{fiducial}} = 977.81\,$pc$\,$cm$^{-3}$ and $\mathrm{Std}(\text{DM}){\mathrm{fiducial}} = 170.10\,$pc$\,$cm$^{-3}$.
For the NoBH model, we find $\mu_{\mathrm{NoBH}} = 0.1712 \pm 0.0017$ and $\sigma_{\mathrm{NoBH}} = 952.46 \pm 1.87\,$pc$\,$cm$^{-3}$. The corresponding mean and standard deviation of DM are $E(\text{DM})_{\mathrm{NoBH}} = 966.52\,$pc$\,$cm$^{-3}$ and $\mathrm{Std}(\text{DM})_{\mathrm{NoBH}} = 166.65\,$pc$\,$cm$^{-3}$.

The median DM values from data, marked by the dashed lines, are 936.40 and 927.25$\,$pc$\,$cm$^{-3}$ for the fiducial and NoBH models, respectively.

Figure~\ref{fig:DM_f_IGM} compares simulations with the Macquart relation (equation~\ref{eq:dm_igm}) for $f_{\mathrm{diff}} = 0.5 - 1.0$. Our simulation results remain consistent with the model predictions.
The fiducial model constrains $f_{\mathrm{diff}}$ at $z=1$ to $0.865^{+0.101}_{-0.165}$, while NoBH gives $0.856^{+0.101}_{-0.162}$. The slightly lower value in NoBH suggests AGN feedback modestly enhances electron density in the IGM. Here, the error bounds of $f_{\mathrm{diff}}$ represent the 16th–84th percentile range across individual sight lines in log space, capturing the intrinsic cosmic variance (i.e., LoS scatter) caused primarily by foreground halo intersections, rather than the standard error on the mean, which is negligible given the large sample size (SEM ≲ 0.001).

Since $\text{DM}_{\rm Halos}$ is not subtracted in Figure~\ref{fig:DM_f_IGM}, inferred $f_{\mathrm{diff}}$ values have larger uncertainties, occasionally exceeding 1 due to dense foreground halos. This overestimation arises because certain sight lines frequently pass through dense and massive foreground halos, which significantly enhance the total DM. As a result, the inferred $f_{\mathrm{Idiff}}$ values are biased upwards.
This bias will be corrected later when we isolate $\text{DM}_{\rm IGM}$ from $\text{DM}_{\rm diff}$.

\section{Halo Profile Analysis for DM Contributions of Foreground Halos} 

We analyze the contributions of foreground halos to the DM observed in FRB observations, comparing simulation results with theoretical models and examining the impact of AGN feedback.

\subsection{Theoretical Model of Halo Density Profile}
To characterize the density profile of dark matter halos, we adopt two widely used models: the Navarro-Frenk-White (NFW) \citep{NFW1997} and modified NFW (mNFW) \citep{Duffy2010} profiles: 
\begin{equation}
\rho(r) = 
\begin{cases} 
\frac{\rho_0}{y(1+y)^2}, & \text{(NFW)} \\[10pt]
\frac{\rho_0}{y^{1-\alpha}(y_0+y)^{2+\alpha}}, & \text{(mNFW)} 
\end{cases}
\label{eq:profile_model}
\end{equation}
where
\begin{align}
& y = c_{200} \frac{r}{R_{200}}, \quad c_{200} = C \left( \frac{M_{\mathrm{halo}, 200}}{10^{14} M_{\odot}} \right)^{-\beta},\notag\\ 
&\rho_0 = \frac{(200 \rho_c)}{3} c_{200}^3 f(c_{200}),
\end{align}
and
\begin{equation}
f(c) = \ln(1+c) - \frac{c}{1+c}.
\end{equation}

Here, $R_{200}$ is the halo radius where density is 200 times the critical density, $\rho_c = 8.6 \times 10^{-30} \, \mathrm{g \, cm}^{-3}$. $M_{\text{halo}, 200}$ is defined as the total mass enclosed within $R_{200}$. The concentration parameter $c_{200}$ depends on halo mass $M_{\text{halo}, 200}$, with $C$ as normalization. $\rho_0$ is set by $c_{200}$ and $f(c_{200})$. 
The mNFW profile modifies the inner slope via $\alpha$ to account for baryonic effects, with $y_0$ controlling the transition scale and $\beta$ describing the mass dependence of $c$.

While the NFW profile provides a standard description of dark matter halos, the mNFW model accounts for baryonic feedback, including AGN-driven outflows. We use these profiles to analyze the DM contributions from foreground halos along FRB sight lines.

\subsection{Density profiles derived from LSS simulations} \label{sec:density_profiles_sim}

\begin{figure*}[htbp]
    \centering
    \begin{overpic}[width=0.48\textwidth]{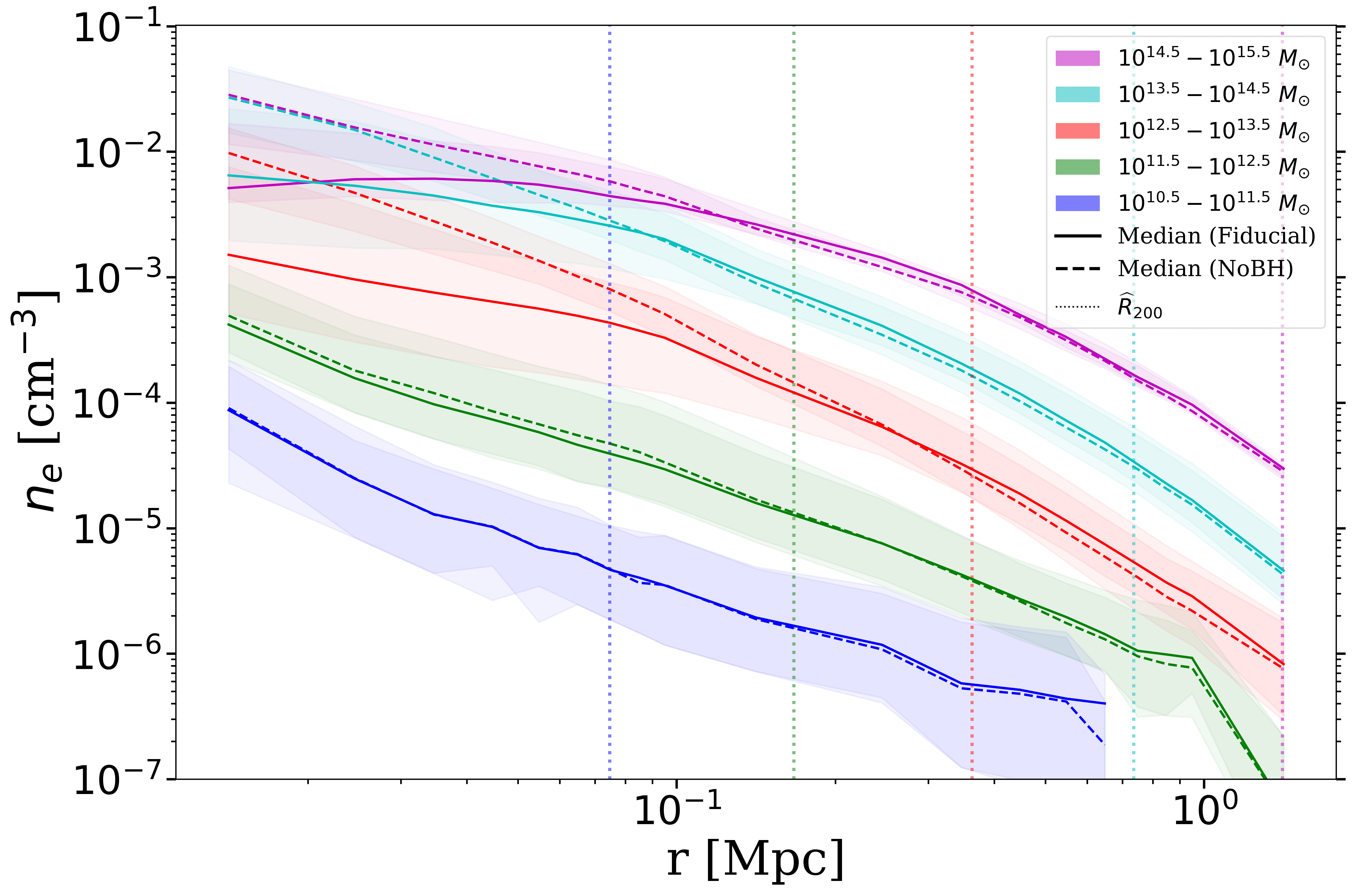}
        \put(5,69){(a)}
    \end{overpic}
    \hfill
    \begin{overpic}[width=0.48\textwidth]{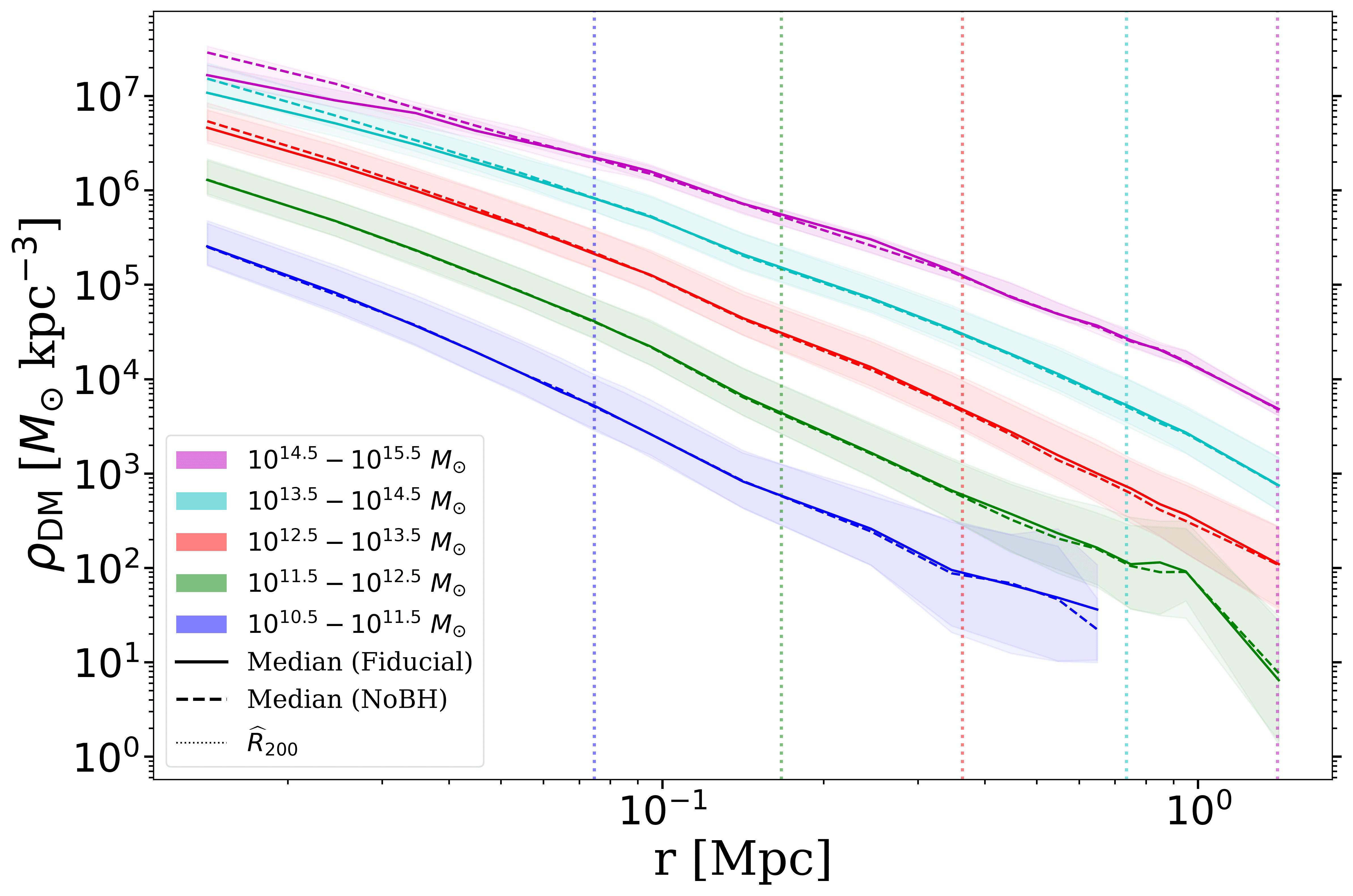}
        \put(5,69){(b)}
    \end{overpic}
    \vspace{5mm} 
    \begin{overpic}[width=0.48\textwidth]{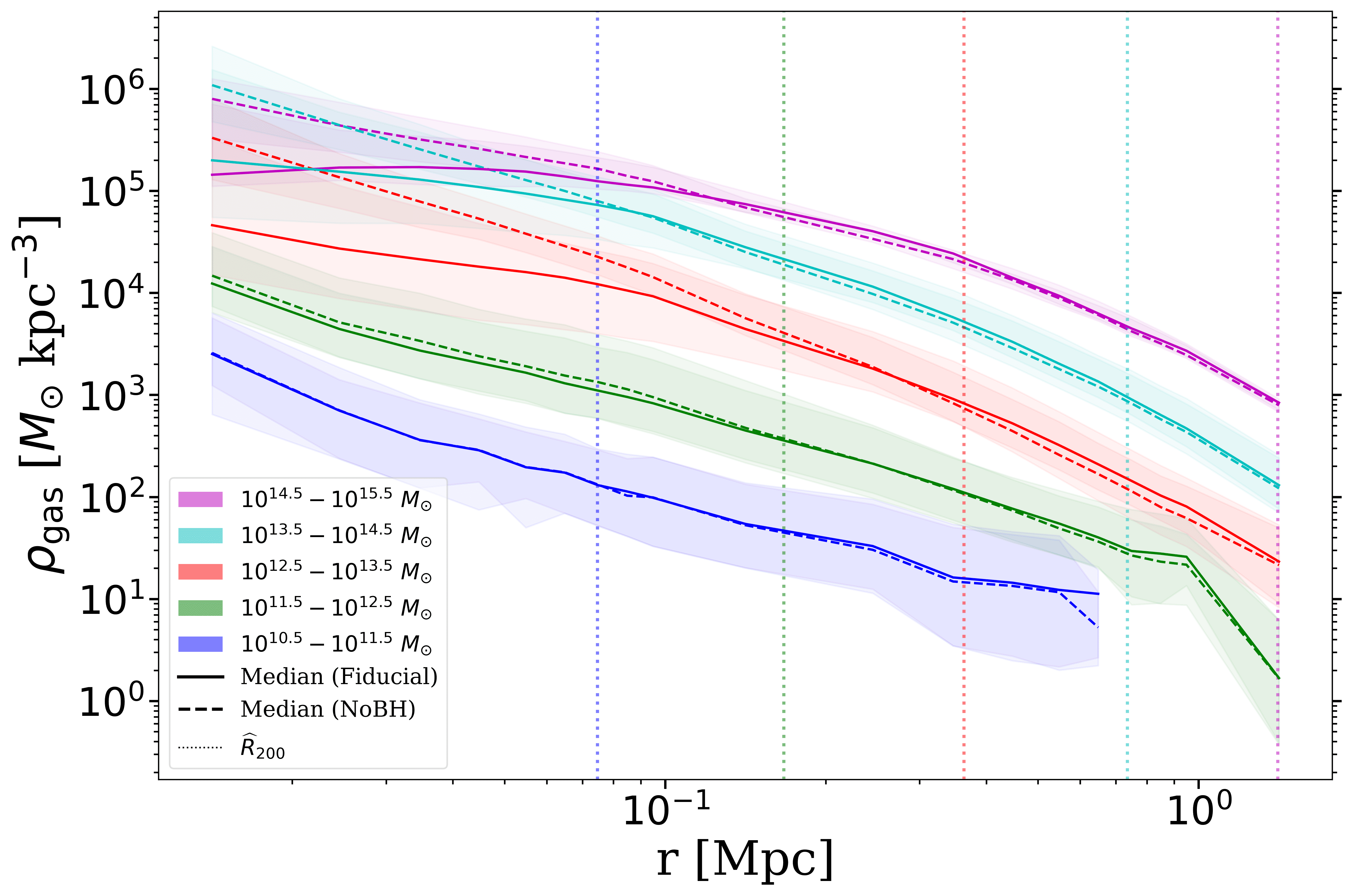}
        \put(5,69){(c)}
    \end{overpic}
    \hfill
    \begin{overpic}[width=0.48\textwidth]{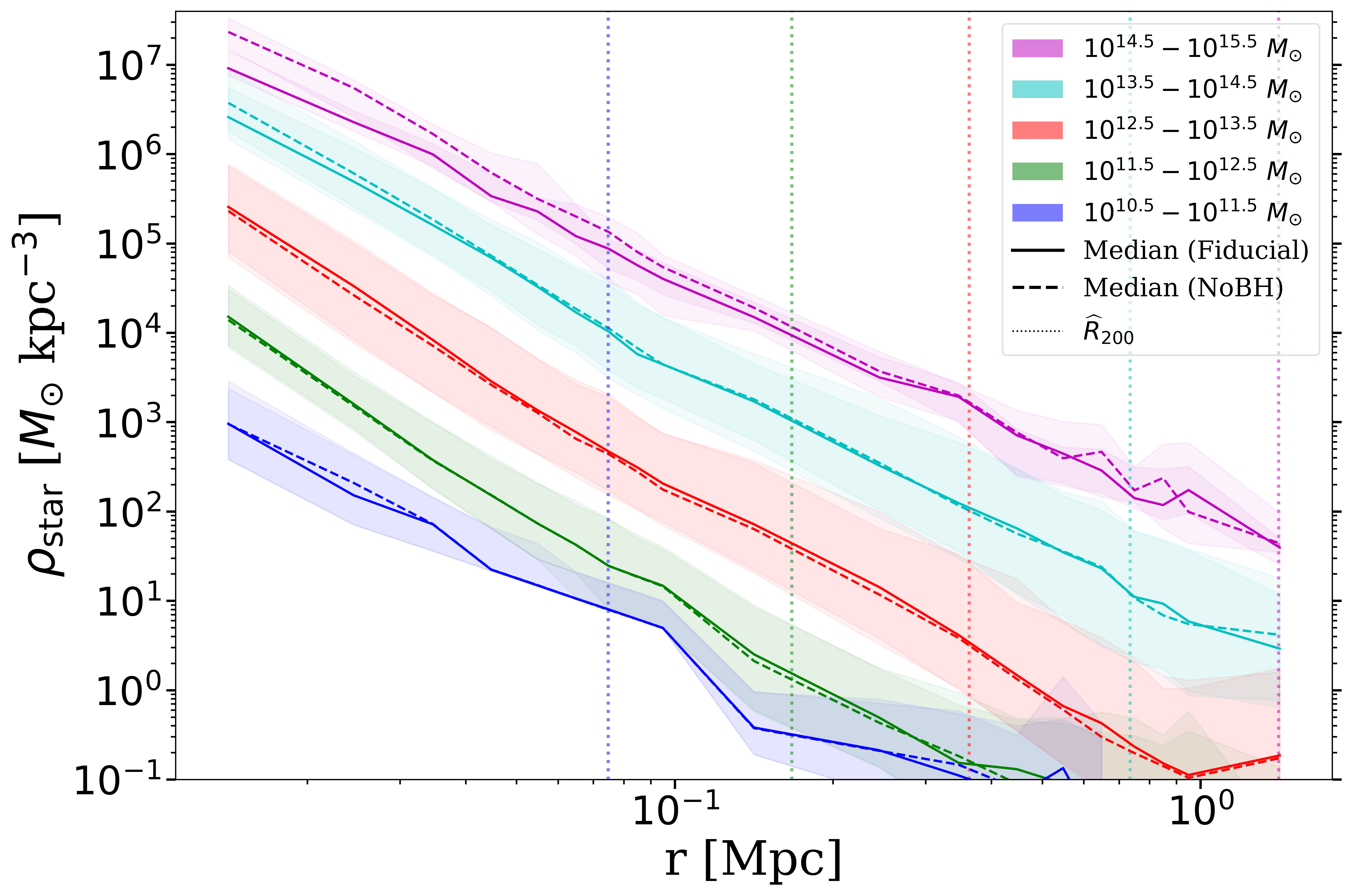}
        \put(5,69){(d)}
    \end{overpic}
    \caption{Comparison of density profiles for fiducial and NoBH simulations in the 100 Mpc box. The panels show the overlay of median density profiles for different components. The shaded regions represent the 1$\sigma$ range of variations in each mass bin. Solid lines represent fiducial simulations, while dashed lines represent NoBH simulations. The dotted lines indicate the median $R_{200}$ for halos in each mass range in the fiducial results, with NoBH results showing nearly no difference.}    
    \label{fig:density_profiles_100Mpc}
\end{figure*}

\begin{figure}
    \centering
    \includegraphics[width=0.47\textwidth]{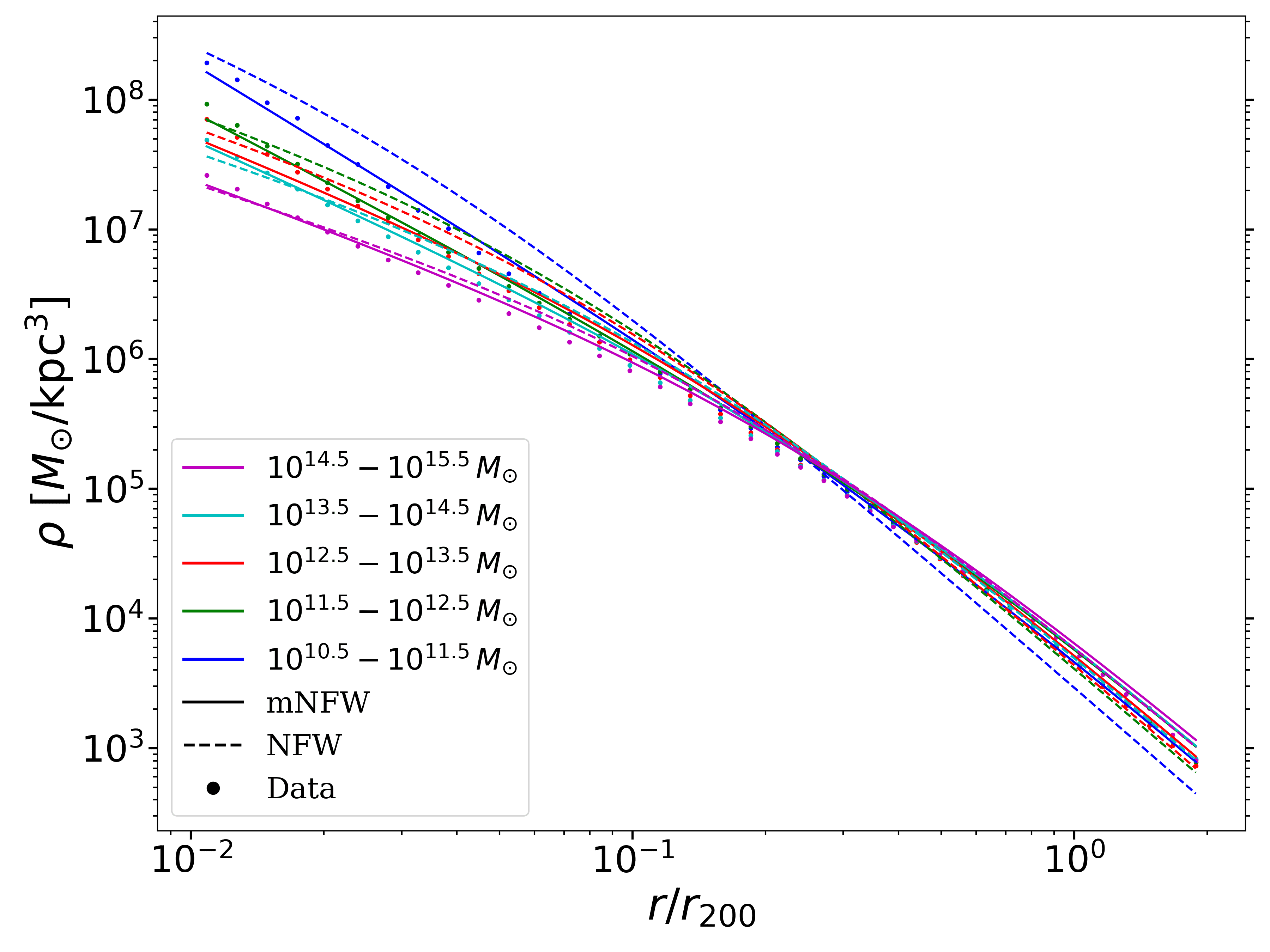}
    \includegraphics[width=0.47\textwidth]{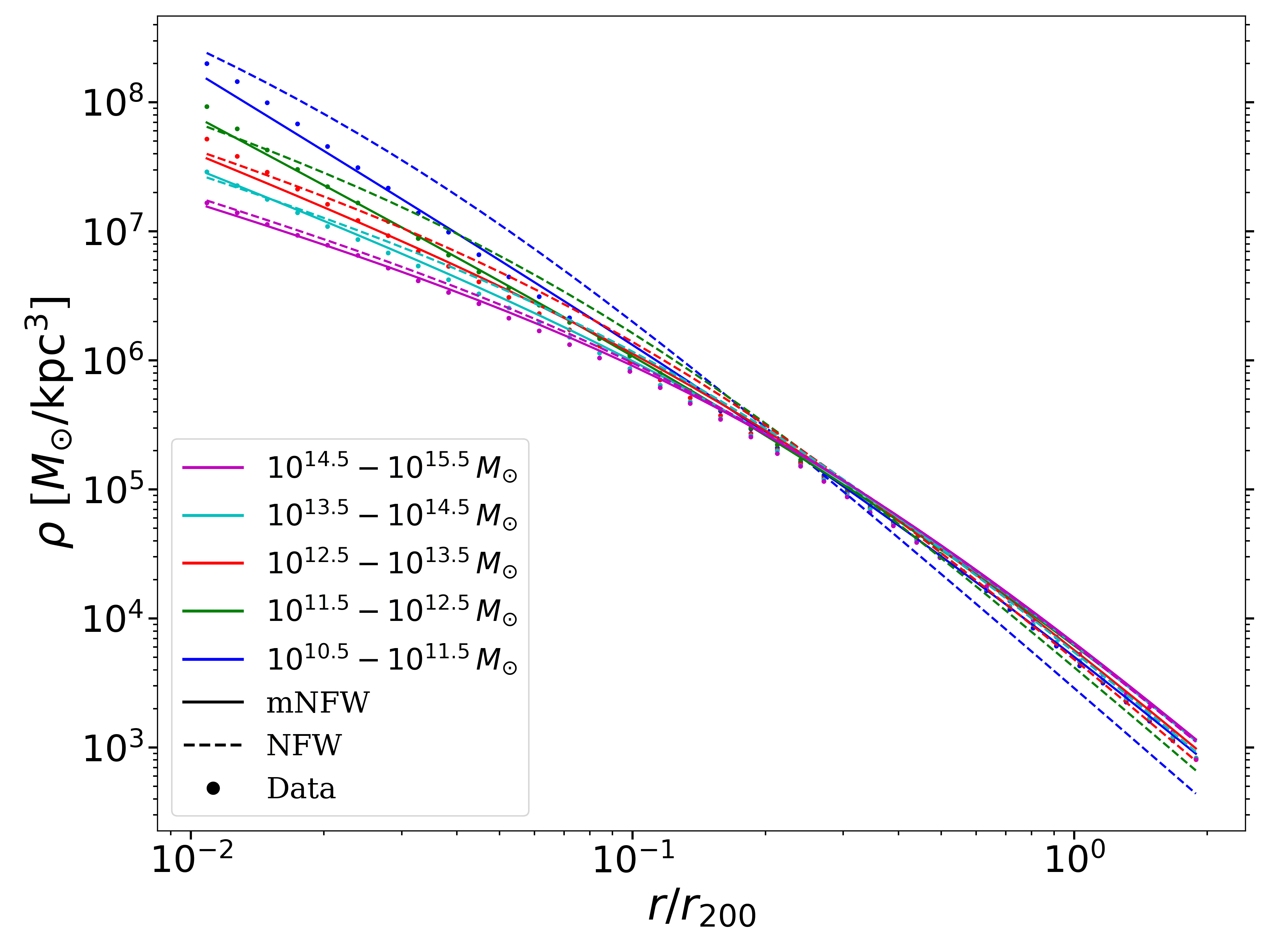}
    \caption{Comparison of the mNFW and NFW mass profiles for fiducial (top) and NoBH (bottom) for 100 Mpc box}
    \label{fig:density_profiles_100Mpc_agn_comparison}
\end{figure}

We present density profiles derived from simulations, including gas, dark matter, stellar mass density, and \( n_e \). To compute these profiles, we first categorize all halos in the LSS simulation into different mass bins, covering the range \(10^{10.5} - 10^{15.5} M_{\odot}\). Within each mass bin, we randomly select 500 halos and compute their individual density profiles for various components. The median density profile is then extracted for each mass range, with the 16th-84th percentile range determined to represent the upper and lower bounds of the distribution. Additionally, we calculate the median \(R_{200}\) for halos in each mass bin. Since the \(R_{200}\) values for the fiducial and NoBH runs show negligible differences, we only retain the fiducial results. These profiles provide insights into the distribution of baryons and dark matter in halos of different masses and serve as an essential component for modeling the DM along FRB sight lines. The resulting density profiles are shown in Figures~\ref{fig:density_profiles_100Mpc} and \ref{fig:density_profiles_100Mpc_agn_comparison}.

Figure~\ref{fig:density_profiles_100Mpc} show the median density profiles from the L100 simulation, respectively, for various mass ranges. (The L500 results are shown in the Appendix \ref{App:boxsize_effect}).  
Panel~{\it (a)} displays gas density, panel~{\it (b)} $n_e$, panel~{\it (c)} dark matter density ($\rho_{\text{DM}}$), and panel~{\it (d)} stellar mass density. The shaded regions indicate the 16th–84th percentile range. range of variations among the selected halos.
Notably, the trends in gas and $n_e$ profiles are consistent, reflecting the high ionization fraction of the gas.

To assess the effects of AGN feedback, we compare the density profiles of fiducial and NoBH simulations in the L100 simulation, as shown in Figure~\ref{fig:density_profiles_100Mpc}. 
Each panel compares the fiducial (solid lines) and NoBH (dashed lines) simulations across different halo mass ranges.
AGN feedback significantly reduces the central densities of gas and $n_e$ in halos, redistributing baryonic matter toward the outskirts. This effect is particularly pronounced in halos with masses between $10^{12.5}\, M_\odot$ and $10^{13.5}\, M_\odot$. 
For lower-mass halos ($< 10^{12.5}\,M_\odot$), AGN feedback has minimal impact, while for more massive halos, gravitational forces limit its effectiveness.

Based on panels~{\it (c)} and {\it (d)}, despite the relatively weak influence of AGN feedback on dark matter and stellar mass, it slightly reduces the central densities of these components in high-mass halos. This subtle effect is likely due to the gravitational drag exerted by gas particles that are expelled by AGN feedback. As the gas is redistributed outward, its gravitational interaction modulates the distribution of non-baryonic components, including dark matter and stars. This effect has been thoroughly analyzed and discussed in the recent work by \citet{Sorini2025}, which provides a detailed characterization of how baryonic processes shape dark matter density profiles. While this effect is not as prominent as the redistribution of gas and $n_e$, it nonetheless underscores the interconnected dynamics of baryonic and non-baryonic matter in high-mass halos influenced by AGN feedback.

We compare the dark matter density profiles for fiducial and NoBH simulations using NFW and mNFW models, as shown in Figure~\ref{fig:density_profiles_100Mpc_agn_comparison}. Both models fit the dark matter density profiles well; however, the mNFW model provides a slightly better fit, capturing the subtle features in our simulation results more effectively.

\begin{figure}
    \centering
    \hspace*{0.28cm}
    \includegraphics[width=0.45\textwidth]{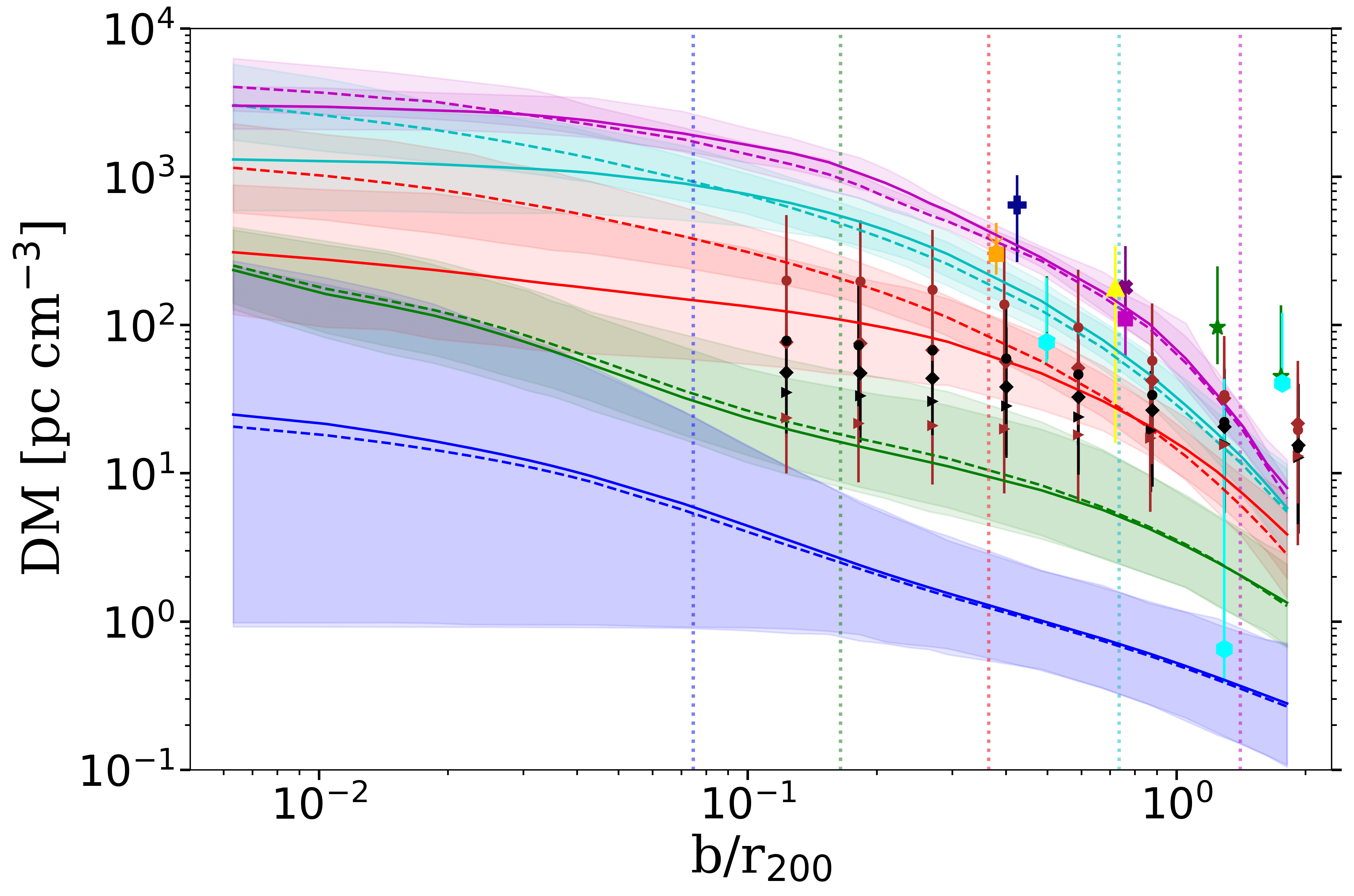}
    \includegraphics[width=0.55\textwidth]{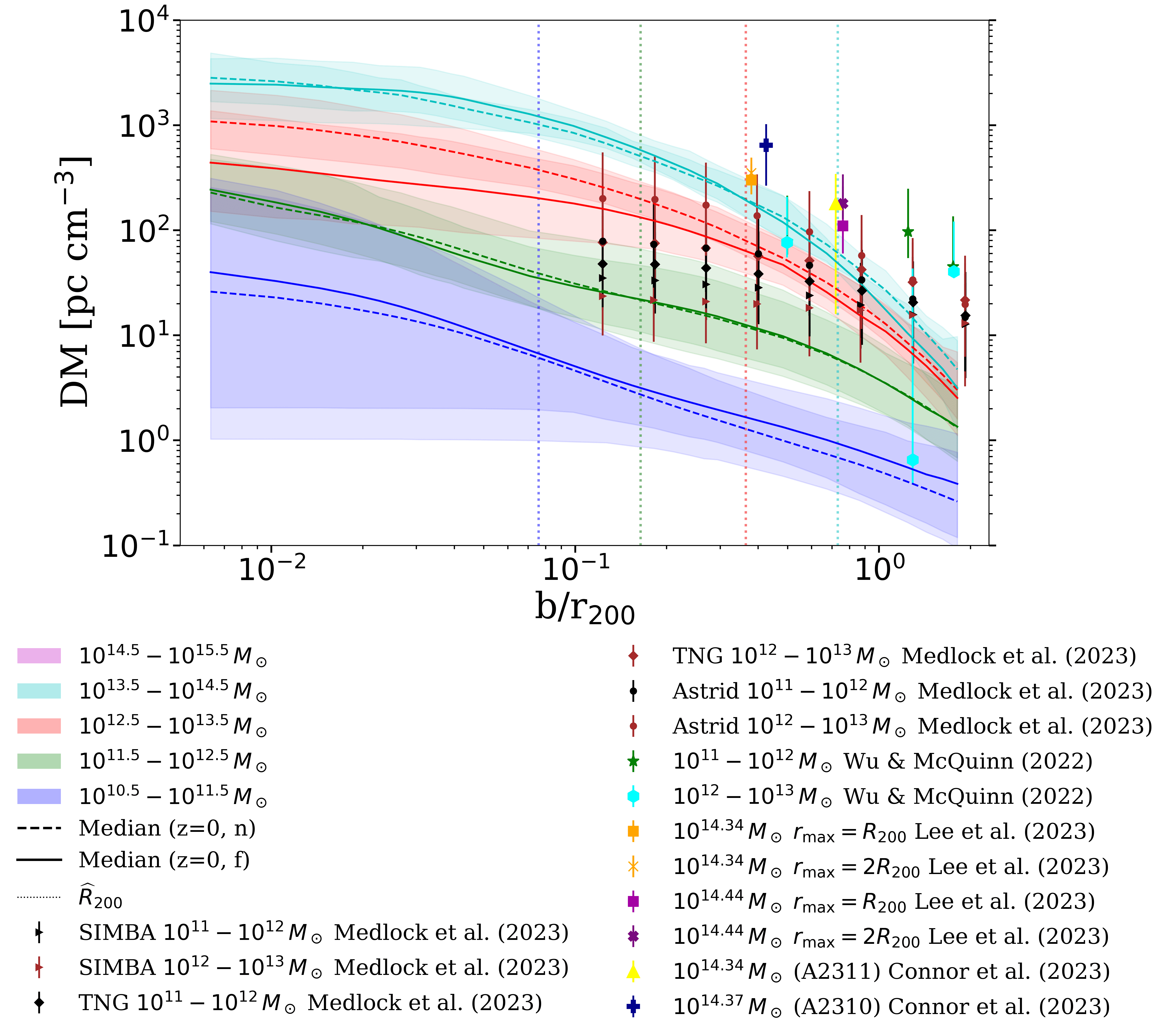}
     \caption{
    Dispersion measure (DM) as a function of impact parameter $b$ for halos in the L100 (top) and L50 (bottom) simulations, categorized by mass ranges. Solid and dashed lines represent our simulations with and without AGN feedback, respectively, while shaded regions denote the 1$\sigma$ range of DM values for 500 halos in each mass range. The central curves correspond to the median DM values at each impact parameter. The DM values are obtained by integrating electron densities along the LoS over a path length of $-2R_{200}$ to $2R_{200}$. Observational constraints from \citet{WuMQ2023}, \citet{Connor2024}, and \citet{Lee2023}, as well as simulation-based fits from \citet{Medlock2024}, are overlaid for comparison. The Medlock data points include results from TNG, Astrid, and SIMBA simulations across different mass bins. The points from \citet{Connor2023} correspond to FRBs intersecting galaxy clusters A2310 and A2311. The \citet{Lee2023} estimates are derived from semi-analytical ICM models for two foreground clusters intersected by FRB~20190520B, assuming cutoff radii ($r_{\mathrm{max}}$) of $R_{200}$ and $2R_{200}$.}
    
    \label{fig:dm_vs_b}
\end{figure}

\subsection{DM profile as a function of impact parameter}
\label{sec:1D_DM_statistic}

To quantify the contribution of foreground halos to the DM, we determined the center of each halo and extracted all gas particles within a sphere of radius $ 2R_{200} $. sight lines were then generated perpendicular to a reference plane passing through the halo center, with the impact parameter $ b $ representing the perpendicular distance from the sightline to the halo center.

The DM for each sightline was calculated using equation~\ref{eq:DM}, which integrates the electron number density $ n_{e,\mathrm{halo}}(l) $ along the sightline over a path length of $ -2R_{200} $ to $ 2R_{200} $. This method was applied to 500 halos in each mass range to obtain the DM distribution as a function of $ b $. Separate analyses were conducted for simulations with and without AGN feedback to investigate its influence.

Figure~\ref{fig:dm_vs_b} illustrates the DM distributions for halos of varying masses. For sight lines passing close to the halo center, DM values range from a few $\mathrm{pc \, cm^{-3}}$ for low-mass halos to several thousand $\mathrm{pc \, cm^{-3}}$ for massive halos. AGN feedback notably reduces the central DM while increasing the outer DM profiles, particularly in halos with masses $ 10^{12.5} \, M_{\odot} < M_H < 10^{13.5} \, M_{\odot} $. 

Observational constraints from \citet{WuMQ2023}, \citet{Connor2023}, and \citet{Lee2023}, along with simulation-based fits from \citet{Medlock2024}, are included for comparison. Although some deviations exist, the simulation results align well with Medlock's fits within similar mass ranges. The estimates from \citet{Lee2023} are derived using a semi-analytical gas profile model for two foreground clusters intersecting the sightline of FRB~20190520B, assuming different cutoff radii ($R_{200}$ and $2R_{200}$). The results from \citet{Connor2023} are based on excess DM values of two FRBs (20220914A and 20220509G) hosted by massive clusters A2310 and A2311, after subtracting Galactic and IGM contributions. These observational points fall within or slightly above the range predicted by our halo DM profiles across the corresponding mass scales. Furthermore, our analysis extends to higher halo masses ($M_H \gtrsim 10^{14} M_{\odot}$), surpassing those considered in CAMELS simulations, allowing us to explore the DM contribution from the most massive halos. This provides new insights into the CGM properties in cluster-scale environments.

\subsection{2D DM Map and 1D Profiles of Individual Halos}
\label{sec:2D_DM_individual}


\begin{figure*}
    \centering
    \begin{tabular}{cc}
        \includegraphics[width=0.4\textwidth]{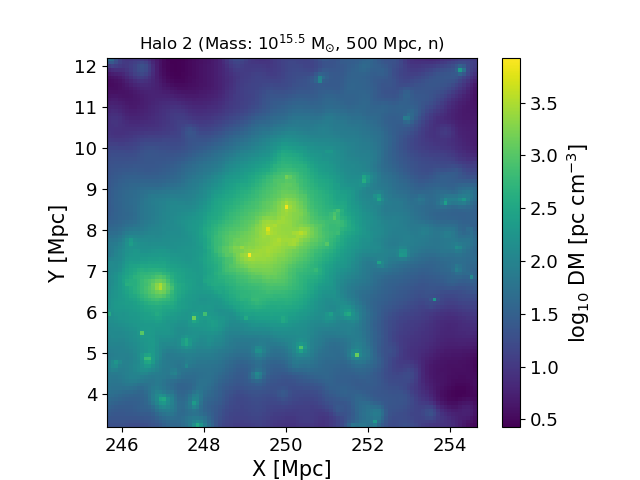} &
        \includegraphics[width=0.35\textwidth]{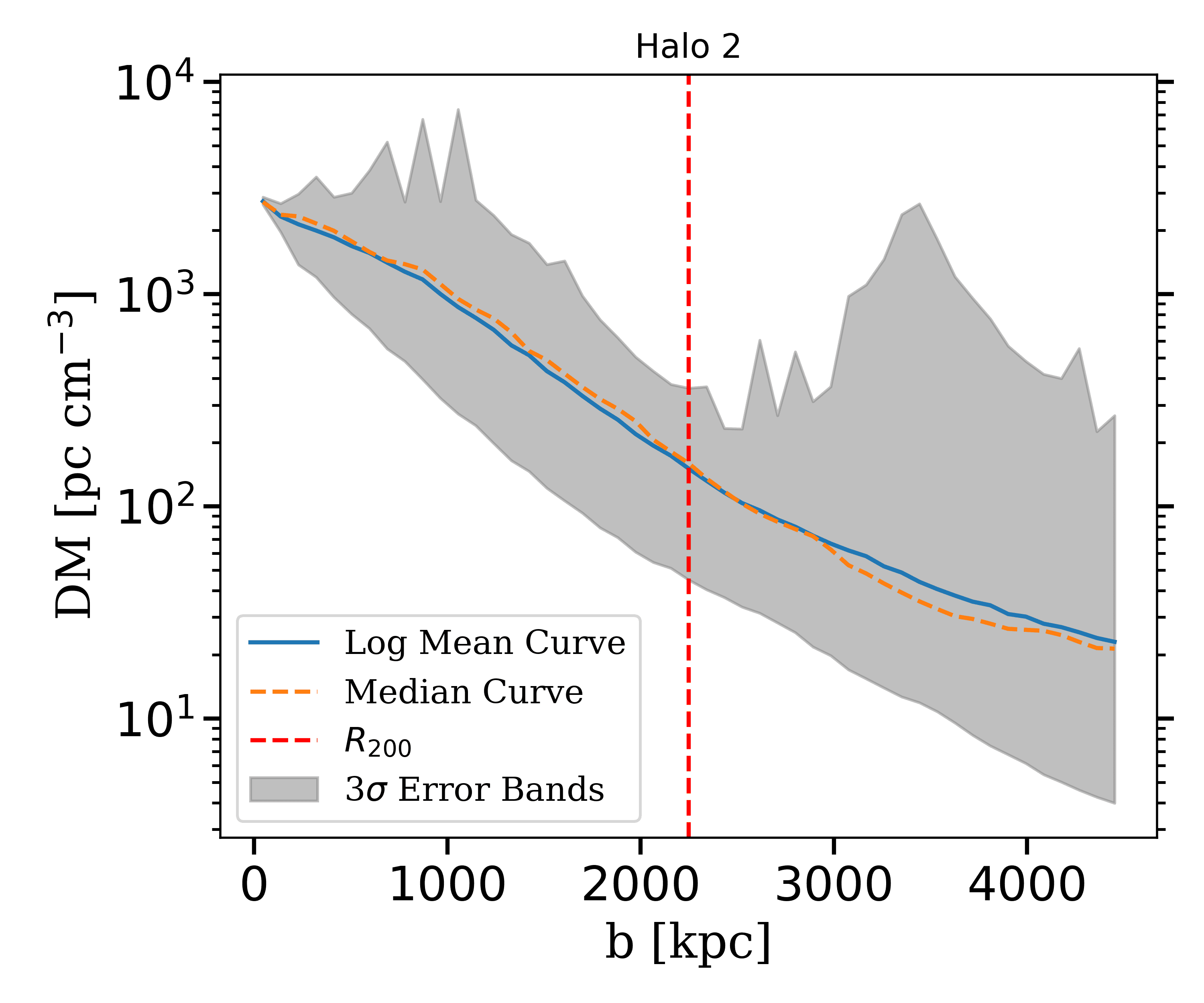}\\
        \includegraphics[width=0.4\textwidth]{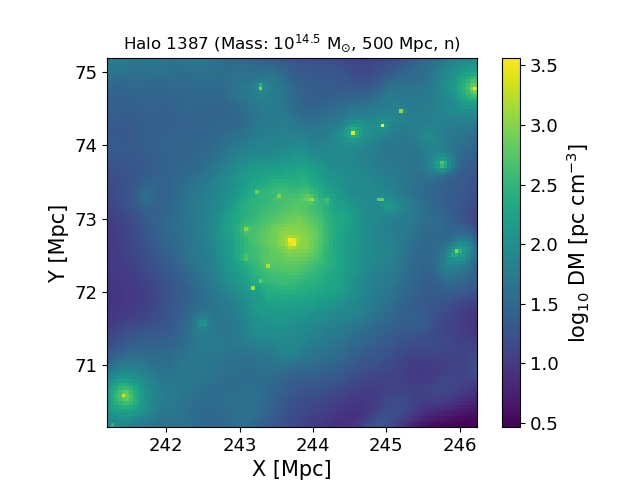} &
        \includegraphics[width=0.35\textwidth]{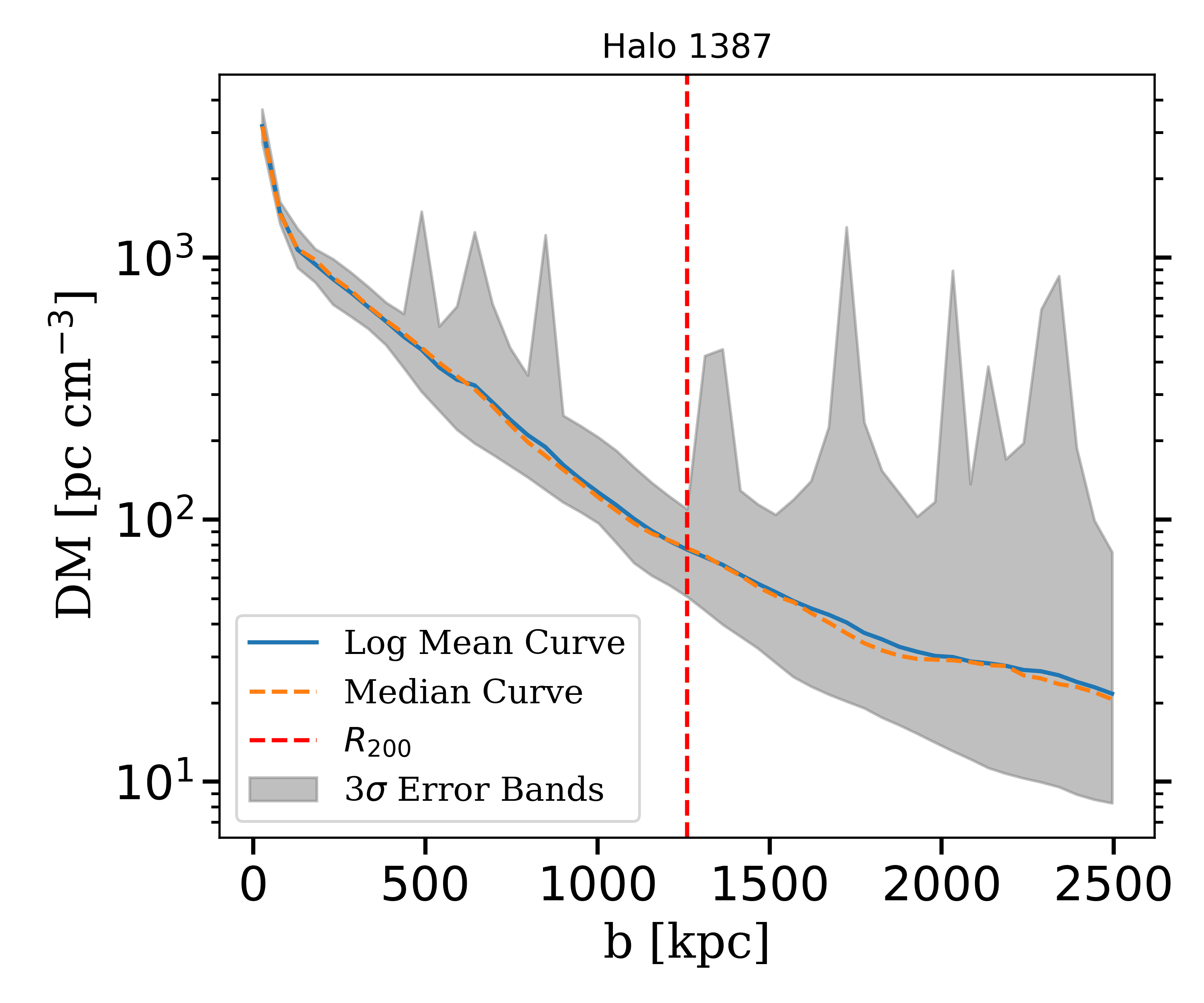} \\
        \includegraphics[width=0.4\textwidth]{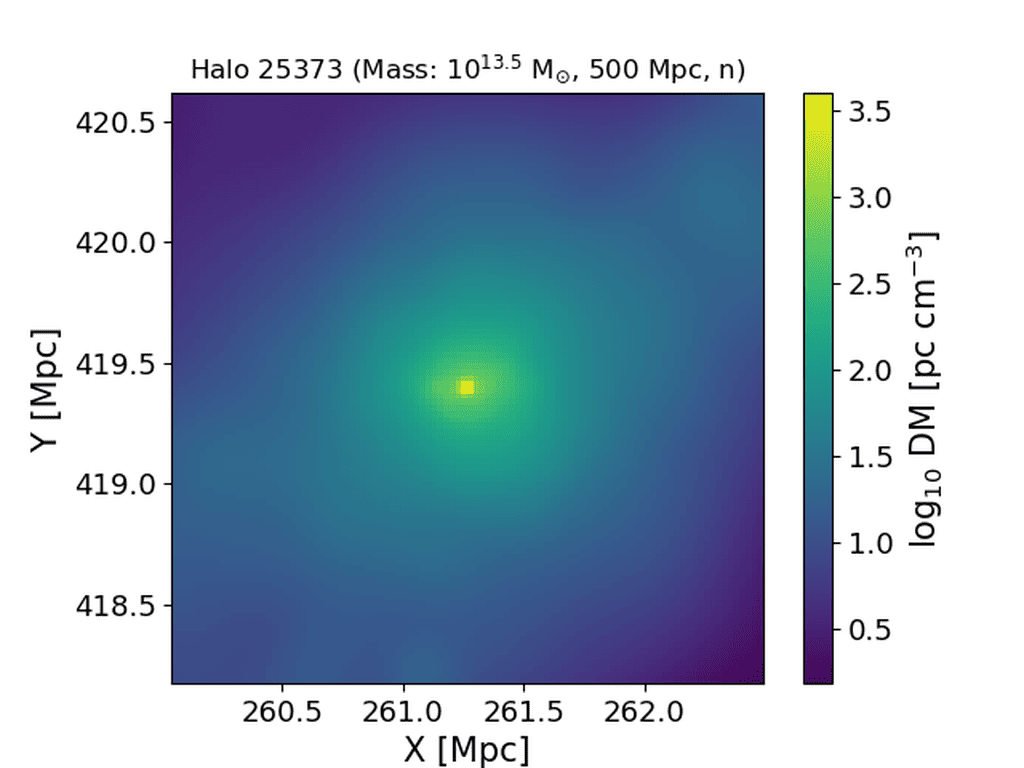} &
        \includegraphics[width=0.35\textwidth]{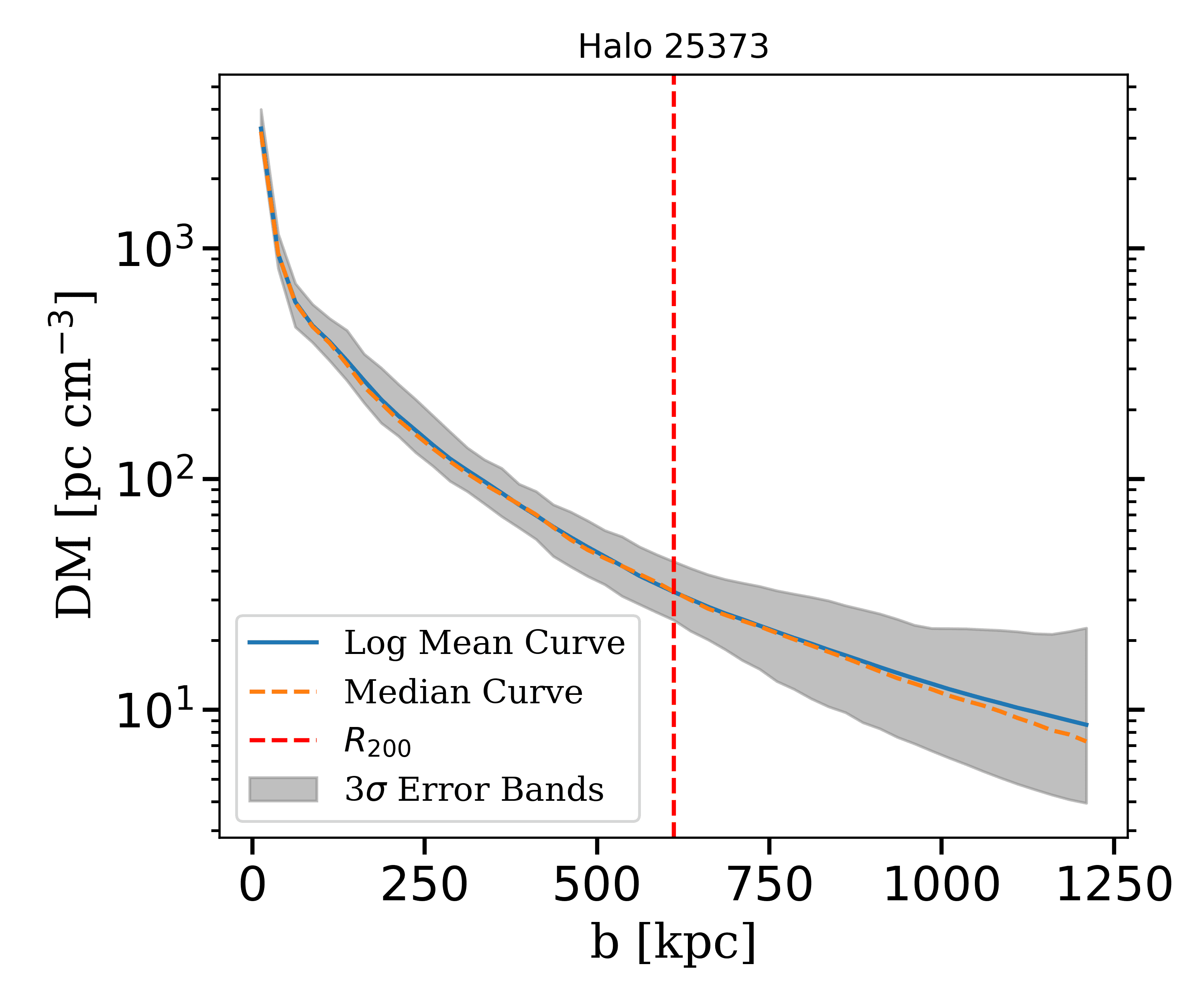} \\
        \includegraphics[width=0.4\textwidth]{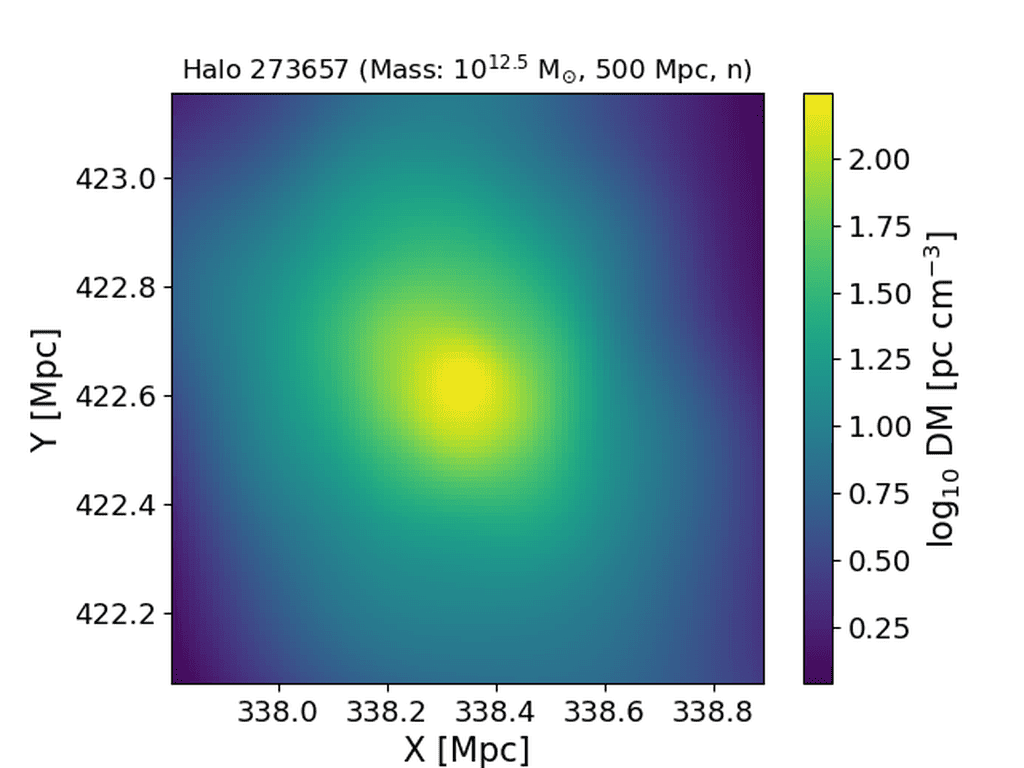} &
        \includegraphics[width=0.35\textwidth]{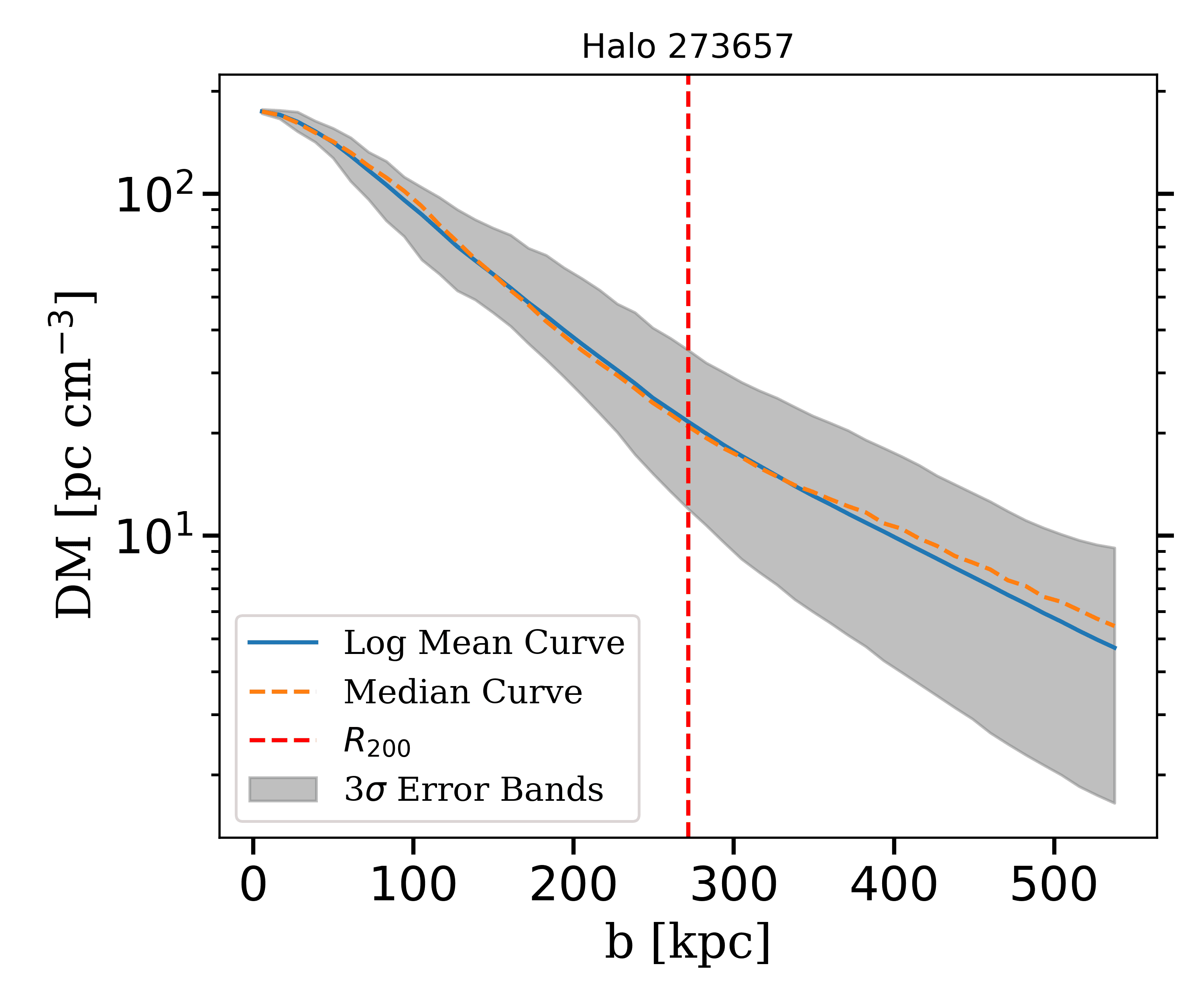} \\

    \end{tabular}
    \caption{2D DM maps (left column) and 1D DM profiles as a function of impact parameter $b$ (right column) for the most massive halos in each representative mass range, as shown in Figure~\ref{fig:density_profiles_500Mpc}, from the L500 simulation. Each row corresponds to a halo, arranged in descending order of mass from top to bottom, with the halo ID and mass indicated in the title of each plot. The vertical red dashed lines in the right panels indicate the virial radius of the halo ($R_{200}$). Shaded regions in the 1D profiles represent the scatter of different sight lines at each $b$. All halos are from simulations without AGN feedback (labeled "n").}
    \label{fig:DM_profiles_500}
\end{figure*}

The two-dimensional DM distribution (i.e. 2D map) of individual halos is often more complex than 1D profiles, especially for high-mass halos, such as galaxy clusters and massive galaxies. These halos can exhibit intricate structures, including satellite galaxies, varying degrees of gas dispersal (potentially influenced by AGN feedback), or even multiple halos in close proximity undergoing merger processes. In these cases, 1D spherically symmetric models may fail to accurately capture the true characteristics of the halos. Using such models requires caution when applied to such complex systems.

To investigate the contributions of dense substructures near halo centers and the surrounding gas environment, as well as to assess their deviations from 1D models, we analyzed the 2D DM map of halos. This analysis also enables us to examine the impact of AGN feedback on these distributions.

Figure~\ref{fig:DM_profiles_500} presents the 2D DM maps and 1D DM profiles of representative halos from different mass ranges in the L500 simulation. For high-mass halos like Halo 2 and Halo 1387, more complex structural distributions are observed. Numerous satellite galaxies surround these halos, leading to noticeable ``peaks'' in their 1D DM profiles at various locations. These fluctuations result in significantly elevated DM values and contribute to variations in the mean profile, although the overall impact on the average remains moderate. Halo 2, compared to Halo 1387, exhibits a larger number of substructures, and its surrounding gas environment shows greater overlap, resulting in a 1D profile with a broader scatter depending on the LoS location. 

In contrast, low-mass halos, such as Halo 25373 and Halo 273657 (two lower sets), tend to be more isolated, lacking complex substructures. As a result, they are better described by 1D spherically symmetric models. Between these two halos, Halo 25373, having a higher mass, shows a sharp density increase near its center, appearing as a distinct peak in the 2D distribution. Conversely, the lower-mass Halo 273657 has a smoother central region, and its surrounding gas environment is more dispersed due to weaker gravitational influence.

To investigate the influence of AGN feedback on the contribution of foreground halos to the DM of FRBs, we compare results from the L100 simulations. Figure~\ref{fig:DM_profiles_14-15} presents the 2D DM maps and 1D DM--$b$ profiles of two representative massive halos ($\sim 10^{14.8} M_{\odot}$) in both the L100N1024$_{\text{fiducial}}$ and L100N1024$_{\text{NoBH}}$ runs. 

An analysis of ten halos with masses exceeding $10^{14.5} M_{\odot}$ in the L100 simulations reveals that AGN feedback generally has a limited impact on the gas distribution within such massive halos, which exhibit strong self-gravitating systems. This is exemplified by Halo 1, where AGN feedback induces only minimal redistribution effects, with slight diffusion observed near the core in both the 2D map and 1D profile. However, among these massive halos, we identify an exceptional case—Halo 0. As seen in its 2D map and 1D profile, the NoBH simulation exhibits a highly dense, compact region approximately 1 Mpc from the halo center, with a gas density exceeding that of the central region, leading to an anomalous DM distribution. Such occurrences are possible in the absence of AGN feedback, where gas can over-concentrate locally. Similar phenomena are frequently observed in massive halos within the L500 simulation (Figure~\ref{fig:DM_profiles_500}). In contrast, the fiducial run of Halo 0 shows that this dense clump is absent, as AGN feedback efficiently disperses the gas, resulting in a more diffuse central gas distribution.

Figure~\ref{fig:DM_profiles_12-14} presents a comparison of two halos in the $10^{12.5}-10^{13.5} M_{\odot}$ mass range, where AGN feedback is expected to be more efficient. Our statistical analysis of over a hundred halos in this range confirms that AGN feedback is particularly effective within this mass interval. We highlight two representative cases: Halo 223 (fiducial) and Halo 222 (NoBH), both with masses of $\sim 10^{13.5} M_{\odot}$, collectively referred to as Halo $\alpha$; and Halo 1000 (fiducial) and Halo 886 (NoBH), with masses in the range of $10^{12.8}-10^{12.9} M_{\odot}$, collectively referred to as Halo $\beta$. 

Halo $\alpha$ exhibits three distinct structures: a primary halo, a massive subhalo in close proximity, and a significantly smaller subhalo. AGN feedback has a pronounced effect on both main components, leading to a substantial reduction in central DM in the fiducial run compared to the NoBH counterpart. Additionally, the highest AGN feedback efficiency is observed near $10^{13} M_{\odot}$, as demonstrated by Halo $\beta$. The NoBH simulation yields a central DM approaching $10^3$ pc cm$^{-3}$, whereas the fiducial counterpart shows a significantly lower central DM, barely exceeding $10^2$ pc cm$^{-3}$. 

To quantify the redistributing impact of AGN feedback on halo gas densities and DM, we compute the AGN-driven central gas expulsion fraction and central DM suppression fraction, defined as:
\begin{align}
   f^{15}_{\rm exp} &= \frac{\widehat{\rho}_{\rm NoBH}(r=15\,\rm kpc) -\widehat{\rho}_{\rm fiducial}(r=15\,\rm kpc)}{\widehat{\rho}_{\rm NoBH}(r=15\,\rm kpc)},\\
   f^c_{\rm suppr} &= \frac{\widehat{\text{DM}}_{\rm NoBH}(b=0) - \widehat{\text{DM}}_{\rm fiducial}(b=0)}{\widehat{\text{DM}}_{\rm NoBH}(b =0)},
\end{align}
where $\widehat{\rho}(r=15\,\rm kpc)$ represents the median gas density at $15\, \mathrm{kpc}$ for different halo mass bins, and $\widehat{\text{DM}}(b =0)$ represents the median DM at impact parameter $b=0$. The computed expulsion and suppression fractions for different halo mass ranges are:

\begin{itemize}
    \item $10^{10.5}-10^{11.5} M_{\odot}$: $f^{15}_{\rm exp}=2.3\%$, $f^c_{\rm suppr}=-20.6\%$,
    \item $10^{11.5}-10^{12.5} M_{\odot}$: $f^{15}_{\rm exp} = 16.2\%$, $f^c_{\rm suppr} = 6.8\%$,
    \item $10^{12.5}-10^{13.5} M_{\odot}$: $f^{15}_{\rm exp} = 86.1\%$, $f^c_{\rm suppr} = 73.0\%$,
    \item $10^{13.5}-10^{14.5} M_{\odot}$: $f^{15}_{\rm exp} = 81.7\%$, $f^c_{\rm suppr} = 57.1\%$,
    \item $10^{14.5}-10^{15.5} M_{\odot}$: $f^{15}_{\rm exp} = 81.9\%$, $f^c_{\rm suppr} = 25.3\%$.
\end{itemize}

This analysis demonstrates that AGN feedback is most efficient at expelling gas from halos in the mass range $10^{12.5}-10^{13.5} M_{\odot}$, significantly reducing central gas densities. The strong correlation between $f^c_{\rm exp}$ and $f^c_{\rm suppr}$ indicates that AGN-driven gas removal directly impacts DM suppression along the LoS.  

It is worth noting that although the AGN feedback in the $10^{13.5}-10^{14.5} M_{\odot}$ and $10^{14.5}-10^{15.5} M_{\odot}$ mass ranges also exhibits strong gas expulsion effects in the central region (at 15 kpc), it does not completely expel the gas from the halo. As shown in Figures~\ref{fig:density_profiles_100Mpc} and~\ref{fig:dm_vs_b}, most of the redistributed gas remains within $R_{200}$, meaning that the expelled material largely stays within the so-called "CGM". Beyond $R_{200}$, the difference between the fiducial and NoBH simulations is relatively small. However, for halos in the $10^{12.5}-10^{13.5} M_{\odot}$ range, we observe that beyond $R_{200}$, the gas density profile, electron number density profile, and DM profile in the fiducial run are all significantly higher than those in the NoBH run. This suggests that halos in this mass range are the primary drivers of baryonic redistribution. In addition to exhibiting the highest AGN feedback efficiency, these halos also play a dominant role in facilitating material exchange between the IGM and CGM.

In the lowest-mass bin ($10^{10.5}-10^{11.5} M_{\odot}$), we observe a very small $f^c_{\rm exp}$ but a significant negative $f^c_{\rm suppr}$. This suggests that rather than AGN-driven expulsion, ionization effects are the primary factor responsible for the observed changes. The increased electron density and corresponding enhancement in DM could be attributed to the following possible mechanisms: (1) collisionally ionized gas due to heating within the halo itself, (2) shock heating effects from nearby massive halos, or (3) thermal energy injected by the SN-driven outflows. A more detailed investigation of these effects is beyond the scope of this paper.

The efficiency of AGN feedback in the $10^{12.5}-10^{13.5} M_{\odot}$ range is further validated in the appendix, where we analyze the relationship between AGN feedback energy and halo mass. As illustrated in Figures~\ref{fig:halo_gas_mass} and~\ref{fig:energy_mass_relations}, the correlation between AGN feedback efficiency and halo mass provides additional evidence supporting our findings.

\section{IGM Baryon Fraction Analysis} \label{sec:Analysis_method}

In Section~\ref{subsec:DM_in_LSS}, we constrained an important parameter, $f_{\mathrm{diff}}$, using the DM--$z$ relation for FRBs. This parameter directly reflects the distribution of the "missing baryons" that we are investigating. As the dark matter halos grow, $f_{\mathrm{diff}}$ evolves over time. Additionally, based on numerical simulations, aside from the DM--$z$ relation, we can also estimate the gas mass fraction belonging to the IGM by excluding the gas in the halos, which is often called CGM. Here, we treat the $f_\mathrm{diff}$ inferred from the DM–$z$ relation as the observational estimate, and the simulation-derived $f_\mathrm{IGM}$ — which excludes the halo (CGM) gas (whose mass dependence is shown in figure \ref{fig:f_gas_evolution0}) — as the theoretical reference or “true value.” In the actual observations, it is usually difficult to subtract the intervening halo contribution, hence $f_\mathrm{diff} = f_\mathrm{IGM} + f_\mathrm{halo}$, as we defined in  Sec~\ref{sec:FRB_DM}.  In some observational papers, the observed $f_\mathrm{diff}$ is written as $f_\mathrm{IGM}$ assuming that $f_\mathrm{Halos}$ is a part of the smooth IGM component. Such an estimate might lead to an overestimate of true $f_\mathrm{IGM}$, and a care is needed in its interpretation. In the present work, we attempt to clarify the distinction between $f_\mathrm{diff}, f_\mathrm{IGM}$, and $f_\mathrm{Halo}$ in the following sections.  

\subsection{``Observed'' $f_{\mathrm{diff}}$}
The calculation of the "observed value" ($f_{\mathrm{diff, obs}}$) follows the same method as in Section~\ref{subsec:DM_in_LSS}, where we record the constrained $f_{\mathrm{diff, obs}}$ at each redshift when constructing the light cone and connecting the simulation box. We emphasize that the observationally constrained $f_{\mathrm{diff,obs}}$—derived through the integration of DM along light cones—is not strictly equivalent to the instantaneous simulation-based definition of $f_{\mathrm{diff}}(z)$ at a given redshift. The subtle differences are elaborated in Appendix~\ref{APP:f_diff_appendix}.

Since most FRB sources are observed at $z<1$, we primarily focus on calculating the evolution of $f_{\mathrm{diff}}$ up to $z=1$, as shown in Figure~\ref{fig:f_IGM_redshift}b, with the redshifts of our simulation snapshots including $z=[0.106, 0.287, 0.508, 0.754, 1.041]$. The black and orange circles represent data points constrained by L100N1024$_{\text{fiducial}}$ and L100N1024$_{\text{NoBH}}$, respectively, with values given as  $f_{\mathrm{diff, obs}}=[0.706, 0.758, 0.810, 0.835, 0.865]$ (fiducial) and [0.685, 0.749, 0.804, 0.829, 0.856] (NoBH). The fiducial results are marginally higher than those of the NoBH case, reflecting the baryon redistribution driven by AGN feedback. A more detailed analysis is presented in Section~\ref{subsec:baryon_redistribution}.

At $z=1$, our fiducial simulation yields $f_{\mathrm{diff, fiducial}}=0.865$ and the NoBH simulation gives $f_{\mathrm{diff, NoBH}}=0.856$, which are slightly lower than the latest FRB observational constraints from \citet{Connor2024}, where the full sample of all localized FRBs (including detections from DSA-110, ASKAP, CHIME, MeerKAT, and VLA realfast) gives $f_{\mathrm{diff}}=0.93^{+0.04}_{-0.02}$. However, our fiducial result is comparable to the constraint from the DSA-110 only sample, which gives $f_{\mathrm{diff}}=0.89^{+0.06}_{-0.06}$. This agreement highlights the robustness of our simulation in capturing the baryonic distribution within large-scale structures and its impact on FRB DM statistics.

\subsection{``True'' $f_{\mathrm{IGM}}$}
The estimation of $f_{\mathrm{IGM, true}}$ (the "true value" of $f_{\mathrm{IGM}}$ which does not include the halo component) depends critically on how we subtract the contributions from intervening halos. Unless we have a sufficiently detailed understanding of the foreground galaxies along the sight lines, which would allow us to identify all intervening halos and their density profiles, we cannot effectively remove their contributions to the IGM fraction. Thus, the calculation of $f_{\mathrm{IGM, true}}$ relies on the subtraction of CGM gas, which involves defining what gas belongs to the CGM. We employed multiple methods to identify the most appropriate approach for subtracting CGM gas fractions. We also find that the constraint on $f_{\text{IGM}} = 0.8^{+0.08}_{-0.09}$ from \cite{Connor2024} falls within this range as well.

\subsubsection{Using the HMF to Calculate $f_{\mathrm{IGM, true}}$} \label{sec:HMF_f}
\begin{figure}[htbp]
    \centering
    \includegraphics[width=0.47\textwidth]{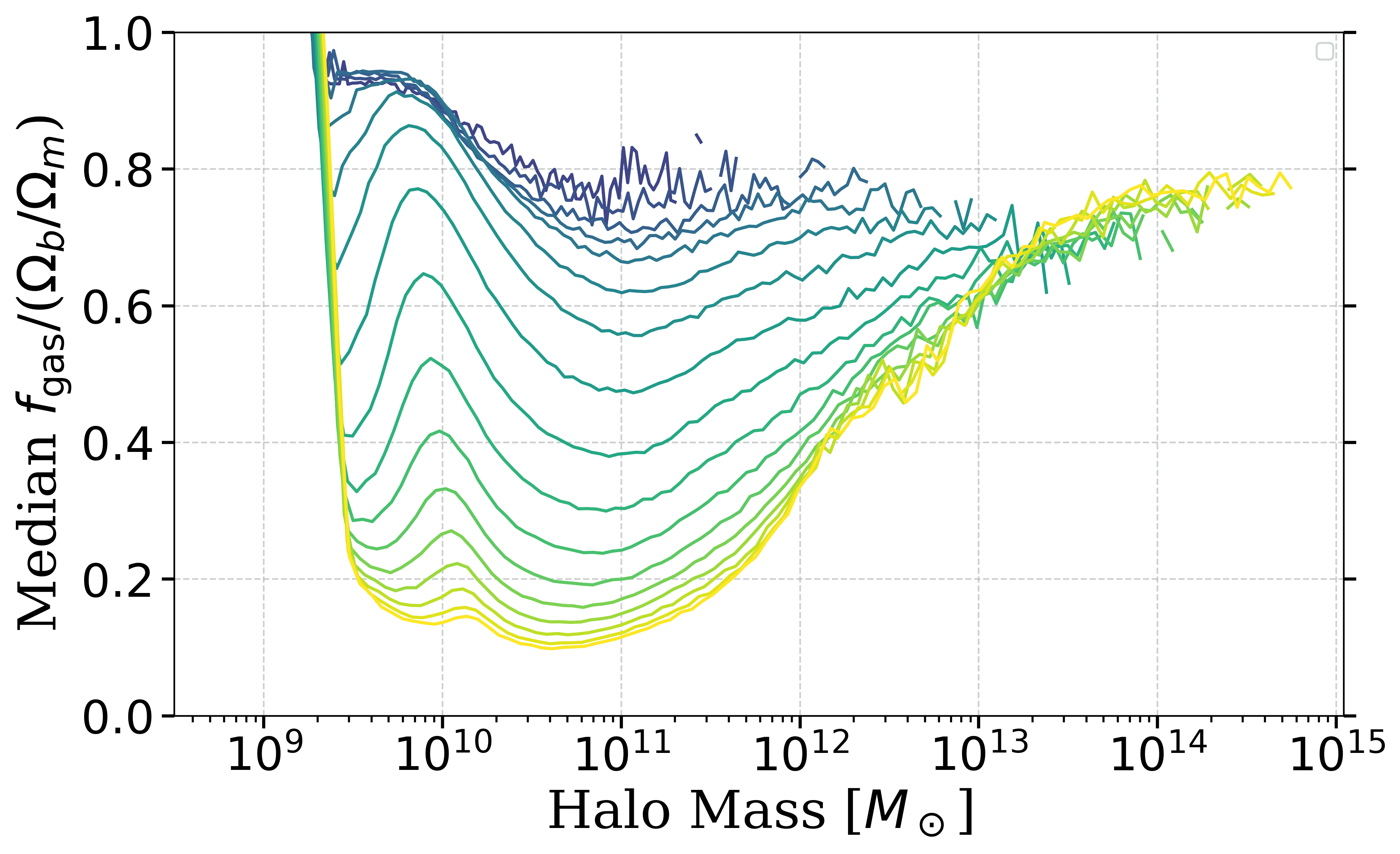} \\
    \includegraphics[width=0.5\textwidth]{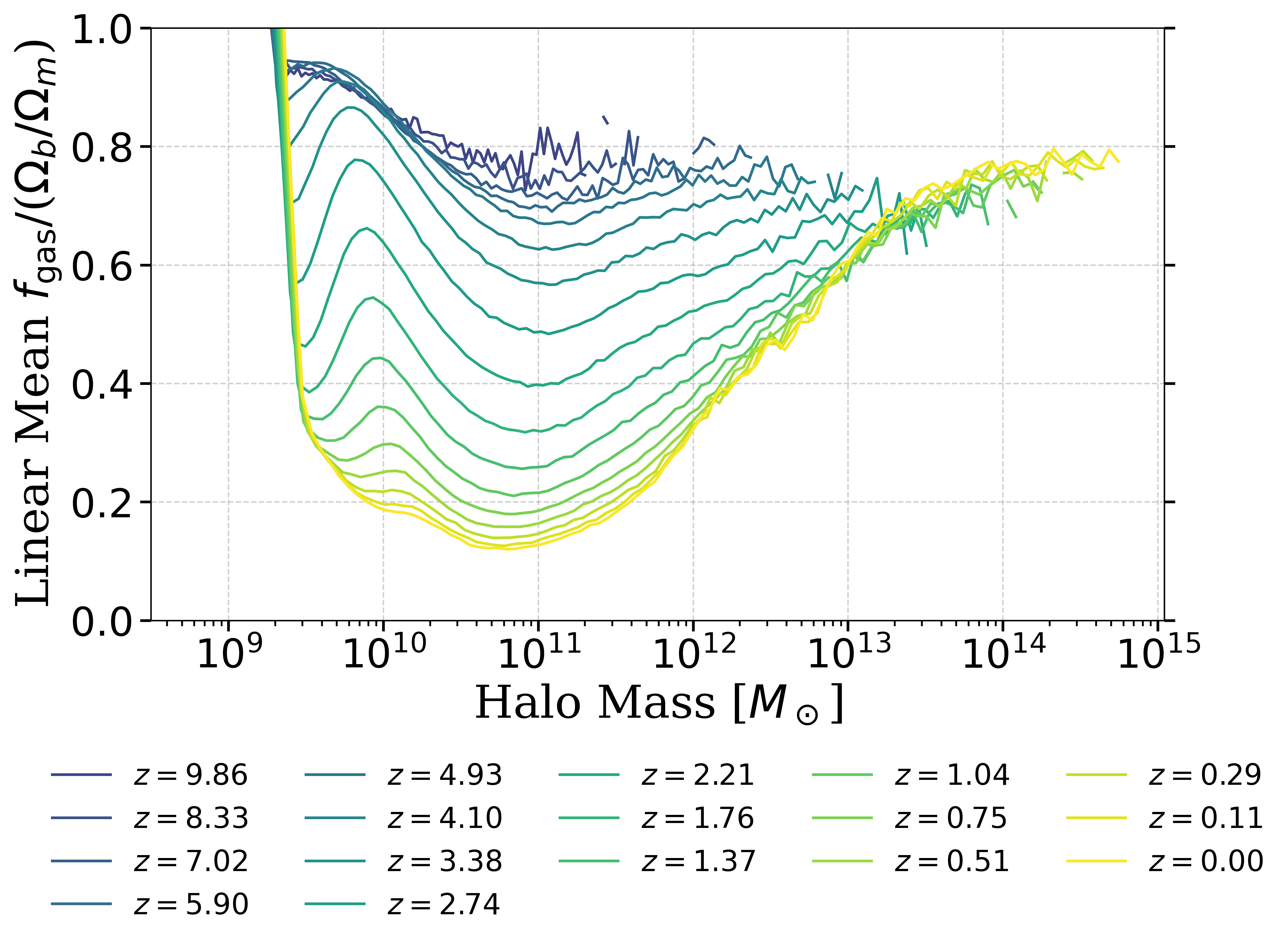}
    \caption{
    Evolution of CGM gas fraction ($f_{\mathrm{gas}}$) as a function of halo mass for the L100N1024$_{\text{fiducial}}$ run. 
    The top panel shows the median gas fractions at different redshifts for various mass ranges. 
    The bottom panel presents the linear mean of $f_{\mathrm{gas}}$ under the same conditions. Both plots reflect the redshift evolution and its dependence on halo mass, highlighting the role of AGN feedback in modifying the gas fraction.}
    \label{fig:f_gas_evolution0}
\end{figure}
\begin{figure*}[htbp]
    \centering
    \begin{overpic}[width=0.78\textwidth]{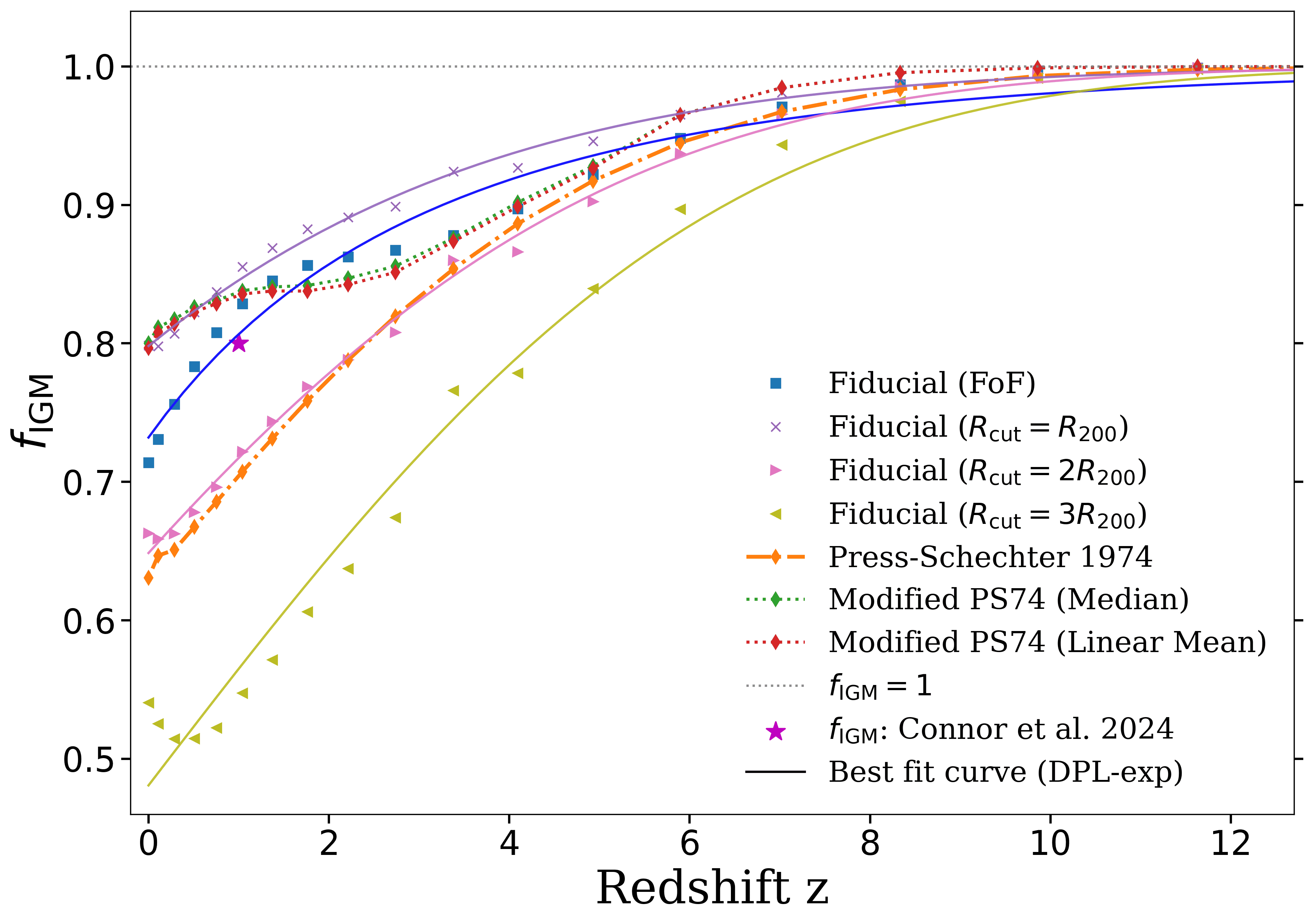}
        \put(2,70){\large(a)}
    \end{overpic}

    \vspace{0.25cm}

    \begin{overpic}[width=0.78\textwidth]{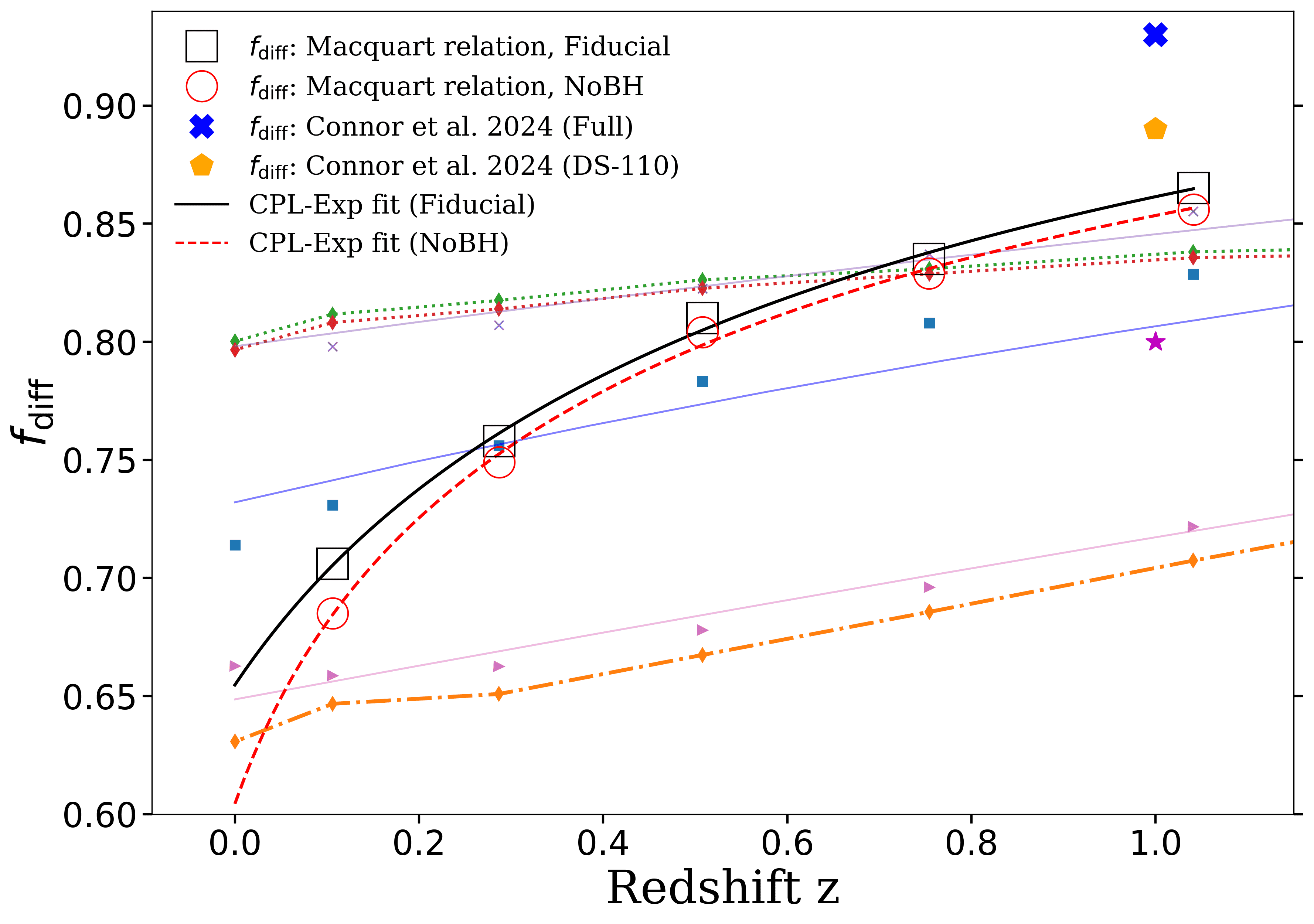}
        \put(2,71){\large(b)}
    \end{overpic}

    \caption{
    Redshift evolution of the baryon mass fractions in the intergalactic and diffuse media. (a) The baryon mass fraction in the intergalactic medium ($f_\text{IGM}$), computed from the fiducial simulation under various CGM gas removal criteria, including radius cutoffs at $R_{\mathrm{cut}} = R_{200}, 2R_{200}, 3R_{200}$ and FoF-based halo boundaries. Solid and dashed lines represent results from the fiducial and NoBH models, respectively. The orange dot-dashed line shows the analytical prediction from the Press-Schechter HMF \citep{Press1974} with $\Omega_b/\Omega_m = 0.16$. Green and red dashed lines indicate “modified PS” estimates derived from the median and linear-mean gas fractions of halos in different mass bins. The orange pentagram shows the constraint on $f_\text{IGM}$ from \citet{Connor2024} based on the full sample of all previously localized FRBs. The grey dashed line at $f_{\mathrm{IGM}}$ indicates the total cosmic baryon budget. (b) The baryon mass fraction in the diffuse medium ($f_{\text{diff}}$) for $z \leq 1$, constrained using the Macquart relation. Black open squares and red open circles show simulation results from the fiducial and NoBH runs, respectively. The blue "X" and orange pentagon correspond to observational constraints from \citet{Connor2024} using the full sample and the DSA-110-only sample. For comparison, the $f_\text{IGM}$ results from panel (a) are overplotted using thinner lines and smaller markers
    }
    \label{fig:f_IGM_redshift}
\end{figure*}

We begin with the simplest approach: utilizing the \citet{Press1974} theoretical model to compute the halo mass function (HMF) at different redshifts (the result from our simulations at $z=0$ is shown in Figure~\ref{fig:HMF}). We assume that the gas mass fraction in the CGM of each halo is a constant $f_{\mathrm{gas}} \sim \Omega_b/\Omega_m \approx  0.16$. Then, based on the HMF, we estimate the CGM gas fraction by summing over halos in different mass bins. This gives us an estimate for $f_{\mathrm{IGM}}$:
\begin{equation}
    f_{\mathrm{IGM}}(z) = 1 - \frac{f_{\mathrm{gas}}\int M \Phi(M,z) dM}{\Omega_b \rho_{\text{crit}}}
\end{equation}
where $\Phi(M, z)$ represents the HMF at redshift $z$. $M$ denotes the halo mass, and the denominator, $\Omega_b \rho_{\mathrm{crit}}$, represents the cosmic mean baryon density, which serves as the total baryonic mass budget of the universe. $f_{\mathrm{gas}}$ accounts for the gas fraction within halos, which we define as the CGM gas fraction: 
\begin{equation} f_{\mathrm{gas}} = \frac{M_{\mathrm{gas}, R_{\mathrm{cut}}}}{M_{\mathrm{halo}, R_{\mathrm{cut}}}}. \end{equation}
Here, $M_{\mathrm{gas}, R_{\mathrm{cut}}}$ is the total gas mass enclosed within a halo up to a cutoff radius $R_\text{cut}$, and $M_{\text{halo}, R_{\text{cut}}}$ is the total halo mass within the same radius. This definition explicitly incorporates the dependence on the chosen truncation radius, which we explore in later sections. 
The Press-Schechter (PS74) prediction is shown by the orange dot-dashed line in Figure~\ref{fig:f_IGM_redshift}a (diamond marker). It is clear that this result is much lower than $f_{\mathrm{IGM}}$ constrained by the Macquart relation, with $f_{\mathrm{IGM, true}} \sim 0.7$ at $z = 1$. This discrepancy likely stems from three factors: (1) we cannot approximate the CGM gas fraction by the cosmic average baryon-to-matter ratio, and (2) $f_{\mathrm{gas}}$ evolves with halo mass, and (3) random LoSs towards FRBs do not go through all halos at each redshift. The redistribution of baryons within halos reduces the CGM gas fraction below the cosmic average, and this effect is mass-dependent, as shown in Fig.~\ref{fig:f_gas_evolution0} and the lower panels of Figure~\ref{fig:halo_gas_mass}. To improve the accuracy of our estimate, We extract the median and mean $f_{\mathrm{gas}}$ evolution across redshifts and mass bins, as illustrated in Figure~\ref{fig:f_gas_evolution0}. We then employ them in a modified calculation:
\begin{equation}
    f_{\mathrm{IGM}}(z) = 1 - \frac{\int f_{\mathrm{gas}}(M, z) M\Phi(M) dM}{\Omega_b \rho_{\text{crit}}}
\end{equation}
where $f_{\mathrm{gas}}(M, z)$ now explicitly accounts for mass- and redshift-dependent evolution. The resulting modified $f_{\mathrm{IGM}}$ is shown by the green and red dotted lines and diamond markers indicated with `Modified PS' in Figure~\ref{fig:f_IGM_redshift}a. This result is closer to the constraints from the Macquart relation, though the evolution in the $z<1$ range is somewhat smoother.
\subsubsection{Directly Subtracting CGM Gas to Calculate $f_{\mathrm{IGM, true}}$} 
\label{subsubsec:CGM_substraction}
Both the DM--$z$ relation and the Press-Schechter HMF-based calculations of $f_{\mathrm{IGM}}$ inherently rely on theoretical assumptions and uniform matter distribution.  
A more direct approach is to identify and precisely remove all CGM-associated gas particles within each halo in the simulation snapshot at each redshift. 
However, this method introduces a key challenge: defining the spatial extent of the CGM. The choice of CGM boundary significantly impacts the computed $f_{\mathrm{IGM}}$. 

\begin{figure*}[htbp]
    \centering
    \includegraphics[width=0.85\textwidth]{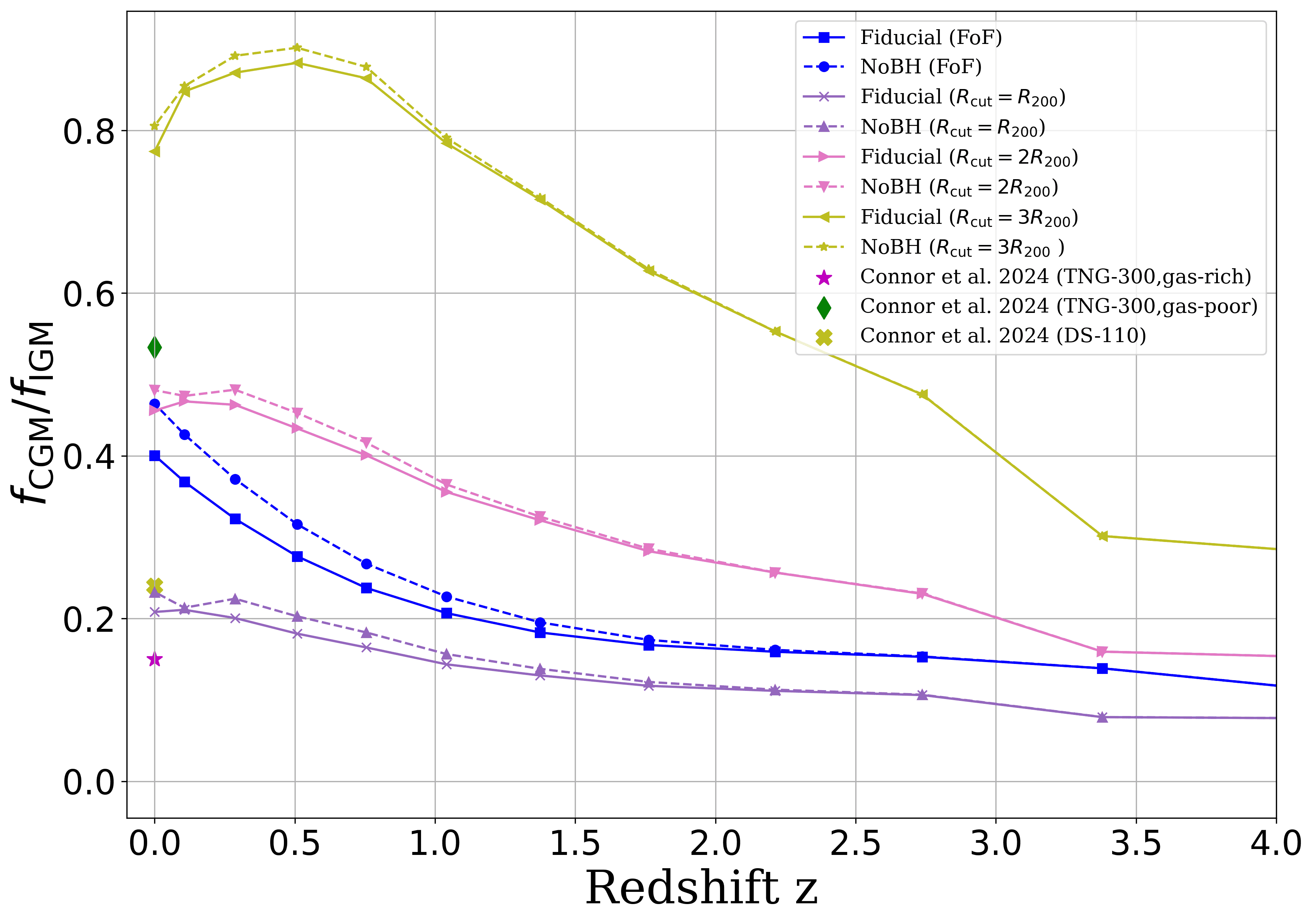}
    \caption{
    Redshift evolution of the baryon fraction ratio between the circumgalactic medium (CGM) and the intergalactic medium (IGM), $f_{\mathrm{CGM}} / f_{\mathrm{IGM}}$, under various CGM gas exclusion methods. Solid and dashed lines represent the fiducial and NoBH simulations, respectively. Observational estimates from \citet{Connor2024} (olive "X" symbols) are included for reference. The purple pentagram and dark green diamond indicate the expected CGM–IGM ratios from the TNG300 gas-rich and gas-poor simulations used in Connor’s study for comparison with observations.
    }
    \label{fig:fdiff_fIGM_fCGM}
\end{figure*}

To address this, we adopt four different halo gas removal criteria. The first method involves directly excluding gas bound within FoF halos. 
The FoF algorithm identifies halos by linking particles within a fixed linking length, set to 0.2 times the mean interparticle separation.
This method effectively captures the total mass and extent of a halo, including diffuse CGM. The results using this approach are shown as the blue solid line (square markers; fiducial) and the blue dashed line (circle markers; NoBH) in Figures~\ref{fig:f_IGM_redshift} and \ref{fig:fdiff_fIGM_fCGM} . 

In addition to the FoF-based method, we further employ three alternative CGM subtraction approaches using different cutoff radii, listed in Table~\ref{tab:markers}.
\begin{table}
    \caption{CGM subtraction methods and their markers}
    \begin{tabular}{cccc}
        \toprule
        $R_{\mathrm{cut}}$ & Type & Line Style & Marker \\
        \midrule
        \multirow{2}{*}{$R_{200}$}  & Fiducial & Light purple solid & $\times$\\
                                    & NoBH & Light purple dashed & $\triangle$\\
        \multirow{2}{*}{$2R_{200}$} & Fiducial & Magenta solid & $\triangleright$\\
                                    & NoBH & Magenta dashed & $\triangledown$\\
        \multirow{2}{*}{$3R_{200}$} & Fiducial & Olive-green solid & $\triangleleft$\\
                                    & NoBH & Olive-green dashed & $\star$\\
        \bottomrule
    \end{tabular}
    \label{tab:markers}
\end{table}

As shown in Figure~\ref{fig:f_IGM_redshift}a, all these methods, along with the $f_{\mathrm{IGM}}$ computed via the Press-Schechter HMF approach (Section~\ref{sec:HMF_f}), converge towards $f_{\mathrm{IGM}} = 1$ (gray dashed line) at high redshifts ($z \sim 10$), indicating their consistency in the early Universe when there are few dark matter halos. 
At lower redshifts, significant differences emerge as halos develop. 
When removing gas only within $R_{200}$, $f_{\mathrm{IGM}}$ remains the highest and exceeds the Macquart relation constraints, 
but at $z = 1$, it reaches $f_{\mathrm{IGM}} = 0.874$ (fiducial) and $0.865$ (NoBH), aligning closely with the Macquart estimate (i.e., using Eq.~(\ref{eq:dm_igm}) and regarding $f_{\mathrm{diff}}$ as $f_{\mathrm{IGM}}$). 
For $R_{\mathrm{cut}} = 2R_{200}$, the resulting $f_{\mathrm{IGM}}$ is lower than the Macquart estimate, yielding $f_{\mathrm{IGM}} = 0.738$ (fiducial) and $0.733$ (NoBH) at $z = 1$. 
When extending $R_{\mathrm{cut}}$ to $3R_{200}$, the lowest $f_{\mathrm{IGM}}$ values are obtained, with $f_{\mathrm{IGM}} = 0.560$ (fiducial) and $0.558$ (NoBH) at $z = 1$.  
These values are obtained only using a single snapshot of CROCODILE simulation at each redshift. 
The results derived from the FoF halo approach lie between the $R_{200}$ and $2R_{200}$ cases, suggesting that the effective radius defined by the FoF method is within this range.
Although the $R_{200}$-based calculation yields $f_{\mathrm{IGM}}$ values closest to the Macquart relation at $z = 1$, 
the overall redshift evolution of $f_{\mathrm{IGM}}$ derived from the FoF halo method exhibits the best agreement with the Macquart relation at low redshifts. This suggests that if CGM is defined within the range of $R_{200}$ to $2R_{200}$, the $f_{\mathrm{diff}}$ constrained by the Macquart relation closely matches the true $f_{\mathrm{IGM}}$ at $z\lesssim 1$. 

However, the observed ``$f_{\mathrm{IGM}}$'' inferred from the FRB $\text{DM}$--$z$ relation is systematically higher than the true $f_{\mathrm{IGM}}$ due to the inclusion of foreground halos, which are not explicitly separated in most previous FRB-based studies (e.g., \citealp{Yang2022ApJ...940L..29Y, Wang2023ApJ...944...50W}). 
Scientifically, our primary interest lies in accurately determining $f_{\mathrm{IGM}}$ rather than $f_{\mathrm{diff}}$, as $f_{\mathrm{IGM}}$ more directly represents the baryon fraction in the IGM. This further underscores the necessity of constructing a foreground galaxy map to refine the Macquart relation's application in constraining $f_{\mathrm{IGM}}$, as emphasized by \citet{Li2019ApJ...884L..26L,Lee2022,Lee2023} and \citet{KHRYKIN2024}.

\subsection{Effect of AGN Feedback on Baryon Redistribution}
\label{subsec:baryon_redistribution}

To more precisely analyze the impact of AGN feedback on baryon redistribution, we examine the evolution of halo gas relative to the IGM. Figure~\ref{fig:fdiff_fIGM_fCGM} shows the redshift evolution of the ratio \footnote{After we define the CGM in Section~\ref{subsubsec:CGM_substraction}, the concept of $\fhalo$ can be naturally replaced by $f_{\mathrm{CGM}}$.}$f_{\mathrm{CGM}} / f_{\mathrm{IGM}}$ for both the fiducial and NoBH models under different $R_{\mathrm{cut}}$ definitions.  In general, we see that this ratio is increasing with decreasing redshift, as the halos grow due to gravitational instability. Notably, the observational constraints from \citet{Connor2024} are broadly consistent with the predictions adopting $R_{\mathrm{cut}} = R_{200}$, suggesting that this structural definition may best capture the CGM scale probed by current FRB observations. At low redshifts, the ratio is systematically lower in the fiducial case, indicating that AGN feedback expels a portion of halo gas into the IGM. This trend confirms that AGN feedback plays a moderate but non-negligible role in redistributing baryons. This ratio therefore serves as a complementary tracer of AGN-induced baryon redistribution.

However, the relatively small gap between fiducial and NoBH simulations suggests that AGN feedback in our CROCODILE simulations has a modest impact on $f_{\mathrm{diff, obs}}$ and $f_{\mathrm{CGM}}/f_{\mathrm{IGM}}$. This result can be attributed to several factors:
\begin{itemize}
    \item Absence of kinetic AGN feedback: Our current implementation of AGN feedback in the CROCODILE simulations primarily relies on thermal feedback and lacks AGN jet (kinetic) feedback. Without this additional kinetic mode, the efficiency of AGN-driven redistribution is likely underestimated.
    
    \item Effects of AGN feedback vary with halo mass: While high-mass halos ($M_{\mathrm{halo}} > 10^{12.5} M_{\odot}$) are expected to exhibit stronger AGN-driven outflows, lower-mass halos ($M_{\mathrm{halo}} < 10^{12} M_{\odot}$) retain a significant fraction of their CGM, leading to only a minor difference in $f_{\mathrm{diff}}$ between the two runs. 
    
    \item Impact of cooling and recycling. Even if AGN feedback expels gas into the IGM, a fraction of this gas can return to the halo through radiative cooling and the gravity of the halo, partially counteracting the redistribution effect. This fallback might reduce the difference in $f_{\mathrm{diff}}$ between the fiducial and NoBH runs.
    
    \item Coupling between $f_{\mathrm{IGM}}$ and $f_{\mathrm{CGM}}$: While $f_{\mathrm{diff}}$ is defined as the sum of $f_{\mathrm{IGM}}$ and $f_{\mathrm{CGM}}$, these two components are intrinsically coupled, especially in the presence of AGN feedback. AGN-driven outflows can increase $f_{\mathrm{IGM}}$ by expelling gas from halos while simultaneously reducing $f_{\mathrm{CGM}}$ by depleting CGM gas. Observationally, constraining $f_{\mathrm{IGM}}$ using FRBs requires careful subtraction of foreground halo contributions along each LoS — a challenging but essential task for FRB-based cosmology. From the simulation perspective, although we have full knowledge of halo properties and provide a detailed comparison of CGM/IGM separation in Figure~\ref{fig:f_structure_evolution}, the precise definition of the CGM–IGM boundary remains model-dependent and nontrivial. This underscores the importance of joint observational and simulation-based efforts.

\end{itemize}

\subsection{Fitting the Redshift Evolution of $f_\text{IGM}$ and $f_\text{diff}$}
\label{subsec:fitting_f}

We provide fitting models for the redshift evolution of $f_\text{IGM}$ and $f_\text{diff}$, which could be useful for semi-analytic models of galaxy formation and IGM. 
A natural first approach is to use a power-law function, as many cosmic-scale evolutionary processes can be well-described by such relations. However, $f_\text{IGM}$ asymptotically approaches 1 at high redshifts, suggesting that a simple power-law function may not be sufficient to capture this behavior. Consequently, we adopt an ``exponential convergence model'', ensuring that as $z$ increases, $f_\text{IGM}$ naturally approaches 1.

To this end, we introduce three models of increasing complexity, each progressively improving the accuracy of the fit:

\begin{enumerate}
    \item Convergence Exponential Model (C-Exp):
    \begin{equation}
        f(z) = 1 - \exp(-\kappa (z + \tau))
        \label{eq:c-exp}
    \end{equation}
    This is the simplest model, assuming that the growth of $f_\text{IGM}$ or $f_\text{diff}$ follows a straightforward exponential process governed by $\kappa$ and $\tau$. While it effectively describes the large-scale growth trend, it lacks flexibility in capturing the detailed evolution at both higher and lower redshifts ($z \lesssim 1$).

    \item Convergence Power-Law Exponential Model (CPL-Exp):
    \begin{equation}
        f(z) = 1 - \exp(-\kappa (z + \tau)^{\zeta})
        \label{eq:cpl-exp}
    \end{equation}
    By introducing the exponent $\zeta$, this model gains additional flexibility, allowing the growth rate to vary with redshift instead of following a fixed exponential form. As a result, it better captures the evolution at lower redshifts while preserving the high-redshift convergence to 1.

    \item Double Power-Law Exponential Model (DPL-Exp):
    \begin{equation}
        f(z) = 1 - \exp(-\kappa (z^\gamma + \tau)^{\zeta})
        \label{eq:dpl-exp}
    \end{equation}
    Further refining the model, we introduce an additional exponent $\gamma$ to allow the redshift dependence to be more flexible. This enables the model to better capture the growth behavior across different redshift ranges. In particular, it significantly improves the fit at lower redshifts, ensuring that the model accounts for early-stage evolution more effectively.
\end{enumerate}

From C-Exp to CPL-Exp to DPL-Exp, each successive model builds upon the previous one by adding degrees of freedom, leading to a progressively better fit for $f_\text{IGM}$ to the data. However, for $f_\text{diff}$ the CPL-Exp model provides the best fit, as indicated by the $\chi^2$/DOF values (table \ref{tab:fitting_parameters} in Appendix~\ref{app:fit_params}).

As seen from the best-fit parameters listed in Table~\ref{tab:fitting_parameters}, the evolution of $f_\text{IGM}$ and $f_\text{diff}$ follows a well-defined trend across different redshifts. Notably, we observe that the observationally inferred $f_\text{diff}$, derived from FRB DM--$z$ constraints, exhibits a significantly higher growth rate compared to the intrinsic $f_\text{IGM}$. This trend aligns with expectations:

- At low redshifts, $f_\text{diff}$ remains low, closely matching the results obtained from applying a halo mass cut of $R_\text{cut} = 2R_{200}$. This suggests that at lower redshifts, the probability of an FRB intersecting a massive foreground halo is low. As a result, the inferred $f_\text{diff}$ is closer to the intrinsic $f_\text{IGM}$, as it receives minimal contributions from foreground halos.

- As redshift increases, the probability of FRBs passing through massive foreground halos increases significantly. Consequently, the inferred $f_\text{diff}$ includes a larger contribution from foreground halos, leading to a steeper increase. By $z \sim 1$, the inferred $f_\text{diff}$ surpasses the FoF-halo-excluded results and closely matches the results obtained from a $1R_{200}$ mass cut at $z\sim 1$.

Furthermore, although direct DM--$z$ constraints cannot determine the exact value of $f_\text{diff}$ at $z=0$, we can use our best-fit models to extrapolate their asymptotic values as $z \to 0$. These extrapolated values are:

\begin{itemize}
    \item C-Exp: $f_\text{diff}(z \to 0)$ = 0.685 (fiducial), 0.667 (NoBH)
    \item CPL-Exp: $f_\text{diff}(z \to 0)$ = 0.655 (fiducial),  0.604 (NoBH)
    \item DPL-Exp: $f_\text{diff}(z \to 0)$ = 0.706 (fiducial), 0.684 (NoBH)
\end{itemize}

For comparison, using the Press-Schechter HMF and the cosmic mean baryon density, the expected value of $f_\text{IGM}$ at $z=0$ is approximately $f_\text{IGM}(z=0) = 0.631$. This value closely matches the extrapolated values from the CPL-Exp model, supporting the validity of its extrapolations.

In summary, our results confirm that the inferred $f_\text{diff}$ from FRB DM--$z$ constraints systematically increases with redshift due to the increasing contribution of foreground halos. Our progressively refined models effectively capture this trend and provide robust extrapolations for $f_\text{diff}$ at $z=0$. These results offer new insights into the distribution of ionized gas in the Universe and its redshift evolution.

\subsection{From FRB Observables to Intrinsic $f_{\mathrm{IGM}}(z)$: A Semi-Analytic Reconstruction Framework}

We now discuss how to observationally constrain the "true" IGM baryon content using FRB sight lines.
The diffuse component of the DM was given by Eq.~(\ref{eq:dm_igm}). From an observational perspective, assuming that FRBs are randomly distributed across the sky, and considering that $f_{\mathrm{diff}}$ evolves with redshift, what we constrain is actually its cumulative, redshift-averaged form as we discussed in Appendix~\ref{APP:f_diff_appendix}:
\begin{equation}
\langle \mathrm{DM}_{\mathrm{diff}}(<z) \rangle \propto \langle f_{\mathrm{diff,\,obs}}(<z) \rangle \int_0^z \frac{1 + z'}{E(z')} \, dz' , 
\label{eq:DM_diff_obs}
\end{equation}
where $E(z) = \sqrt{\Omega_m (1+z)^3 + \Omega_\Lambda}$.  Here we assume a constant ionization fraction $\chi_e\approx0.875$ for fully ionized diffuse medium. 
Explicitly considering the contribution from intersecting halos, the average DM contribution from both IGM and halos can be expressed as:
\begin{equation}
\langle \mathrm{DM}_{\mathrm{diff}}(<z) \rangle \propto \int_0^z \left[ f_{\mathrm{IGM}}(z') + \langle f^{\mathrm{intersect}}_{\mathrm{Halos}}(z') \rangle \right] \frac{1 + z'}{E(z') } \, dz',
\label{eq:DM_theory}
\end{equation}
where the term $\langle f^{\mathrm{intersect}}_{\mathrm{Halos}}(z) \rangle$ represents the average contribution from halos intersected by FRB sight lines at redshift $z$. To incorporate the extended gas density profiles of halos in a physically motivated manner, we express this halo term as an integral over a halo mass range [$M_{\mathrm{low}}$, $M_{\mathrm{up}}$], with the projected baryonic column density:
\begin{align}
\left\langle f^{\mathrm{intersect}}_{\mathrm{Halos}}(z) \right\rangle 
&= \frac{1}{\rho_b(z)} \int^{M_\mathrm{up}}_{M_\mathrm{low}} n(M_h, z) \times \notag\\
&\quad\left[ \int_0^{R_{\mathrm{cut}}(M_h)} 2\pi b \, \Sigma_g (b; M_h, z) \, db \right]\, dM_h ,
\label{eq:intersect_fhalo}
\end{align}

where
\begin{equation}
\Sigma_g(b) = 2 \int_b^{R_\mathrm{cut}(M_h)} \frac{rf_{\mathrm{gas}}\rho_{\mathrm{model}}(r)}{\sqrt{r^2 - b^2}} \, dr.
\label{eq:sigma_projection}
\end{equation}
Here $\Sigma_g (b; M_h, z)$ is the projected gas mass column density of a halo with mass $M_h$ at impact parameter $b$, $n(M_h, z)$ is the HMF, $\rho_{\rm model}(r)$ is the modeled density profile of the dark matter halo (e.g., Equation \ref{eq:profile_model}), and $\rho_b(z)$ is the mean cosmic baryon density at redshift $z$. The upper limit $R_{\mathrm{cut}}(M_h)$ is the assumed maximum cutoff radius of halo gas (e.g., $R_{200}$ or $2R_{200}$). This formulation properly accounts for both the extended gas structure and the geometric probability of intersecting a halo, making it a physically motivated expression for the average halo contribution to the FRB dispersion measure.

By differentiating both sides of Equations~\ref{eq:DM_diff_obs} and \ref{eq:DM_theory}, we derive the following expression for the IGM baryon fraction:

\begin{align}
f_{\mathrm{IGM}}(z) 
&= \left\langle f_{\mathrm{diff,obs}}(<z) \right\rangle - \left\langle \fhalo^{\mathrm{intersect}}(z) \right\rangle
\notag \\
&+  \left( \frac{d}{dz} \left\langle f_{\mathrm{diff,obs}}(<z) \right\rangle \right)
\cdot \left[ \frac{1}{X(z)} \int_0^z X(z')\, dz' \right] 
\label{eq:f_IGM}
\end{align}

where  
\begin{equation}
    \displaystyle X(z) = \frac{(1 + z)}{E(z)}. 
\end{equation}
Here, $\langle f_{\mathrm{diff,obs}}(<z) \rangle$ can be inferred from the observation of the Macquart relation up to $z$, which statistically links FRB DMs and redshift. Its analytic form can be approximated using the empirical fitting functions (e.g., C-exp) described in Section~\ref{subsec:fitting_f}, allowing us to compute the derivative term $\frac{d}{dz} \langle f_{\mathrm{diff,obs}}(<z) \rangle$.

The sightline-weighted halo baryon fraction $\langle \fhalo^{\mathrm{intersect}}(z) \rangle$ can also be constrained observationally, by modeling the density field of foreground galaxies along FRB sight lines \citep[e.g.,][]{Lee2022}.

\begin{figure*}[htbp]
    \centering
    \includegraphics[width=0.8\textwidth]{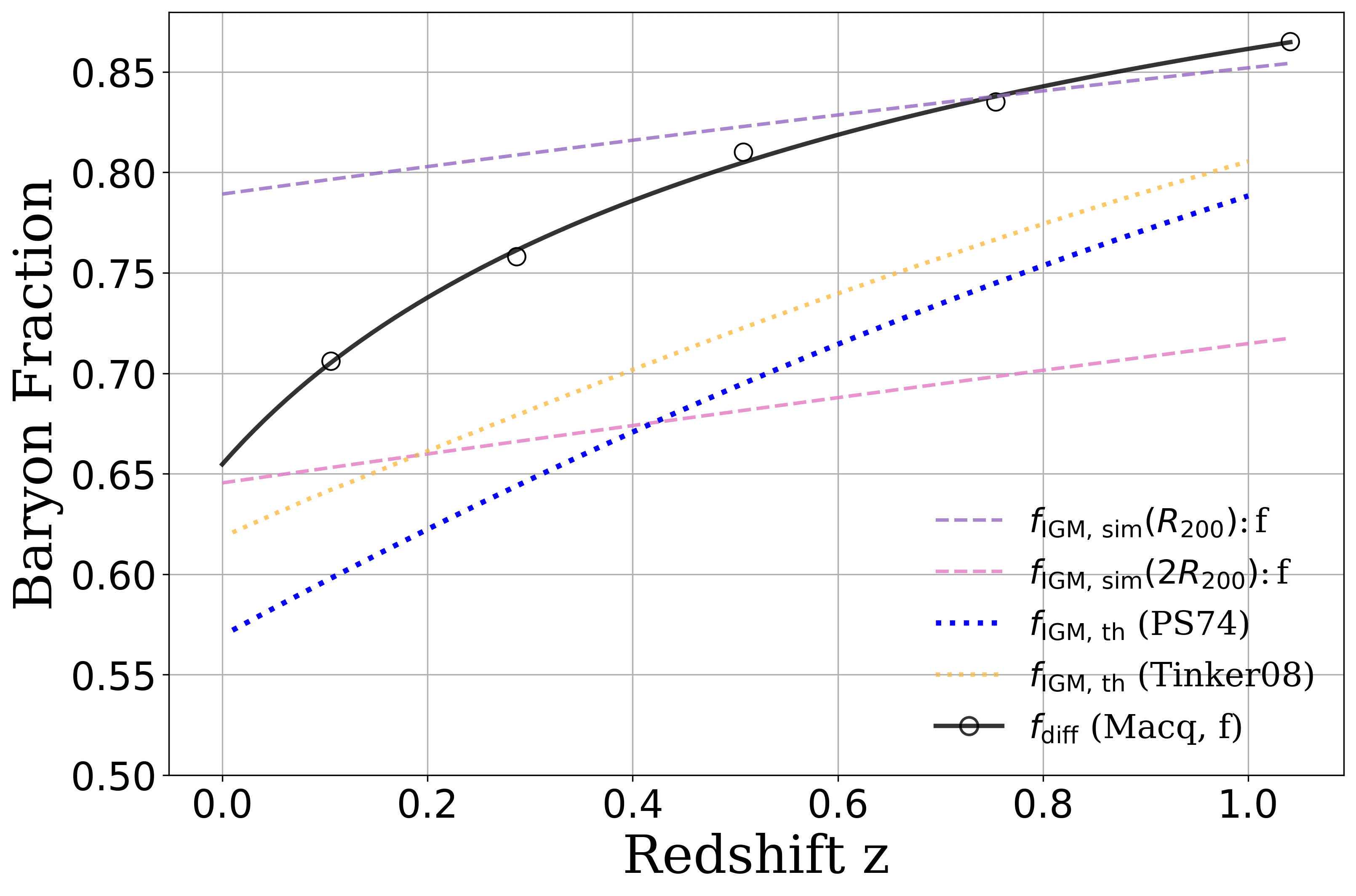}
    \caption{
    Comparison of the redshift evolution of the average diffuse baryon fraction $\langle f_{\mathrm{diff}}^{\mathrm{obs}} \rangle$ (black) obtained from the DM--$z$ relation against our theoretical modeling. 
    The dashed lines represent simulation results of $f_{\mathrm{IGM}}$ under two different CGM boundary definitions: $R_{200}$ (light purple) and $2R_{200}$ (pink). 
    The dotted lines show the reconstructed IGM baryon fraction recovered from $f_{\mathrm{diff}}$ using PS74 (blue) and Tinker08 (yellow) HMFs.
    }
    \label{fig:f_diff_obs}
\end{figure*}

Figure \ref{fig:f_diff_obs} compares the reconstructed IGM baryon fraction $f_{\mathrm{IGM, th}}(z)$—recoverd from the cumulative observational constraint $\langle f_{\mathrm{diff,obs}}(<z) \rangle$—with the baryon fraction directly computed from simulations, $f_{\mathrm{IGM, sim}}(z)$. The black curve in the figure corresponds to the CPL-Exp fit to the fiducial $\langle f_{\mathrm{diff,obs}}(<z) \rangle$ shown in Table \ref{tab:fitting_parameters}.

To calculate $\langle \fhalo^{\mathrm{intersect}}(z) \rangle$ (Equation \ref{eq:intersect_fhalo}), we adopt two standard HMFs—PS74 and Tinker08. For the gas density profiles, we use the fitted parameters of the mNFW profile at $z = 0$, evaluated in different halo mass bins (listed in Table \ref{tab:mnfw_vertical_compact}), assuming a gas distribution of the form $f_{\mathrm{gas}} \rho_{\mathrm{model}}(r)$. The resulting values of $\langle \fhalo^{\mathrm{intersect}}(z) \rangle$ are shown in Table \ref{tab:fhalo_vs_fCGM}.

As shown in Figure \ref{fig:f_diff_obs}, $f_{\mathrm{IGM, th}}(z)$ lies closer to the simulated $f_{\mathrm{IGM, sim}}(z)$ obtained with a CGM exclusion radius of $2R_{200}$. Compared to the $1R_{200}$ case, it shows a steeper evolution with redshift and lies inbetween the $R_{200}$ and $2R_{200}$ definitions at $z \sim 1$.

It is important to note several sources of systematic uncertainties in this reconstruction:
\begin{enumerate}
    \item
    The analytic form of $\langle f_{\mathrm{diff,obs}}(<z) \rangle$ significantly affects the inferred IGM fraction and depends on the validity of the chosen fitting function.
    
    \item
    The HMF used in our model is based on analytical (PS74) and empirical (Tinker08) relations.
    However, as shown in Figure~\ref{fig:HMF}, these HMFs can deviate from the simulation results, especially at the high-mass end. In principle, one could interpolate the HMF directly from simulation outputs at each redshift for higher accuracy, but this would undermine the purpose of using FRBs to test theoretical models — as in reality, we do not know the true HMF of the Universe. Therefore, we opt for established analytic HMFs instead of interpolated simulation values.

    \item 
    The mNFW fitting parameters are derived from median density profiles at $z = 0$ in each mass bin. These do not represent individual halos and are used as characteristic profiles for population-level estimation.

    \item 
    The concentration parameter $c_{200}$ evolves with redshift and alters the relationship between $M_{200}$ and $R_{200}$, thereby modifying the density profile shape of the halo. This is particularly important for gas, whose distribution is further affected by baryon–DM interactions (see \citet{Ludlow2016, Sorini2025}).

    \item
    The gas density profiles are more complex than simply scaling dark matter profiles by a constant $f_{\mathrm{gas}}$ — baryonic processes such as feedback and thermal pressure modulate the gas structure differently from dark matter \citep{Sorini2025}.
\end{enumerate}

Despite the limitations from these sources of uncertainty, this semi-analytic framework provides a direct bridge between FRB cosmological observables, such as $f_{\mathrm{diff,\,obs}}$, and physically motivated quantities predicted by theoretical models (e.g., in the standard $\Lambda$CDM cosmology), enabling improved interpretation of FRB-derived constraints on the cosmic baryon distribution.

\section{Zoom-in Cosmological Hydrodynamic Simulations and their results} 
\label{sec:Host_Galaxy}
To examine the contribution of the HHs of FRBs (DM\textsubscript{host}), it is crucial to analyze FRB host environments with high-resolution simulations. This allows for a detailed separation of DM\textsubscript{host} contributions and helps in understanding how different types of host galaxies and the specific locations of FRBs within these galaxies (e.g., central regions versus outskirts) impact the total DM. 
To this end, we employ zoom-in cosmological hydrodynamic simulations performed by both {\Gthree} and {\Gfour} codes (simply due to the historical timeline).  We employ three different zoom simulations to probe host galaxies of different mass scales and different environments: dwarf galaxy, MW (MW)-like galaxy, and galaxy cluster.  
The initial conditions and parameters are summarized in Table~\ref{tab:zoom-in}.

\begin{figure*}[htbp]
    \centering
    \begin{tabular}{cc}
        \includegraphics[width=0.5\textwidth]{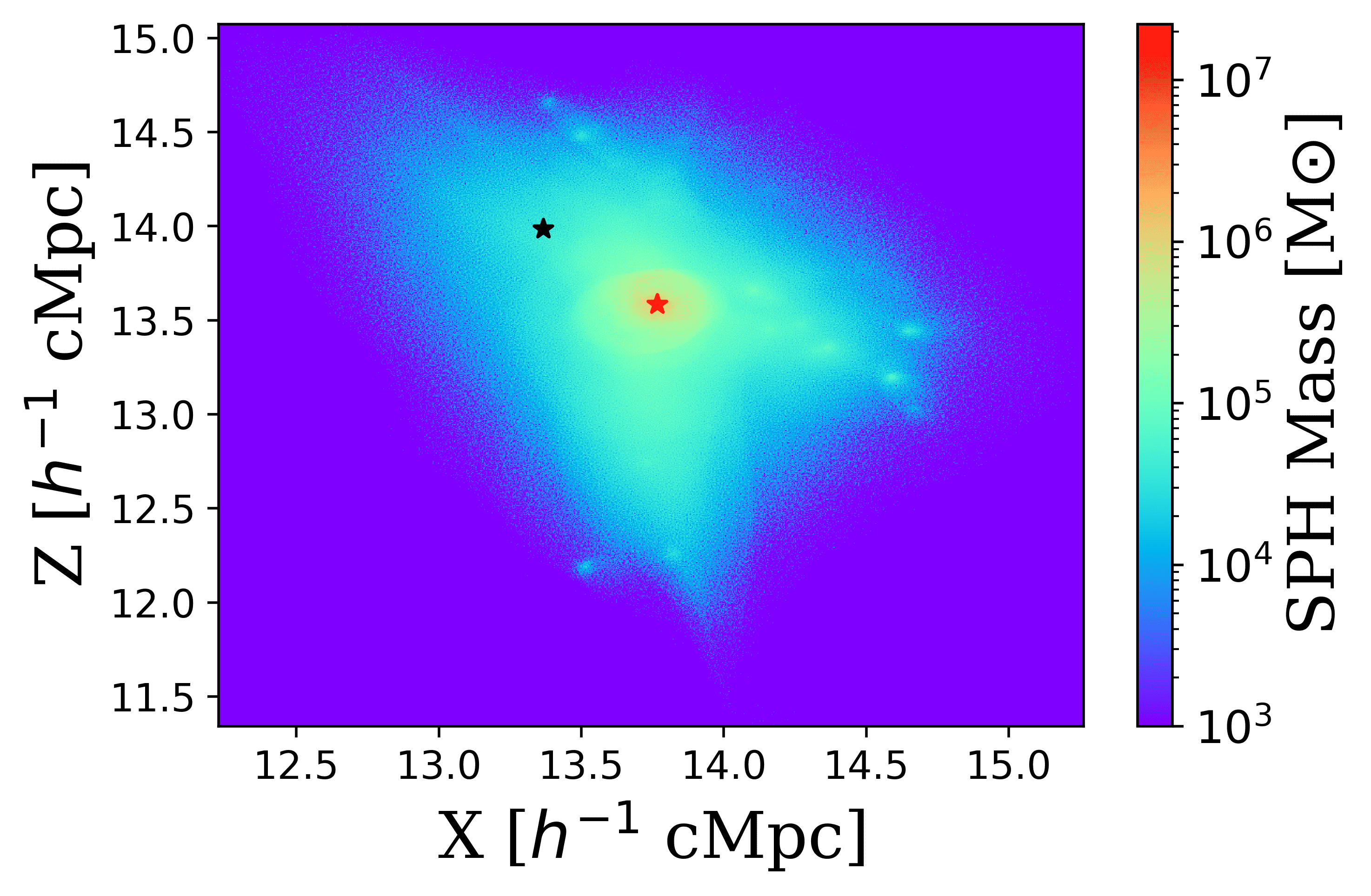} &
        \includegraphics[width=0.5\textwidth]{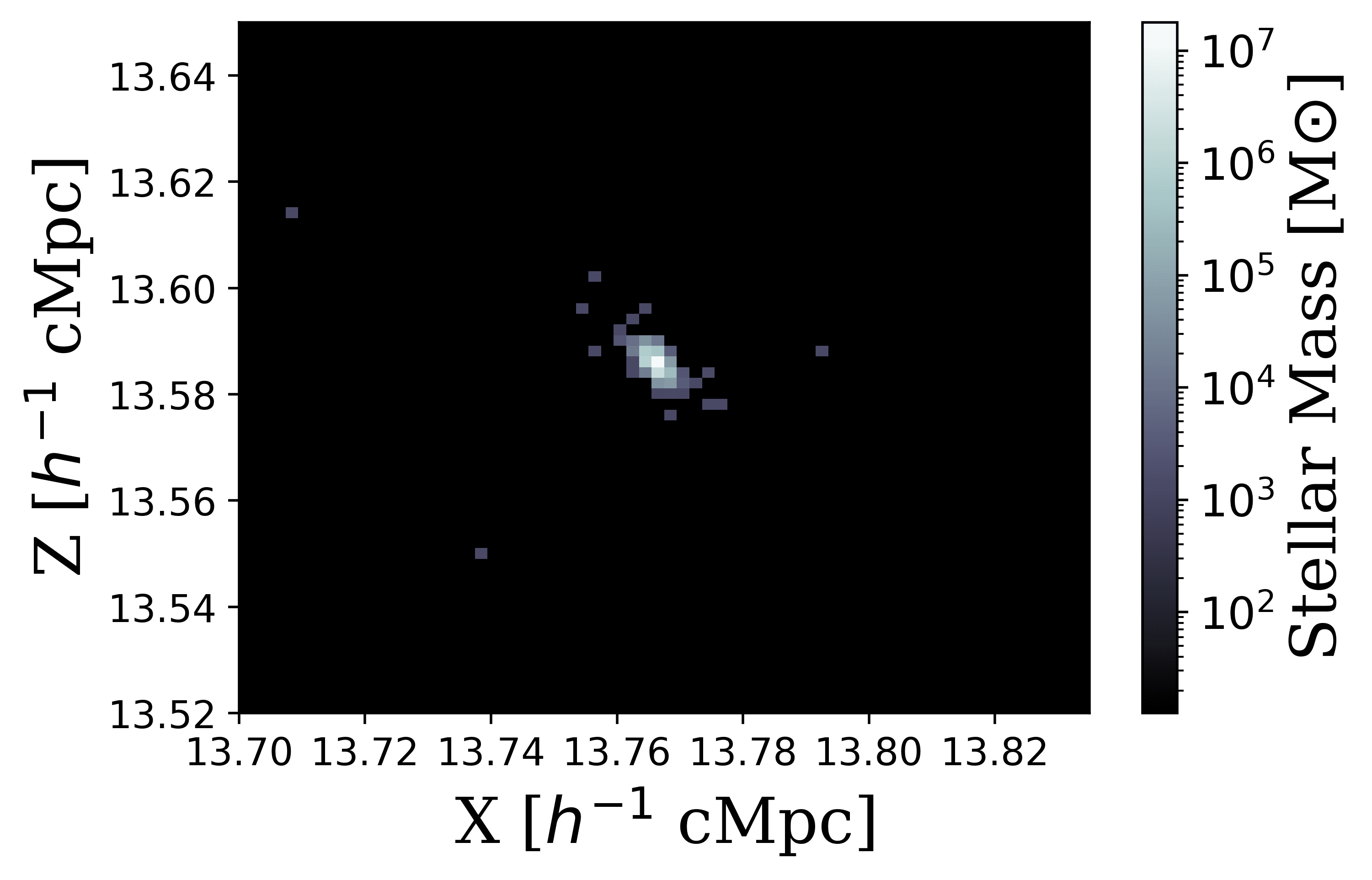} \\
        \includegraphics[width=0.5\textwidth]{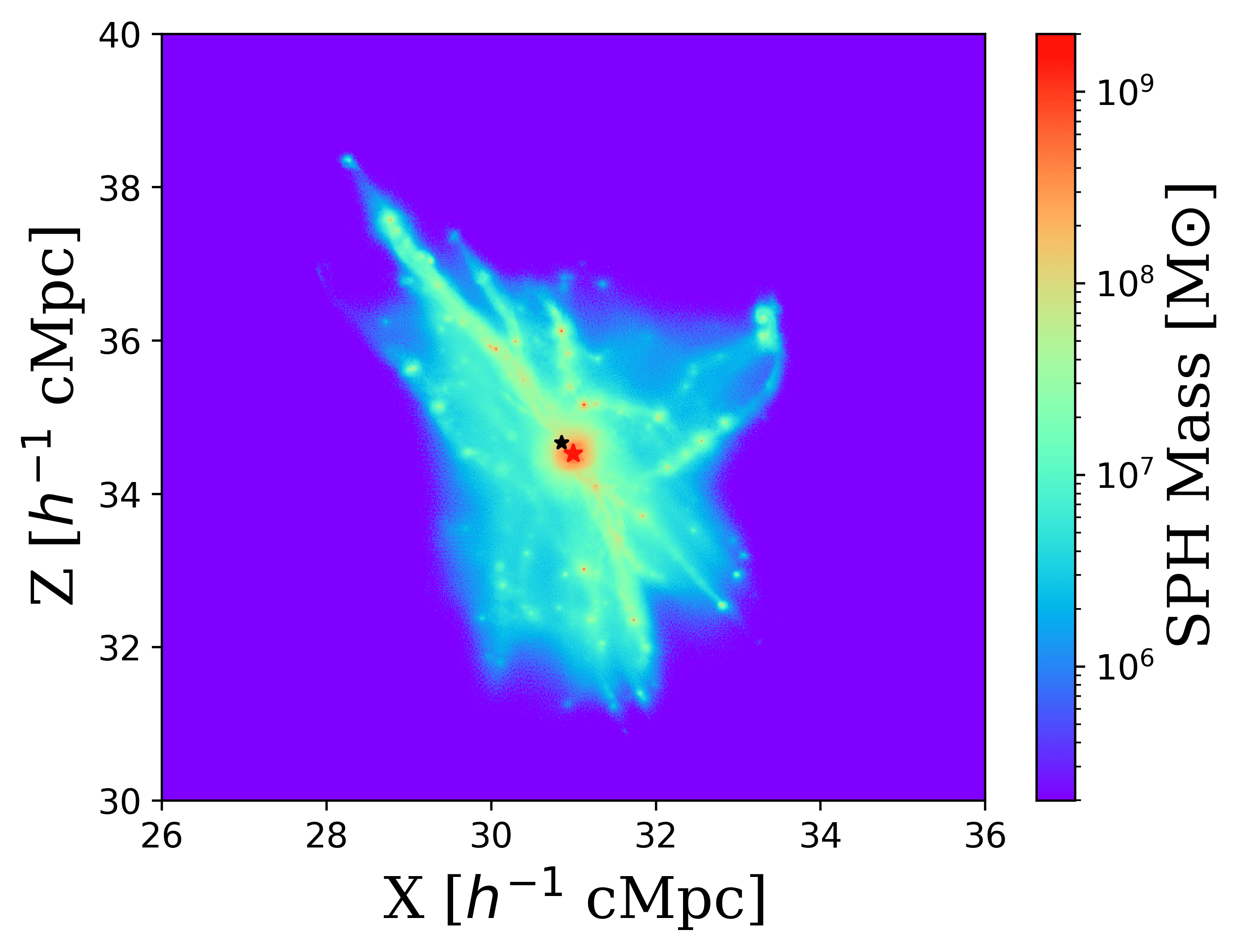} &
        \includegraphics[width=0.5\textwidth]{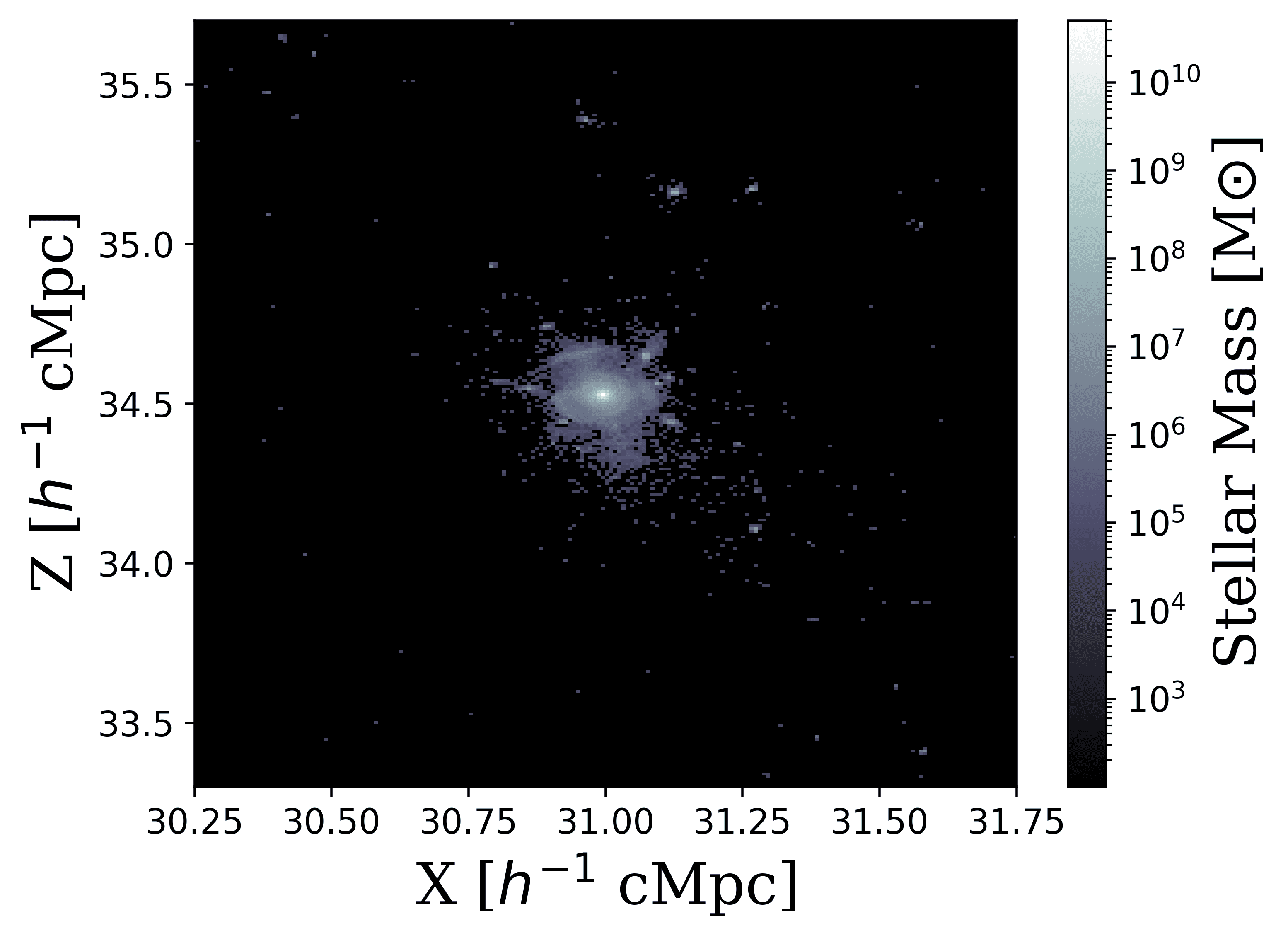} \\
        \includegraphics[width=0.5\textwidth]{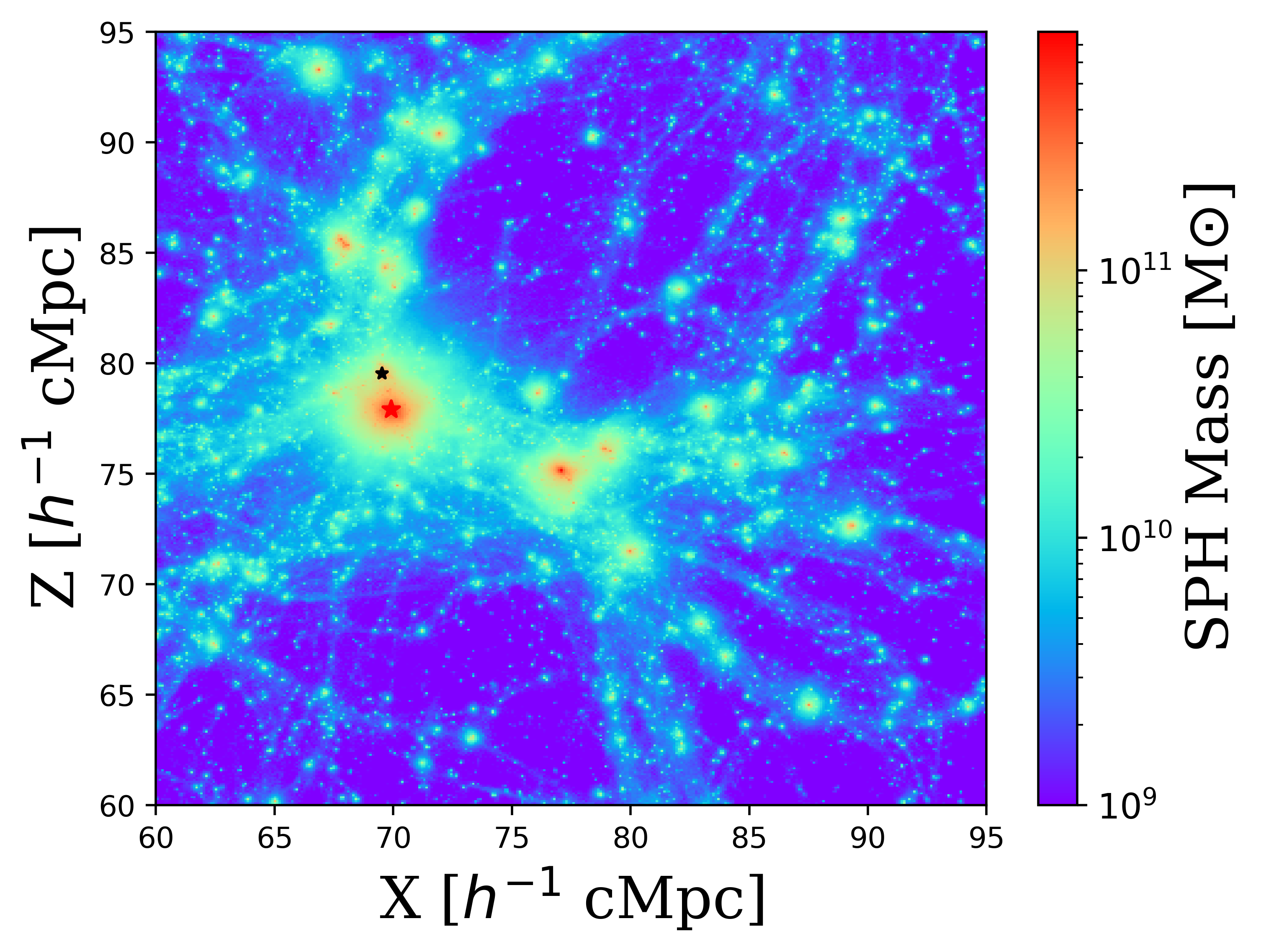} &
        \includegraphics[width=0.5\textwidth]{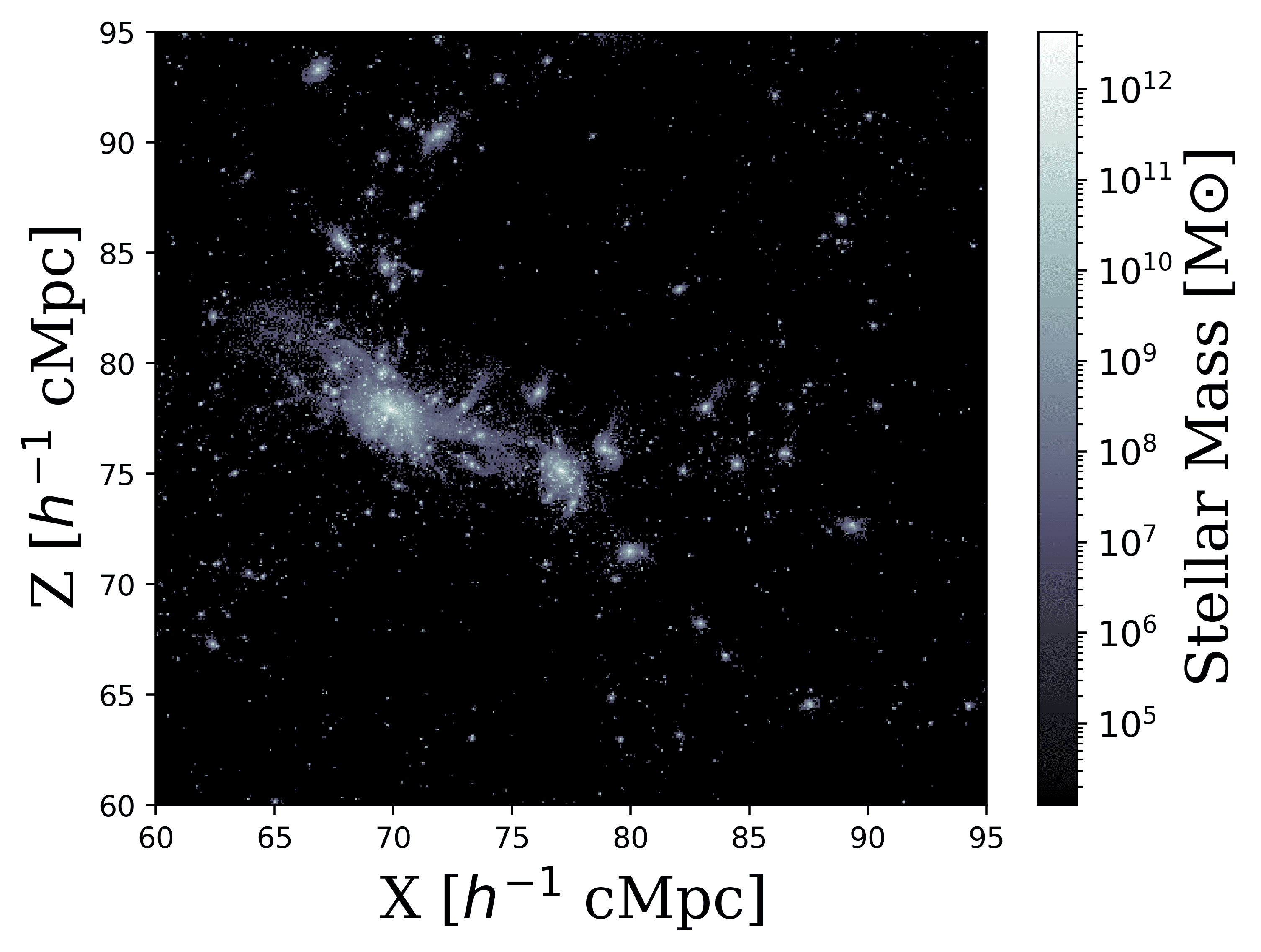} \\
    \end{tabular}
    \caption{Distribution of gas mass (left panels) and stellar mass (right panels) for a dwarf galaxy (top row; Tomaru et al. in prep.), a MW-like galaxy from the AGORA simulation \citep[middle row][]{Santi2021ApJ, Santi2024ApJ}, and a galaxy cluster (bottom row) from \citet{Fukushima2023MNRAS.525.3760F}. The red stars indicate the center-of-mass of each target halo.}
    \label{fig:zoomin_sims}
\end{figure*}

Figure~\ref{fig:zoomin_sims} presents the gas mass (left panels) and stellar mass (right panels) distributions for three different types of galaxies (dark matter halos) simulated using zoom-in techniques. 

\begin{itemize}
    \item Dwarf Galaxy: In the top row of Figure~\ref{fig:zoomin_sims}, the results for a dwarf galaxy (K. Tomaru et al. in prep.) from GADGET4-Osaka are displayed. The gas mass distribution shows a concentrated core, with some surrounding substructures. Given the small overall mass of this galaxy ($M_{200} = 9.3 \times 10^{9}\,\Msun$), the gas distribution is primarily centered around the core, and the star formation is also concentrated in this region, reflecting the compact nature of this system. The stellar mass is $2.2 \times 10^7\,\Msun$, further emphasizing the compactness of the galaxy.

    \item MW-like Galaxy: The bottom row shows the results from a MW-like galaxy in the AGORA GADGET3 simulation \citep{Santi2021ApJ, Santi2024ApJ}. The gas mass distribution reveals a well-defined spiral structure, typical of a disk galaxy. The stellar mass is concentrated in the disk and bulge regions, with significant star formation occurring throughout the spiral arms. With $M_{200} = 1.24 \times 10^{12} M_\odot$, this structure highlights the characteristic organization of stars and gas in spiral galaxies.
    
    \item Galaxy Cluster: The middle row displays the gas and stellar mass distributions for a galaxy cluster, also from {\Gthree} \citep{Fukushima2023MNRAS.525.3760F}. The gas mass is widely distributed, showing the large scale of this system ($M_{200} = 9.23 \times 10^{14}\,\Msun$). The gas distribution covers a vast region, with the stellar mass being more concentrated in the central region, reflecting the dense core of the cluster. This extensive distribution and the filamentary LSS are characteristic of galaxy clusters, with star formation centered in the core but more dispersed compared to the dwarf galaxy.
\end{itemize}

\begin{table*}[tb]  
\centering
\caption{Zoom-in Simulation Parameters} 
\label{tab:zoom-in}
\renewcommand{\arraystretch}{1.1} 
\setlength{\tabcolsep}{3pt} 
\resizebox{1\textwidth}{!}{ 
\begin{tabular}{lccc}
\hline
Parameter & Dwarf Galaxy$^\dagger$ & MW-like Galaxy$^\sharp$ & Galaxy Cluster$^\ddagger$ \\
\hline
Simulation Code & GADGET4-Osaka & AGORA-GADGET3 & GADGET3-Osaka \\
$H_0$ [km\,s$^{-1}$\,Mpc$^{-1}$] & 70.2 & 70.2 & 67.74 \\
$\Omega_m$ & 0.272 & 0.272 & 0.3099 \\
$\Omega_\Lambda$ & 0.728 & 0.728 & 0.6901 \\
$\Omega_b$ & 0.0455 & 0.0455 & 0.04889 \\
Box Size [$h^{-1}$\,cMpc] & 20 & 60 & 100 \\
$m_{\text{DM}}$ [$h^{-1}\,M_\odot$] & $7.32\times 10^3$ & $2.8\times10^5$ & $6.73\times10^{7}$ \\
$m_{\text{gas}}$  [$h^{-1}\,M_\odot$] & $1.47 \times 10^3$ & $5.65\times10^4$ & $1.26\times10^{7}$ \\
$\epsilon_{\text{grav}}$ [$h^{-1}$\,kpc] & 0.195 & 0.080 & 3.26 \\
$z_{\text{snapshot}}^\S$ & 0 & 0.3 & 0\\ 
$R_{200}$ [kpc] & 43.35 & 141.3 & 2052.94 \\
$M_{\text{halo}, 200}$ [$M_\odot$]& $9.32\times10^{9}$ & $1.24\times10^{12}$ & $9.23\times10^{14}$\\
$M_{\text{FoF}}$ [$M_\odot$] & $1.13\times10^{10}$ & $1.33\times10^{12}$ & $1.20\times10^{15}$\\
Dark Matter Mass [$M_\odot$]  & $1.11\times10^{10}$ & $1.17\times10^{12}$ & $1.05\times10^{15}$ \\
Gas Mass [$M_\odot$] & $2.10\times10^{8}$ & $1.02\times10^{11}$ & $1.22\times10^{14}$ \\
Stellar Mass [$M_\odot$] & $2.2\times10^{7}$ & $6.58\times10^{10}$ & $3.70\times10^{13}$ \\
Stellar Feedback & $\checkmark$ & $\checkmark$ & $\checkmark$ \\
AGN Feedback & \ding{53} & \ding{53} & \ding{53}\\
\hline
\end{tabular}
}
\vspace{5pt} 
\noindent \\
$^\dagger$ Tomaru et al. in prep. 
$^\sharp$ \citet{Santi2021ApJ, Santi2024ApJ} 
$^\ddagger$ \citet{Fukushima2023MNRAS.525.3760F} \\
$^\S$ $z_{\text{snapshot}}$ is the redshift at which host galaxy properties are analyzed. 
All halo-related quantities  \\
($R_{200}$, $M_{\text{halo},200}$, $M_{\text{FoF}}$, and total masses of dark matter, gas, and stars) are computed at this redshift.
\end{table*}
\subsection{FRBs in Dwarf Galaxies} 
\label{sec:Host_Dwarf}

Many repeating FRBs, such as FRB 121102 and FRB 20190520B, are located in low-metallicity, star-forming dwarf galaxies \citep{tendulkar2017, Niu2022}. These FRBs are believed to originate from highly magnetized environments, possibly driven by young neutron stars or magnetars as discussed in Appendix \ref{APP:FRB-Origin}. The stellar masses of these dwarf galaxies range approximately from $10^7 M_{\odot}$ to $10^9 M_{\odot}$, with star formation rates not exceeding $1 M_{\odot} \text{yr}^{-1}$. Here, we select a relatively low-mass dwarf galaxy to reflect its characteristic features, with a stellar mass of only $2.2 \times 10^7 M_{\odot}$ and a star formation rate of $0.03 M_{\odot} \text{yr}^{-1}$. On the other hand, although BNS mergers are no longer considered a dominant FRB formation mechanism, they may still contribute a minor subset of the FRB population, as discussed in Appendix \ref{APP:FRB-Origin}. We compute the DM distribution along 10000 different sight lines originating from both the central and off-center locations of the host galaxies (Figure~\ref{fig:Combined-curves}). For the central case, the DM integration length endpoint for each galaxy type is set to $2R_{200}$, and the LoS resolution length is chosen as $\epsilon_{\text{grav}}$, both listed in Table~\ref{tab:zoom-in}. For the off-center case, the DM integration length is set to 2 Mpc, and the LoS resolution length is chosen as 1 kpc. Unlike the central case, where the dominant contribution comes from the dense interstellar medium (ISM), the off-center DM primarily originates from the CGM and the extended halo. A shorter integration length might lead to an artificially narrow or underestimated DM distribution, failing to capture the full contribution from the CGM and halo gas. 

To account for off-center FRB sources caused by binary neutron star (BNS) mergers, we consider that these sources can be offset due to the kick velocity imparted to the system during supernova explosions. Here we simply assume that the source is located at an off-center location of 580 kpc for this dwarf galaxy.

%
%
%

In Figure~\ref{fig:Combined-curves}, the blue component represents the dwarf galaxy throughout the figure, while the orange and pink components correspond to the MW-like galaxy and galaxy cluster, respectively, which will be discussed in later sections. Panel~{\it (a)} shows that the electron number density in the central region of the dwarf galaxy can exceed $10^{-3}\,\text{cm}^{-3}$ but rapidly declines within tens of kpc. The corresponding DM profile in panel~{\it (c)} shows that the median DM rapidly converges in the high-density region within a few tens of kpc and is primarily distributed (16th-84th percentile) in the range of $2.3\text{-}9.6 \, \text{pc cm}^{-3}$. The DM distribution at \(2R_{200}\) (\(\text{DM}_{2R_{200}}\)) in the dwarf galaxy is best described by a Gaussian Mixture Model (GMM), indicating that multiple underlying components contribute to the observed distribution rather than a single Gaussian or log-normal function. The GMM is a probabilistic model that represents the data as a weighted sum of multiple Gaussian distributions:
\begin{equation}
P(x) = \sum_{i=1}^{n} w_i \cdot \mathcal{N}(x | \mu_i, \sigma_i^2),
\label{eq:gmm}
\end{equation}
where \(n\) is the number of Gaussian components, \(w_i\) is the weight of each Gaussian component, and \(\mathcal{N}(x | \mu_i, \sigma_i^2)\) is a Gaussian distribution defined as:

\begin{equation}
\mathcal{N}(x | \mu, \sigma^2) = \frac{1}{\sqrt{2\pi\sigma^2}} \exp \left( -\frac{(x - \mu)^2}{2\sigma^2} \right).
\end{equation}

We apply a three-component GMM (\(n=3\)) to fit the DM distribution of the dwarf galaxy, effectively capturing the observed structure, as shown in panel~{\it (e)}. The best-fit parameters, including the means (\(\mu_i\)), covariances (\(\sigma_i^2\)), and weights (\(w_i\)), are summarized in Table~\ref{tab:GMM}.

\begin{table*}[!htbp]
    \centering
    \caption{Gaussian Mixture Model Fit for DM$_{2R_{200}}$ in Dwarf and MW-like Galaxies.}
    \label{tab:GMM}
    \renewcommand{\arraystretch}{1.1} 
    \setlength{\tabcolsep}{3pt} 
    \scriptsize 
    \resizebox{0.9\textwidth}{!}{ 
    \begin{tabular}{c|ccc|ccc}
        \hline
        & \multicolumn{3}{c|}{Dwarf Galaxy} & \multicolumn{3}{c}{MW-like Galaxy (AGORA)} \\
        \hline
        Component & Mean $\mu_i$ & Variance $\sigma_i^2$ & Weight $w_i$ & Mean $\mu_i$ & Variance $\sigma_i^2$ & Weight $w_i$ \\
        & (pc cm$^{-3}$) & (pc$^2$ cm$^{-6}$) &  & (pc cm$^{-3}$) & (pc$^2$ cm$^{-6}$) &  \\
        \hline
        1 & $3.65^{+8.42}_{-0.02}$  & $2.32^{+16.56}_{-0.05}$ & $0.80^{+0.01}_{-0.66}$ & 
            $120.03^{+323.91}_{-1.48}$  & $2571.98^{+52246.88}_{-140.31}$ & $0.60^{+0.01}_{-0.33}$ \\
        2 & $25.50^{+0.31}_{-13.67}$  & $9.84^{+1.76}_{-3.10}$ & $0.070^{+0.07}_{-0.01}$ & 
            $448.16^{+1423.77}_{-37.11}$  & $56537.28^{+5.38\times10^4}_{-5.38\times10^4}$ & $0.27^{+0.32}_{-0.11}$ \\
        3 & $12.11^{+3.25}_{-0.52}$  & $19.13^{+1.57}_{-11.58}$ & $0.1340^{+0.01}_{-0.06}$ & 
            $1847.64^{+1420.32}_{-326.08}$  & $917426.72^{+8.67\times10^5}_{-1.11\times10^5}$ & $0.15^{+0.11}_{-0.01}$ \\
        \hline
    \end{tabular}
    }
\end{table*}
Among the three Gaussian components, the dominant one is centered at $\mu_1 = 3.65^{+8.42}_{-0.02}\, \text{pc cm}^{-3}$, with the highest weight of $w_1 = 0.80^{+0.01}_{-0.66}$, indicating that this component contributes the most to the overall distribution. This suggests that the majority of sight lines in the dwarf galaxy have DM values close to this peak, likely corresponding to the bulk of the ISM within the galaxy. The second and third components (see Table~\ref{tab:GMM}), which correspond to higher DM values with relatively small weights, likely represent more extreme sight lines. These may trace regions with higher gas densities or outer halo structures 

The median DM value for the dwarf galaxy is found to be $\text{DM}_{\mathrm{median}} = 4.13 \, \text{pc cm}^{-3}$, aligning closely with the primary Gaussian component, further supporting the interpretation that this component dominates the overall distribution.

For the off-center case, the electron number density profile is more dispersed, hovering around the cosmic mean electron density of $2.2 \times 10^{-7} \, \text{cm}^{-3}$. Consequently, its contribution to the FRB's $\text{DM}_{\text{Host}}$ is negligible and does not fit a log-normal distribution. Its symmetry is weaker, with most values concentrated near the lower boundary. As a result, the $\text{DM}_{\text{2Mpc}}$ distribution exhibits a shape characterized by a rapid rise followed by a slow decline. Therefore, we applied the Generalized Pareto Distribution (GPD) to fit the distribution.
\begin{equation}
f(x; k, \sigma, \theta) =
\begin{cases}
\frac{1}{\sigma} \left( 1 + k \frac{x - \theta}{\sigma} \right)^{-\frac{1}{k} - 1}, & k \neq 0 \\
\frac{1}{\sigma} e^{-\frac{x - \theta}{\sigma}}, & k = 0
\end{cases}
\label{GPD}
\end{equation}

Where the shape parameter is 
$ k = 0.65^{+0.01}_{-0.02}, $ and the scale parameter is 
$ \sigma = 0.02^{+0.0004}_{-0.0001}$.  The location parameter is tightly constrained at $ \theta = 0.01 $ with negligible uncertainty. The median DM for the off-center case is only $ 0.02 \, \text{pc cm}^{-3}$.

This result indicates that, if the FRB source is located in the star-forming region of this dwarf galaxy, its DM contribution would generally remain within $4 \, \text{pc cm}^{-3}$. In contrast, for the off-center case, the DM contribution can be considered negligible. 

Although our analysis indicates that the DM contribution from this dwarf galaxy is quite limited, it is important to note that the stellar mass of our selected FRB host galaxy is only $2.2 \times 10^7 \, M_{\odot}$, which is relatively low even among dwarf galaxies. This mass is also close to the lower end of the dwarf galaxy sample studied in \cite{Zhang2020}. While even lower-mass dwarf galaxies exist, FRB hosts below this mass range are rarely identified, likely due to both observational limitations and intrinsically lower FRB occurrence rates. Their faintness makes precise localization challenging, and their lower star formation rates may result in fewer FRB events. Considering that some more massive dwarf galaxies can reach up to $10^9 \, M_{\odot}$, the DM contribution from this galaxy can be regarded as a lower bound for identifiable FRB host dwarf galaxies.

\subsection{FRBs in Spiral Galaxies} 
\label{sec:host_MW_like}

Repeating FRBs such as FRB 180916 and FRB 20201124A are also believed to be magnetar-driven, but their host galaxies differ from FRB 121102. These FRBs are located in the star-forming regions of spiral galaxies, whose stellar masses range from $10^9 - 10^{10} \, M_{\odot}$ and star formation rates exceeding $0.01 \, M_{\odot} \, \text{yr}^{-1}$. 

For this study, we select a MW-like galaxy with a stellar mass of $1.21 \times 10^{11} \, M_{\odot}$ and a star formation rate of $6.6 \, M_{\odot} \, \text{yr}^{-1}$, as shown in Table~\ref{tab:zoom-in}, to represent its key characteristics. The analysis method follows the same procedure as in Section~\ref{sec:Host_Dwarf}. Results are presented in Figure~\ref{fig:Combined-curves}. However, due to the higher mass of the spiral galaxy compared to the dwarf galaxy, the BNS merger timescale is expected to be shorter. Therefore, we adopt a smaller off-center distance ~ $ 194 \, \mathrm{kpc}$.


In Figure~\ref{fig:Combined-curves}, panel~{\it (a)} shows that the differences in electron density profiles between central and off-center sight lines are within two orders of magnitude. In the central region, the electron density exceeds $10^{-2} \, \mathrm{cm}^{-3}$, while in the off-center case, it is around $10^{-4} \, \mathrm{cm}^{-3}$. Despite this, the electron density curves exhibit noticeable differences: in the central case, $n_e$ decreases more steeply, whereas in the off-center case, the decline is more gradual. Additionally, the off-center profiles exhibit larger deviations near the source. Both cases display numerous peaks in their profiles, indicative of the complex structures within spiral galaxies.

Panel~{\it (c)} presents the $\text{DM}_{\text{2Mpc}}$ distribution. The DM for the central case is primarily distributed in $83\text{-}813 \, \mathrm{pc \, cm}^{-3}$,  we also use the GMM (equation~\ref{eq:gmm}) with $n = 3$, to fit the distribution, as shown in panel~{\it (e)}. The best-fit parameters are listed in Table~\ref{tab:GMM}. The dominant component is centered at $\mu_1 = 120.03^{+323.91}_{-1.48} \, \text{pc cm}^{-3}$, with the highest weight of $w_1 = 0.60^{+0.01}_{-0.33}$, indicating that this component contributes the most to the overall distribution.

For the off-center case (panel~{\it e}), the DM values range between $6.38\text{-}28.42\, \mathrm{pc \, cm}^{-3}$. The distribution's shape is similar to the off-center case for dwarf galaxies, with the values more concentrated near the lower boundary. Consequently, we use the GPD (Equation~\ref{GPD}) to fit the distribution.  The best-fit parameters are: $k$ = $0.21^{+0.03}_{-0.01}$, $\sigma$ = $11.43^{+0.37}_{-0.87}$, and $\theta=3.91^{+0.19}_{-0.20}$.

The median DM values are $163.53 \, \mathrm{pc \, cm}^{-3}$ and $11.59\, \mathrm{pc \, cm}^{-3}$ for the central and off-center cases, respectively. The median value for the central case is comparable to the Galactic ISM contribution $\text{DM}_\text{MW}\approx 140\,\mathrm{pc \, cm}^{-3}$ or $\text{DM}_\text{MW}\approx 200\,\mathrm{pc \, cm}^{-3}$, according to the NE2001 model\citep{Cordes_Lazio2002} and the YMW16 model\citep{Yao2017}. 

\subsection{FRBs in Galaxy Clusters} \label{sec:host_GalCluster}

Galaxy clusters represent the most massive gravitationally bound structures in the Universe, and their vast gravitational potential significantly impacts the DM distribution of any hosted FRB. Unlike spiral or dwarf galaxies, the massive gravitational potential of a galaxy cluster confines its BNS mergers to regions very close to the cluster center. This results in a negligible difference between central and off-center sources. Therefore, instead of studying the off-center case, we analyze the DM contributions from a satellite galaxy located within the galaxy cluster. 

\begin{figure*}[htbp]
    \centering
    \begin{tabular}{cc}
        \includegraphics[width=0.48\textwidth]{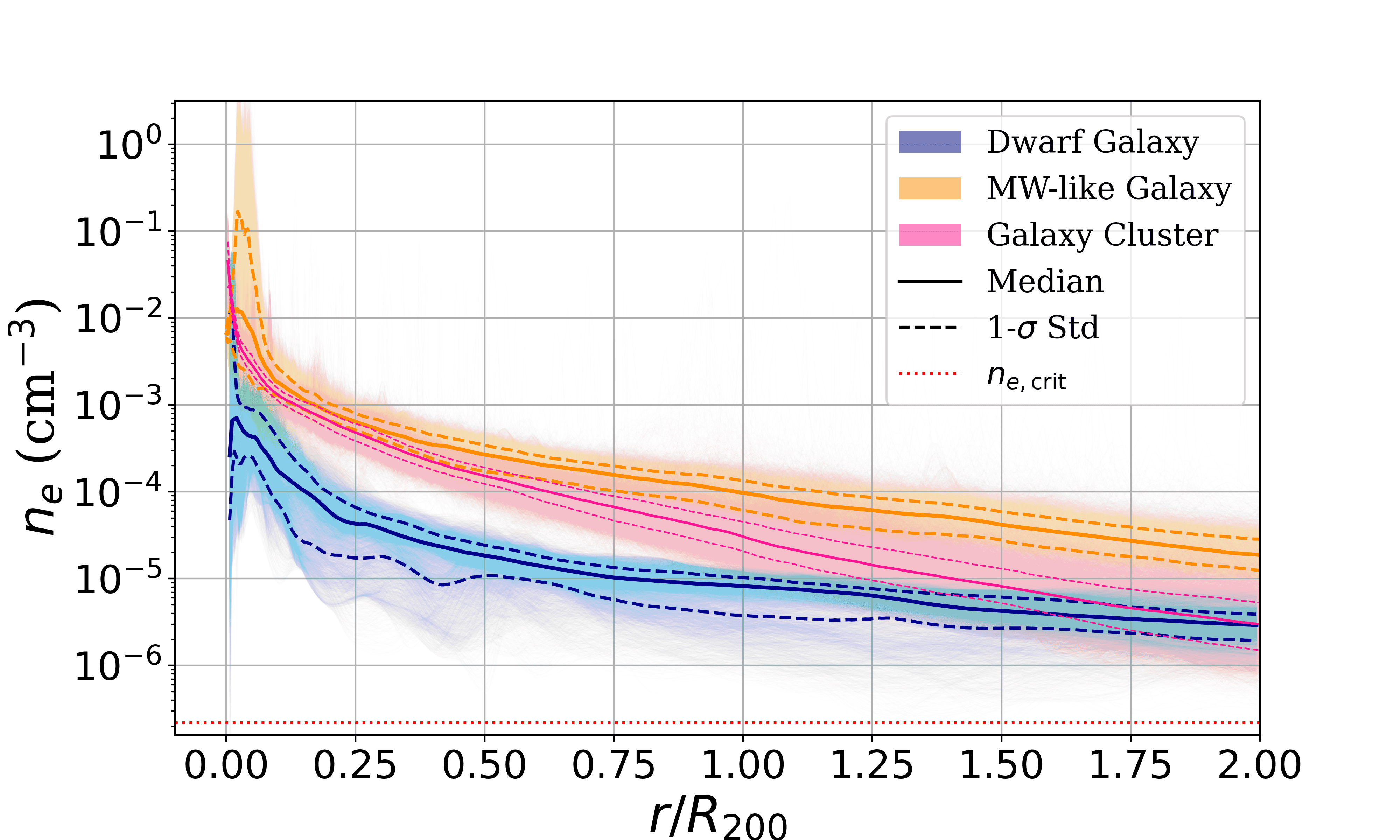} & 
        \includegraphics[width=0.48\textwidth]{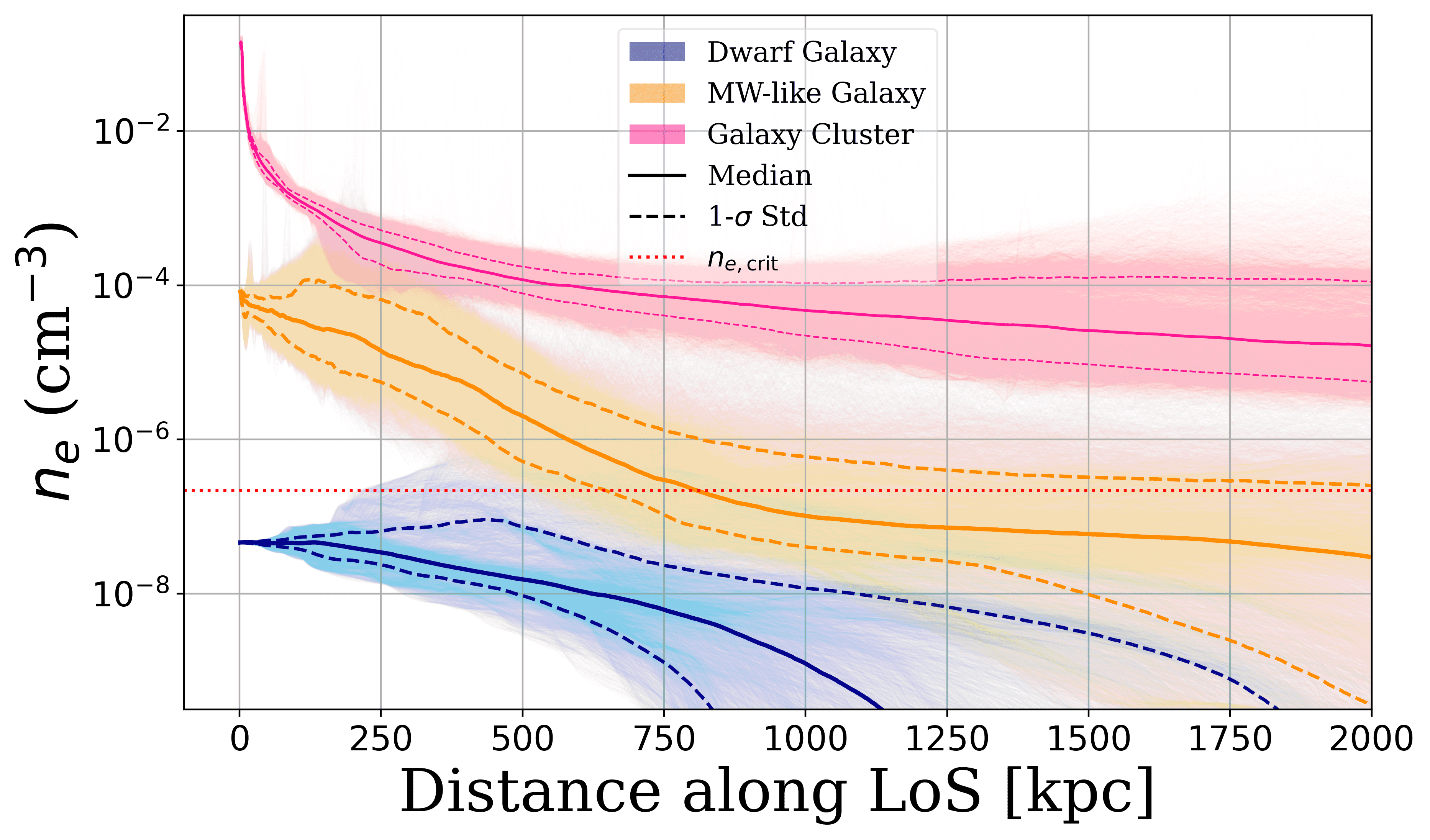} \\
        (a) Electron density profile (central) & (b) Electron density profile (off-center/satellite) \\
        \includegraphics[width=0.48\textwidth]{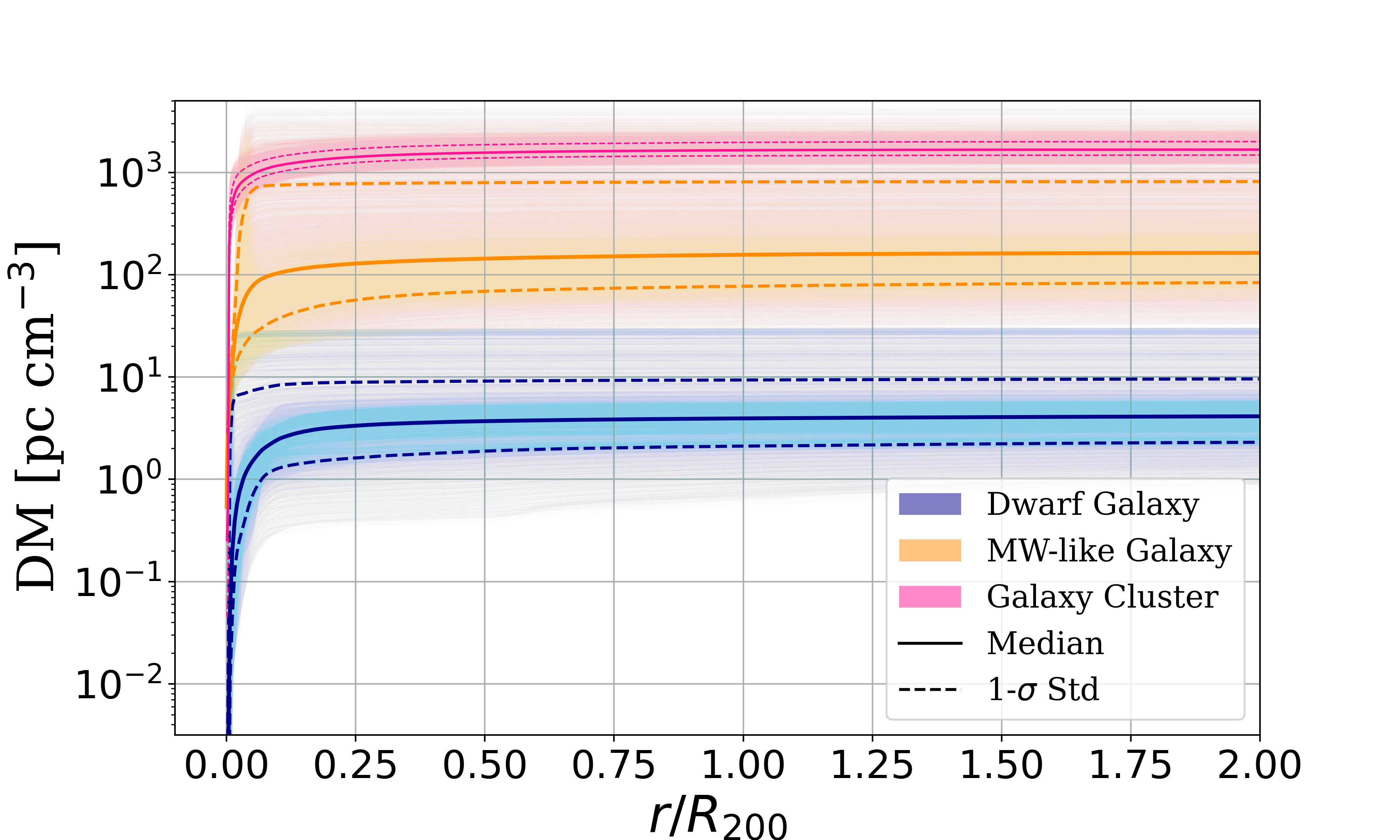} & 
        \includegraphics[width=0.48\textwidth]{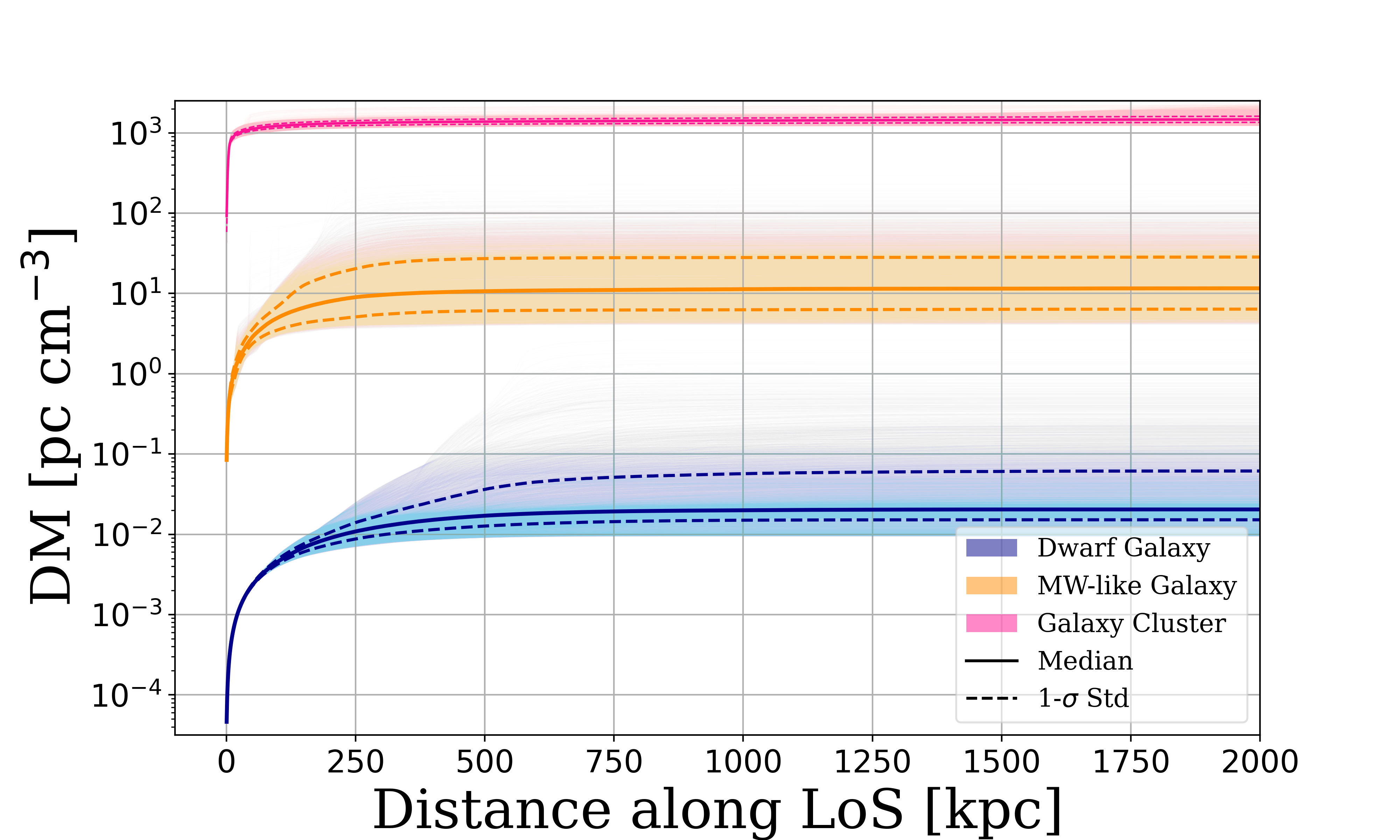} \\
        (c) DM profile (central) & (d) DM profile (off-center/satellite) \\
        \includegraphics[width=0.45\textwidth]{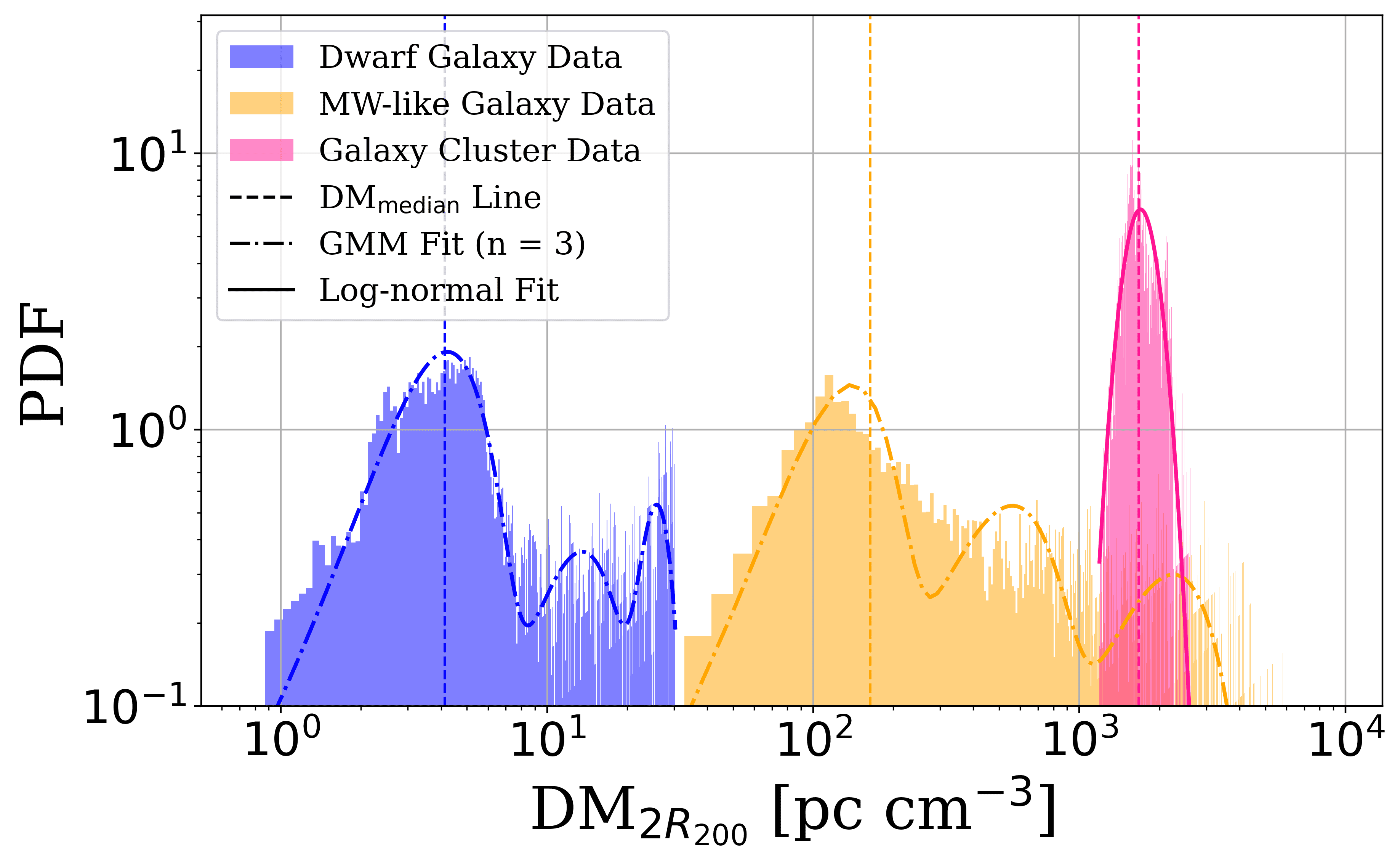} & 
        \includegraphics[width=0.45\textwidth]{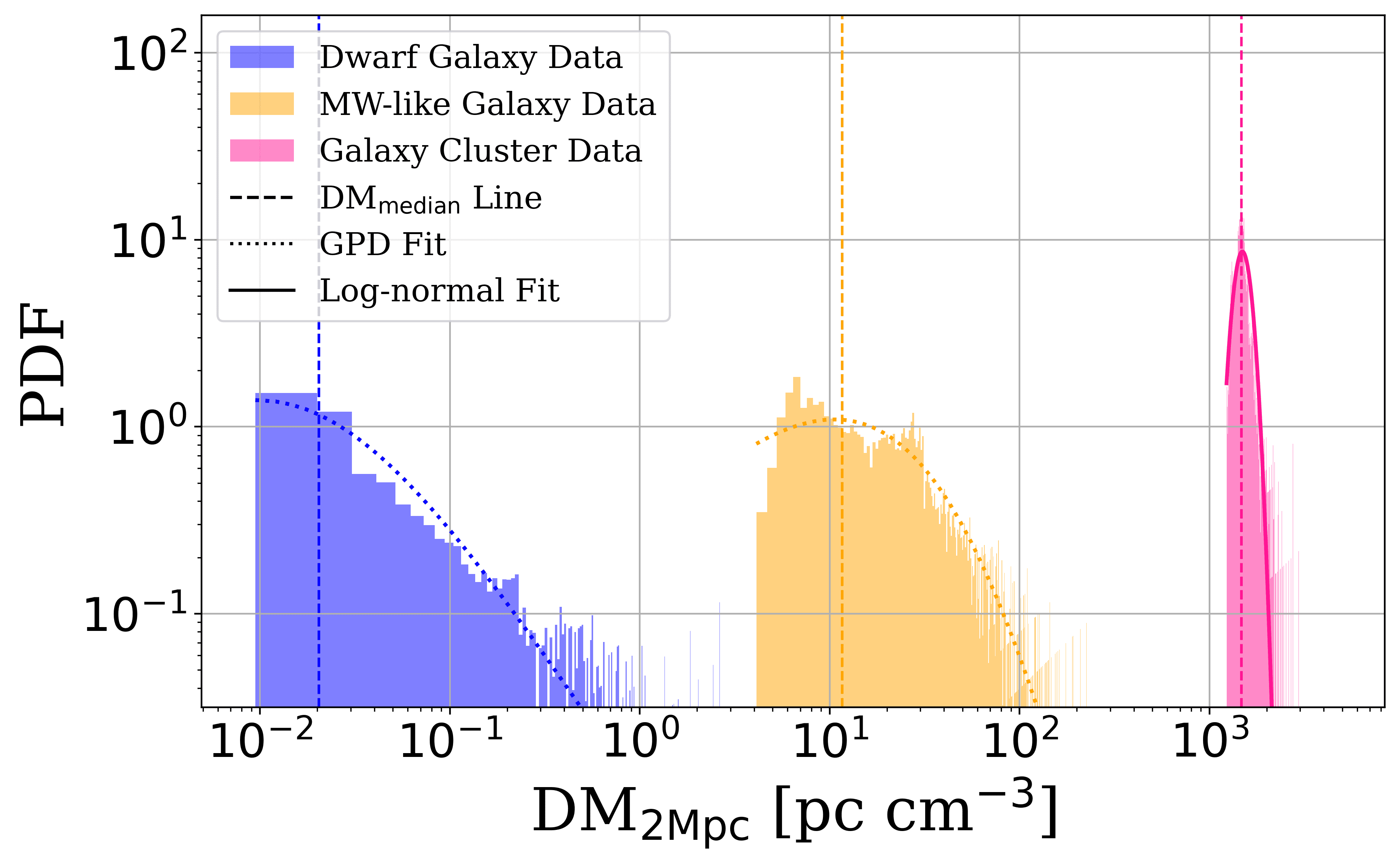} \\
        (e) Log-Normal Fit for DM distribution (center)& (f) Exponential Fit for DM distribution (off-center/satellite)\\
    \end{tabular}
    \caption{
  Consolidated analysis for different galaxy environments. The blue, orange, and pink colors represent the dwarf galaxy, MW-like galaxy, and galaxy cluster cases, respectively. We utilize 10000 sight lines, with different integration limits: for the central cases (panels~{\it a, c, e}), the integration extends only up to $2R_{200}$ for higher resolution, while for the off-center/satellite cases (panels~\textit{b, d, f}), the integration extends to 2000 kpc.  The solid lines in panels~{\it (a)} and {\it (b)}  indicate the median values, while the dashed lines represent the uncertainty range of 16th–84th percentile. The red dotted line in panels~{\it (a)} and {\it (b)} denotes the critical electron density ($n_{e,\text{critical}}$). Panels~{\it (c)} and {\it (d)} display the DM profiles along the LoS. 
   For the central cases (panel {\it e}), the DM distributions are fitted using a Gaussian Mixture Model (GMM) with 3 components $n=3$ for the dwarf and MW-like Galaxies, and a log-normal distribution for the galaxy cluster. For the off-center/satellite cases (panel \textit{f}), the dwarf and MW-like galaxies represent off-center DM values, with DM$_\text{median}$ values of $0.02 \, \mathrm{pc \, cm^{-3}}$ and $11.59 \, \mathrm{pc \, cm^{-3}}$, respectively, both fitted with a Generalized Pareto Distribution (GPD). The galaxy cluster instead represents a satellite halo near the cluster center, fitted with a skewed log-normal distribution, with a DM$_\text{median}$ of $1472.86 \, \mathrm{pc \, cm^{-3}}$. 
    }
    \label{fig:Combined-curves}
\end{figure*}

For this analysis, we select the largest subhalo near the galaxy cluster halo center as the representative satellite galaxy. This subhalo is located approximately $1.86 \, \mathrm{Mpc}$ from the cluster center. The analysis follows the same methodology as that used for spiral galaxies, as outlined in ~\ref{sec:host_MW_like}, but the off-center case is replaced by the satellite case. The results are presented using pink component in Figure~\ref{fig:Combined-curves}. 

In Figure~\ref{fig:Combined-curves}, panel~{\it (a)}) shows the electron density profile along central sight lines, while panel~{\it (b)} displays the profile for the satellite galaxy. The central case exhibits electron densities approaching $10^{-1} \, \mathrm{cm}^{-3}$ near the cluster center, gradually declining over a distance of several thousand kiloparsecs. The satellite case, while showing lower electron densities near the center, still reflects a complex density structure, including peaks indicative of substructures within the cluster. These profiles suggest that galaxy clusters possess extremely dense and complex baryonic distributions.

Panels~{\it (c)} and {\it (d)} show the DM profiles for the central and satellite cases, respectively. 
The central case demonstrates a steep increase in DM near the center, and the 16th-84th percentile DM range is distributed between 
$1477\text{-}2003 \, \mathrm{pc \, cm}^{-3}$, with a median DM value of $1670.16 \, \mathrm{pc \, cm}^{-3}$. 
The log-normal distribution (Equation~\ref{eq:lognormal}) provides a good fit to the DM distribution. The best-fit parameters for the central case are $\sigma = 0.1466^{+0.0008}_{-0.0010}$ and $\mu = 1698.05^{+2.31}_{-2.18} \, \mathrm{pc \, cm}^{-3}$.  The corresponding mean and standard deviation (Equation \ref{eq:lognorm_mean_std}) of DM are $E(\text{DM}) = 1716.40 \, \mathrm{pc \, cm}^{-3}$ and $\mathrm{Std}(\text{DM}) = 252.98 \, \mathrm{pc \, cm}^{-3}$.

For the satellite case, the DM distribution exhibits a more gradual increase, with a median value of $1472.86 \, \mathrm{pc \, cm}^{-3}$. 
The log-normal fit for this case gives $\sigma = 0.1063^{+0.0014}_{-0.0015}$ and $\mu = 1490.15^{+1.34}_{-1.50} \, \mathrm{pc \, cm}^{-3}$.  The corresponding mean and standard deviation of DM are $E(\text{DM}) = 1498.59 \, \mathrm{pc \, cm}^{-3}$ and $\mathrm{Std}(\text{DM}) = 159.75 \, \mathrm{pc \, cm}^{-3}$.

The results reveal that if a galaxy cluster serves as the host for an FRB, its DM contribution ($\mathrm{DM}_{\mathrm{HG}}$) is expected to dominate the total $\mathrm{DM}_{\mathrm{FRB}}$, potentially exceeding the contributions from all other components combined. Even for FRBs originating in satellite galaxies within the cluster, the $\mathrm{DM}_{\mathrm{HG}}$ remains significant, highlighting the importance of considering the environment of galaxy clusters in FRB studies.

\section{Discussion} \label{sec:discussion}

Our analysis highlights the potential variations in FRB DM statistics when applying our methodology to numerical simulations that incorporate different physical processes. In this section, we discuss the impact of different physical processes and modeling choices on DM estimation, as well as future directions for improving both numerical simulations and observational constraints.

\begin{itemize}
    \item \textit{AGN Feedback and DM Estimation:} 
    AGN feedback plays a crucial role in redistributing baryons between the CGM and IGM, influencing the DM observed along FRB sight lines. Strong feedback expels more gas into the IGM, reducing CGM’s DM contribution, while weaker feedback retains more baryons in the CGM, increasing its impact on FRB sight lines.
    
    Different simulations implement AGN feedback differently, leading to variations in DM statistics. For example, SIMBA employs strong kinetic jet feedback that efficiently removes gas from halos, whereas IllustrisTNG and EAGLE \citep{Schaye2015} use different combinations of thermal and kinetic feedback, resulting in distinct CGM structures. These variations directly affect the free electron column density and DM measurements. 
    
    Our CROCODILE simulations currently lack AGN jet feedback, which may contribute to the relatively small difference in DM statistics between the fiducial and NoBH runs. This limitation is consistent with the results from \citet{Medlock2025a}, who analyzed feedback energetics in CAMELS and found that different AGN feedback implementations lead to varying baryon redistribution efficiencies. In particular, kinetic feedback models tend to push gas farther into the IGM compared to thermal models, significantly affecting the fraction of CGM gas. The regulation of kinetic feedback activation (e.g., the black hole mass threshold) is a key factor controlling its impact at low redshift ($z=0$). These results suggest that future improvements to the CROCODILE simulations should incorporate kinetic AGN feedback to better capture the redistribution of gas.

    Additionally, \citet{Medlock2025b} explored the impact of $f_{\mathrm{IGM}}$ on the suppression of the matter power spectrum, showing that baryon redistribution due to feedback processes alters the clustering of matter on different scales. This effect is particularly relevant for FRB-based constraints on the baryon fraction, as it influences the statistical properties of DM fluctuations along sight lines. Future work should examine whether such suppression is observable in FRB samples and how it compares with large-scale simulations like CROCODILE.

    \item \textit{$F$-parameter and Baryon Spread:}
    The $F$-parameter provides a statistical characterization of feedback strength based on the scatter in the Macquart relation, following the formalism introduced in \cite{McQuinn2014} and later developed in \cite{Macquart2020Nature}. The parameter is derived from the fractional standard deviation of the DM ($\sigma_{\text{DM}}$) and is given by:
    \begin{equation}
    \sigma_{\text{DM}}(\Delta) = F z^{-1/2}
    \end{equation}
    where $\sigma_{\text{DM}}$ represents the standard deviation of the DM at a given redshift $z$, and $F$ quantifies the amount of feedback-driven baryon redistribution. A lower $F$ value corresponds to stronger feedback (under fixed $\Omega_m$ and $\sigma_8$), where more baryons are expelled into the IGM, leading to reduced DM scatter. Conversely, higher $F$ values indicate weaker feedback, where baryons remain in CGM, resulting in increased DM scatter along FRB sight lines.
    
   \citet{Medlock2024, Medlock2025a} have used CAMELS simulations to study how different feedback implementations affect $F$. Their results suggest that SIMBA’s kinetic feedback produces lower $F$ values, indicating stronger baryon redistribution, while IllustrisTNG and Astrid exhibit weaker feedback with higher $F$. They also highlight that box size limitations constrain their analysis, as small volumes may not fully capture the impact of large-scale structures on the DM distribution, similar to the findings of \citet{Batten2020}, who analyzed box size effects using the EAGLE simulation.

    Moreover, observational constraints on $F$ remain limited. The recent analysis by \citet{Baptista2024} provided an empirical lower limit on $F$ using localized FRBs. Comparing these results with simulations can offer insights into the dominant feedback mechanisms shaping the IGM. The relationship between $F$ and baryon spread, as shown in Figure~1 of \citet{Medlock2025b}, underscores the need for refined models incorporating AGN-driven gas expulsion.
    
    \item \textit{Impact of Initial Conditions.} 
    Variations in cosmological parameters, such as $\sigma_8$ and $\Omega_m$, influence LSS formation and thus affect CGM and IGM gas distribution \citep{Khrykin2024MNRAS.529..537K, KHRYKIN2024} (the $F$-parameter is also affected). These differences may introduce systematic shifts in DM measurements.
    
    Additionally, stochastic initial perturbations cause sight-line-to-sight-line variations in DM, leading to scatter in statistical results. The formation history of galaxies, including merger rates and cooling timescales, also shapes CGM structure, further influencing FRB DM contributions. Future simulations with refined treatments of these factors will improve the accuracy of DM-based baryon fraction estimates.

    \item \textit{Numerical Methodology and Computational Effects.} 
    Numerical methods contribute to variations in DM statistics. Smoothed Particle Hydrodynamics (SPH) produces smoother CGM distributions, especially in low-density regions, which may lead to underestimating  DM.  
    Similarly, differences in radiative transfer models, ultraviolet background intensity, and AGN radiation feedback can impact CGM ionization and alter DM estimates. While these computational effects are relevant, our focus remains on the physical processes governing baryon distribution.
    
    \item \textit{Bridging Simulations and Observations.} 
    Future improvements in both simulations and observations will be crucial for refining DM-based baryon fraction estimates. Larger simulation volumes, such as TNG-300 and CAMELS, will allow for a more comprehensive statistical analysis and improve our understanding of DM variability. Higher resolution and more refined galaxy formation models can enhance halo density profile calculations, leading to better estimates of foreground galaxy contributions.

    From an observational perspective, upcoming large-scale FRB surveys, such as DSA-2000 \citep{Hallinan2019} and BURSTT \citep{Lin2022}, will provide significantly larger FRB samples, enabling precise constraints on CGM’s DM contribution. Additionally, deep spectroscopic surveys, such as those conducted with the Prime Focus Spectrograph on the Subaru Telescope \citep{PFS_reference}, will help characterize foreground galaxies along FRB sight lines. These observations will constrain CGM gas distribution, ionization state, and metallicity, improving DM modeling.

    The synergy between simulations and observations is essential for resolving the missing Baryon Problem. Simulations provide physically motivated models for baryon distribution, while new observations offer critical constraints to refine them. As both approaches progress, their interplay will continue to enhance our understanding of baryon distribution in the Universe.
\end{itemize}

\section{Summary}\label{sec:summary}
	
We have conducted a comprehensive investigation into the cosmic baryon distribution using FRBs within the framework of the CROCODILE simulations. These simulations, performed with the {\small GADGET3/4-OSAKA} SPH code, incorporate star formation, SN feedback, and AGN feedback to model LSSs in the Universe.

By generating light cones from our simulations, we computed gas density profiles and DMs along FRB sight lines. Our findings indicate that AGN feedback plays a crucial role in modulating baryon distribution, reducing the central gas densities in halos and reshaping the transition between the CGM and IGM. The impact of AGN feedback is particularly pronounced in halos within the mass range of $10^{12.5} - 10^{13.5} M_{\odot}$, where it efficiently redistributes gas and central gas density at $15\,\mathrm{kpc}$ is reduced by $86.1\%$.

We derived the DM--$z$ relation from our light-cone data and constrained the diffuse baryon mass fraction at $z = 1$ to be $f_{\mathrm{diff}} = 0.865^{+0.101}_{-0.165}$ for our fiducial model and $f_{\mathrm{diff}} = 0.856^{+0.101}_{-0.162}$ for the NoBH model. These results align with observational constraints derived from FRB data and confirm the robustness of the Macquart relation when halo-associated gas contributions are properly accounted for.
It also provides a nice cross-check of the $\Lambda$CDM model. 

We quantified the values of $f_{\mathrm{diff}}$ and $f_{\mathrm{IGM}}$ as a function of redshift using different cutoff radius for the halo gas (CGM). We provided fitting results to these results, which might be useful for semianalytic models of galaxy formation and IGM.  

To further explore the role of foreground halos in shaping FRB DMs, we examined the density profiles of dark matter halos and their contributions to DM along different impact parameters. Our results confirm that AGN feedback significantly lowers the DM contributions from the central regions of halos while increasing contributions at larger radii. Specifically, in the $10^{12.5}-10^{13.5} M_{\odot}$ mass range, AGN feedback suppresses the median central DM contribution by approximately 73\% compared to the NoBH case, highlighting its strong impact on baryon redistribution. The modified NFW profile provides a better fit to our simulated density distributions than the standard NFW profile, capturing the effects of AGN-driven gas expulsion more accurately.

Finally, we investigated the DM contributions from FRB host galaxies using higher-resolution zoom-in cosmological hydrodynamic simulations of dwarf galaxies, MW-like galaxies, and galaxy clusters. Our analysis reveals that the host DM varies significantly depending on the galaxy type and FRB location. For example, in dwarf galaxies, the central DM is typically below $100\,\text{pc cm}^{-3}$, whereas in galaxy clusters, the central DM can exceed $1300\,\text{pc cm}^{-3}$, dominating the total FRB DM budget.

Our study provides a framework for understanding the cosmic baryon distribution and the role of AGN feedback in redistributing gas across cosmic structures. By leveraging FRBs as probes of the IGM and CGM, we demonstrate the power of combining simulations with observational data to refine our understanding of baryonic physics in the universe. Future observational advancements in FRB surveys, coupled with high-resolution simulations, will further enhance our ability to constrain baryon fraction evolution and feedback processes.

\section*{acknowledgments}

We are grateful to Volker Springel for providing the original version of GADGET-3/4, on which the GADGET3/4-OSAKA codes are based. Our numerical simulations and analyses were carried out on the XC50 systems at the Center for Computational Astrophysics (CfCA) of the National Astronomical Observatory of Japan (NAOJ), and SQUID at the D3 Center, The University of Osaka, as part of the HPCI system Research Project (hp230089, hp240141, hp250119). Z.J.Z. appreciates the helpful assistance and constructive discussions from Nicolas Ledos, Ke Xu, Daisuke Toyouchi, Zijian Zhang, Qing Wu, and Pingzheng Chen. This work is supported by the JSPS KAKENHI grant No. 19H05810, 20H00180, 24H00002, 24H00241 (K.N.), and the JSPS International Leading Research (ILR) project, JP22K21349. 
K.N. acknowledges support from the Kavli IPMU, the World Premier Research Center Initiative (WPI), UTIAS, the University of Tokyo.

\section*{Data Availability}

The data presented in this paper is available to the research community upon request to the authors. 

\clearpage	
\newpage
\bibliography{ms}{}
\bibliographystyle{aasjournal}

\newpage

\appendix
\section{Overview of LSS Simulations}
\label{App:LSS_overview}
\begin{figure*}[ht]
\centering
\includegraphics[width=0.45\textwidth]{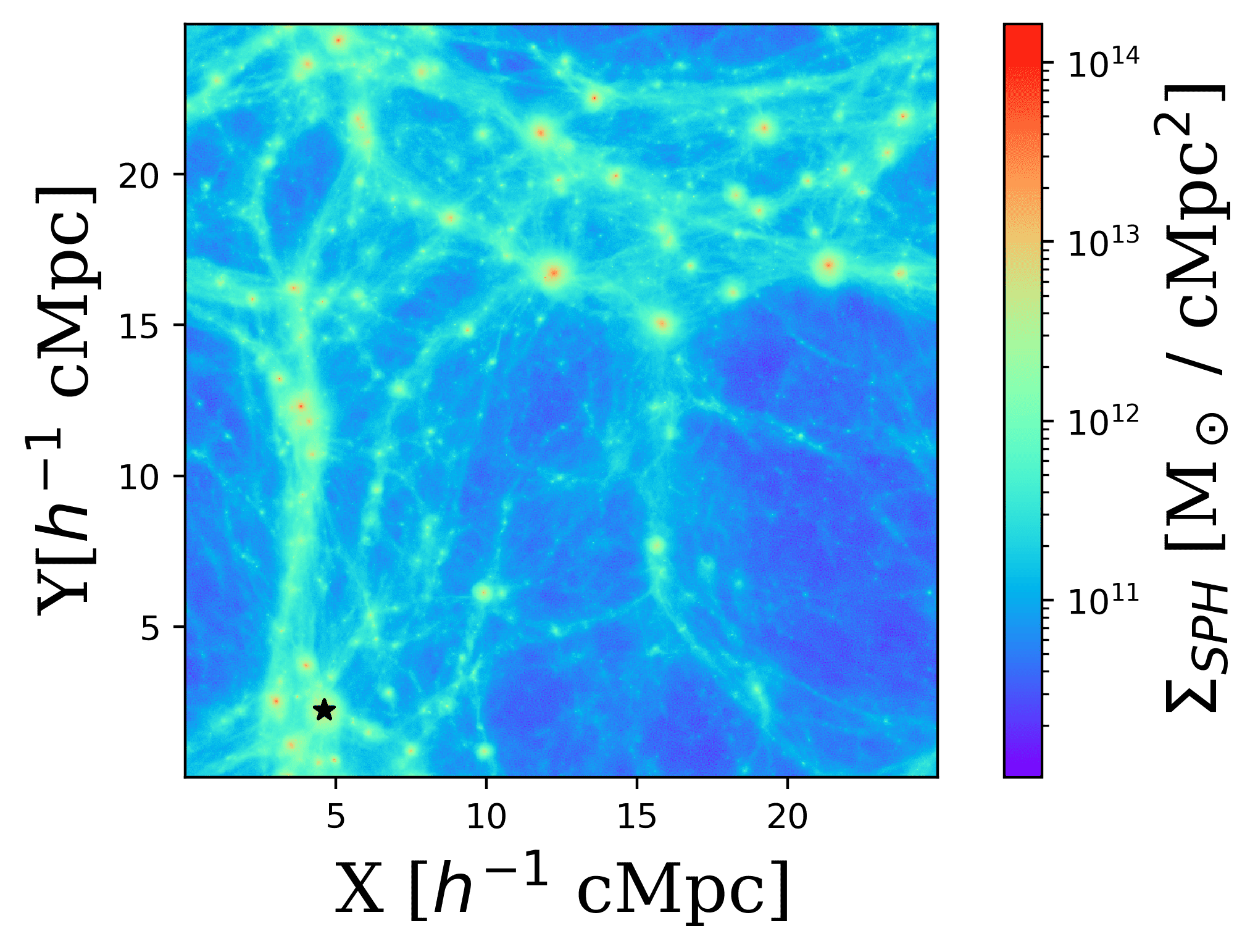}
\includegraphics[width=0.45\textwidth]{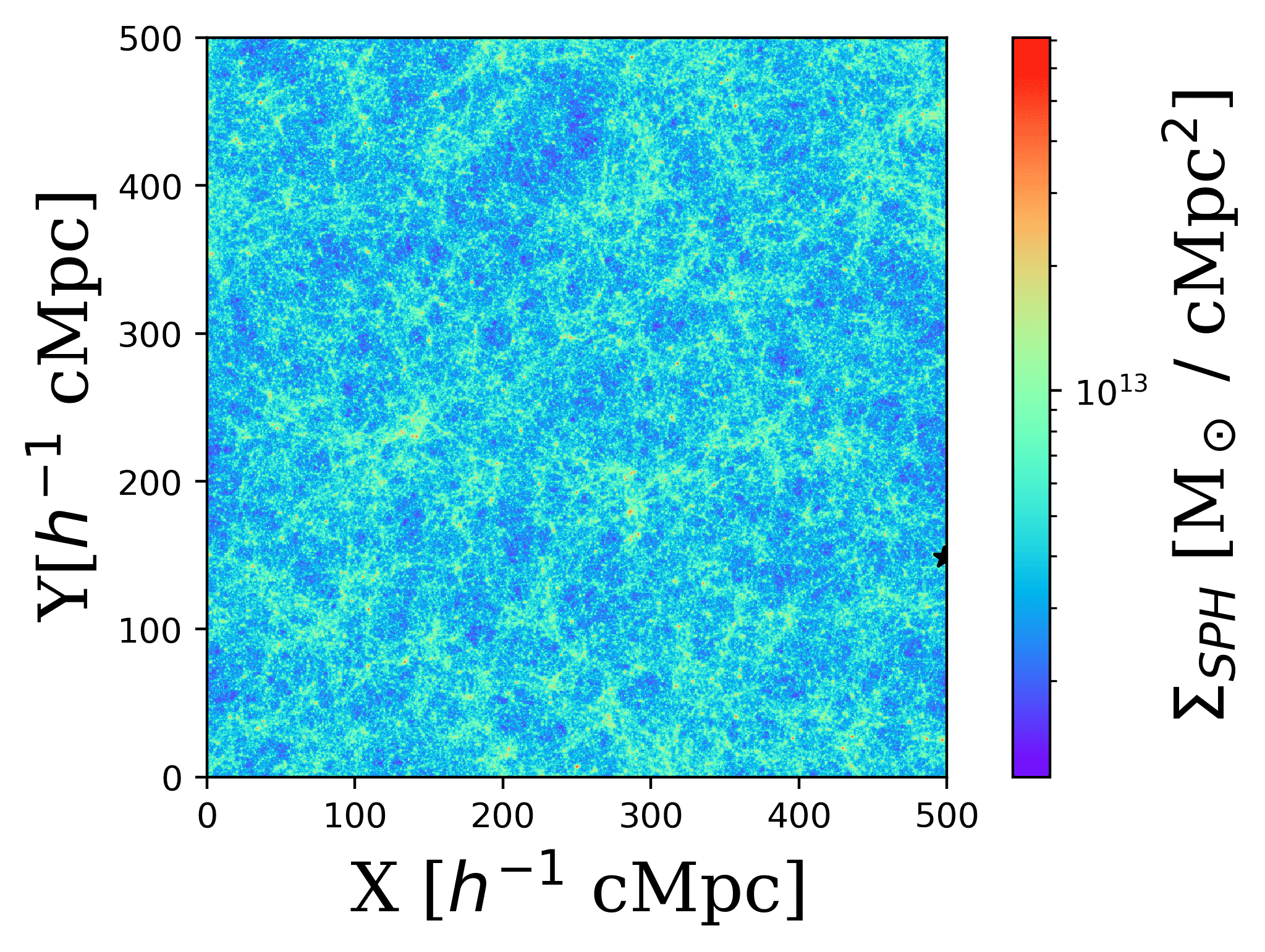}
\includegraphics[width=0.45\textwidth]{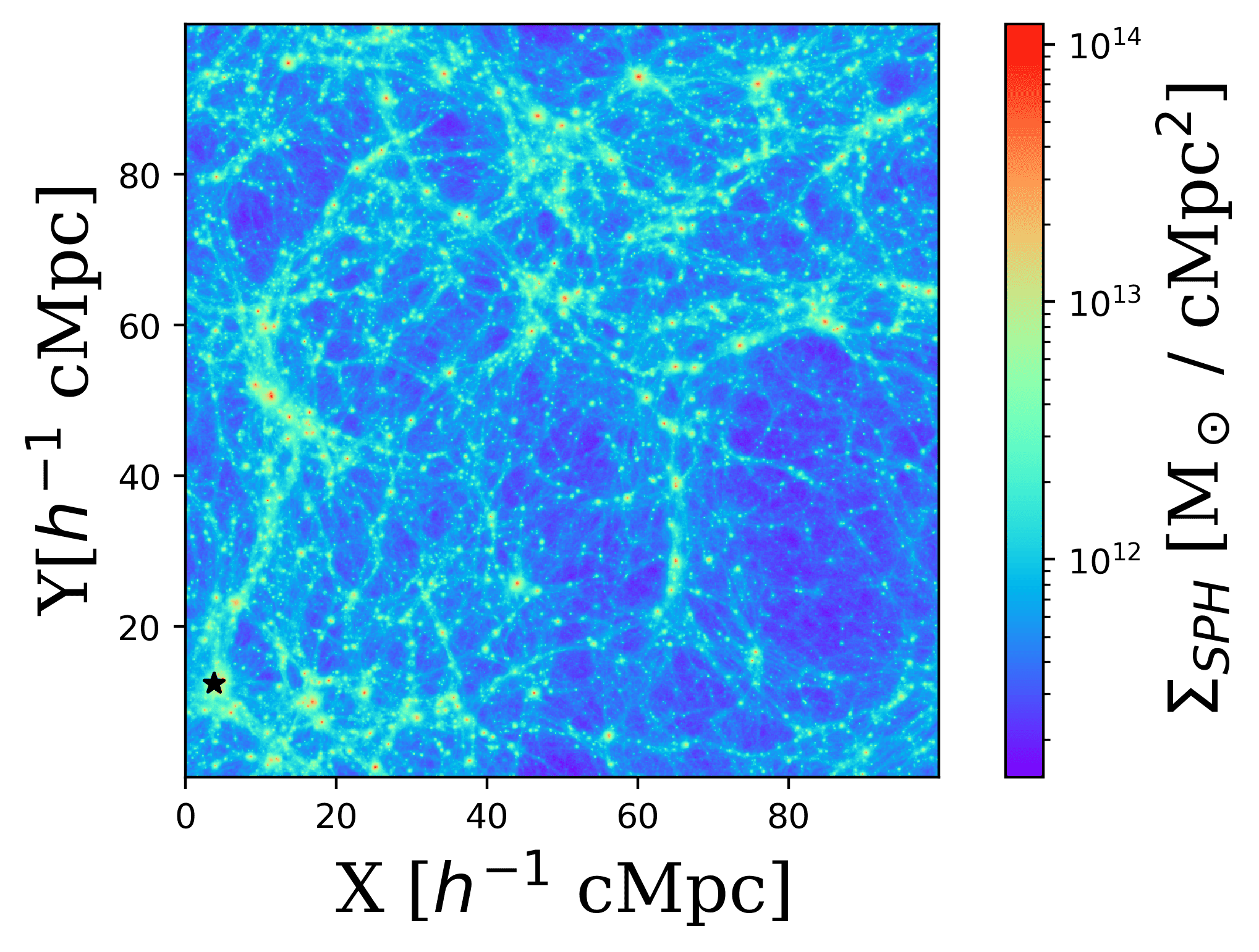}
\includegraphics[width=0.45\textwidth]{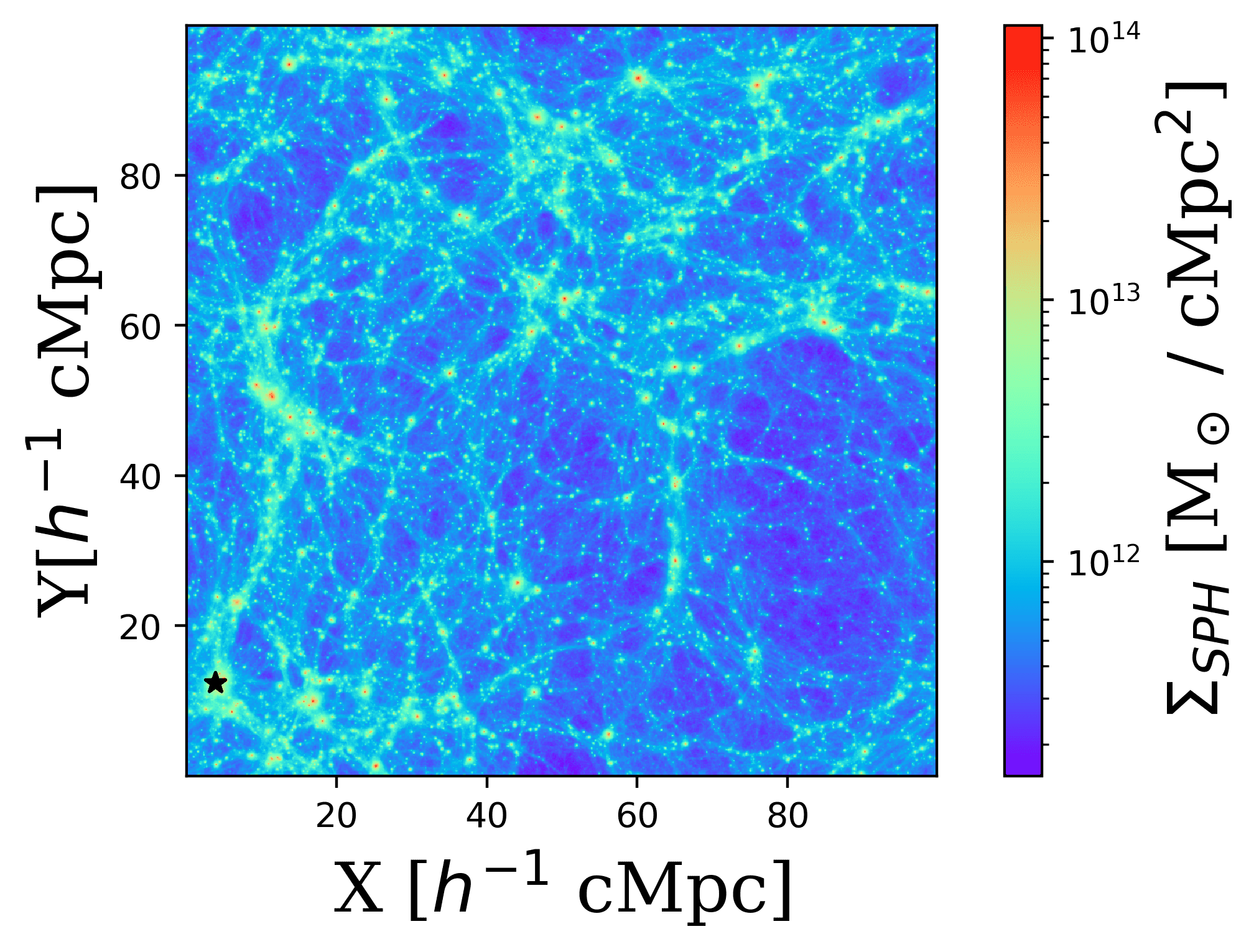}
\includegraphics[width=0.45\textwidth]{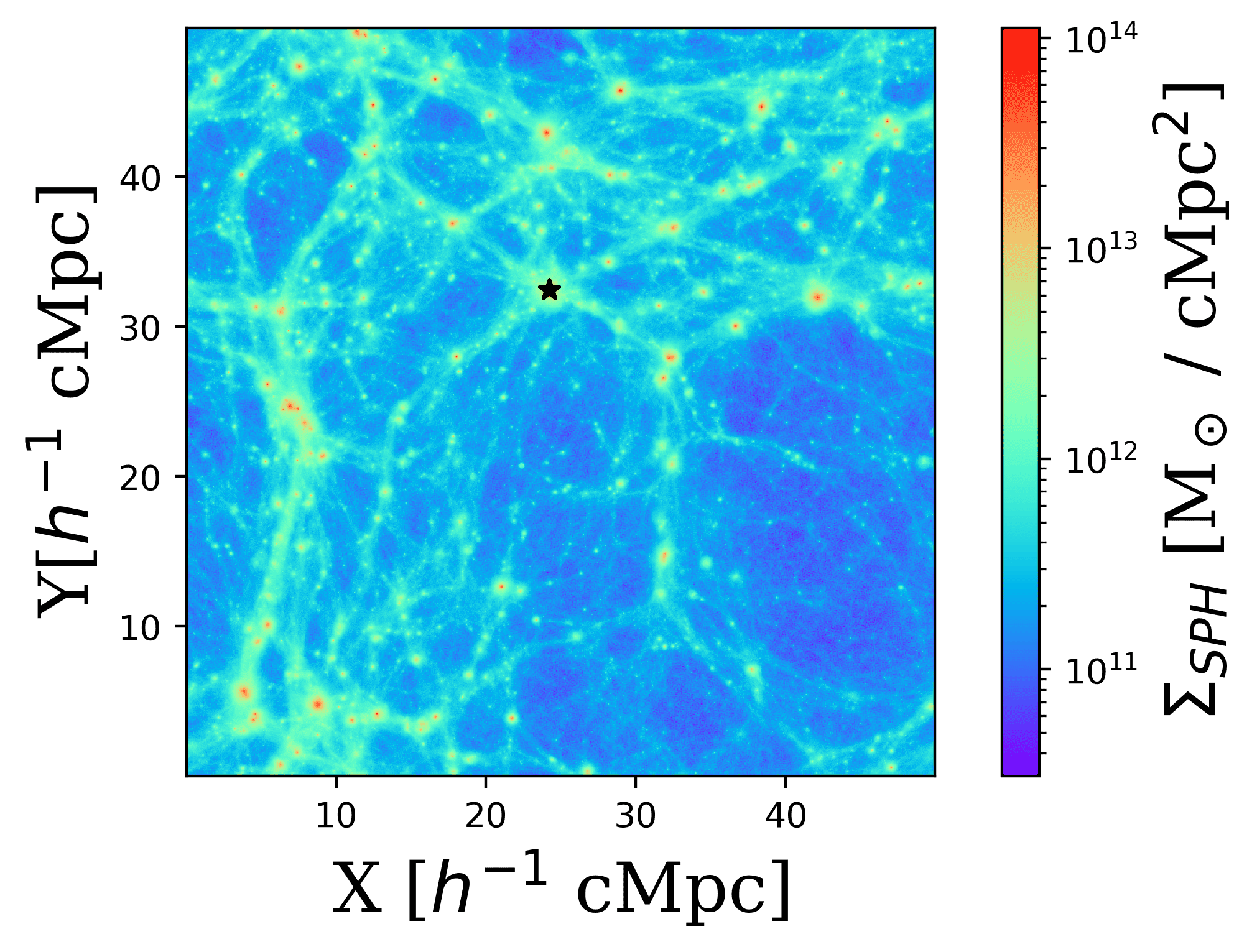}
\includegraphics[width=0.45\textwidth]{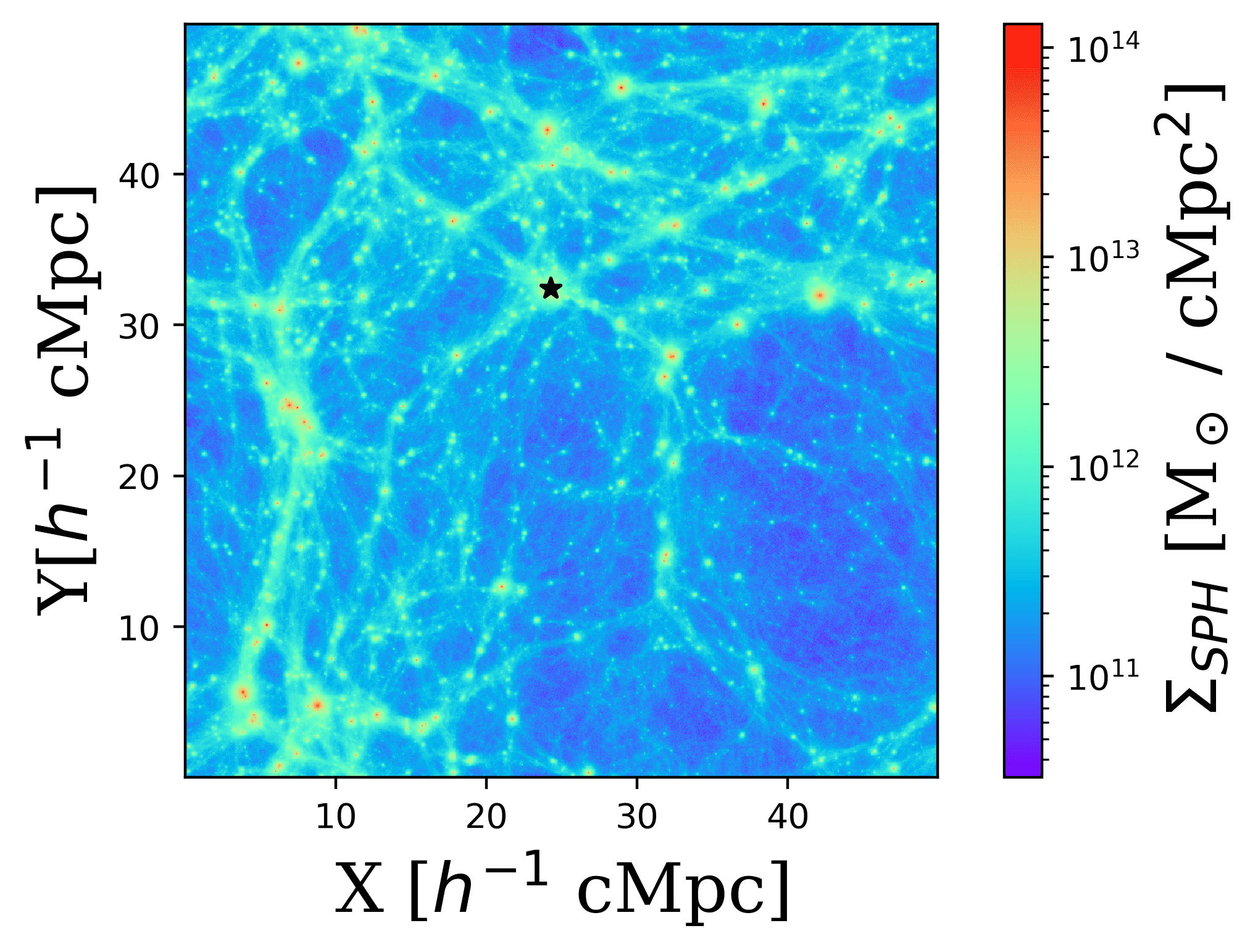}
\caption{Projected gas mass density distributions for different simulation box sizes: L25N512 (top left), L500N1024 (top right), L100N1024$_{\rm fiducial}$ (middle left), L100N1024$_{\rm NoBH}$ (middle right), L50N512$_{\rm fiducial}$ (bottom left) and L50N512$_{\rm NoBH}$ (bottom right). The black stars indicate the locations of the most massive halos.}
\label{fig:SPHCD_combined}
\end{figure*}
\begin{figure*}[ht]
\centering
\includegraphics[width=0.49\textwidth]{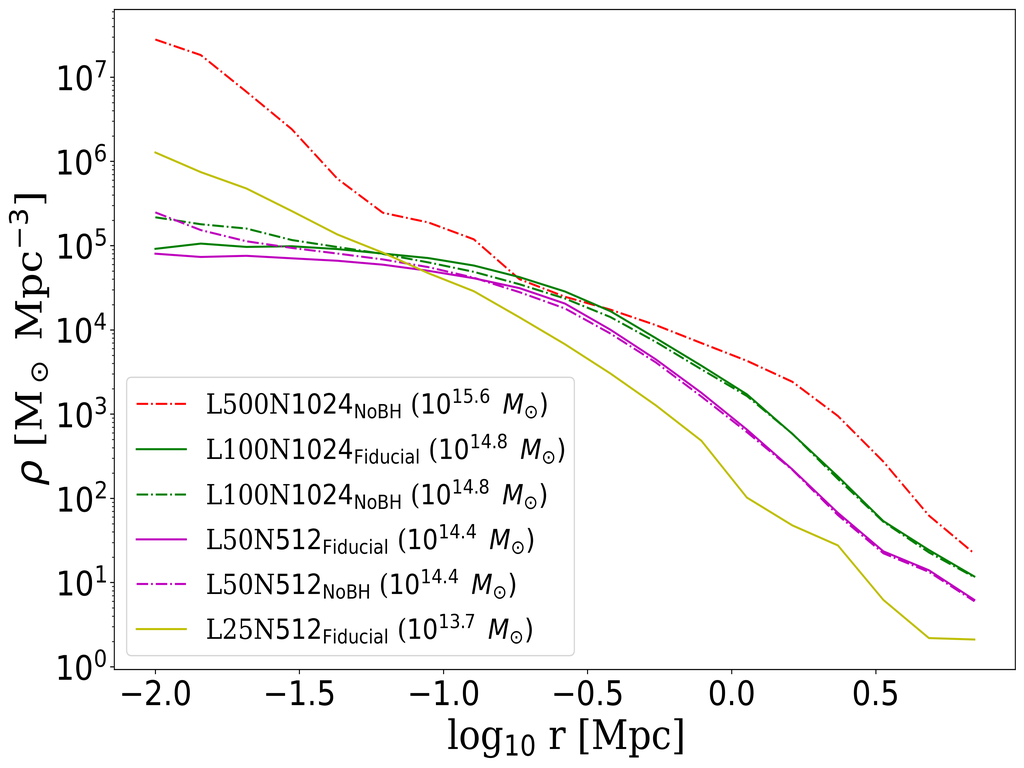}
\includegraphics[width=0.49\textwidth]{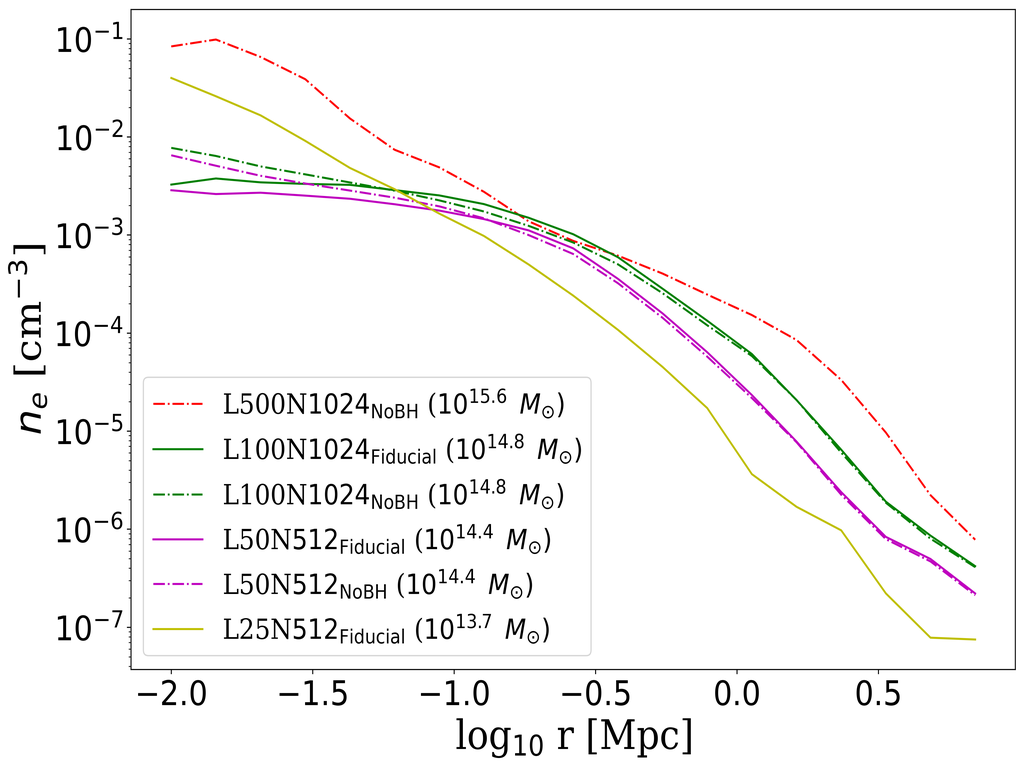}
\caption{Radial profiles of gas mass density (left) and electron number density (right) for the most massive halos in simulations with different box sizes, as summarized in Table~\ref{tab:simulation}.} 
\label{fig:shell}
\end{figure*}

Figure~\ref{fig:SPHCD_combined} shows the projected gas mass density distributions for different simulations.  The L50N512 and L100N512 simulations include runs both with and without AGN feedback. In the present paper, we only consider the thermal AGN feedback without the jet (radio) mode.
We plot the radial profiles of mass density and electron number density of the most massive halos in Figure~\ref{fig:shell}.  The simulation without AGN feedback (L50/L100 NoBH) shows a significantly higher central density, indicating that in the absence of AGN feedback, there is no mechanism to drive material outward, resulting in a more concentrated mass distribution at the center. 
In contrast, the simulation with AGN feedback (L50/L100 fiducial) has a lower central density but higher densities in the outer regions. This suggests that AGN feedback is effective in expelling material outwards into the CGM and even into the IGM, leading to a more extended distribution of mass and electron density.

\begin{figure}[ht]
\centering
\includegraphics[width=0.7\textwidth]{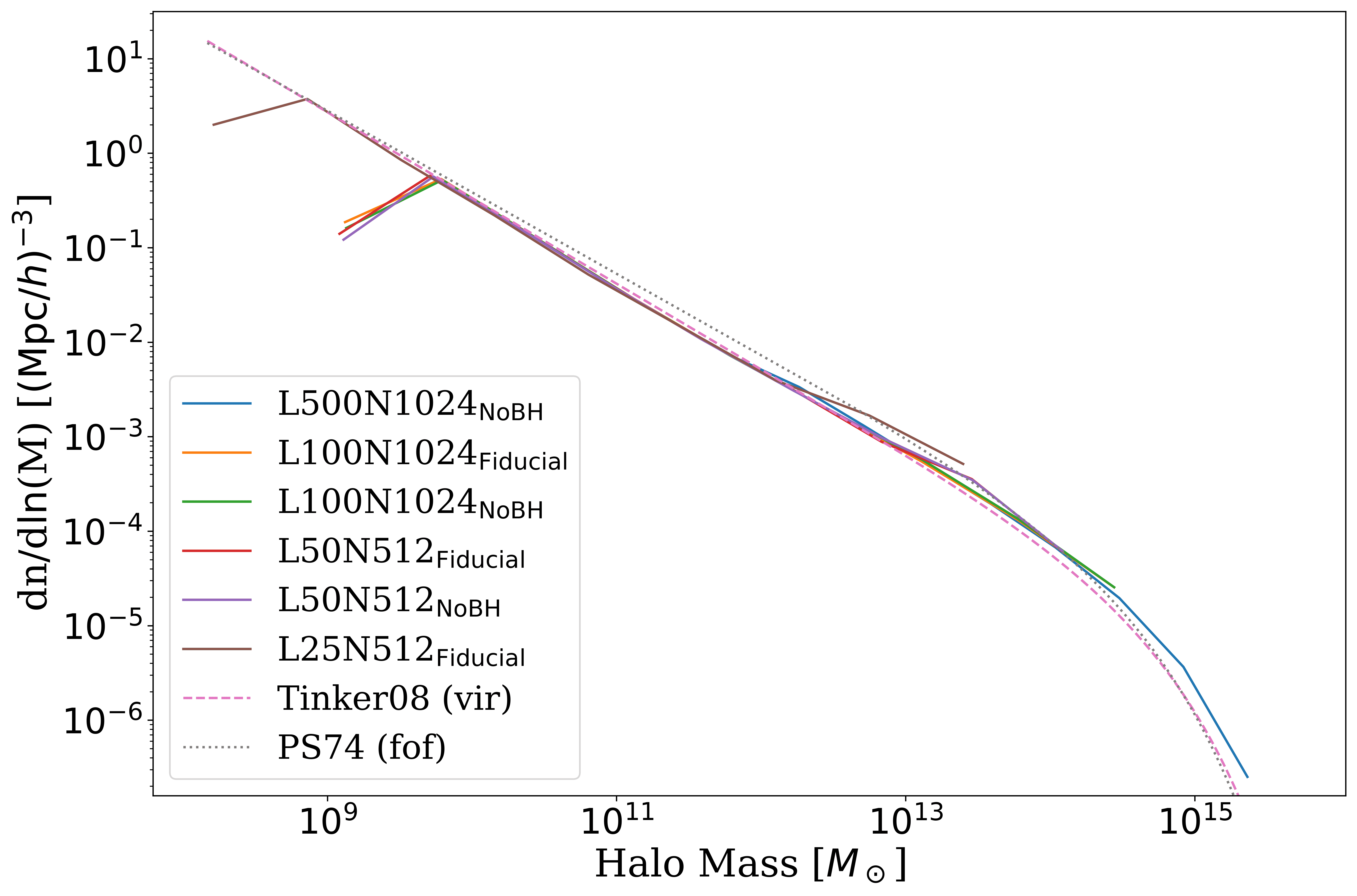}
\caption{HMF for different simulations. The solid lines represent the simulation data, while the dashed and dotted lines indicate the theoretical models from \citet{Tinker2008} and \citet{Press1974}.}
\label{fig:HMF}
\end{figure}

We also perform statistical analysis on all the halos in each simulation data to examine the HMF. 
Figure \ref{fig:HMF} shows the HMF for simulations with different box sizes, comparing the results with theoretical models from \citet{Press1974, Tinker2008}.

The results agree well with the theoretical HMFs, indicating accurate halo formation processes. Notably, the L25 simulation shows a larger number of low-mass halos at $M < 10^{9.5}\,M_{\odot}$ due to higher resolution than in other runs. 
It also results in the absence of halos above $10^{14}\,M_{\odot}$ in the L25 run due to its small box size. 
In contrast, in larger simulation volumes, the effects of AGN feedback are more likely to be averaged out across a larger number of halos, leading to a more balanced distribution of halo masses.
For the L50/L100 runs, differences in the HMF between simulations with and without AGN feedback are relatively small. 


\begin{figure*}[htbp]
    \centering
    \begin{minipage}[b]{0.45\textwidth}
        \includegraphics[width=\textwidth]{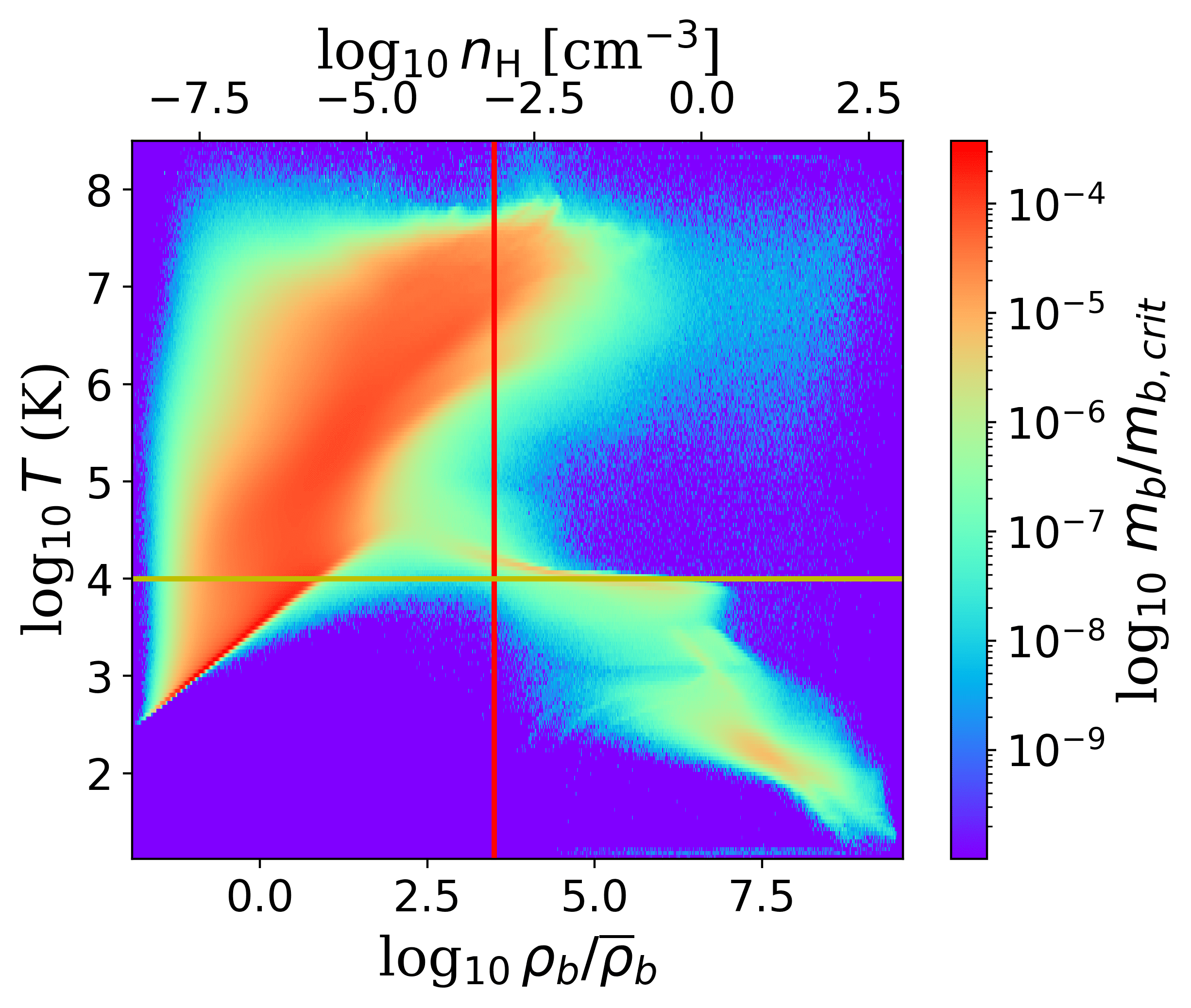}
        \label{fig:rho_T_fiducial}
    \end{minipage}
    \begin{minipage}[b]{0.45\textwidth}
        \includegraphics[width=\textwidth]{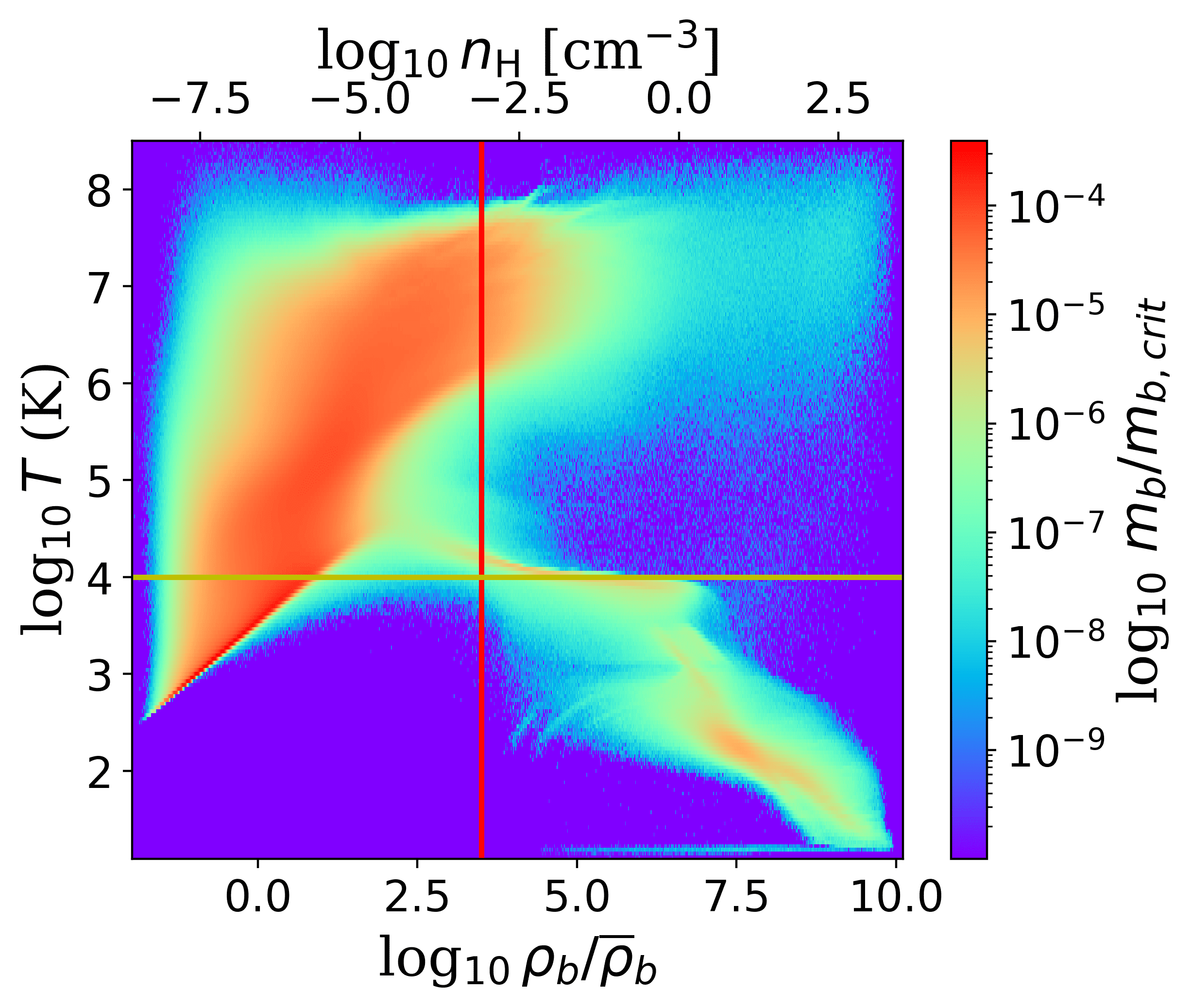}
        \label{fig:rho_T_noAGN}
    \end{minipage}

    \vspace{-0.4cm} 
    
    \begin{minipage}[b]{0.45\textwidth}
        \includegraphics[width=\textwidth]{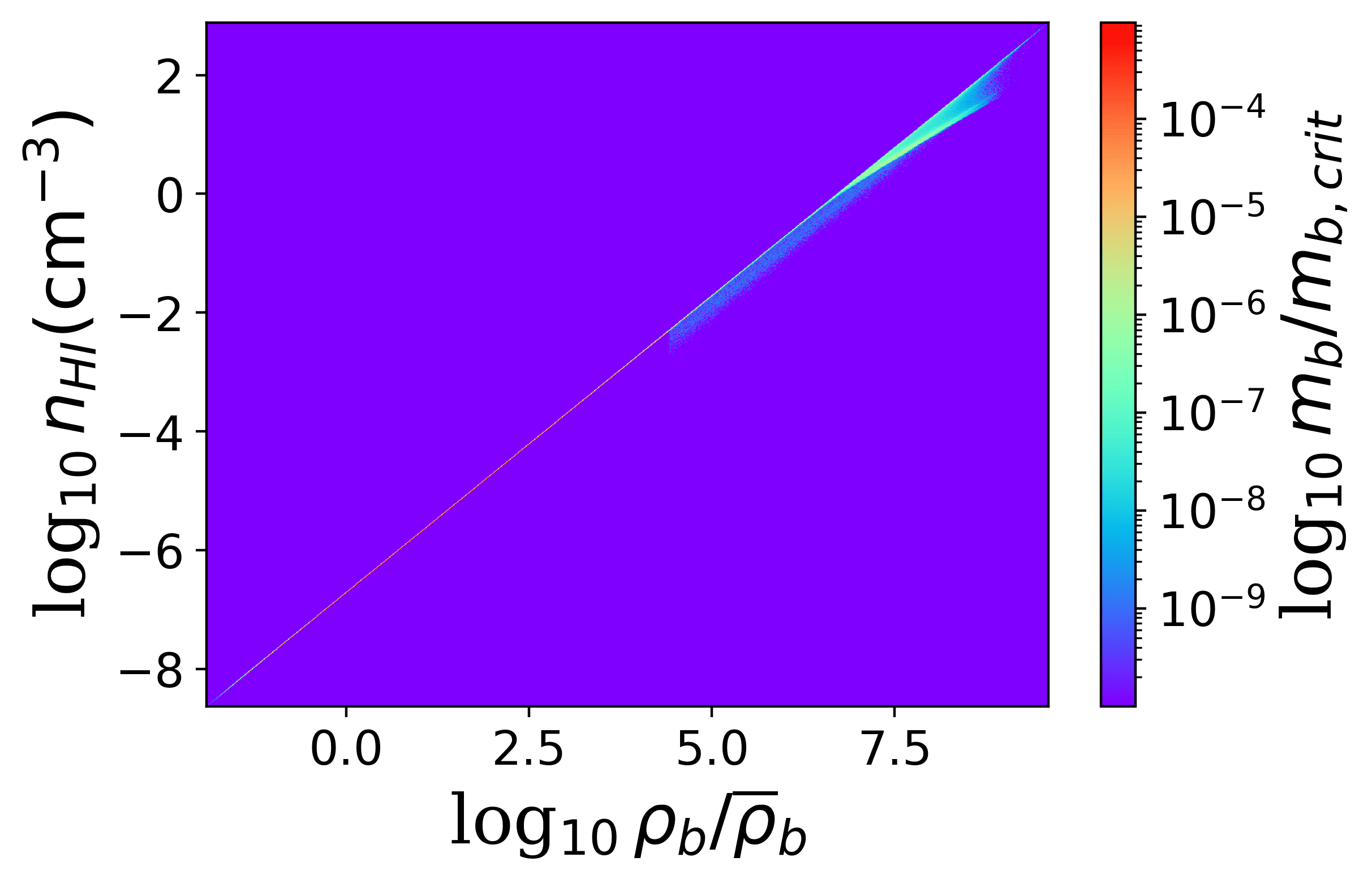}
        \label{fig:rho_nH_fiducial}
    \end{minipage}
    \begin{minipage}[b]{0.45\textwidth}
        \includegraphics[width=\textwidth]{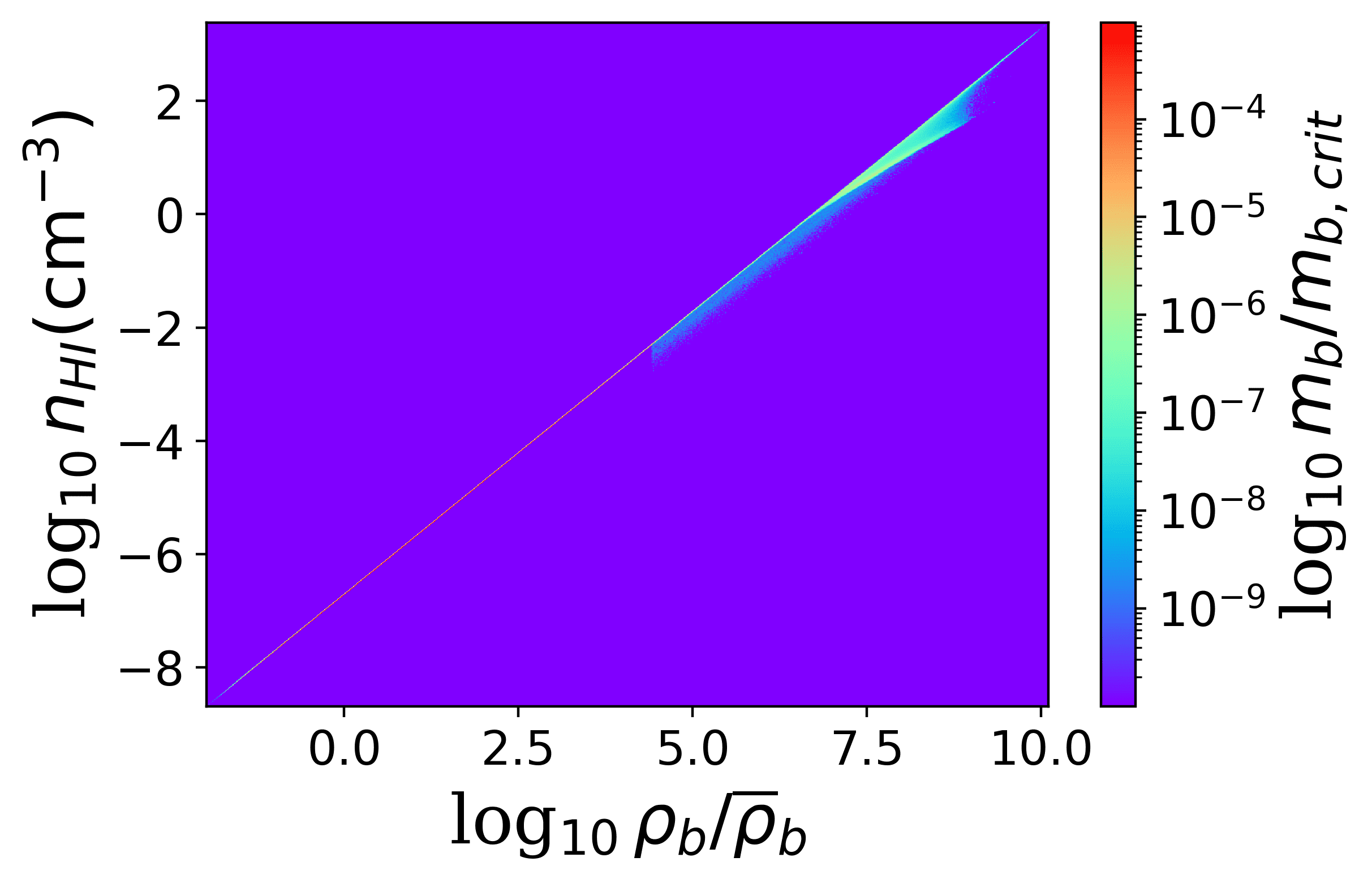}
        \label{fig:rho_nH_noAGN}
    \end{minipage}

    \vspace{-0.4cm} 
    
    \begin{minipage}[b]{0.45\textwidth}
        \includegraphics[width=\textwidth]{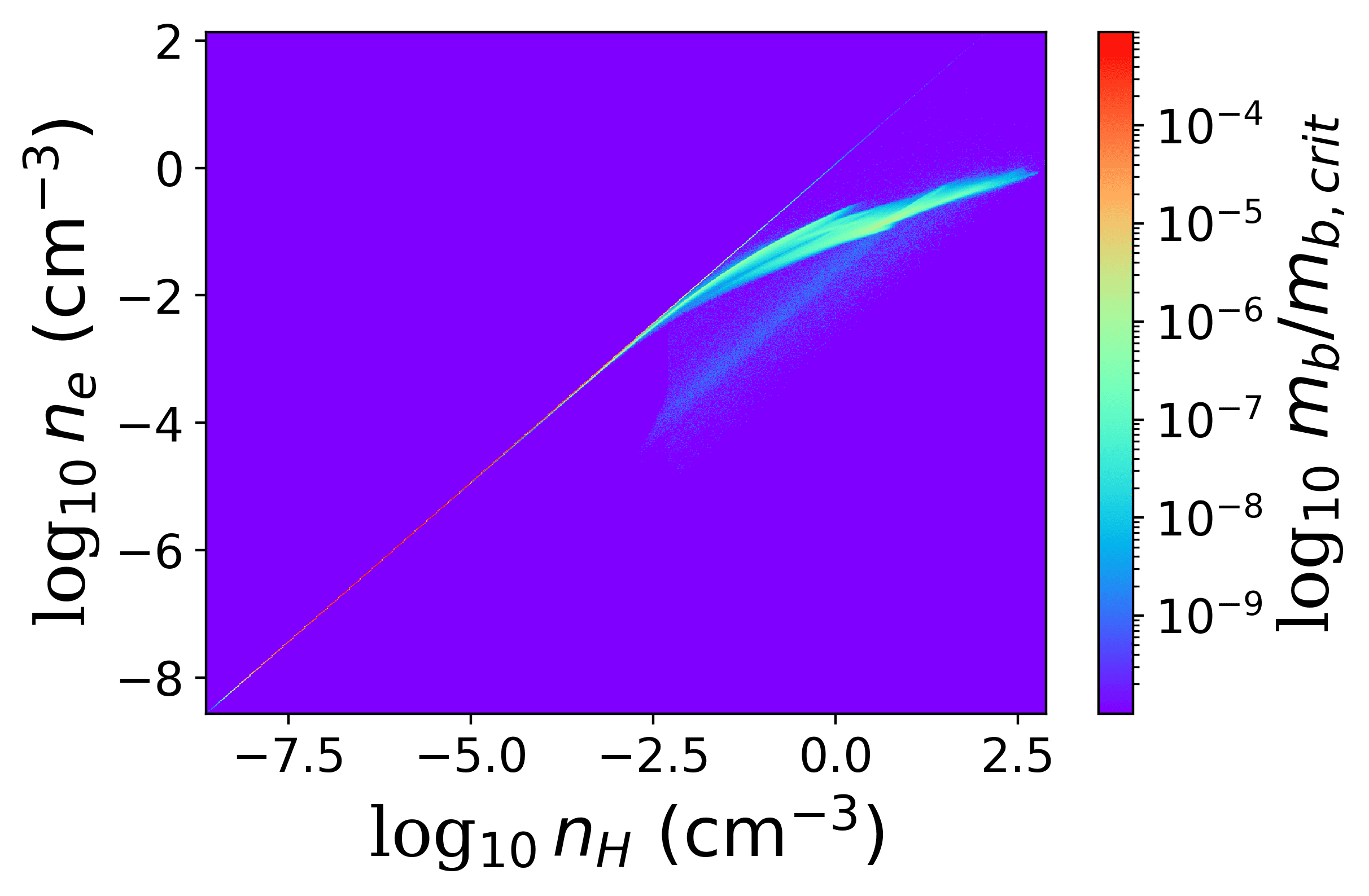}
        \label{fig:nH_ne_fiducial}
    \end{minipage}
    \begin{minipage}[b]{0.45\textwidth}
        \includegraphics[width=\textwidth]{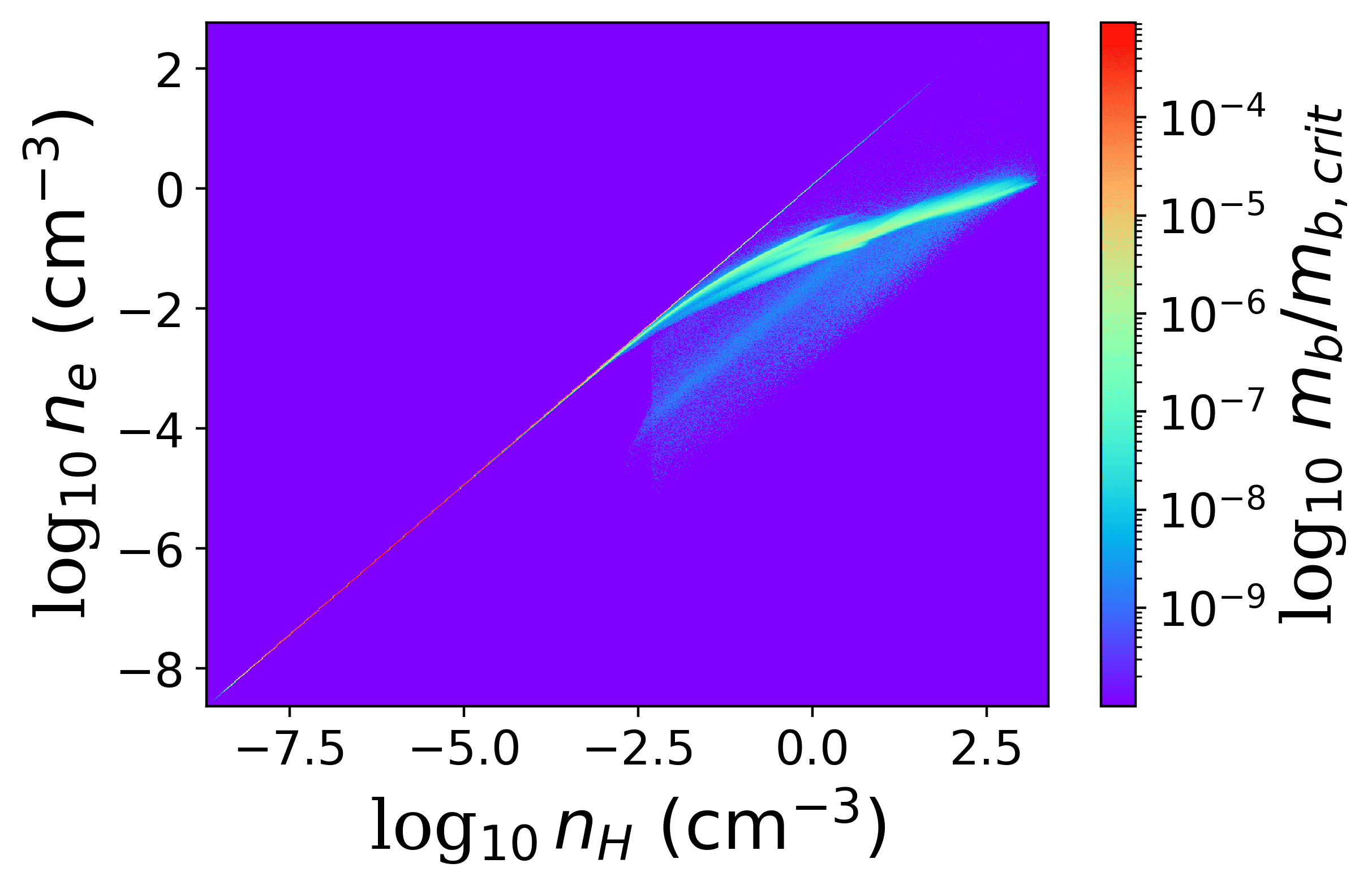}
        \label{fig:nH_ne_noAGN}
    \end{minipage}
    
    \vspace{-0.4cm} 
    \begin{minipage}[b]{0.45\textwidth}
        \includegraphics[width=\textwidth]{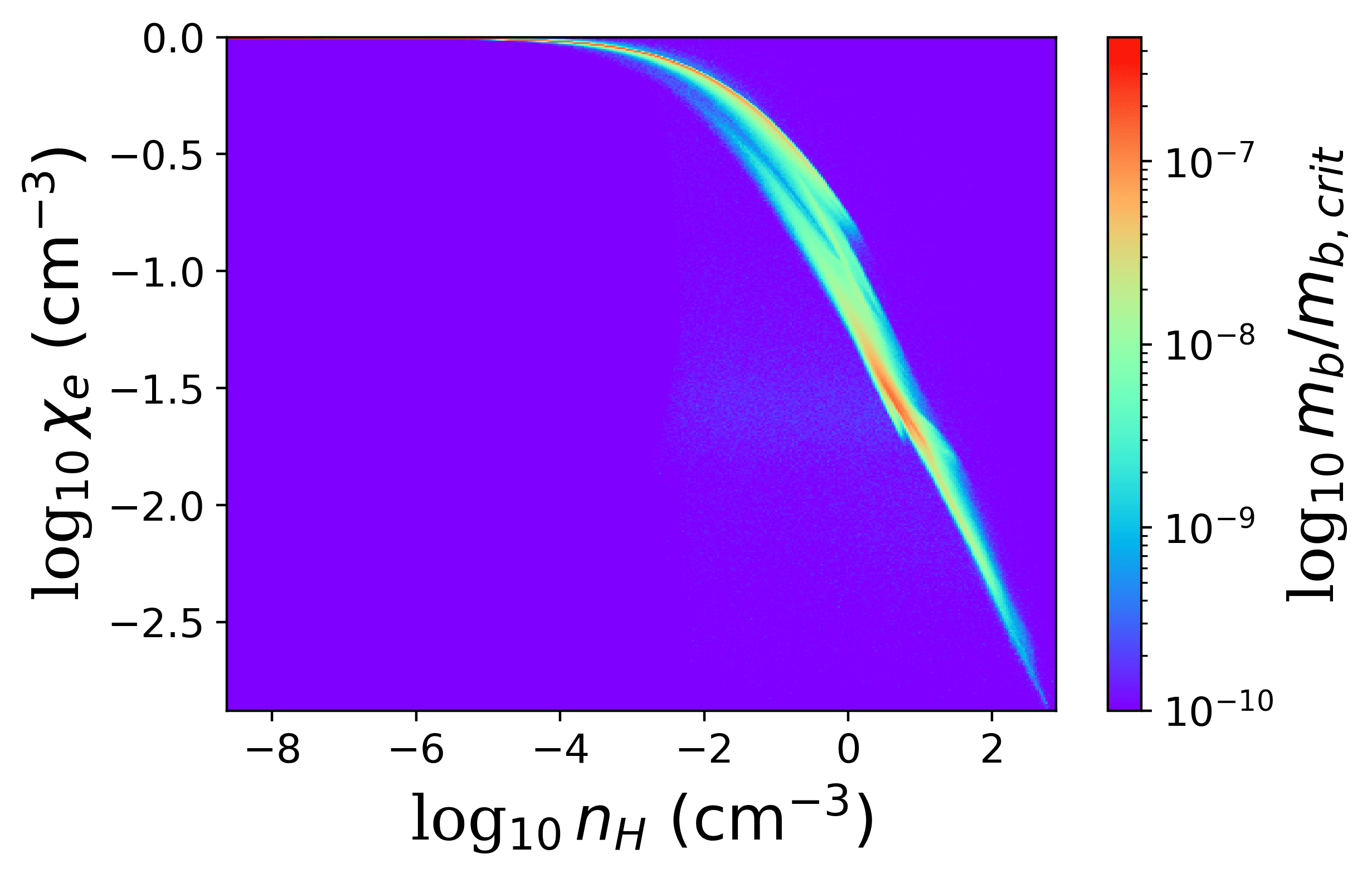}
        \label{fig:nH_Xe_fiducial}
    \end{minipage}
    \begin{minipage}[b]{0.45\textwidth}
        \includegraphics[width=\textwidth]{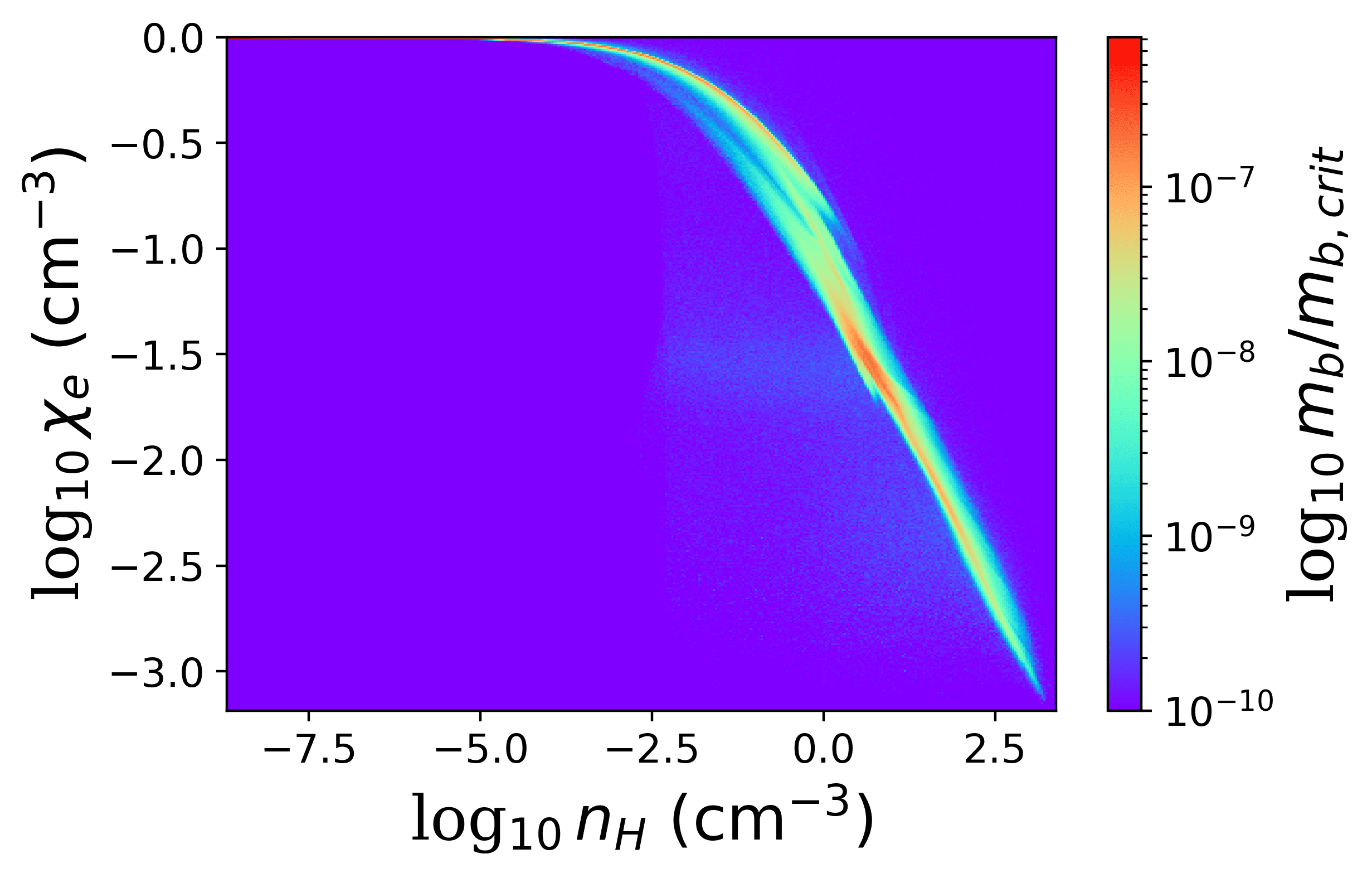}
        \label{fig:nH_Xe_noAGN}
    \end{minipage}
    \caption{Comparison of various properties in fiducial and NoBH simulations for the L100 run.  Each row represents a different set of properties: temperature vs. density (top row), baryon overdensity vs. hydrogen number density (second row), hydrogen number density vs. electron number density (third row), and hydrogen number density vs. ionization fraction (bottom row). The results from the fiducial simulation (left panels) are compared with those from the NoBH simulation (right panels).}
    \label{fig:comparison_fiducial_noAGN}
\end{figure*}

\section{Phase Diagram and Ionization Fraction}

To investigate the impact of AGN feedback, 
Figure~\ref{fig:comparison_fiducial_noAGN} presents the results from the L100 simulations, highlighting the relationship between gas temperature and density, as well as other key properties such as electron number density and hydrogen ionization state.

In the presence of AGN feedback, the phase diagram (left) exhibits a distinct difference from the NoBH scenario. In the temperature--density diagram (top row), AGN feedback leads to more hot gas with $T\gtrsim 10^8$\,K in the low-density, high-temperature regions, while the NoBH case exhibits lower temperatures in these areas. 
In the high-density regions, the NoBH simulation shows a higher degree of condensation, with particles distributed more uniformly and extensively. This indicates that AGN feedback is actively expelling gas from the dense, collapsed regions, heating it to higher temperatures and disrupting the condensation process observed in the NoBH scenario.

In the hydrogen ionization fraction diagram (bottom row), while low-density hydrogen remains almost fully ionized in both scenarios, the NoBH simulation exhibits a broader distribution of highly ionized regions at high densities. This suggests that in the absence of AGN feedback, ionized gas remains more concentrated in dense environments, leading to a more extended ionization structure. 

These differences highlight the role of AGN feedback in regulating the thermal and ionization states of gas, as it redistributes baryons from dense regions into a more diffuse, hotter intergalactic medium. This underscores the significant impact of AGN activity in shaping the balance between the circumgalactic medium (CGM) and the intergalactic medium (IGM).

\section{Halo, Black Hole, and Stellar Mass Relations}

To understand the broader impact of AGN feedback, we next examine how it influences the relationships between halo mass and both black hole mass and stellar mass. By exploring this connection, we can further elucidate the role of AGN feedback in the broader context of galaxy evolution, particularly in regulating black hole growth and quenching star formation.

\begin{figure}[htbp]
    \centering
    \includegraphics[width=0.48\textwidth]{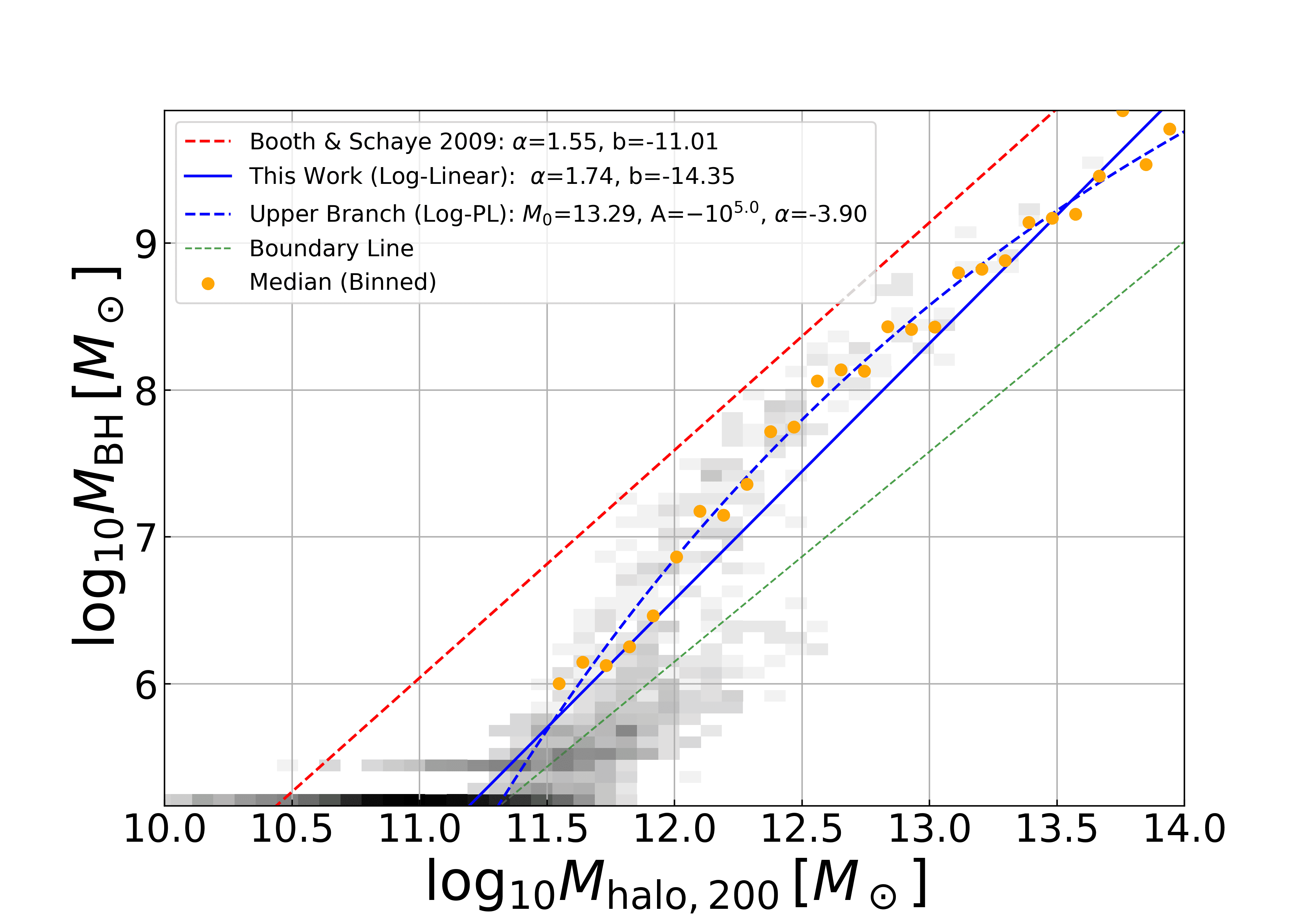}
    \includegraphics[width=0.48\textwidth]{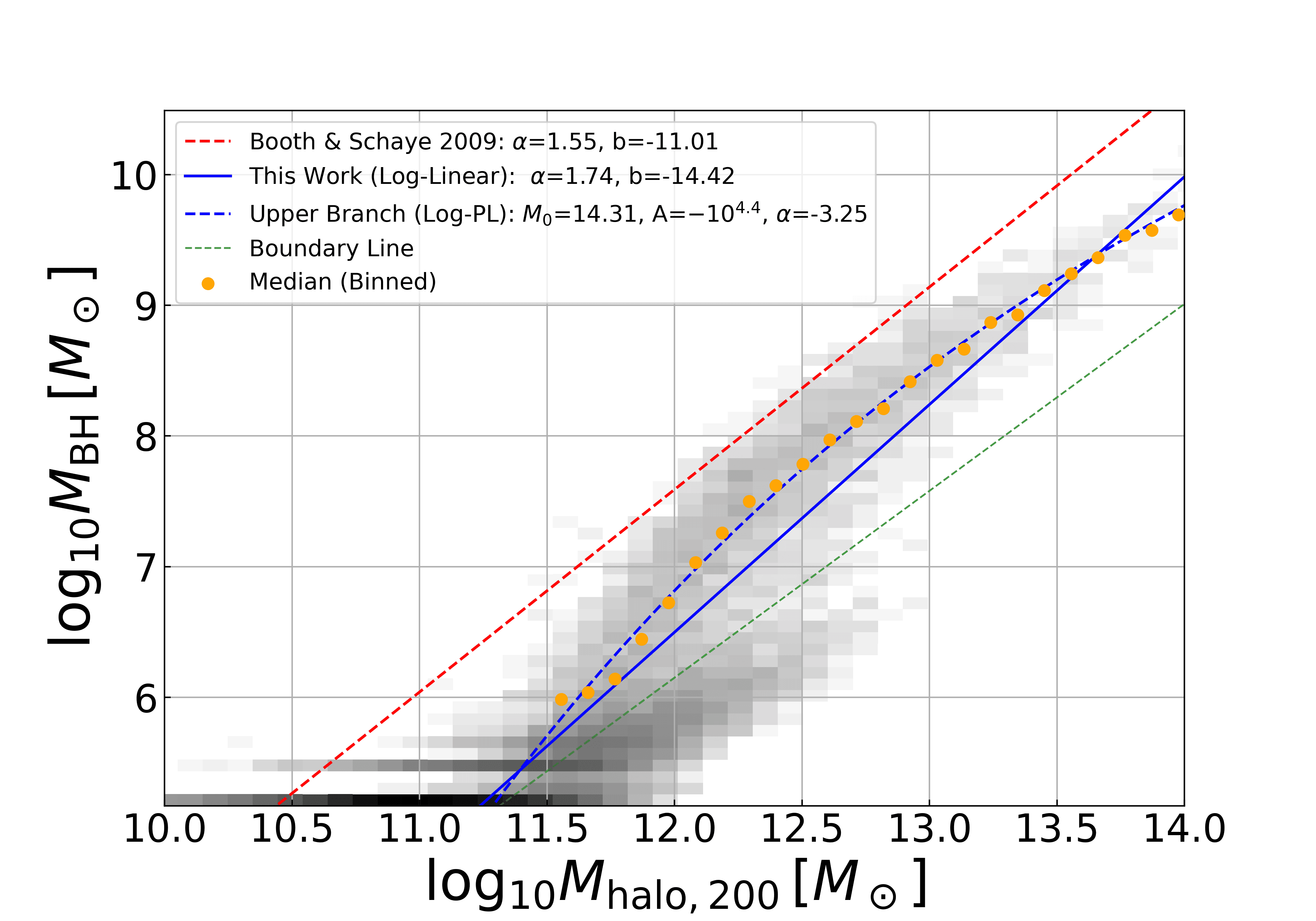}
    \caption{Black hole mass vs. dark matter halo mass relations for the L50N512$_{\mathrm{fiducial}}$ (upper) and L100N1024$_{\mathrm{fiducial}}$ (lower) fiducial simulations. The dashed red line represents the relation from \cite{Booth2010}, while the solid blue line represents the best-fit relation from this work using a log-linear fit. The green dashed line represents the manual boundary line separating the upper and lower branches. The upper branch is fitted separately using a log-power-law (Log-PL) function. The best-fit values for each fitting method are annotated in each panel. The orange points indicate the binned median values used for fitting the upper branch.}

\label{fig:MBH_Mhalo}
\end{figure}
Figure~\ref{fig:MBH_Mhalo} compares the black hole mass vs. halo mass relations in the L50 (upper) and L100 (lower) simulations. In this context, $M_{\mathrm{halo},200}$ refers to the total halo mass enclosed within a radius $R_{200}$. The dashed red line represents the relation from \cite{Booth2010}, while the solid blue line indicates the best-fit log-linear relation derived from this work:
\begin{equation}
    \log_{10} M_{\mathrm{BH}} = \alpha \log_{10} M_{\mathrm{halo},200} + b,
\end{equation}
where $\alpha$ and $b$ represent the slope and intercept of the log-linear fit.The observed differences might stem from the smaller sample size in the L50 box, which could enhance the impact of stochastic effects and deviations from the mean relation. 

Additionally, the presence of two distinct black hole branches in both simulations might be attributed to the different merger histories of the black holes and growth histories. 
To characterize the upper branch, we apply a log-power-law (Log-PL) fit to its median values:
\begin{equation} 
M_{\mathrm{BH}} = 10^{M_0} + A M_{\mathrm{halo},200}^{\alpha}, \end{equation} 
where $M_0$ accounts for the shift in normalization, $A$ is the amplitude, and $\alpha$ is the power-law exponent.
The upper branch potentially results from major mergers, while the lower branch could be due to more isolated black hole systems that have experienced fewer mergers, which requires further analysis in our follow-up paper (D. Nishihama et al. 2025, in preparation).



\section{Relation between gas mass and dark matter halo mass }

\begin{figure*}[htbp]
    \centering
    \begin{tabular}{cc}
        \includegraphics[width=0.5\textwidth]{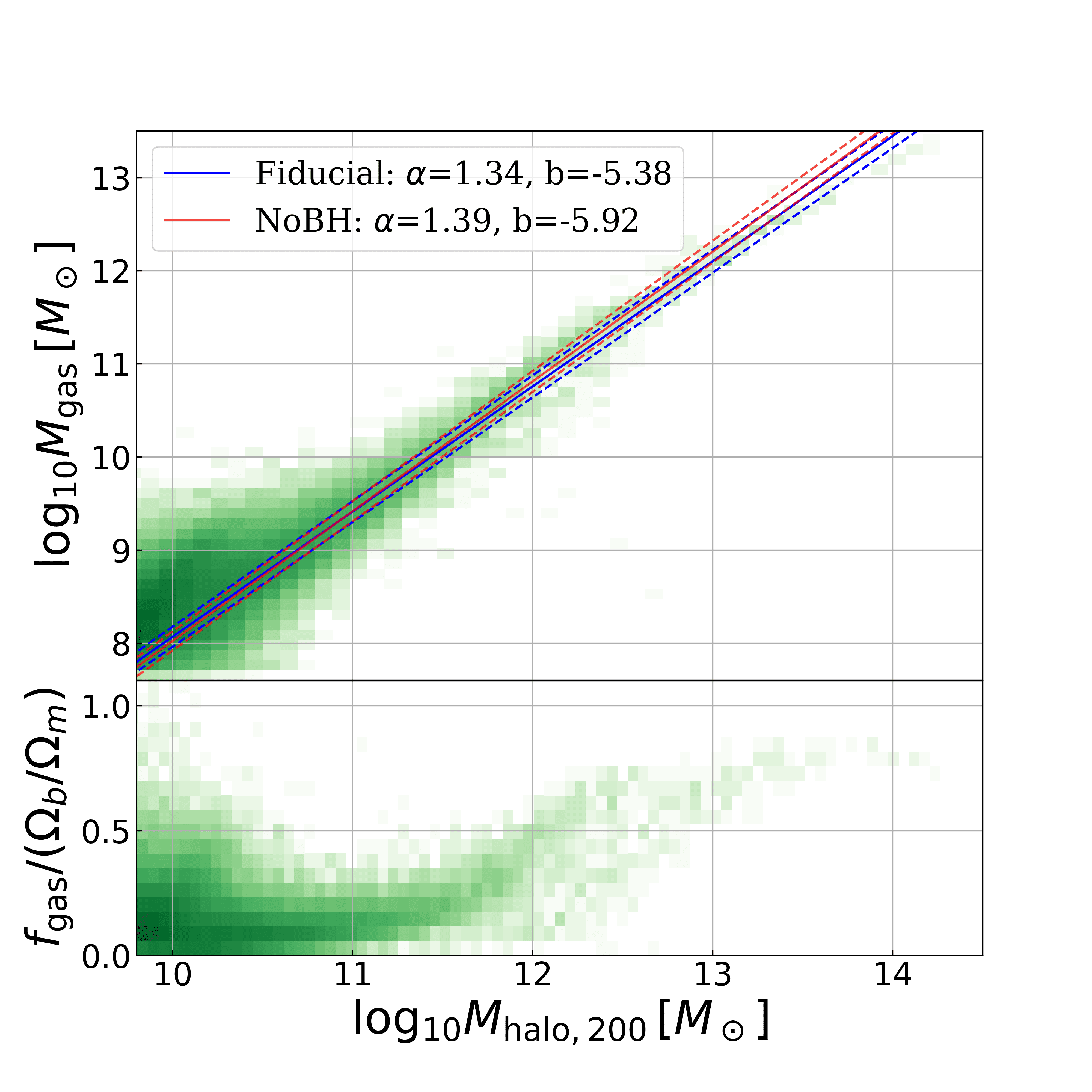} &
        \includegraphics[width=0.5\textwidth]{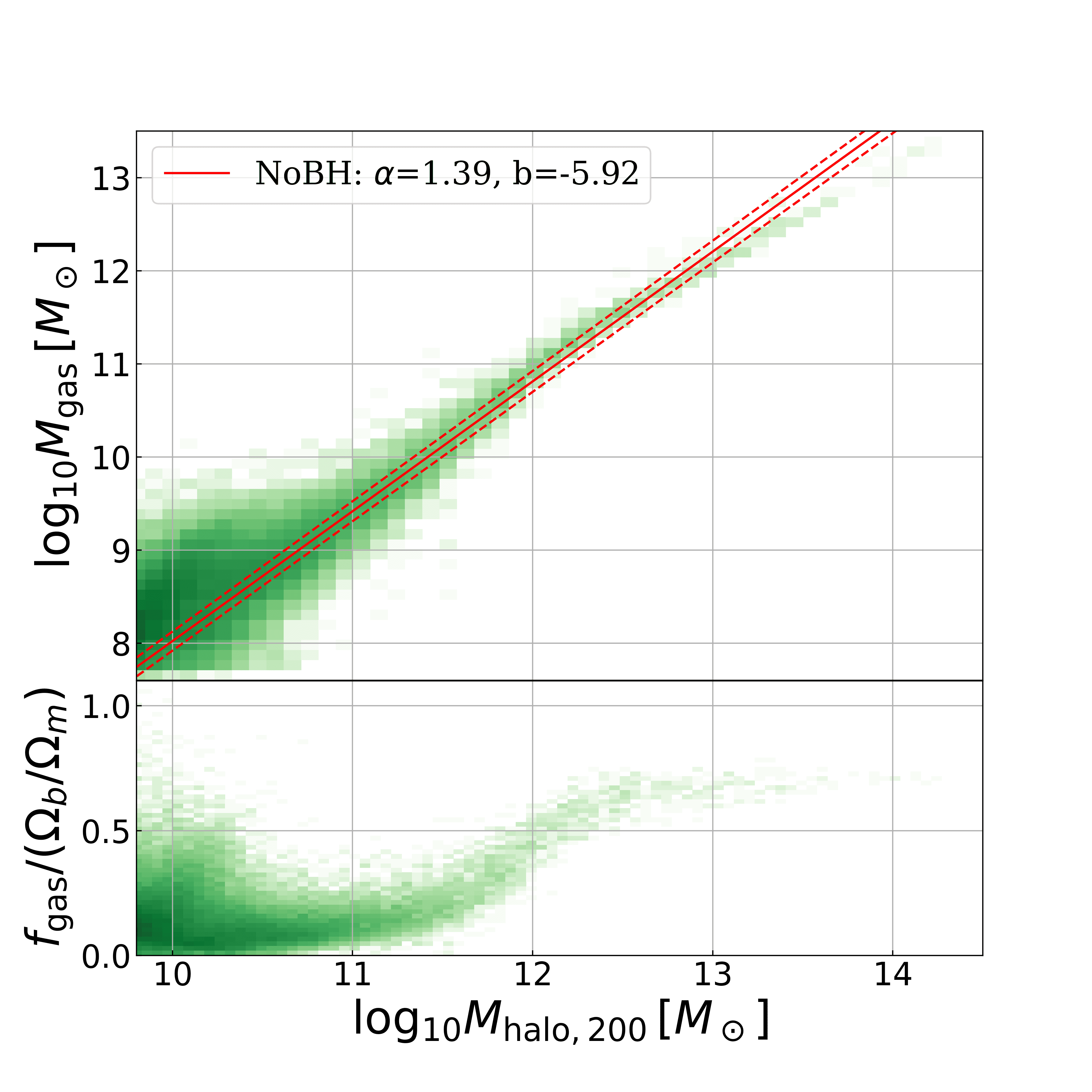} \\
        \includegraphics[width=0.5\textwidth]{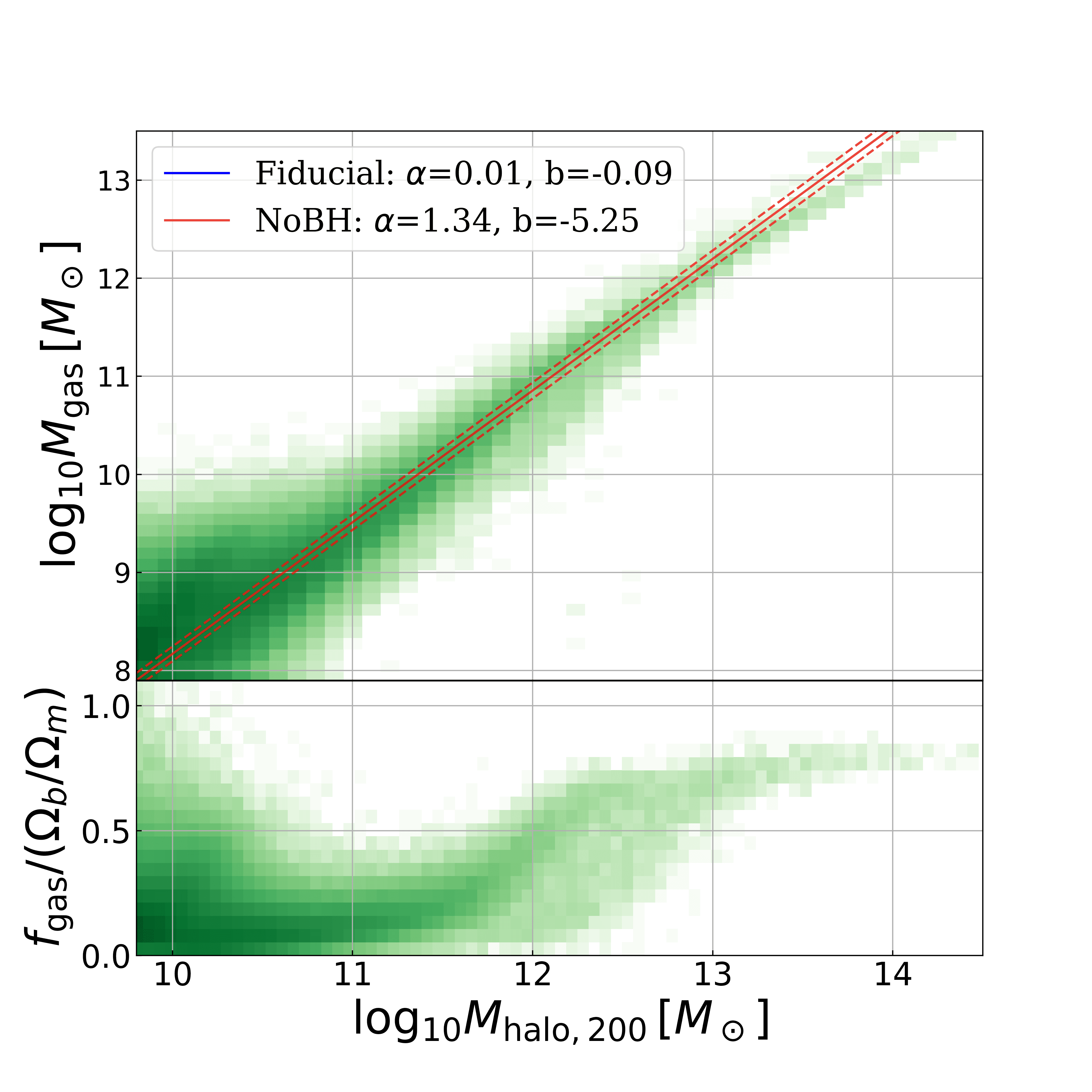} &
        \includegraphics[width=0.5\textwidth]{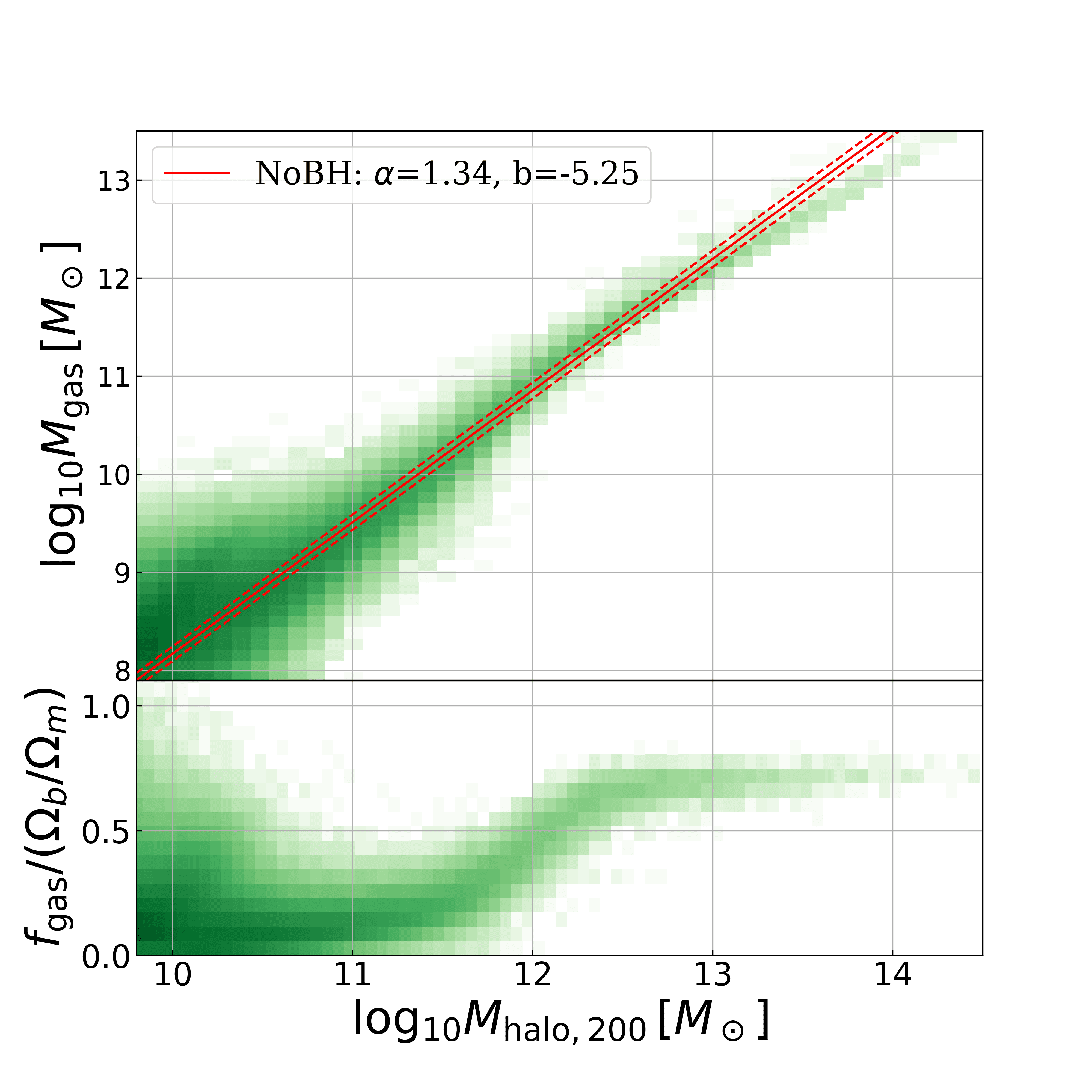} \\
    \end{tabular}
\caption{Comparison of dark matter halo mass vs. gas mass relations. The top row corresponds to the L50N512$_{\text{fiducial}}$ (left) and L50N512$_{\text{NoBH}}$ (right), while the bottom row corresponds to the L100N1024$_{\text{fiducial}}$ (left) and L100N1024$_{\text{NoBH}}$ (right). The color gradient represents the number of halos, and the solid lines indicate the best-fit relations for each case.}
\label{fig:halo_gas_mass}
\end{figure*}

Figure~\ref{fig:halo_gas_mass} presents the comparison between the halo mass ($M_{\rm halo}$) and gas mass ($M_{\rm gas}$) relations across different simulations with and without AGN feedback. The gas mass generally increases with halo mass across all simulations, as expected. However, AGN feedback tends to reduce the gas mass at a given halo mass, especially in more massive halos. 

In all the upper panels of Figure~\ref{fig:halo_gas_mass}, the relationship between $M_{\rm gas}$ and $M_{\text{halo,200}}$ shows a linear fit in log--log space, with minor differences in the slope $\alpha$. The fiducial run has a slightly shallower slope, particularly in the high-mass region ($M_{\text{halo}} > 10^{12}M_\odot$), likely due to stronger AGN feedback expelling gas more efficiently in massive halos.

In the bottom panels of Figure~\ref{fig:halo_gas_mass}, gas fraction $f_{\text{gas}}$ is shown as a function of $M_{\text{halo,200}}$.
As halo mass increases, the overall trend of $f_{\text{gas}}$ shows a gradual increase. 
In the low-mass region ($M_{\text{halo}} < 10^{11} M_{\odot}$), the $f_{\text{gas}}$ distribution is relatively dispersed, with most $f_{\text{gas}}$ below 0.1, while some retain higher gas fractions (exceeding 0.2). This suggests that lower-mass halos tend to lose more gas due to stellar feedback and supernova-driven winds, but there are still some halos capable of retaining a significant fraction of their baryonic content.

In the intermediate mass range ($M_{\text{halo}} \in 10^{12}-10^{13.5} M_{\odot}$), AGN feedback becomes the dominant mechanism in redistributing baryons between CGM and IGM. The efficiency of AGN-driven outflows in this range is high enough to expel gas from the CGM, effectively lowering the gas fraction of halos and causing a more dispersed distribution in $f_{\text{gas}}$. This indicates that AGN feedback is most efficient at removing gas entirely from the halo potential well in this mass regime.

However, for massive halos beyond the galaxy group scale ($M_{\text{halo}} > 10^{13.5} M_{\odot}$), the scatter in $f_{\text{gas}}$ becomes significantly smaller. This suggests that despite stronger AGN feedback at higher masses, it is no longer strong enough to expel gas beyond the deep gravitational potential well of these halos. Instead, AGN feedback redistributes gas within the halo rather than fully removing it, likely pushing gas from the central regions to the outskirts of the CGM. This expelled gas remains gravitationally bound and will eventually reaccrete back into the halo center over time, contributing to the stable gas fraction observed in massive halos.

\begin{figure*}[htbp]
    \centering
    \includegraphics[width=0.7\textwidth]{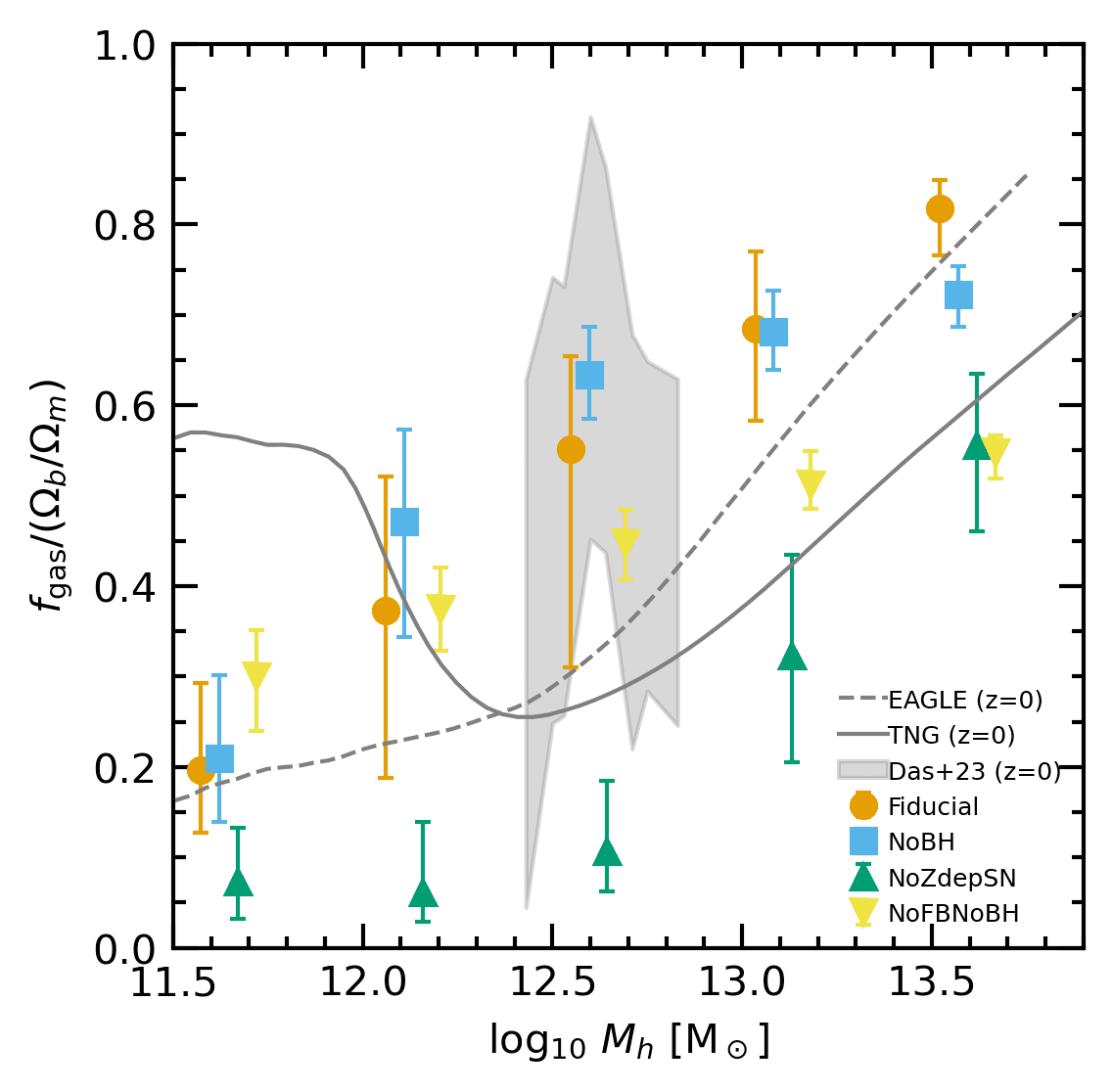}
    \caption{Comparison of CGM gas mass fraction ($f_{\text{gas}}$) as a function of halo mass at $z=0$. The gray dashed and solid lines indicate the median relations from the EAGLE-Ref and TNG-100 simulations, obtained from \citet{Davies2020}. The shaded gray region represents observational constraints from \citet{Das2023}. Data points represent the median values, with error bars indicating the 16th and 84th percentile. Different colors and symbols correspond to various feedback implementations in \citet{Oku2024}: fiducial (orange circles), no BHs (blue squares), fixed Chabrier IMF across different metallicities (green triangles), and no BHs and no SN feedback cases (yellow triangles). Our fiducial model better reproduces the baryon fraction in intermediate-mass halos compared to EAGLE-Ref and TNG-100.}
    \label{fig:gas_fraction_comparison}
\end{figure*}

To further evaluate our results, we compare the baryon mass fraction ($f_{\text{baryon}}$) in our simulations to previous theoretical models and observational constraints. Figure~\ref{fig:gas_fraction_comparison} illustrates the $f_{\text{baryon}}$ as a function of halo mass at $z=0$, including results from our fiducial and various feedback models described in \citet{Oku2024}, along with comparisons to EAGLE-Ref and TNG-100 simulations. Observational constraints from \citet{Das2023} are also shown.
We can see a better agreement to the observed gas fraction in the intermediate halo mass range for the CROCODILE simulation compared to EAGLE-Ref and TNG-100, which could be pushing the gas too much by AGN feedback.

\section{Feedback Energy, Potential Energy, vs. Halo Mass Relations}
\begin{figure*}[htbp]
    \centering
    \includegraphics[width=0.49\textwidth]{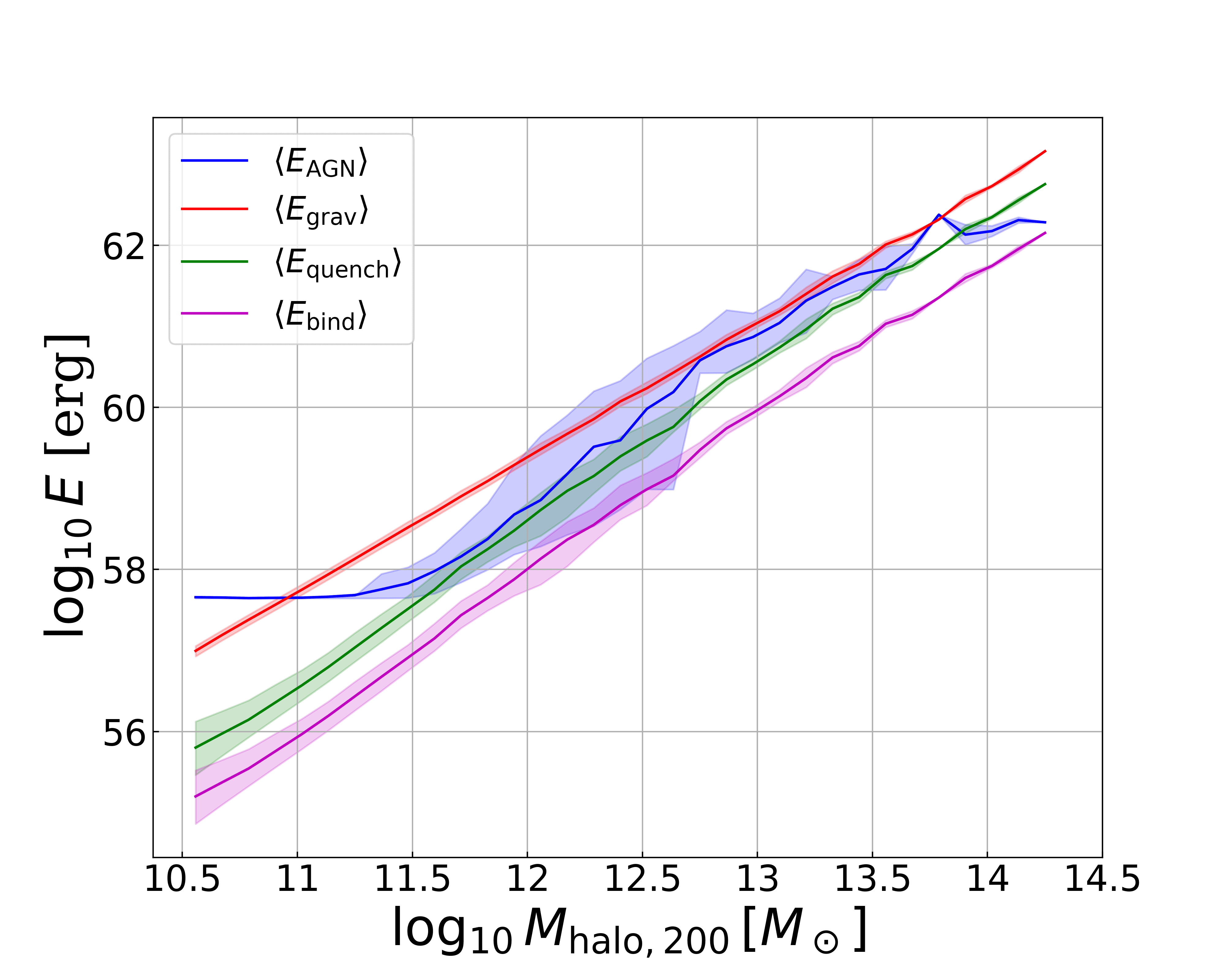}
    \includegraphics[width=0.49\textwidth]{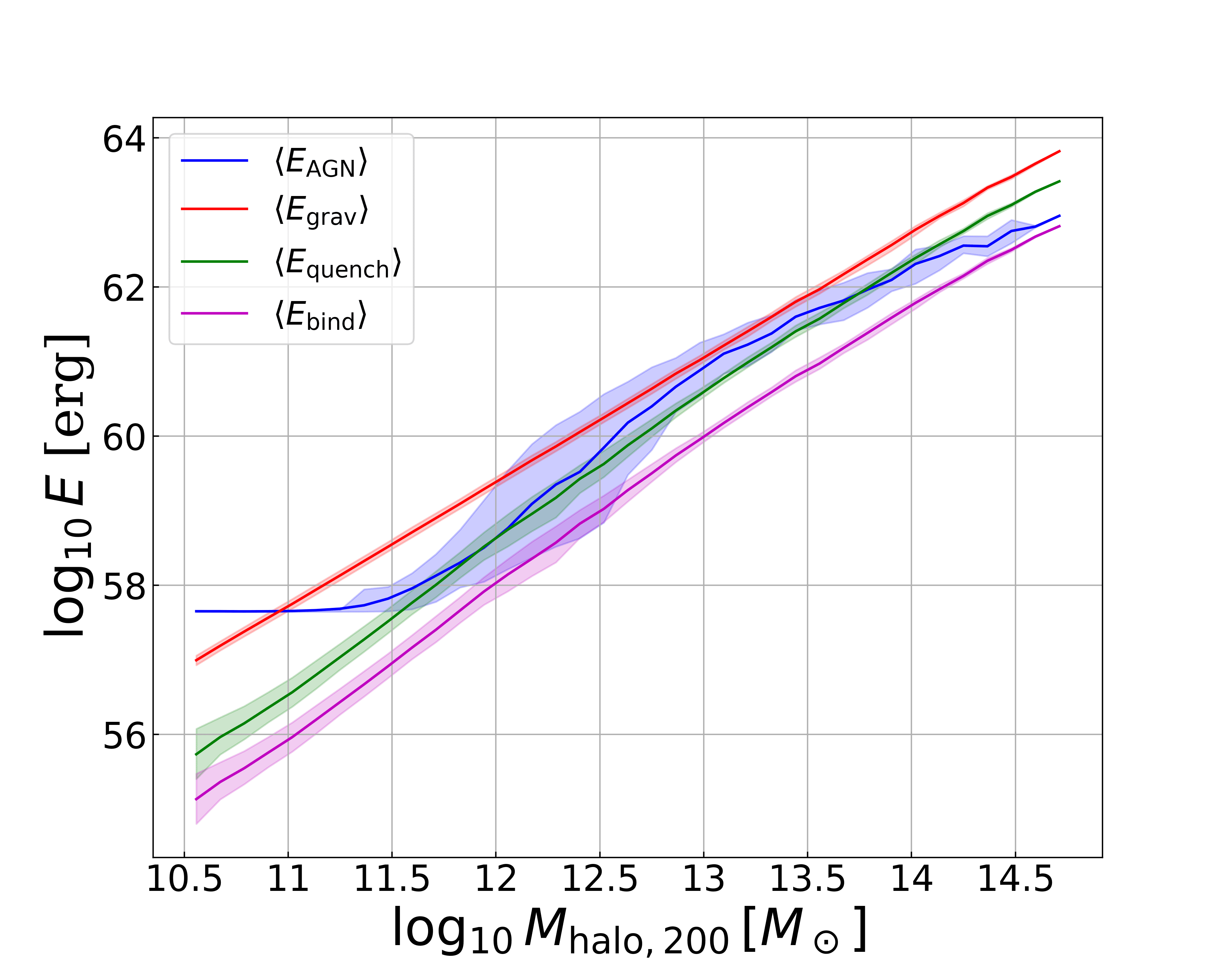}
    \caption{Energy vs. dark matter halo mass relations for the L100N1024$_{\text{fiducial}}$ runs. The top panel shows results from the 50 $h^{-1}$\,cMpc simulation, and the bottom panel shows results from the 100 $h^{-1}$\,cMpc simulation. The different curves represent the average AGN energy $\langle E_{\text{AGN}} \rangle$, gravitational energy $\langle E_{\text{grav}} \rangle$, binding energy $\langle E_{\text{quench}} \rangle$, and total binding energy $\langle E_{\text{bind}} \rangle$.}
    \label{fig:energy_mass_relations}
\end{figure*}
To identify the halo mass range where AGN feedback is most effective, we compare the energy from AGN feedback ($E_{\mathrm{AGN}}$) with the gravitational potential energy ($E_{\mathrm{grav}}$), the binding energy ($E_{\mathrm{bind}}$), and the quench energy ($E_{\mathrm{quench}}$) across different halo masses, as shown in Figure~\ref{fig:energy_mass_relations}. We can compute each energy component by the  following equations:
\begin{align}
\dot{E}_{\mathrm{AGN}} &= \epsilon_f \epsilon_r \dot{m}_{\mathrm{acc}} c^2 = \frac{\epsilon_f \epsilon_r}{1 - \epsilon_r} \dot{m}_{\mathrm{BH}} c^2, \\
E_{\mathrm{grav}} &= \frac{3GM_{200}^2}{5R_{200}}, \\
E_{\mathrm{bind}} &= \frac{1}{2} f_{\mathrm{gas}} \frac{GM_{200}^2}{R_{\mathrm{200}}}, \\
E_{\mathrm{quench}} &= f_B \times E_{\mathrm{bind}}.
\end{align}
Here, the parameter $\epsilon_f$ is the AGN feedback efficiency, where we adopt  $\epsilon_f = 0.15$, as calibrated by \cite{Booth2010}.
$\epsilon_r$ is the radiative efficiency, typically derived from the standard model by \cite{Shakura1973}, and in our work, we adopt $\epsilon_r = 0.1$.
$\dot{m}_{\mathrm{acc}}$ represents the accretion rate of gas onto the black hole.
$\dot{m}_{\mathrm{BH}}$ is the rate of mass accretion rate onto the black hole.
$f_{\mathrm{gas}}$ is the gas mass fraction within the halo.
$f_B$ is a coupling efficiency factor defined as 
\begin{equation}
f_B = \frac{\eta E_{\text{AGN}}}{E_{\text{bind}}},
\end{equation}
where $\eta$ encapsulates the inefficiencies in energy transfer, including radiative losses and the interaction between AGN feedback and halo gas. $f_B$ serves as a calibration factor to align theoretical models with observational constraints \citep{Chen2020ApJ...897..102C}, reflecting the effective energy transfer required to overcome the binding energy and quench star formation or drive gas outflows.

The comparison between $E_{\mathrm{AGN}}$ and $E_{\mathrm{quench}}$ highlights the halo mass range where AGN feedback is most effective. As shown in Figure \ref{fig:energy_mass_relations}, the largest difference between $E_{\mathrm{AGN}}$ and $E_{\mathrm{quench}}$ occurs in halos with masses around $10^{12.5} - 10^{13.5}\,M_\odot$, where AGN feedback efficiently overcomes the gravitational binding energy, facilitating the quenching of star formation and the removal of gas. In contrast, in halos with masses exceeding $10^{14}\,M_\odot$, the gravitational binding energy becomes dominant, significantly reducing the effectiveness of AGN feedback. Although AGN energy may still contribute to heating and driving outflows, gravity emerges as the primary factor regulating the gas and limiting the feedback's capacity to quench star formation.


\section{Impact of Resolution and Box Size on Density Profiles} \label{App:boxsize_effect}

\begin{figure*}[htbp]
\centering
\begin{overpic}[width=.48\textwidth]{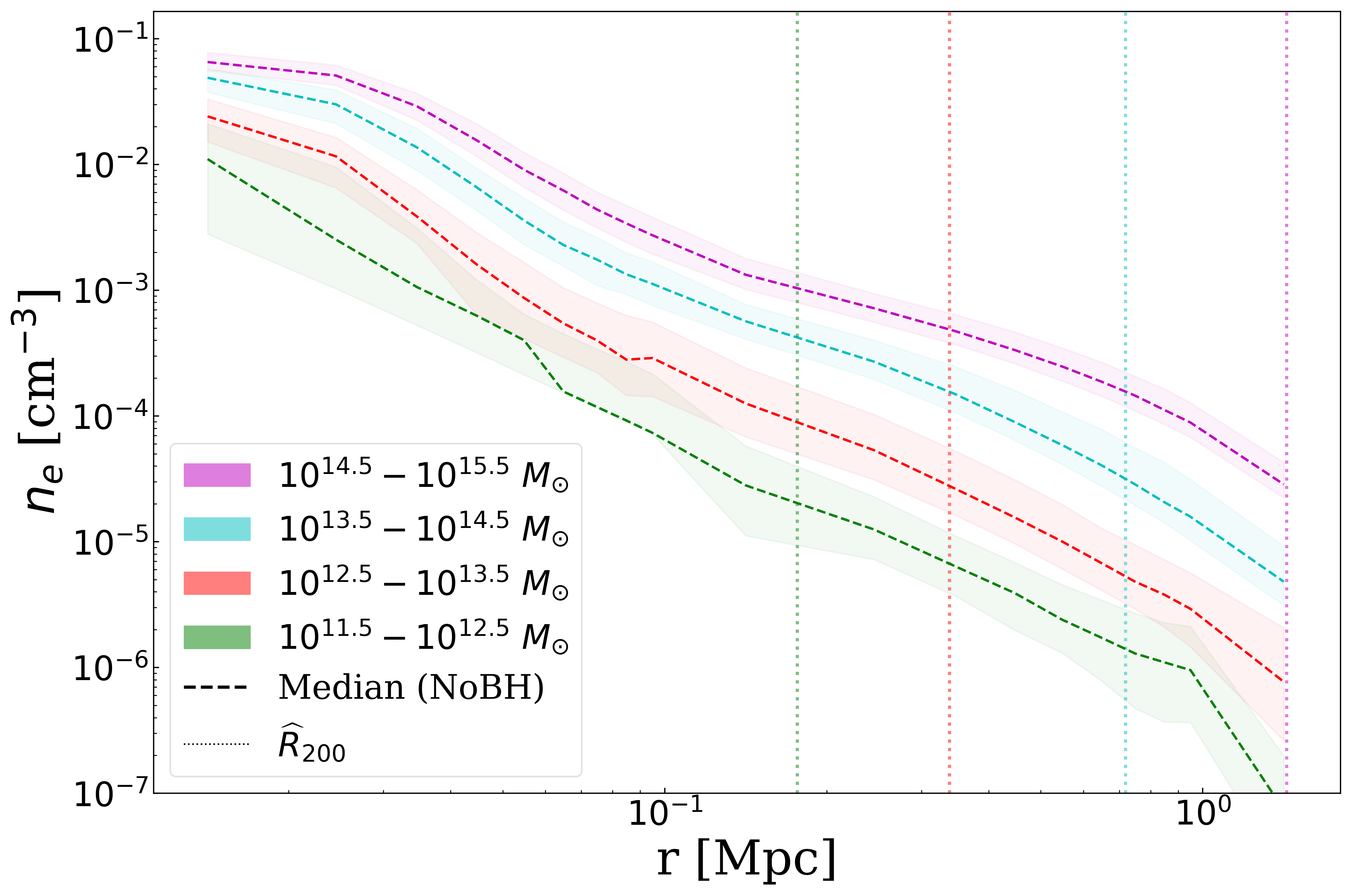}
\put(5,68){(a)}
\end{overpic}
\hfill 
\begin{overpic}[width=.48\textwidth]{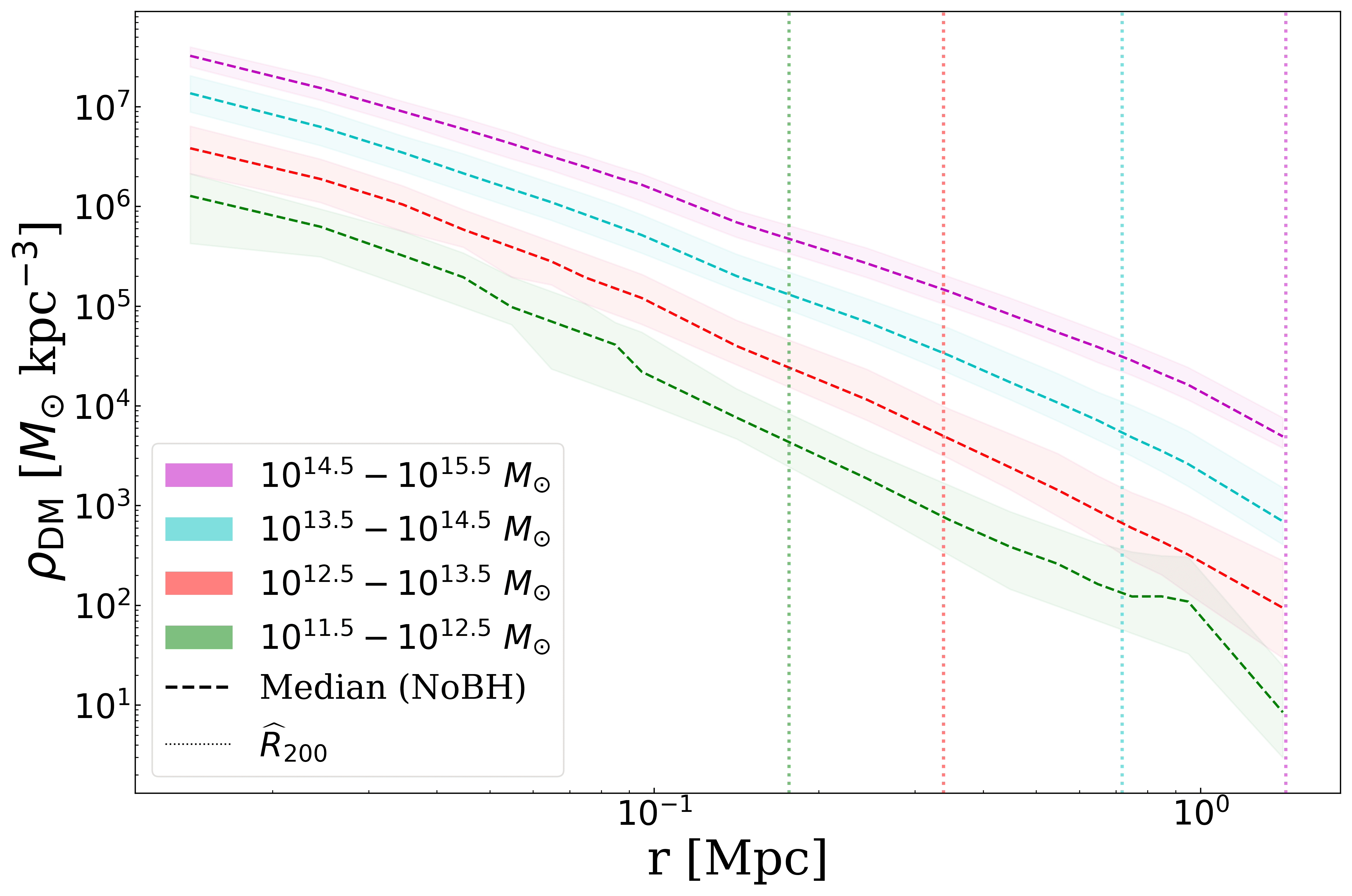}
\put(5,68){(b)}
\end{overpic}
\\ 
\vspace{5mm} 
\begin{overpic}[width=.48\textwidth]{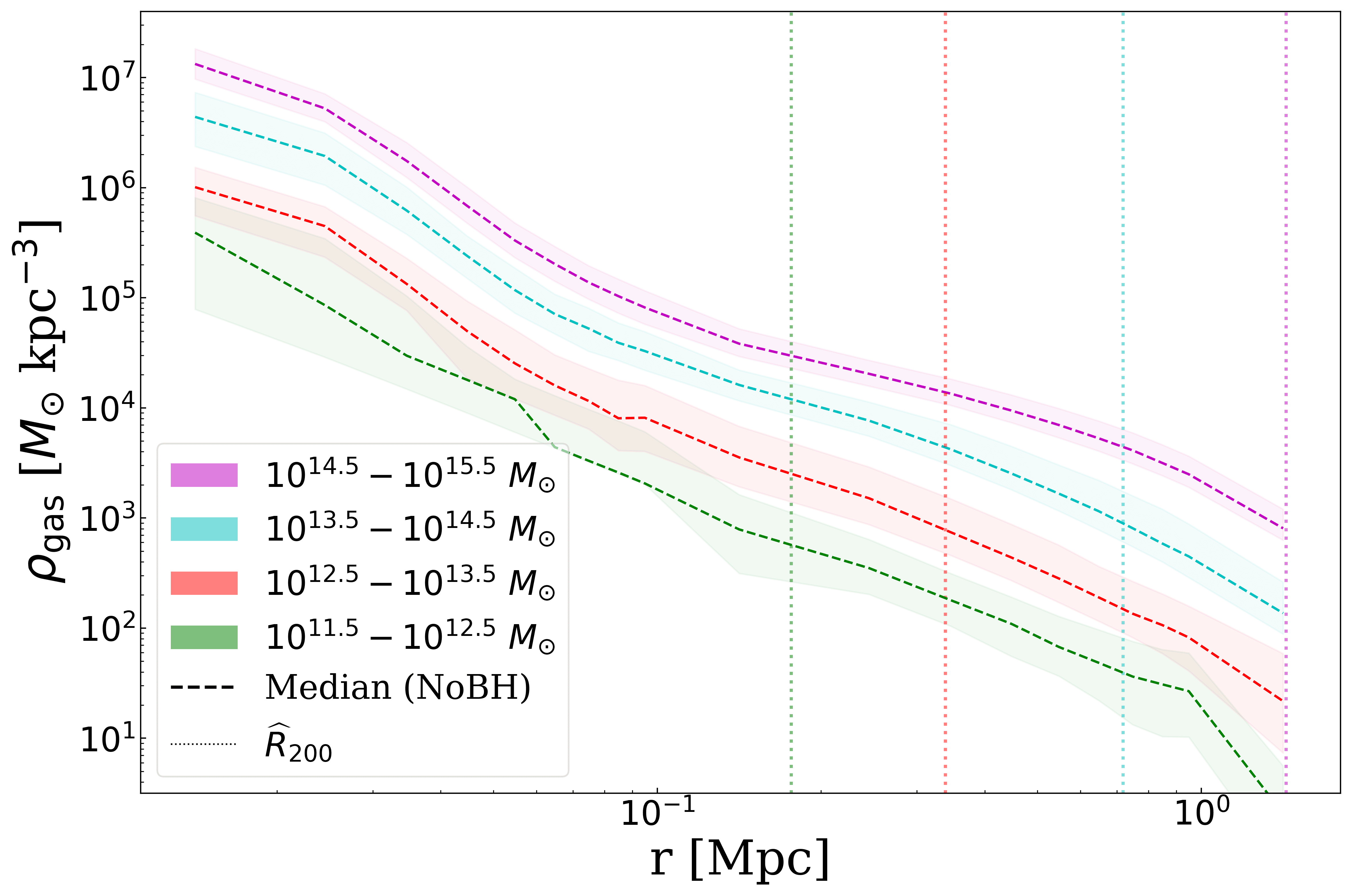}
\put(5,69){(c)}
\end{overpic}
\hfill 
\begin{overpic}[width=.48\textwidth]{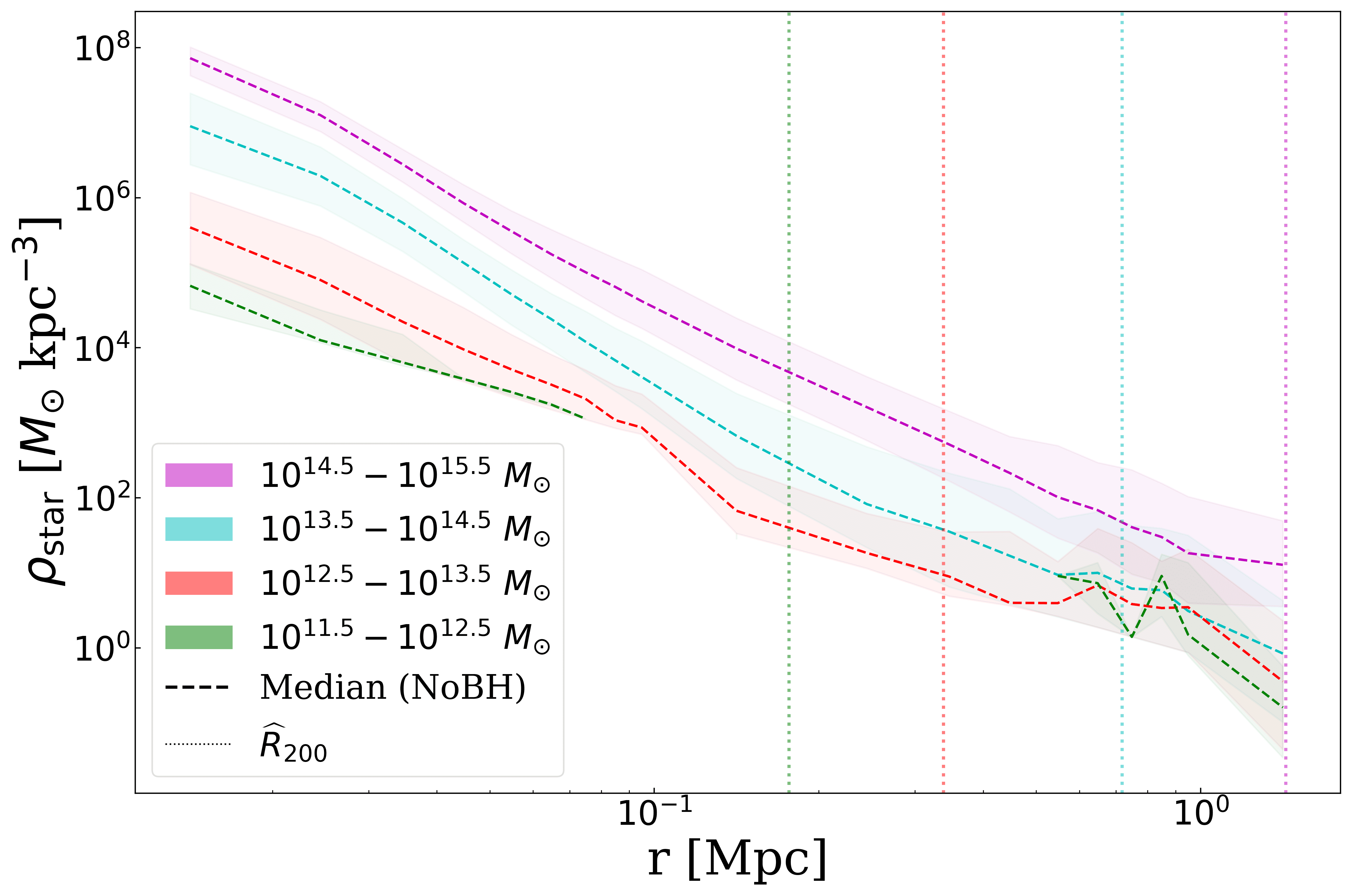}
\put(5,69){(d)}
\end{overpic}
\caption{Median density profiles for halos in the L500NoBH simulation, categorized by mass ranges:  electron number density $n_e$ (panel {\it a}), dark matter density $\rho_{\mathrm{DM}}$ (panel {\it b}),  gas density $\rho_{\mathrm{gas}}$ (panel {\it c}), and stellar mass density $\rho_{\mathrm{star}}$ (panel {\it d}). The shaded regions indicate $1\sigma$ uncertainties for 500 halos in each mass range. The vertical dotted lines represent the median $\widehat{R}_{200}$ of halos in each mass range.}
\label{fig:density_profiles_500Mpc}
\end{figure*}

Figure~\ref{fig:density_profiles_500Mpc} show the median density profiles from the L500 simulation for various mass ranges, similar to Figure~\ref{fig:density_profiles_100Mpc}. 

\begin{figure*}[htbp]
    \centering
    \begin{overpic}[width=0.48\textwidth]{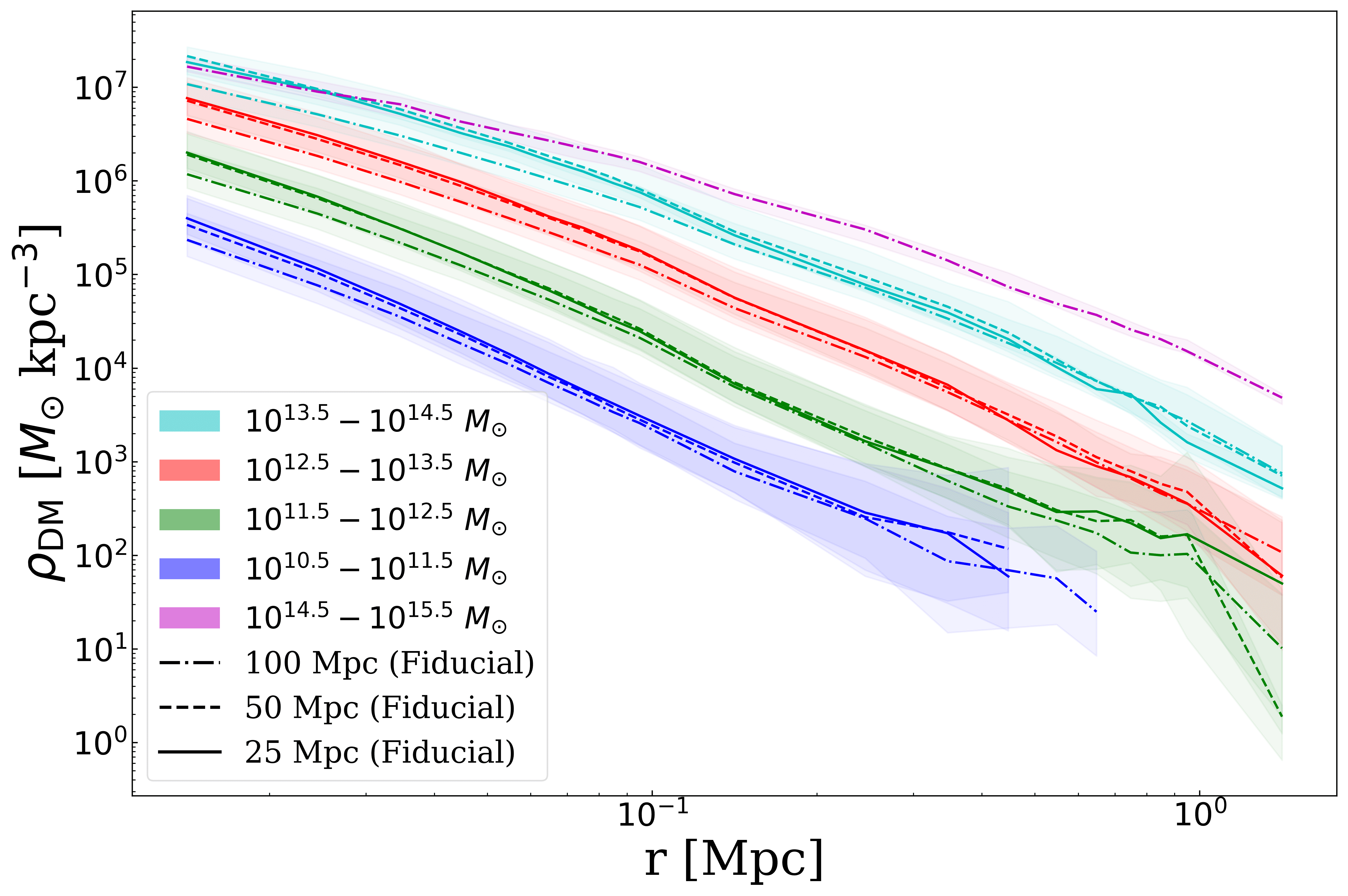}
        \put(5,69){(a)}
    \end{overpic}
    \hfill
    \begin{overpic}[width=0.48\textwidth]{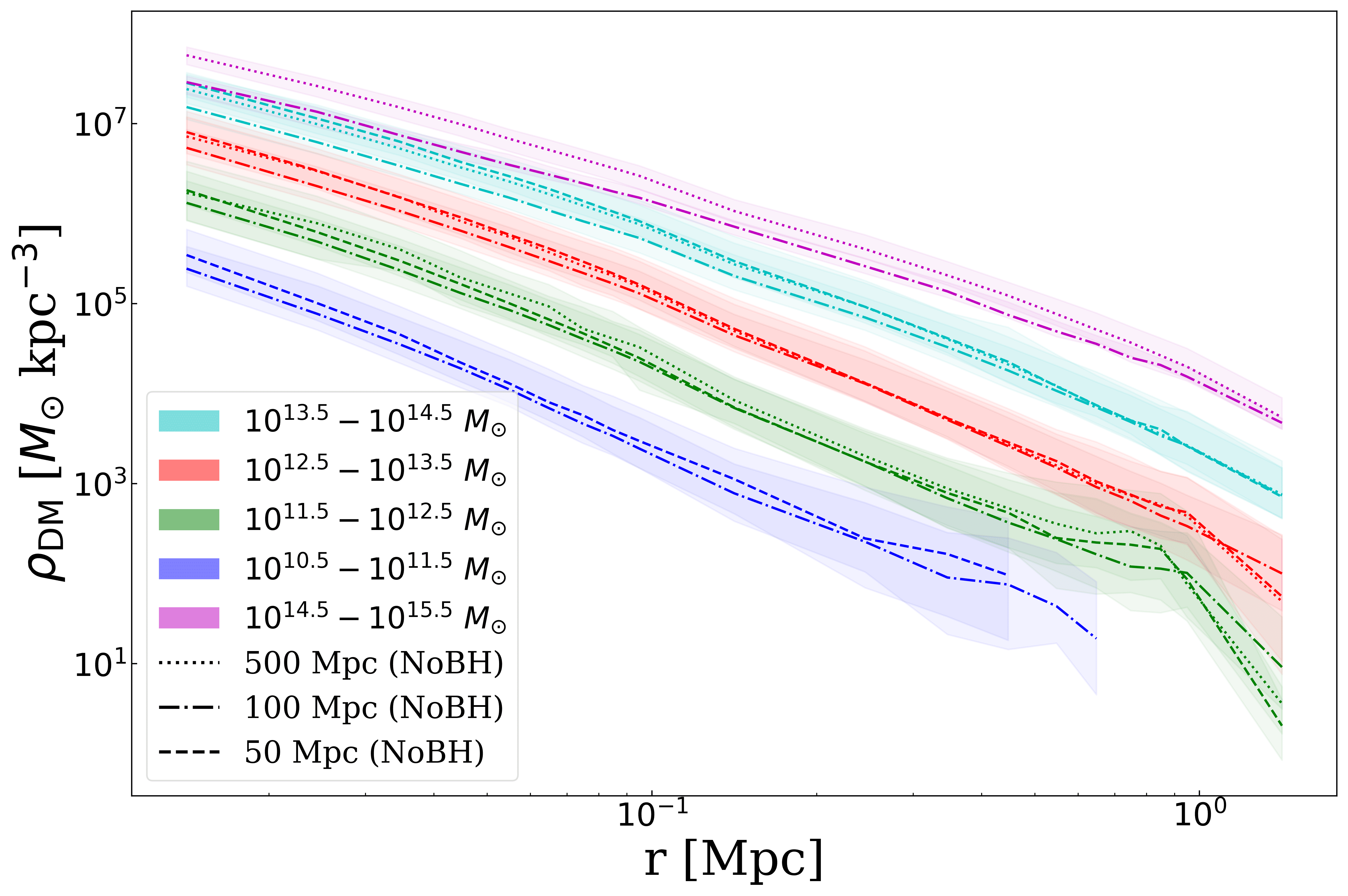}
        \put(5,69){(b)}
    \end{overpic}
    \vspace{5mm} 
    \begin{overpic}[width=0.48\textwidth]{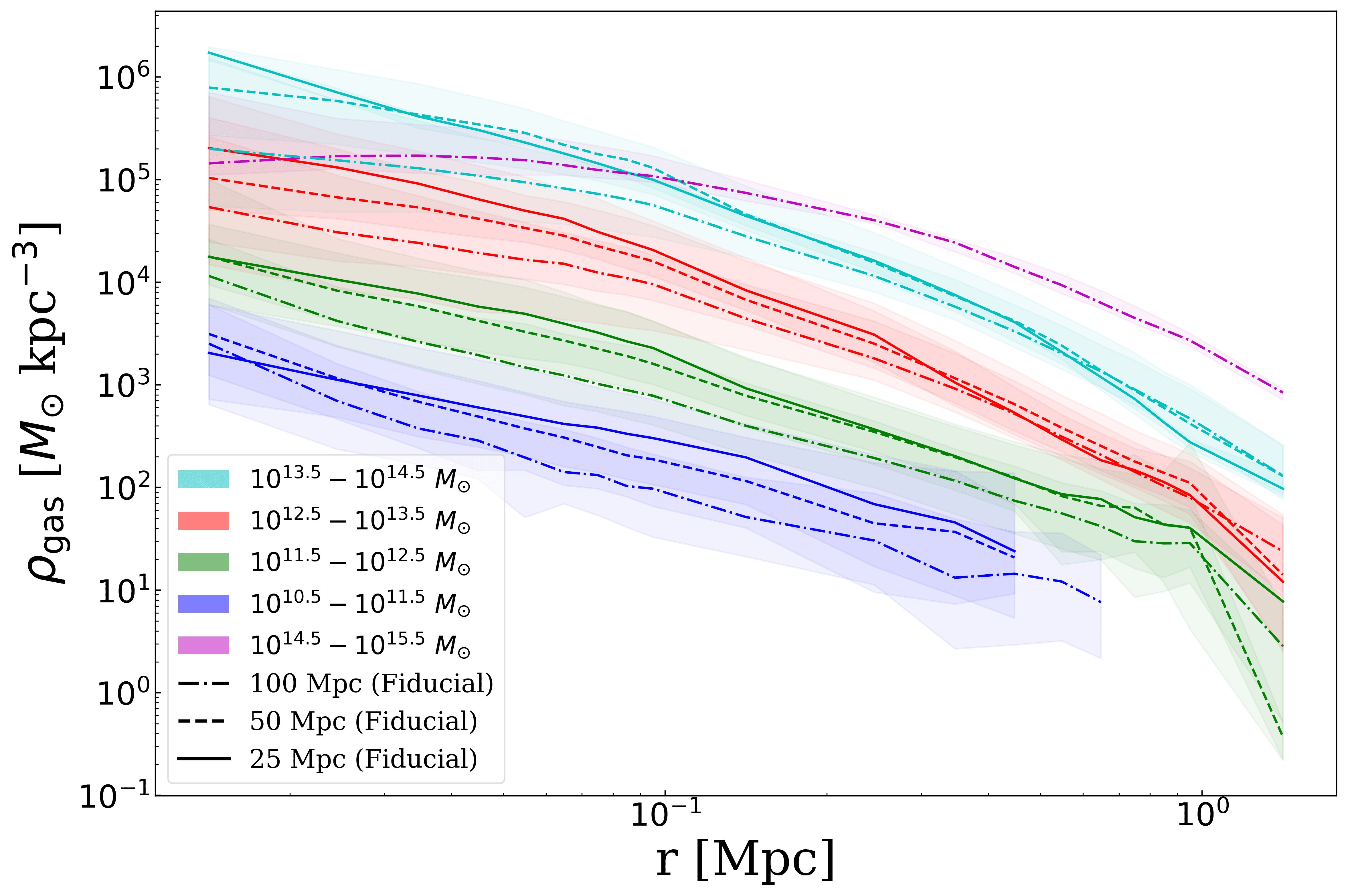}
        \put(5,67){(c)}
    \end{overpic}
    \hfill
    \begin{overpic}[width=0.48\textwidth]{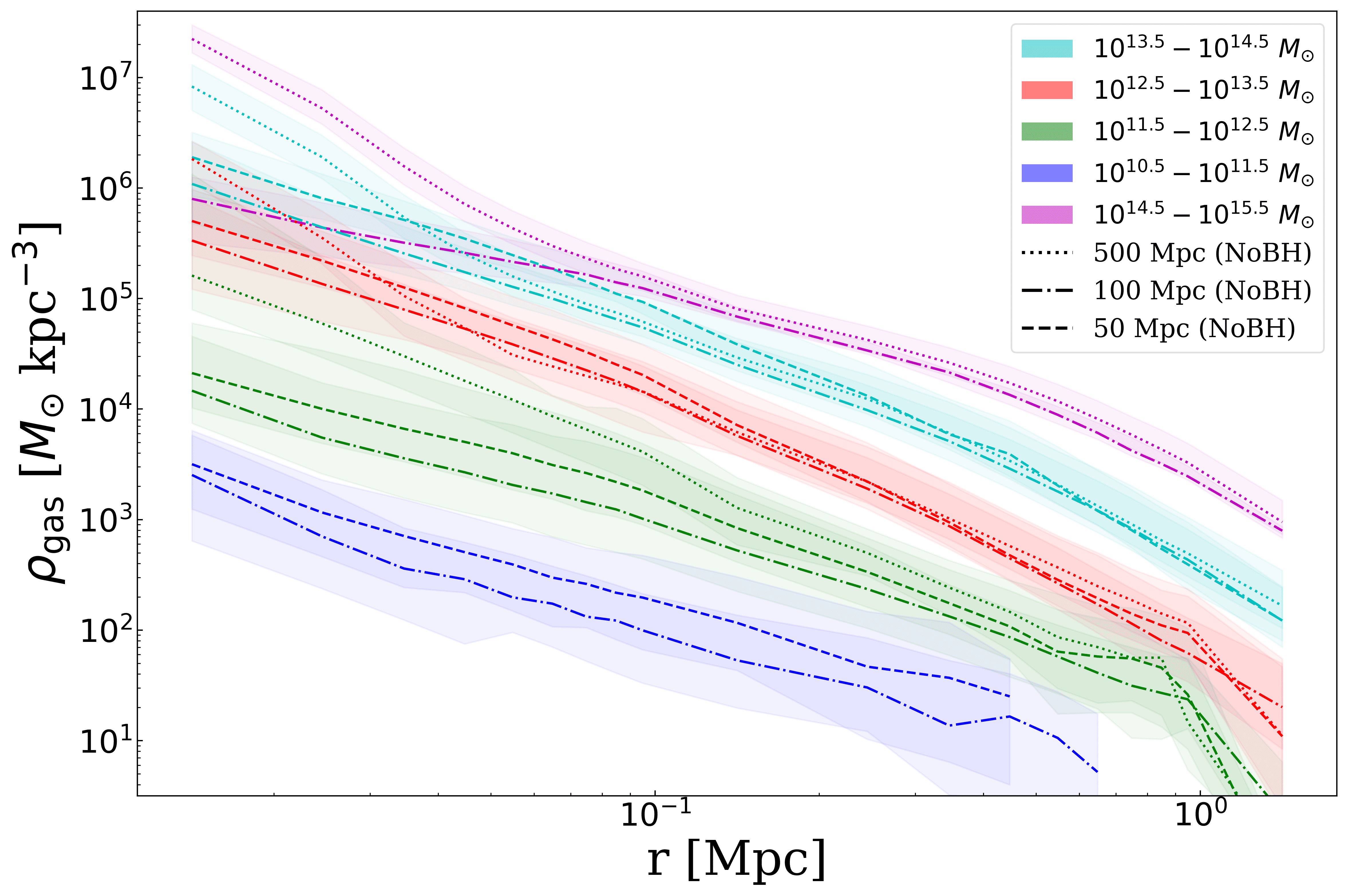}
        \put(5,67){(d)}
    \end{overpic}
    \caption{Comparison of density profiles for halos in different box sizes (25, 50, 100, and 500 Mpc). The left column compares fiducial simulations in 25, 50, and 100 Mpc boxes, and the right column compares NoBH simulations in 50, 100, and 500 Mpc boxes.}
    \label{fig:boxsize_density_profiles}
\end{figure*}

Figure~\ref{fig:boxsize_density_profiles} compares the density profiles of gas and dark matter across simulations with different box sizes and resolutions. Panels~{\it (a)}  and (b) illustrate the fiducial L25, L50, L100 simulations, while panels {\it (c)} and {\it (d)} display the NoBH L50, L100, L500 simulations. This comparison highlights the effects of box size and resolution on the density profiles of halos within similar mass ranges.

In panels~{\it (a)} and {\it (v)}, the gas and dark matter profiles in the L25 and L50 simulations exhibit slight deviations from the L100 results. For dark matter, the L25 and L50 results are relatively close and slightly higher than the L100 simulation. This discrepancy likely arises from sample variance due to the smaller box sizes, where fewer massive halos are available for statistical averaging. For gas, the profiles show distinct trends across mass ranges. In low-mass halos (\(10^{10.5} \, M_{\odot} < M_H < 10^{12.5} \, M_{\odot}\)), the L25 and L50 simulations yield similar results, deviating from the L100. However, for halos in the \(10^{12.5} \, M_{\odot} < M_H < 10^{13.5} \, M_{\odot}\) range, the profiles near the halo center differ significantly across all box sizes, highlighting the crucial role of AGN feedback in this mass regime. In the highest mass range (\(M_H > 10^{13.5} \, M_{\odot}\)), the L25 and L50 profiles remain closer, with larger deviations observed for the L100 simulation.

In panels~{\it (c)}) and {\it (d)}, the NoBH simulations show different trends. For dark matter, the L50 and L500 profiles are more aligned, while the L100 profiles deviate slightly. This consistency between L50 and L500 simulations suggests that sample variance in the L100 box may contribute to the differences. 
For gas, the L500 simulation exhibits a steep rise in the central density, distinct from the L50 and L100 results. This steep rise is attributed to the lower resolution of the L500 simulation, where the gravitational softening length (\(16 \, h^{-1} \, \mathrm{ckpc}\)) coincides with the scale of the steep density increase. 
The limited resolution smooths gas distributions and artificially concentrates gas particles near the halo center, leading to an overestimation of central densities.

The differences in gas and dark matter profiles highlight the impact of resolution and box size on simulations. Gas profiles are more sensitive to resolution due to the interplay of gravitational forces, hydrodynamics, and cooling processes. In contrast, dark matter profiles, driven solely by gravitational interactions, remain relatively robust against resolution changes. The findings highlight the necessity of using higher-resolution simulations for accurately studying baryonic processes, particularly in large-scale simulations with low mass resolution.

\section{Light-cone generation for FRB sight lines}
\label{App:LC}
The Light-cone generation process involves:
\begin{itemize}
    \item Initial setup and LoS Selection: A simulation box at a given redshift (e.g., $z=0$) is positioned at the origin of one plane (e.g., $xy$-plane). A random direction vector $\mathbf{d}$ is generated to define the initial LoS, ensuring uniform sampling in three-dimensional space. The components of $\mathbf{d}$ are given by:
    \begin{equation}
        d_i = 
        \begin{cases} 
            \text{random}(0,1) & \text{with probability } 0.5, \\
            \text{random}(-1,0) & \text{otherwise}.
        \end{cases}
    \end{equation}
    One of the three components is set to $\pm 1$ to normalize the major axis. The corresponding LoS vector is:
    \begin{equation}
        \mathbf{LoS} = \left(\frac{L}{m}, \frac{L}{n}, \frac{L}{o}\right),
    \end{equation}
    where $m=1/d_x$, $n=1/d_y$, and $o=1/d_z$. The total length of the light cone in each single box is computed as:
    \begin{equation}
        L_{\text{LC}} = \sqrt{\sum \mathbf{LoS}^2}.
    \end{equation}
    
    \item DM Calculation within a single snapshot:
    The electron number density $n_e$ along the light cone is computed using a gridding approach. The simulation box is divided into cells based on hierarchical levels: lv7 ($128^3$), lv8 ($256^3$), lv9 ($512^3$), and lv10 ($1024^3$). Each gas particle's mass and electron number density are distributed among the overlapping cells according to its smoothing length. The LoS is subdivided into bins (e.g., 400 bins for lv8). The electron density of a given LoS bin $i$ is taken from the corresponding cell $j$, and the DM contribution for that bin is computed as:
    \begin{equation}
        \text{DM}_i = n_{e,j} \frac{l_{i}}{1+z}.
    \end{equation}
    The total DM along the LoS is obtained by summing over all bins. Tests show that for cell levels lv8 and higher (cell size $\sim 400$ kpc), DM results are numerically converged. Thus, we adopt lv8 in our study.
    
    \item Extending and shifting across snapshots:
    The light cone is extended beyond a single snapshot by assigning a new random angle for the next segment. This ensures diverse structure sampling while preserving continuity in the cosmic web. The process is repeated until the total comoving path length matches the required distance to the next snapshot (e.g., $z=0.1$).

    \item Handling Snapshot Transitions and Smoothing:
    When transitioning between snapshots at different redshifts, an overlapping region of $5\,h^{-1}\,$cMpc is introduced to prevent artificial discontinuities. The electron density in this overlap region is interpolated using a weighted linear approach:
    \begin{equation}
        n_{e, \text{overlap}}(i) = \frac{w_{\text{low}} n_{e, \text{low}}(i) + w_{\text{up}} n_{e, \text{up}}(i)}{w_{\text{low}} + w_{\text{up}}},
    \end{equation}
    where the weights are defined as:
    \begin{equation}
        w_{\text{low}} = \frac{N_{\text{overlap}} - i}{N_{\text{overlap}}}, \quad
        w_{\text{up}} = \frac{i}{N_{\text{overlap}}}.
    \end{equation}
    Here, $N_{\text{overlap}}$ represents the number of bins within the overlapping region (typically 13 bins for lv8). This interpolated region is then inserted at the interface between snapshots, and the DM calculation is repeated following the method described above.
\end{itemize}

\section{Implications of Random Sightline Placement}
\label{APP:halo_los}

In our large-scale light cone simulation, FRBs are modeled as occurring along random sight lines through the cosmic web, without explicitly fixing their origin or termination within halos.
This assumption allows for efficient statistical sampling but does not guarantee a ``halo-to-halo'' configuration. 
As a result, individual sight lines may start or end inside, near, or outside of halos.

In the real Universe, FRBs originate in galaxies and terminate in the MW.
Our use of random sight lines may therefore lead to a modest underestimation of total DM, compared to an idealized configuration in which all sight lines pass through both a HH and the MW halo.
This effect is partially mitigated by averaging over a large number of sight lines, which collectively sample diverse environments.

To assess this potential offset, we refer to our dedicated HH analysis (see Section~\ref{sec:Host_Galaxy}, Figure \ref{fig:Combined-curves}e, f), where the local DM contribution from FRBs varies by environment:
\begin{itemize}
    \item Galaxy Cluster (central): $\sim$1477--2003~pc~cm$^{-3}$ (median: 1670.16)
    \item MW-like Galaxy (central): $\sim$83--813~pc~cm$^{-3}$ (median: 163.53)
    \item MW-like Galaxy (off-center): $\sim$6.38--28.42~pc~cm$^{-3}$ (median: 11.59)
    \item Dwarf Galaxy (central): $\sim$2.3--9.6~pc~cm$^{-3}$ (median: 4.13)
    \item Dwarf Galaxy (off-center): $\sim$0.01--0.02~pc~cm$^{-3}$
\end{itemize}

\noindent
These values estimate the systematic DM contribution omitted by random-LOS sampling.
Accordingly, we caution that our DM results represent only the cosmic contribution outside the MW and host galaxies, which are analyzed separately in Section \ref{subsec:DM_in_LSS}.

\section{Impact of AGN Feedback on 2D DM Distributions in Massive Halos} \label{sec:AGN_DM_mapping}

\begin{figure*}[h]
    \centering
    \begin{tabular}{cc}
        \includegraphics[width=0.4\textwidth]{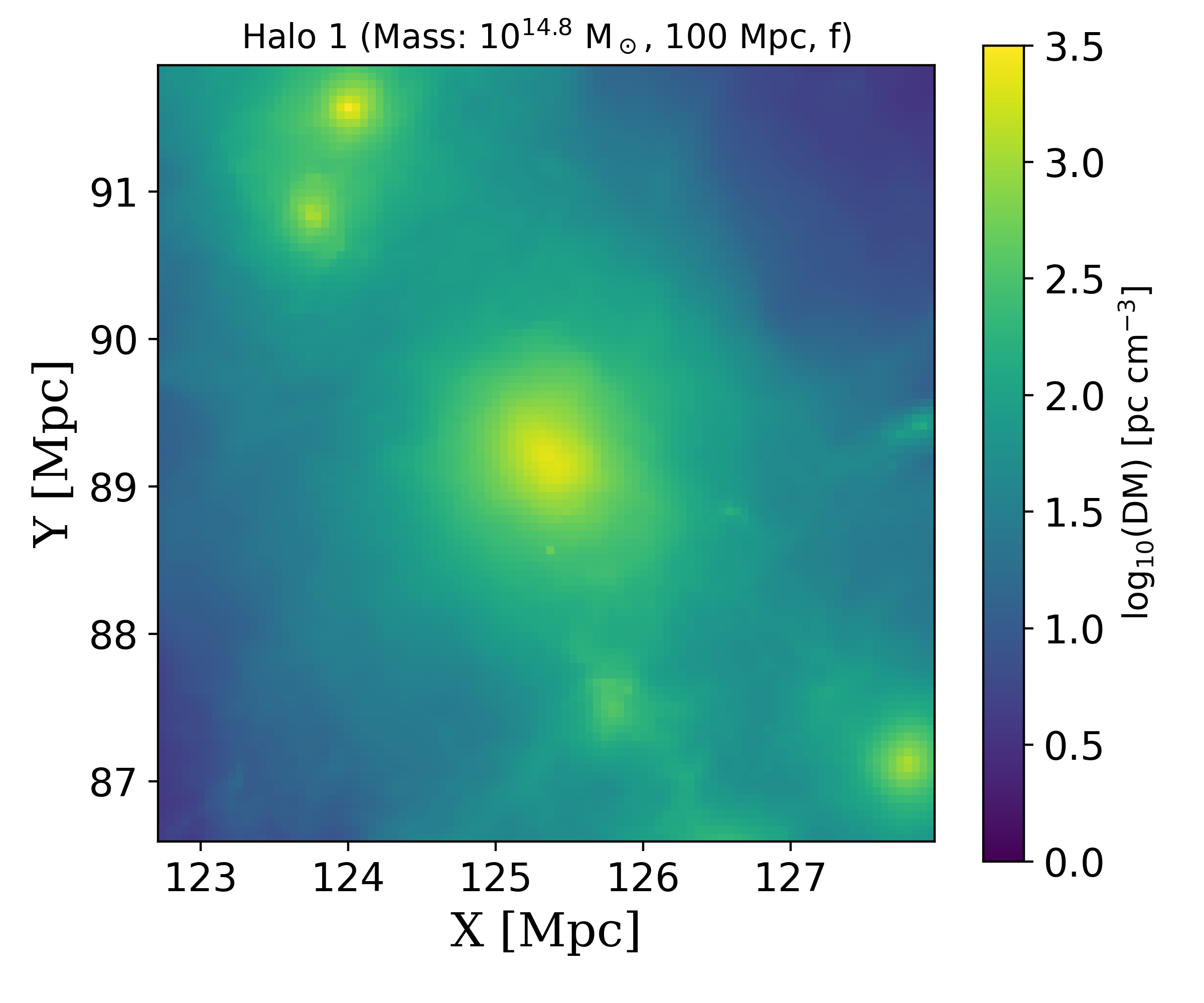} &
        \includegraphics[width=0.4\textwidth]{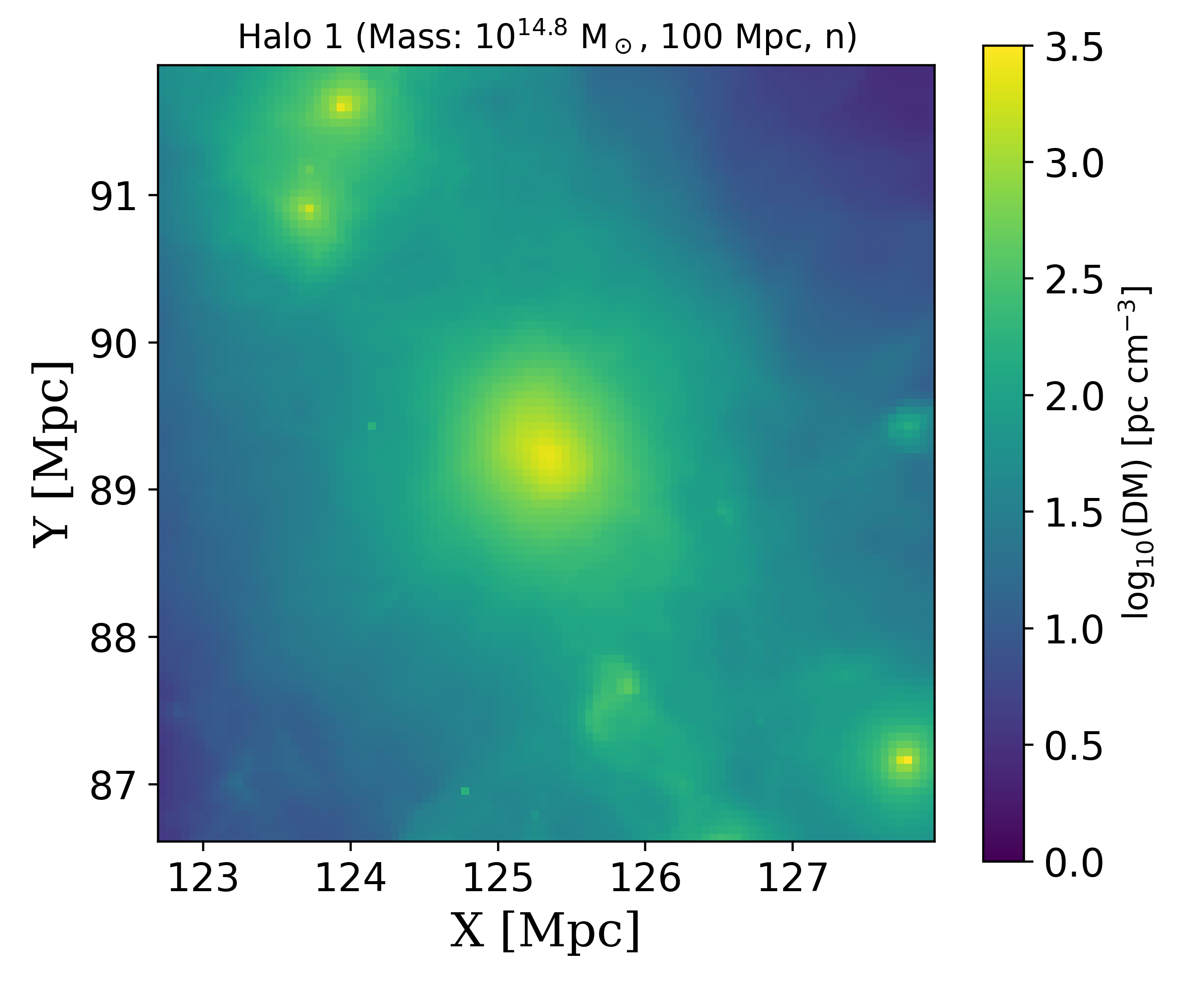} \\
        \includegraphics[width=0.35\textwidth]{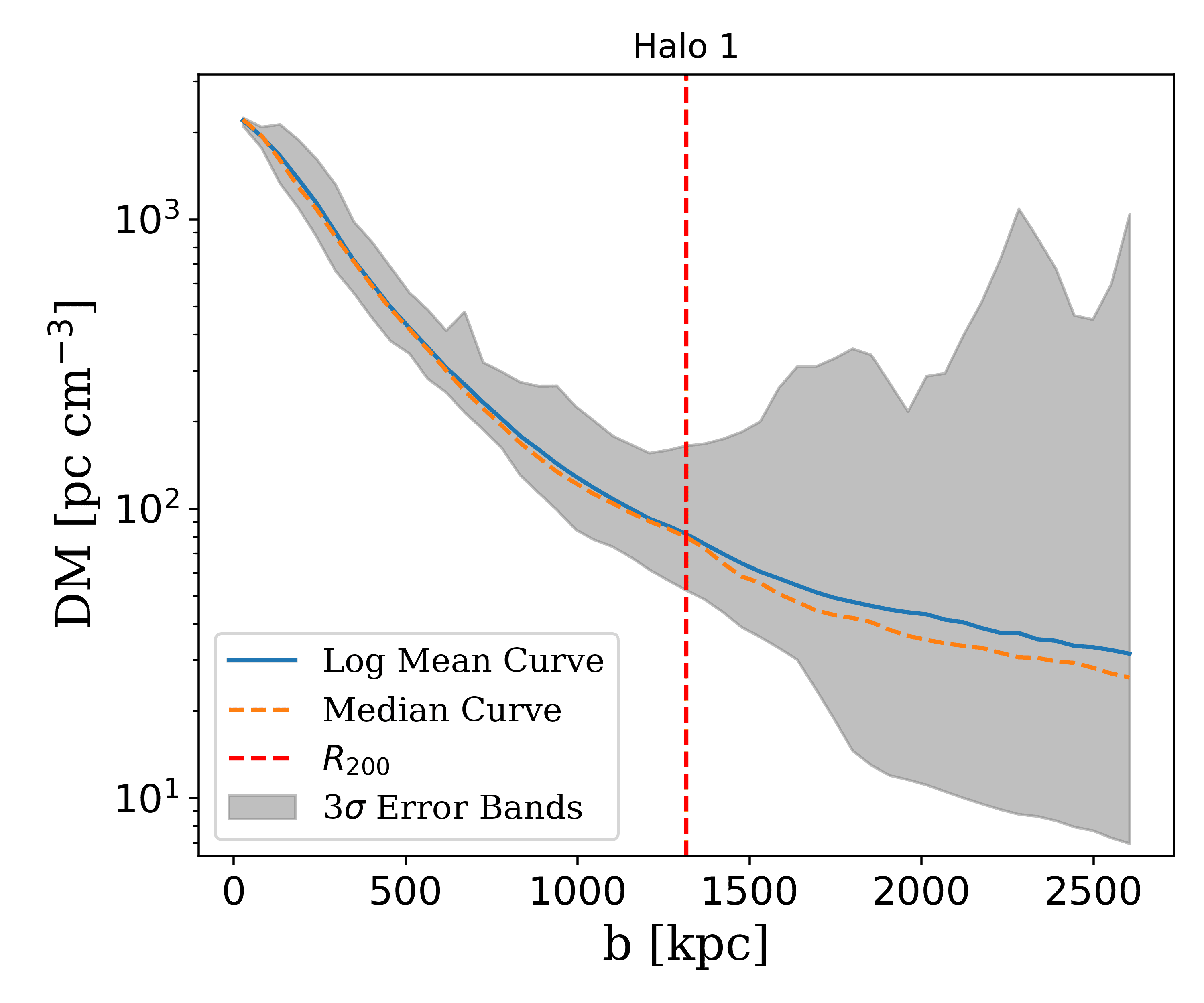} &
        \includegraphics[width=0.35\textwidth]{1_100_f_dm_b_relation.png} \\
        \includegraphics[width=0.4\textwidth]{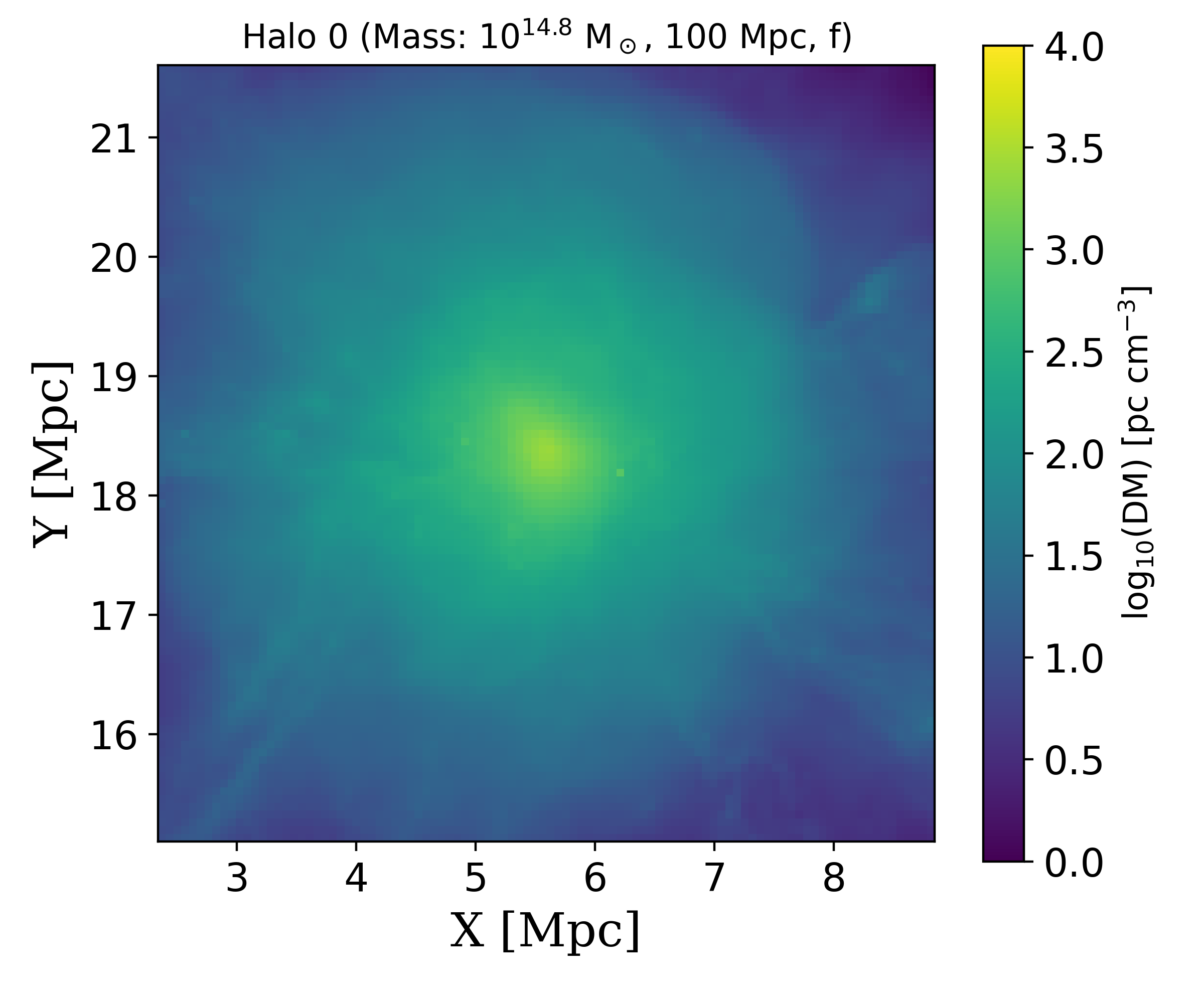} &
        \includegraphics[width=0.4\textwidth]{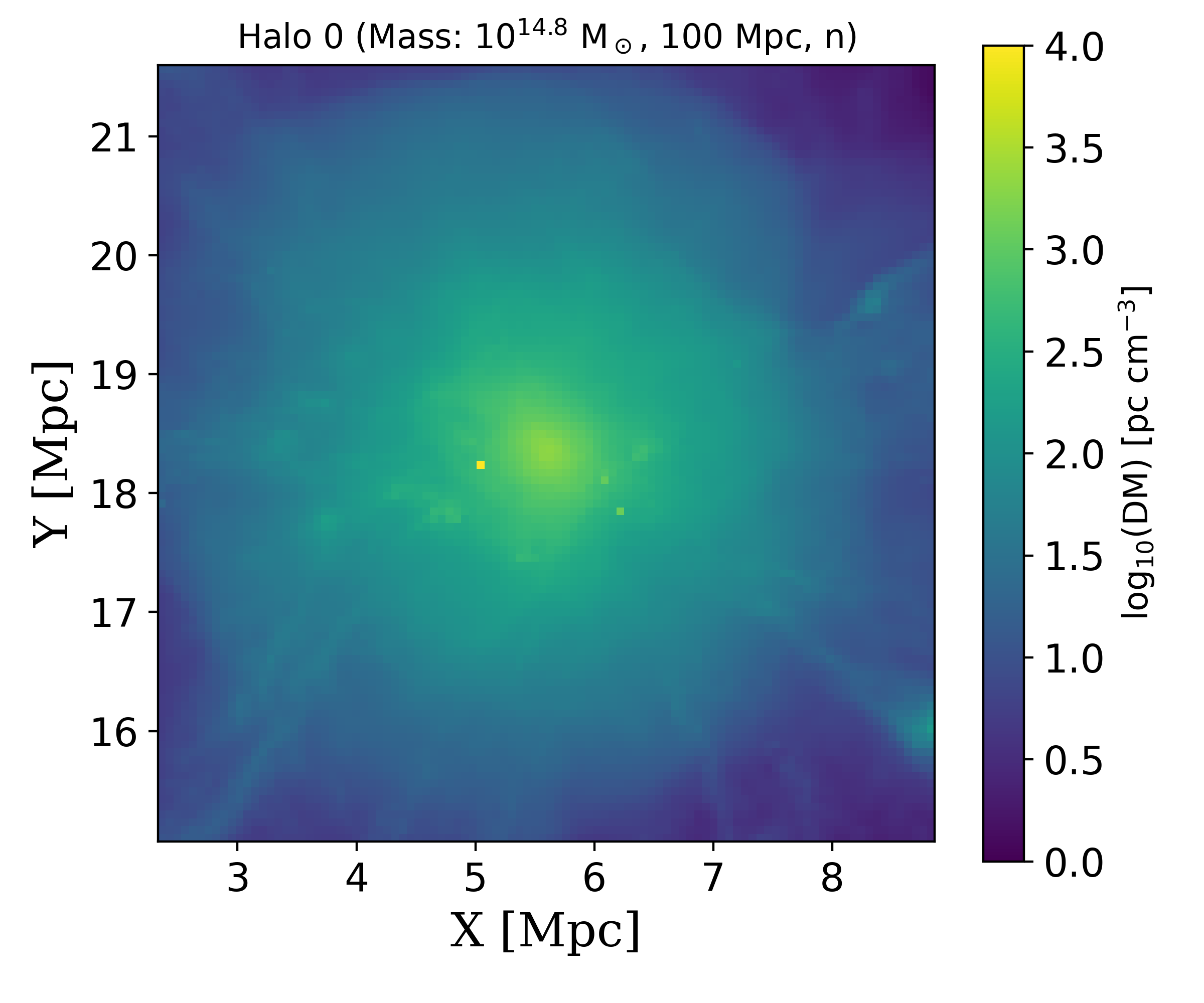} \\
        \includegraphics[width=0.35\textwidth]{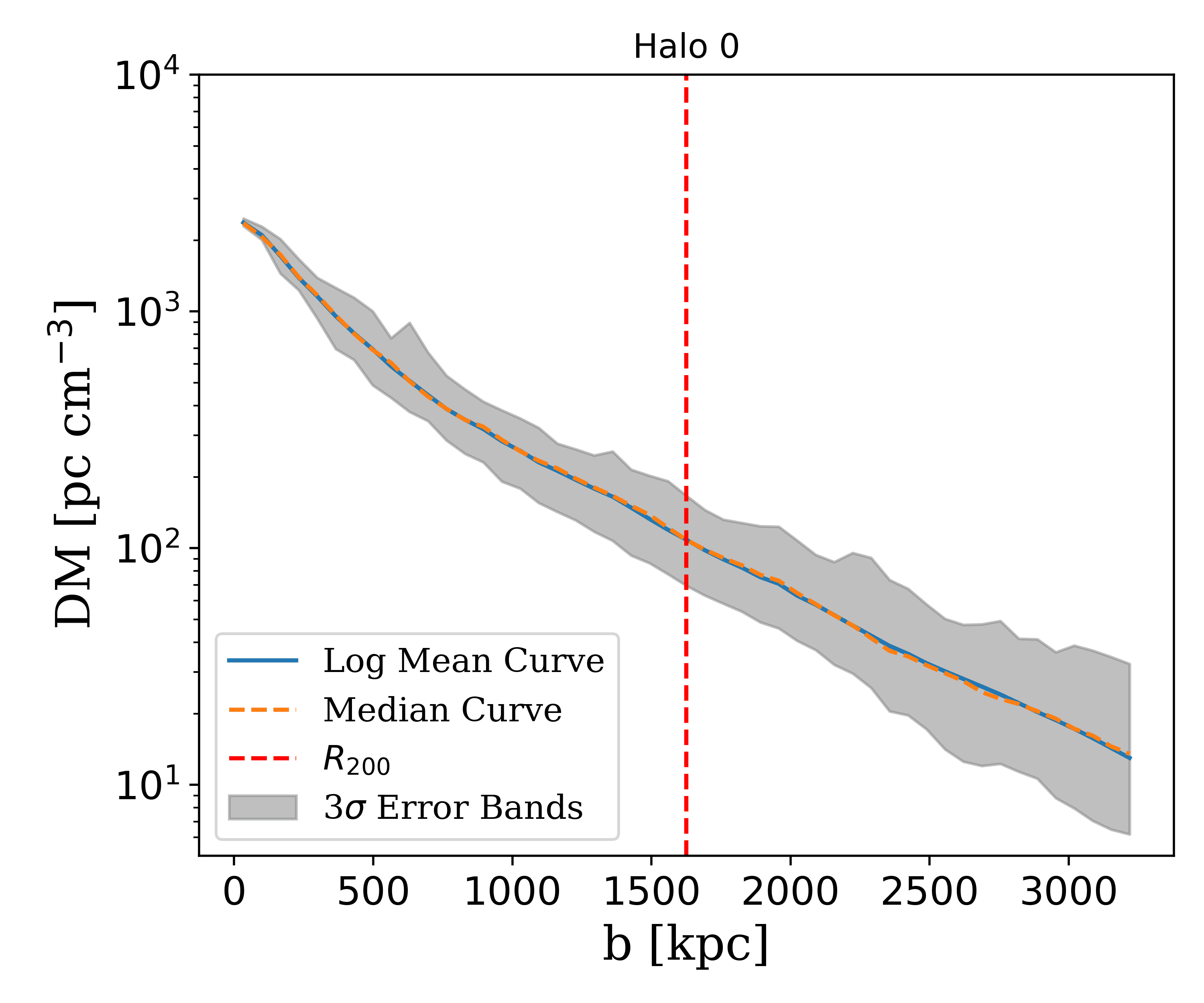} &
        \includegraphics[width=0.35\textwidth]{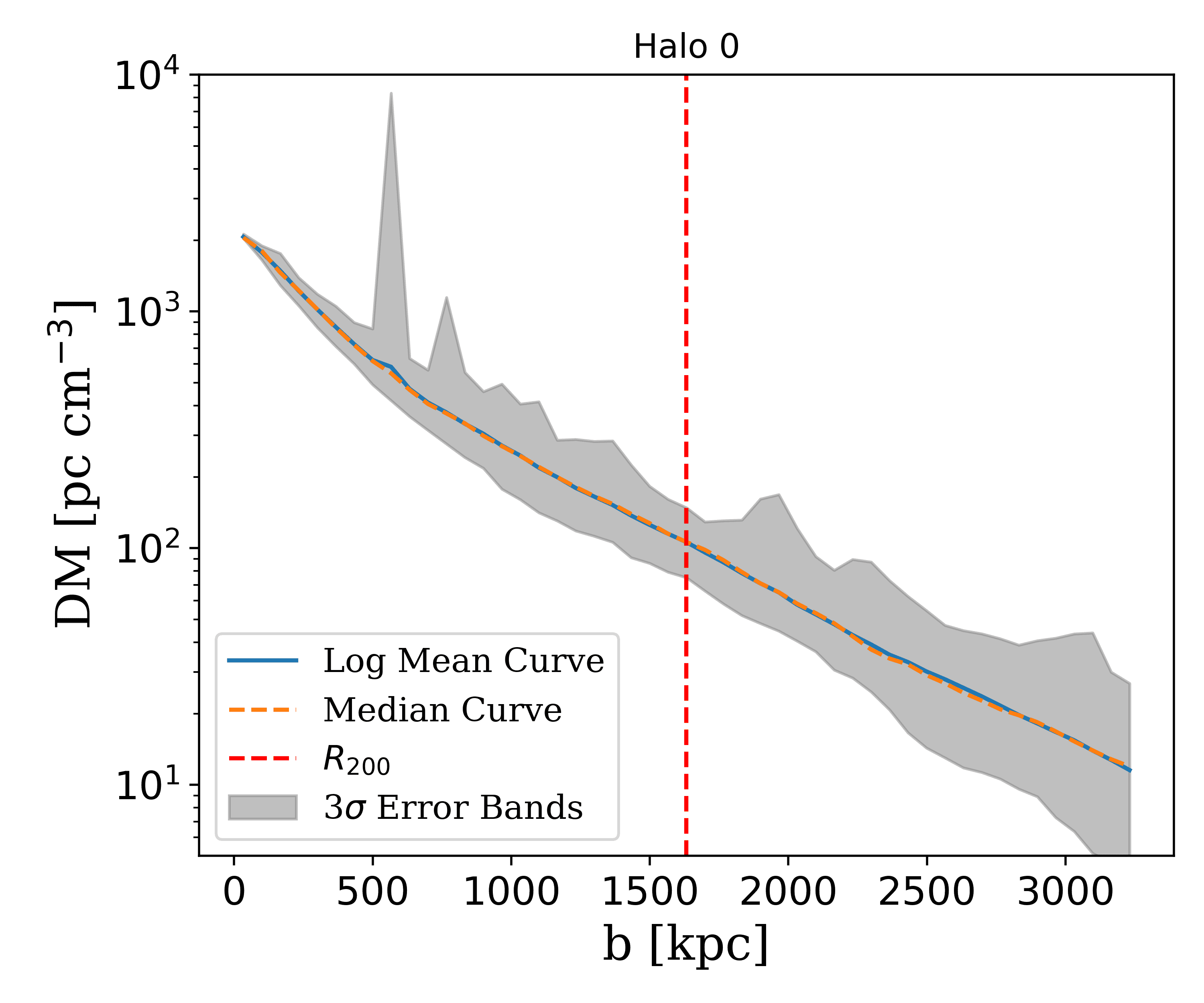} \\
    \end{tabular}
    \caption{Similar to Figure~\ref{fig:DM_profiles_500}, this figure compares 2D DM maps (top row) and 1D DM--$b$ profiles (bottom row) for halos in the highest mass range, focusing on Halo 0 and Halo 1. Results with AGN feedback (L100N1024$_{\text{fiducial}}$) are shown in the left column, and results without AGN feedback (L100N1024$_{\text{NoBH}}$) are shown in the right column.}
    \label{fig:DM_profiles_14-15}
\end{figure*}

\begin{figure*}[h!]
    \centering
    \begin{tabular}{cc}
        \includegraphics[width=0.4\textwidth]{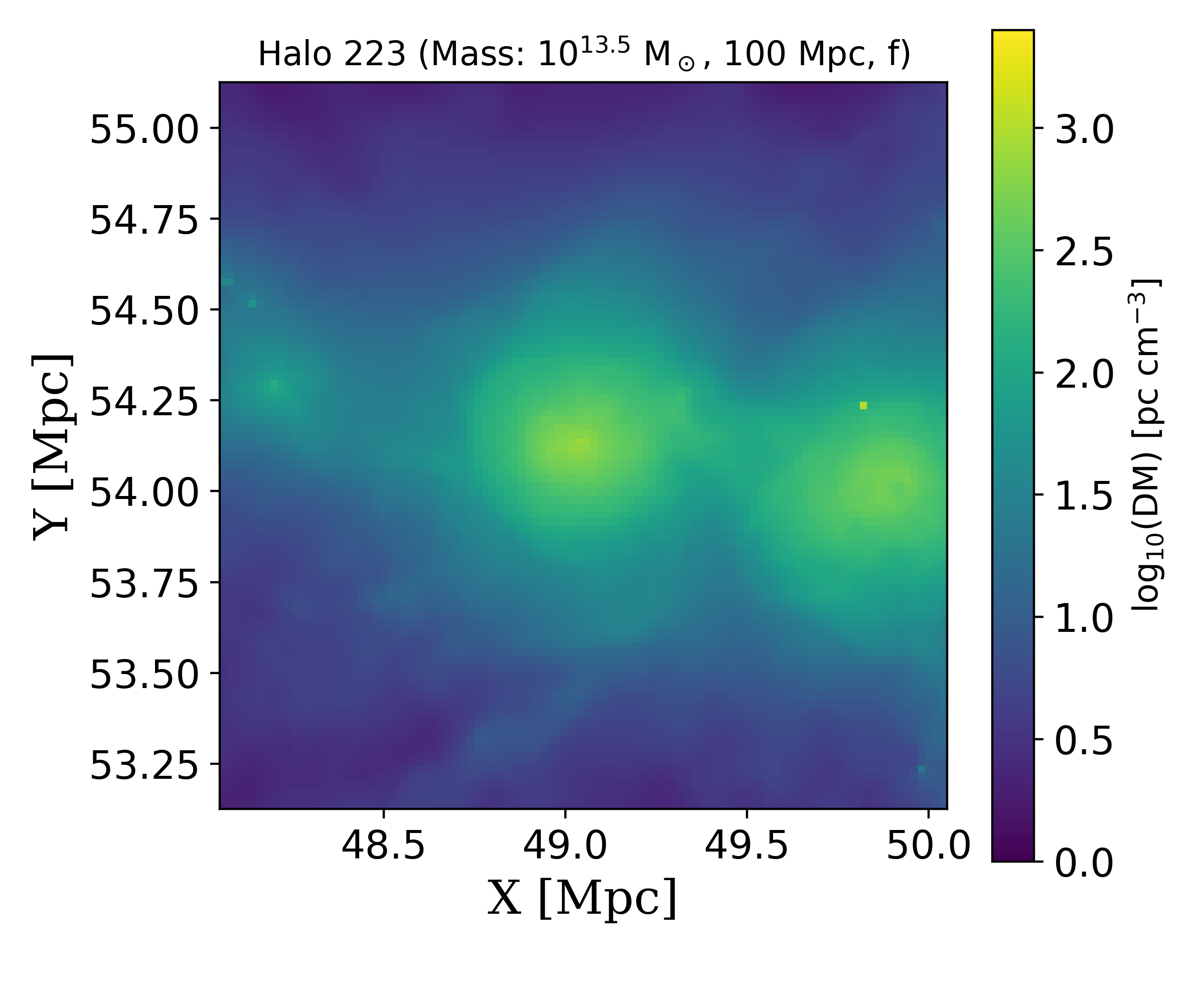} &
        \includegraphics[width=0.4\textwidth]{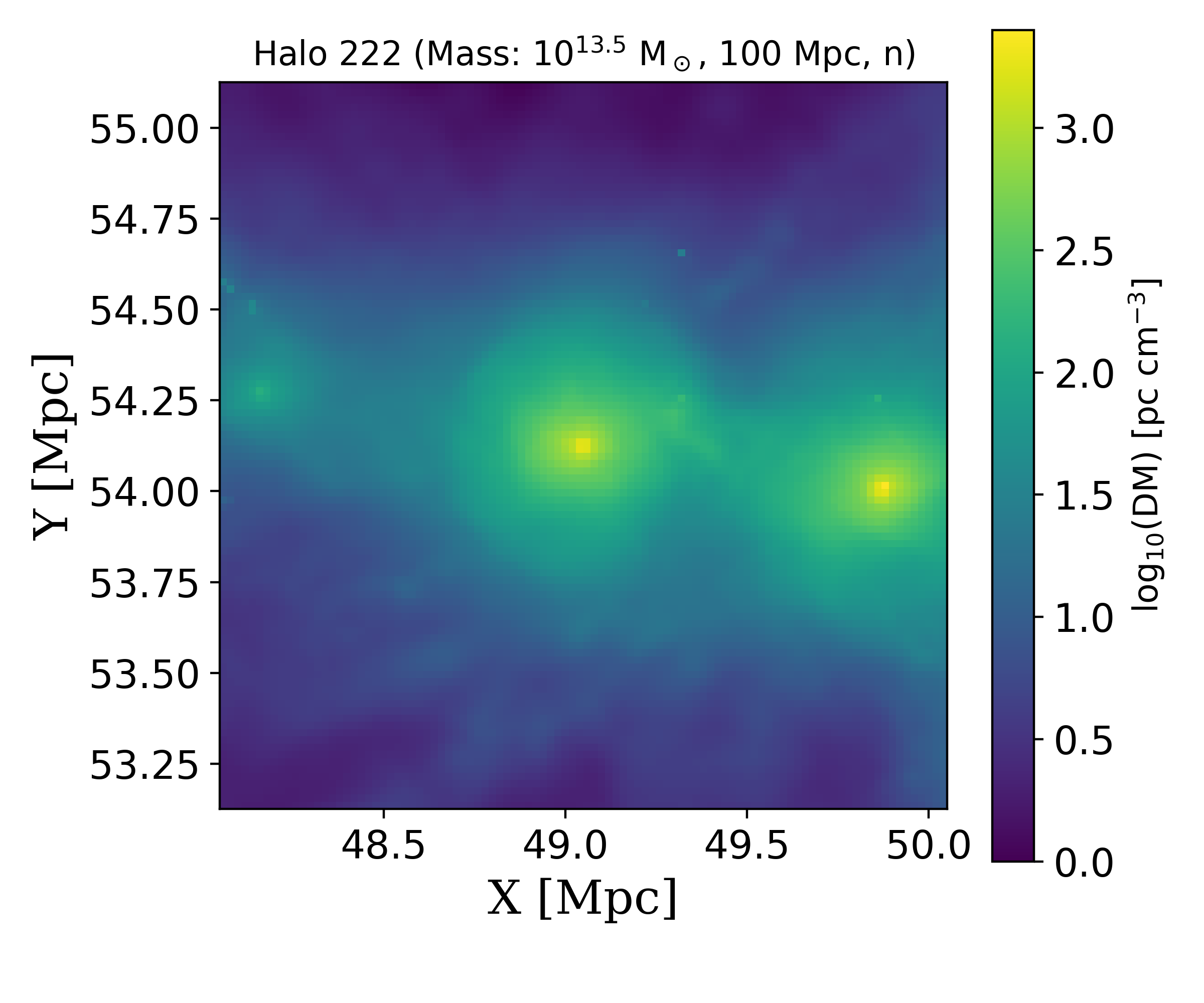} \\
        \includegraphics[width=0.35\textwidth]{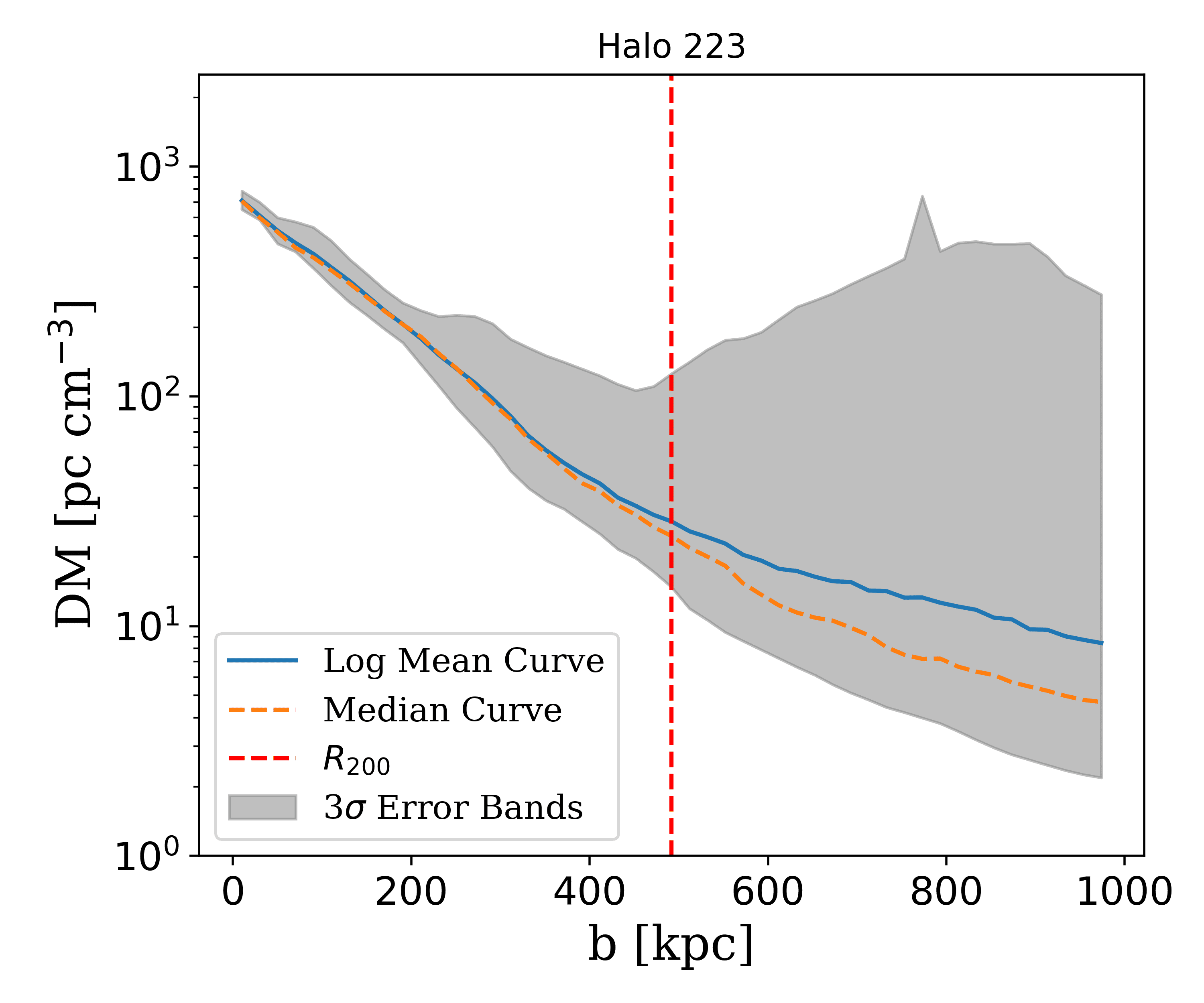} &
        \includegraphics[width=0.35\textwidth]{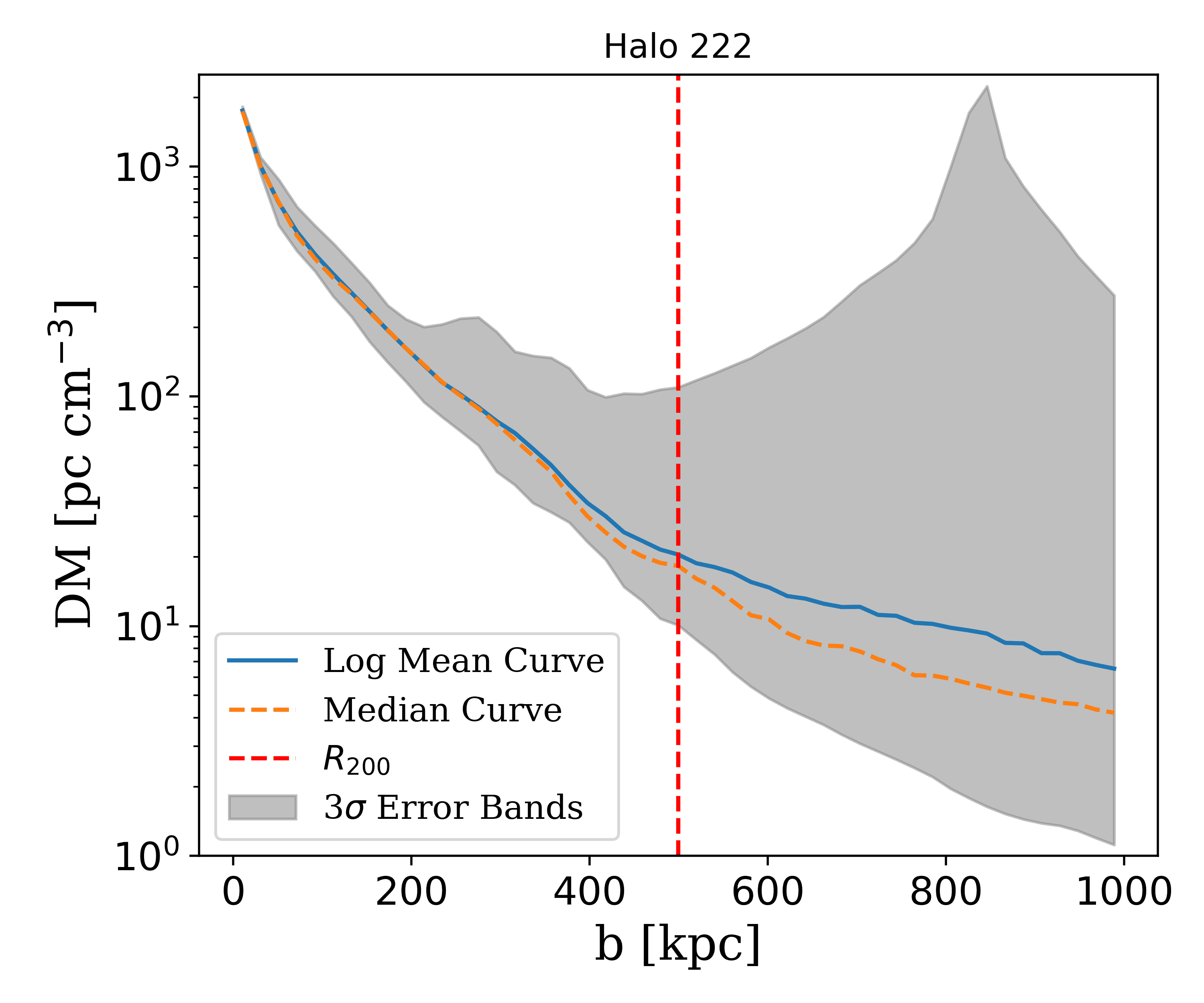} \\
        \includegraphics[width=0.4\textwidth]{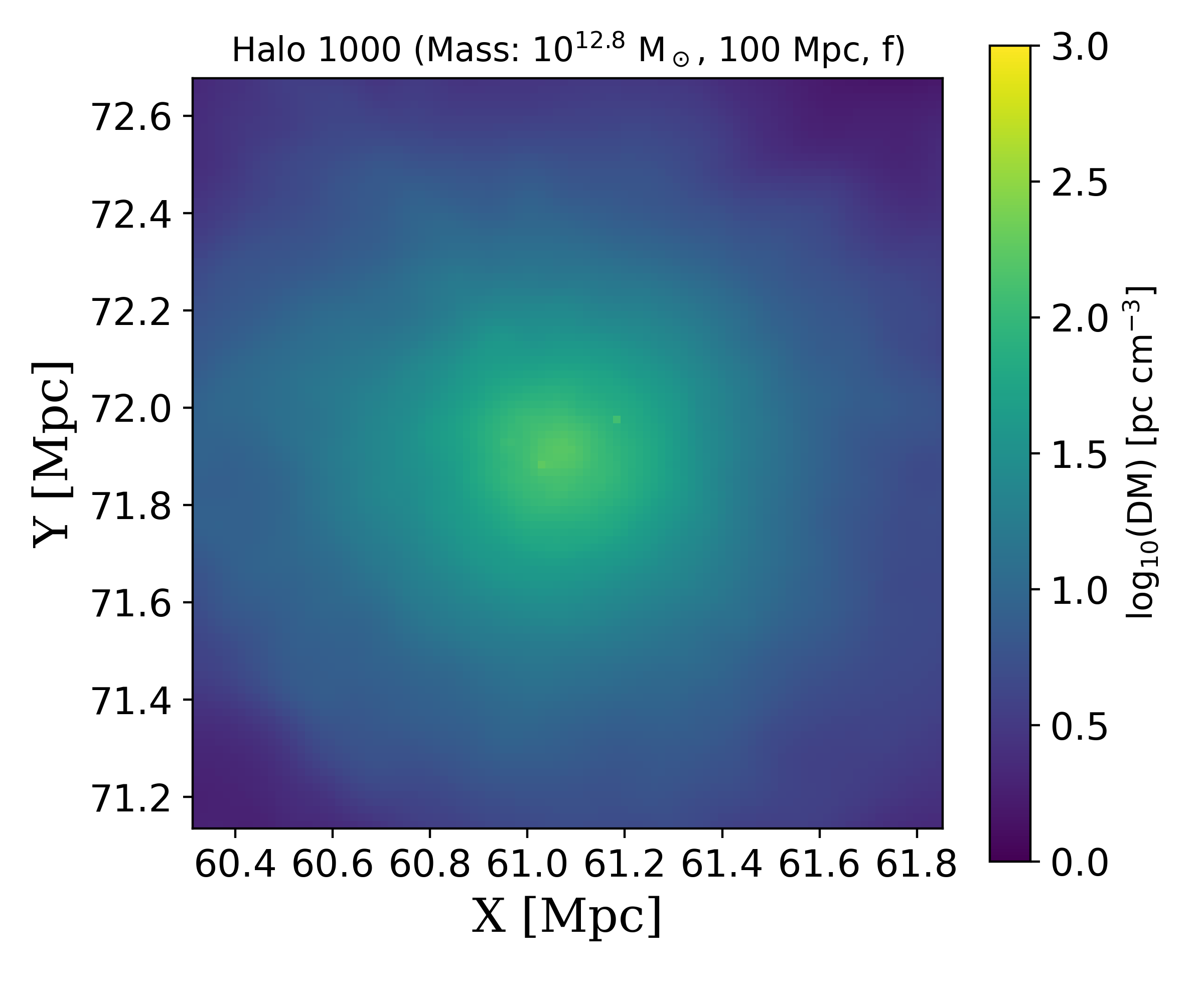} &
        \includegraphics[width=0.4\textwidth]{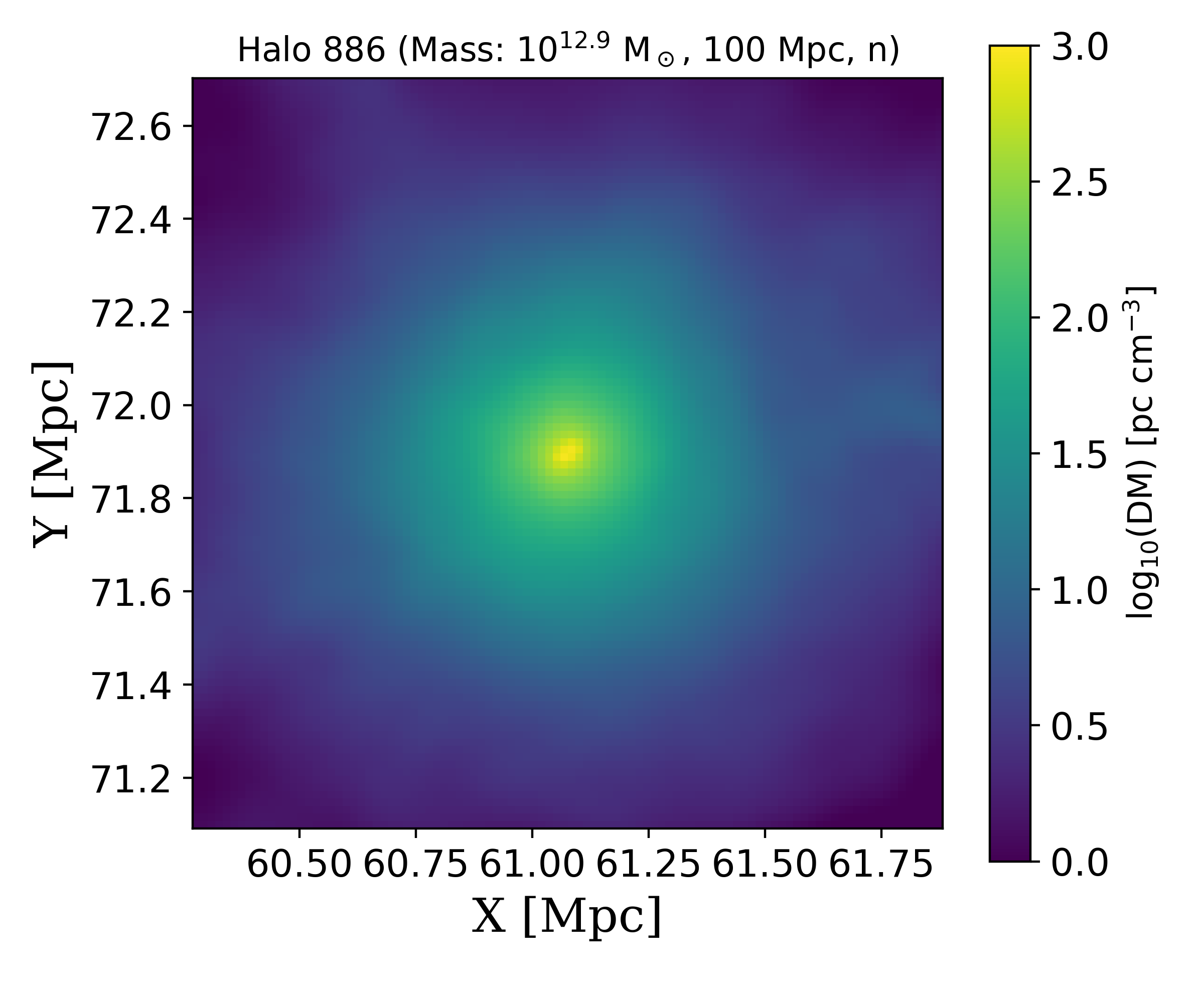} \\
        \includegraphics[width=0.35\textwidth]{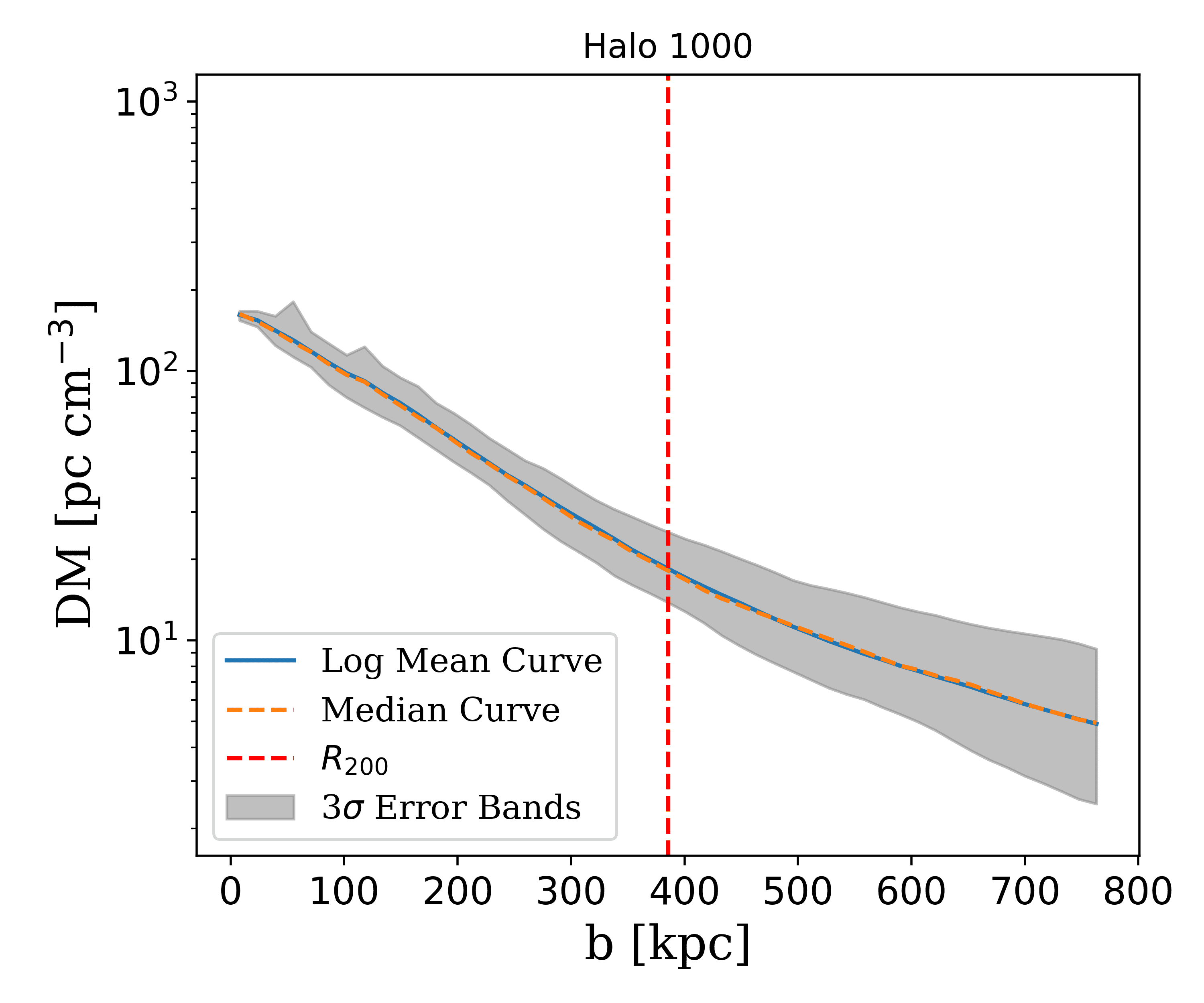} &
        \includegraphics[width=0.35\textwidth]{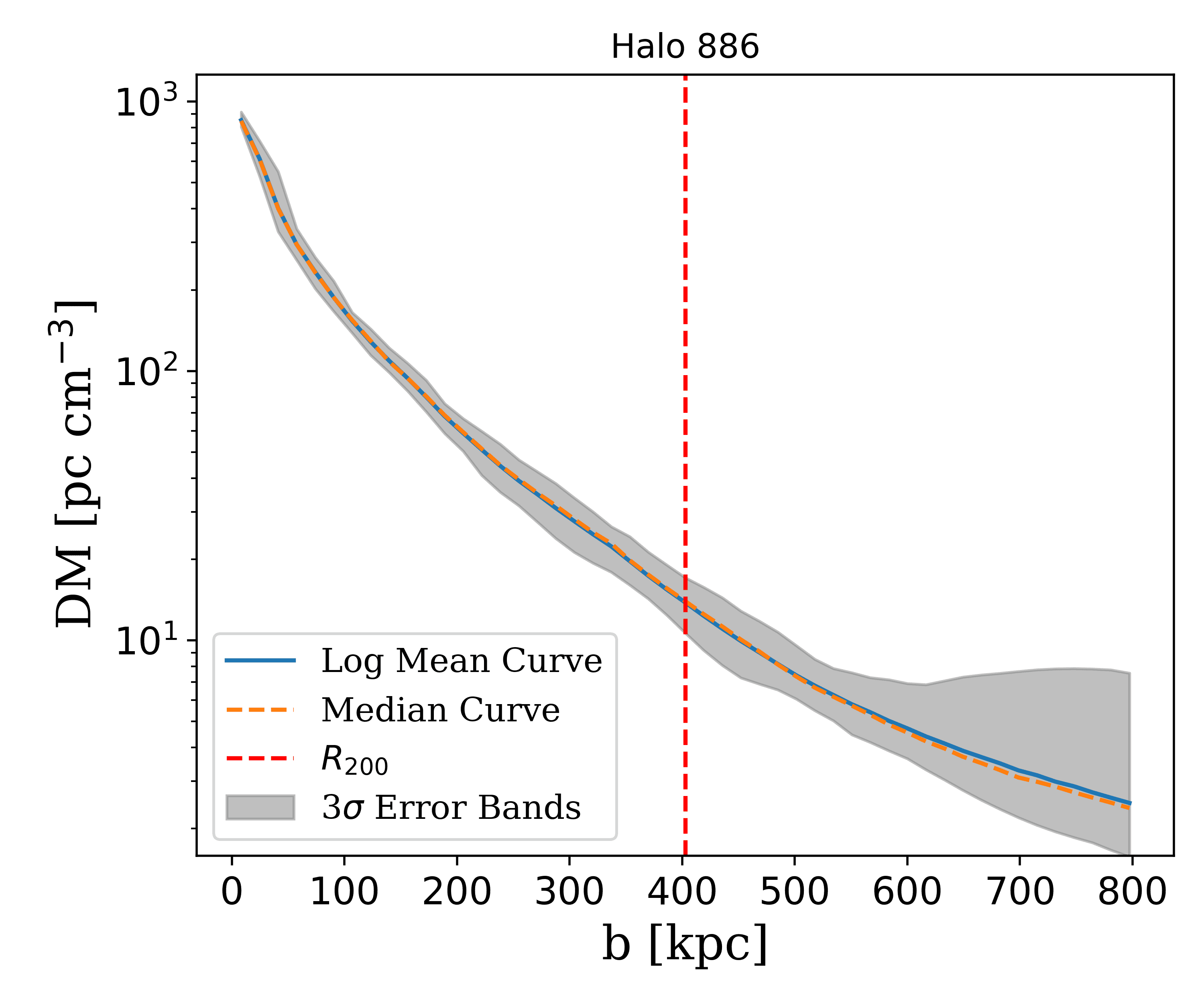} \\
    \end{tabular}
     \caption{Similar to Figure~\ref{fig:DM_profiles_500}, this figure is for halos in the mass range of $10^{12.5} \text{-} 10^{13.5} \, M_{\odot}$, where AGN feedback is expected to be more efficient. The left column shows results from the fiducial simulation (L100N1024$_{\text{fiducial}}$), while the right column corresponds to the NoBH simulation (L100N1024$_{\text{NoBH}}$). The impact of AGN feedback is evident, particularly in reducing central DM contributions.}
    \label{fig:DM_profiles_12-14}
\end{figure*}

Here we present additional visualizations of the DM distributions in halos from the L100 simulations. Figures~\ref{fig:DM_profiles_14-15} and~\ref{fig:DM_profiles_12-14} compare the effects of AGN feedback in two different halo mass ranges.

Figure~\ref{fig:DM_profiles_14-15} shows the 2D DM maps and 1D DM--$b$ profiles of two representative massive halos ($\sim 10^{14.8} M_{\odot}$) in both the fiducial and NoBH simulations. The analysis of these halos suggests that AGN feedback has a limited impact on the central gas distribution in the most massive halos but may still affect their outskirts.

Figure~\ref{fig:DM_profiles_12-14} illustrates the effect of AGN feedback in the $10^{12.5}-10^{13.5} M_{\odot}$ mass range, where feedback mechanisms are expected to be more effective. The comparison highlights the redistribution of gas and its impact on DM profiles due to AGN feedback.

These results support the findings discussed in Section~\ref{sec:2D_DM_individual} of the main text, emphasizing the varying efficiency of AGN feedback across different halo mass ranges.

\section{Further Clarification on the Definition and Interpretation of \fdiffmath}
\label{APP:f_diff_appendix}
In Section~\ref{sec:FRB_DM}, we defined the diffuse baryon fraction, $f_{\mathrm{diff}}$, as the sum of those in the IGM and halos. However, both components are redshift-dependent, and thus $f_{\mathrm{diff}}$ evolves with redshift as $f_{\mathrm{diff}}(z) = f_{\mathrm{IGM}}(z) + \fhalo(z)$ (see Figure~\ref{fig:fdiff_fIGM_fCGM}). This relation can be rigorously obtained from cosmological simulations. However, the value of $f_{\mathrm{diff,\,obs}}$ constrained observationally via Equation~\ref{eq:dm_igm} represents a distinct concept. It reflects the integrated contribution of diffuse gas and halos along the entire sightline up to a given redshift $z$, weighted by the DM contribution of intersected structures.

More precisely, for an FRB located at redshift $z=z_i$ providing a sightline $i$, the observed fraction can be written as
\begin{equation}
    f^{(i)}_{\mathrm{diff,\,obs}}(z<z_i) = f^{(i)}_{\mathrm{IGM}}(z<z_i) + f^{(i)}_{\mathrm{Halos}}(z<z_i),
\end{equation}
which depends not only on the redshift integration but also on the intervening halo distribution along the LoS. When we take a median or logarithmic mean over many FRBs, this yields a statistical measure $\langle f_{\mathrm{diff,\,obs}}(<z) \rangle$  that reflects both cosmic structure and observational selection effects. To achieve this goal, one would need a well-defined sample of localized FRBs with good redshift measurements.

\begin{figure*}
    \centering
    \includegraphics[width=0.8\textwidth]{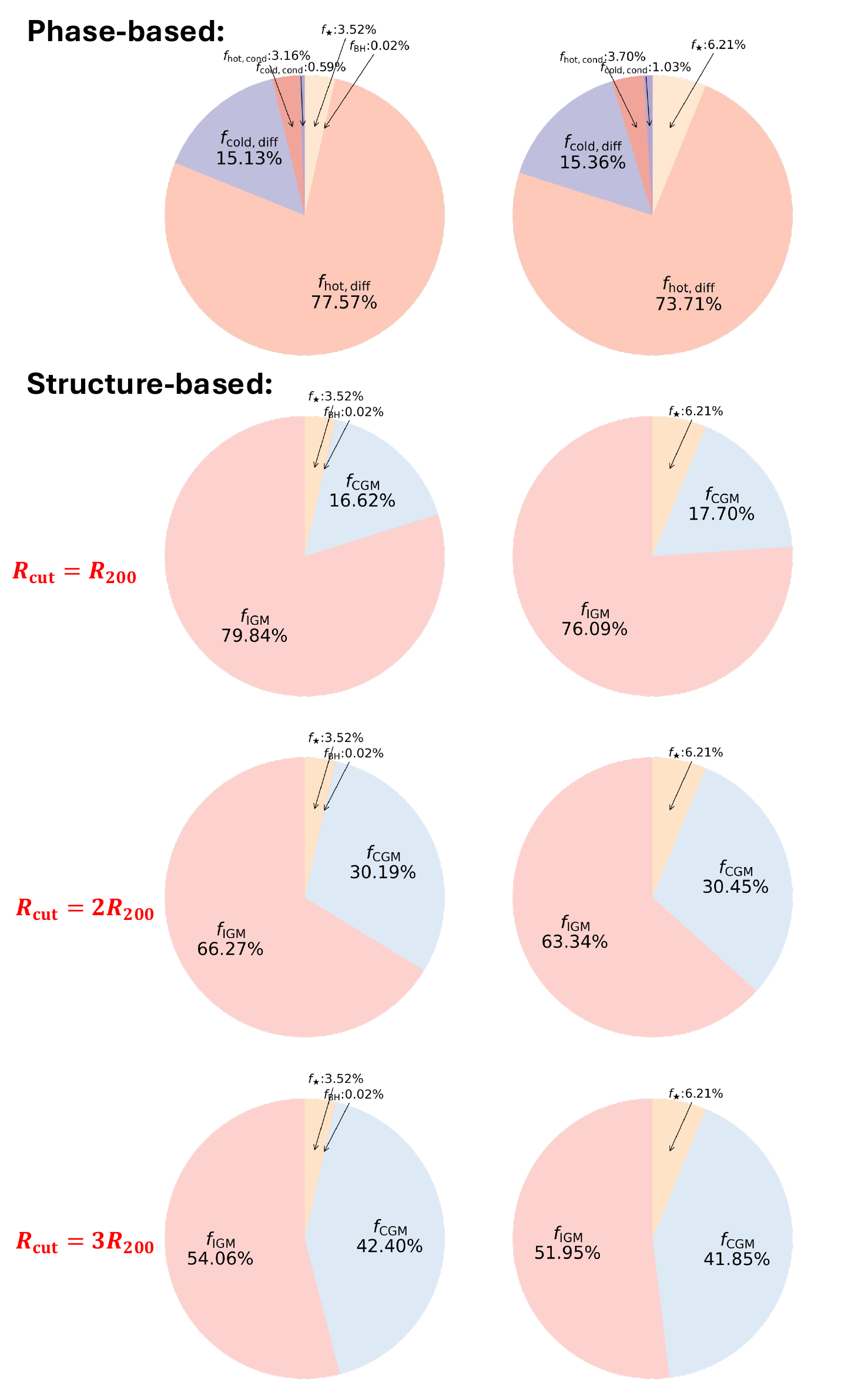}
    \caption{
    Pie chart comparison of baryon composition at $z = 0$ under two classification schemes: gas phase-based (top row) and structure-based definitions (bottom three rows).
    The top row shows the distribution between cold/hot IGM/ phases.
    The remaining panels show baryon fractions under different CGM definitions based on $R_{\mathrm{cut}} = R_{200}, 2R_{200}, 3R_{200}$ for both fiducial (left) and NoBH (right) simulations.
    }
    \label{fig:piecharts}
\end{figure*}

Furthermore, Figure~\ref{fig:piecharts} illustrates the composition of cosmic baryons at $z = 0$ in our CROCODILE simulation in terms of both physical phases and large-scale structures. Baryons are primarily composed of gas, with only minor contributions from stars and black holes.

In the {\it phase-based} classification, gas is divided using a temperature threshold of $10^4$ K to separate {\it cold} and {\it hot} components, and a baryon overdensity threshold of $\log_{10}(\rho_{\rm gas}/\bar{\rho}_b) = 3.5$ to distinguish {\it diffuse} from {\it dense}  phases. In contrast, the {\it structure-based} classification is based on the CGM subtraction method (Section~\ref{subsubsec:CGM_substraction}), in which gas within halos is excluded based on a cutoff radius $R_{\mathrm{cut}}$.

This leads to two distinct definitions of the diffuse baryon fraction:
\begin{align}
f_{\mathrm{diff,\,phase}} &= f_{\mathrm{cold,\,diff}} + f_{\mathrm{hot,\,diff}} \nonumber\\ 
                          &= 1 - f_{\mathrm{cold,\,dense}} - f_{\mathrm{hot,\,dense}} - f_\star - f_{\mathrm{BH}}, \\
f_{\mathrm{diff,\,struct}} &= f_{\mathrm{CGM}} + f_{\mathrm{IGM}} \nonumber\\
                            &= 1 - f_\star - f_{\mathrm{BH}}.
\end{align}
Each component can be measured directly from simulations, but cannot be independently constrained from FRB DM observations. For $f_{\mathrm{diff,\,struct}}$, we do not distinguish between the ISM and intra-cluster medium (ICM), as our CGM subtraction method does not depend on halo mass or internal structure --- these components are effectively included in the CGM term.

By comparing fiducial and NoBH simulations, we find that the fiducial model yields more hot and diffuse gas in the phase-based decomposition, reflecting AGN-driven heating and gas ejection. Similarly, the structure-based decomposition shows a larger IGM fraction and a smaller CGM fraction in the fiducial case, indicating baryon redistribution due to AGN feedback. Additionally, the stellar mass fraction is clearly lower in the fiducial model, confirming suppression of star formation by AGN.

We find $f_{\mathrm{diff,\,phase}} = f_{\mathrm{cold,\,diff}} + f_{\mathrm{hot,\,diff}} \approx 93\%$ at $z = 0$, in agreement with estimates from \citet{Connor2024}. Assuming that the IGM is primarily composed of hot diffuse gas (with $f_{\mathrm{hot,\,diff}} \approx 78\%$ in fiducial and $74\%$ in NoBH), our values closely match those from the structure-based method with $R_{\mathrm{cut}} = R_{200}$, where $f_{\mathrm{IGM}} \approx 80\%$ (fiducial) and $76\%$ (NoBH) at $z=0$. As $R_{\mathrm{cut}}$ increases, the balance between CGM and IGM varies significantly, reaffirming the sensitivity of these components to the adopted halo boundary.

The quantity $\langle f_{\mathrm{diff,\,obs}}(<z) \rangle$ represents a redshift-integrated and sightline-averaged observable derived from many FRB sight lines. {\em It is not a snapshot of the instantaneous baryon fraction at a given redshift, but rather a statistical measure shaped by both the redshift integration and the distribution of intervening halos along the LoS.}  For example, to interpret an FRB observed at $z=1$, one must sample many sight-lines in a light-cone data up to $z=1$, effectively stacking the contributions from all structures within that redshift range. As such, $\langle f_{\mathrm{diff,\,obs}}(<z) \rangle$ can be understood as a weighted average over the redshift-dependent curves of (instantaneous or differential) $f_{\mathrm{CGM}}(z)$ and $f_{\mathrm{IGM}}(z)$ shown in Figure \ref{fig:f_structure_evolution} (Appendix~\ref{app:baryon_evolution}).

While the IGM contribution dominates due to its nearly universal presence in all directions, the halo contribution varies significantly depending on redshift and sightline geometry. At low redshift, many FRBs may not intersect any foreground halos, leading to an underestimated $\langle f_{\mathrm{diff,\,obs}} \rangle$, as seen in the behavior of the Macquart relation fit toward $z=0$ in Figure~\ref{fig:f_IGM_redshift}b and Table~\ref{tab:fitting_parameters} ($f_{\mathrm{diff}}(z\rightarrow0)$). At higher redshifts, although the number of intersecting halos increases, their relative DM contributions decrease due to smaller physical sizes and increased redshift dilution. Thus, $\langle f_{\mathrm{diff,\,obs}} \rangle$ encodes both the underlying cosmic gas distribution and observational sample bias effects.

\begin{table*}[h]
    \centering
    \caption{Fitting parameters and $\chi^2$ values for C-Exp, CPL-Exp, and DPL-Exp.}
    \begin{tabular}{llccccc}
        \hline
        Model & Case & $\kappa$ & $\tau$ & $\zeta$ & $\gamma$ & $\chi^2$/DOF \\
        \hline
        \multirow{12}{*}{C-Exp} 
        & $f_{\mathrm{IGM}}$:& $\kappa$ & $\tau$ & - & - & $\chi^2$/DOF\\
        & Fiducial (FoF) & $0.27^{+0.02}_{-0.02}$ & $5.44^{+0.56}_{-0.51}$ & - & - & 0.142 \\
        & NoBH (FoF) & $0.28^{+0.02}_{-0.02}$ & $5.03^{+0.54}_{-0.39}$ & - & - & 0.127 \\
        & $R_{\mathrm{cut}} = R_{200}$ (Fiducial) & $0.30^{+0.01}_{-0.01}$ & $5.37^{+0.21}_{-0.24}$ & - & - & 0.017 \\
        & $R_{\mathrm{cut}} = R_{200}$ (NoBH) & $0.33^{+0.02}_{-0.02}$ & $4.51^{+0.28}_{-0.28}$ & - & - & 0.037 \\
        & $R_{\mathrm{cut}} = 2R_{200}$ (Fiducial) & $0.27^{+0.02}_{-0.01}$ & $3.76^{+0.20}_{-0.30}$ & - & - & 0.107 \\
        & $R_{\mathrm{cut}} = 2R_{200}$ (NoBH) & $0.28^{+0.01}_{-0.01}$ & $3.42^{+0.17}_{-0.23}$ & - & - & 0.106 \\
        & $R_{\mathrm{cut}} = 3R_{200}$ (Fiducial) & $0.24^{+0.02}_{-0.02}$ & $2.67^{+0.34}_{-0.33}$ & - & - & 1.371 \\
        & $R_{\mathrm{cut}} = 3R_{200}$ (NoBH) & $0.24^{+0.02}_{-0.01}$ & $2.53^{+0.24}_{-0.33}$ & - & - & 1.284 \\
        \cline{2-7}
        & $f_{\mathrm{diff}}$ (Fiducial) & $0.87^{+0.05}_{-0.09}$ & $1.32^{+0.23}_{-0.10}$ & - & - & 0.057 \\
        & $f_{\mathrm{diff}}$ (NoBH) & $0.90^{+0.11}_{-0.14}$ & $1.22^{+0.36}_{-0.16}$ & - & - & 0.129 \\
        \hline
        & $f_{\mathrm{diff}}(z\rightarrow0)$ (Fiducial) & \multicolumn{5}{c}{$0.685$} \\
        & $f_{\mathrm{diff}}(z\rightarrow0)$ (NoBH) & \multicolumn{5}{c}{$0.666$} \\
        \hline\hline

        \multirow{12}{*}{CPL-Exp} 
        & $f_{\mathrm{IGM}}$:& $\kappa$ & $\tau$ & $\zeta$ & $\gamma$ & $\chi^2$/DOF\\
        & Fiducial (FoF) & $0.51^{+0.48}_{-0.51}$ & $3.74^{+14.91}_{-1.86}$ & $0.78^{+1.68}_{-0.23}$ & - & 0.143 \\
        & NoBH (FoF) & $0.62^{+0.37}_{-0.61}$ & $3.01^{+10.30}_{-1.31}$ & $0.72^{+1.24}_{-0.18}$ & - & 0.123 \\
        & $R_{\mathrm{cut}} = R_{200}$ (Fiducial) & $0.05^{+0.11}_{-0.05}$ & $9.56^{+7.33}_{-2.74}$ & $1.56^{+0.95}_{-0.35}$ & - & 0.016 \\
        & $R_{\mathrm{cut}} = R_{200}$ (NoBH) & $0.25^{+0.25}_{-0.20}$ & $5.07^{+3.69}_{-1.62}$ & $1.09^{+0.51}_{-0.23}$ & - & 0.039 \\
        & $R_{\mathrm{cut}} = 2R_{200}$ (Fiducial) & $1.00^{+0.03}_{-0.07}\times 10^{-3}$ & $13.61^{+0.25}_{-0.32}$ & $2.66^{+0.02}_{-0.02}$ & - & 0.034 \\
        & $R_{\mathrm{cut}} = 2R_{200}$ (NoBH) & $1.00^{+0.05}_{-0.01}\times 10^{-3}$ & $12.86^{+0.27}_{-2.97}$ & $2.71^{+0.01}_{-0.49}$ & - & 0.049 \\
        & $R_{\mathrm{cut}} = 3R_{200}$ (Fiducial) & $1.00^{+0.01}_{-0.05} \times 10^{-3}$ & $10.88^{+0.48}_{-0.54}$ & $2.72^{+0.04}_{-0.04}$ & - & 0.651 \\
        & $R_{\mathrm{cut}} = 3R_{200}$ (NoBH) & $1.00^{+0.05}_{-0.07} \times 10^{-3}$ & $10.58^{+0.0Ω39}_{-0.47}$ & $2.74^{+0.04}_{-0.03}$ & - & 0.560 \\
        
        \cline{2-7}
        & $f_{\mathrm{diff}}$ (Fiducial) & $1.83^{+0.12}_{-0.16}$ & $0.24^{+0.19}_{-0.23}$ & $0.37^{+0.11}_{-0.11}$ & - & 0.014 \\
        & $f_{\mathrm{diff}}$ (NoBH) & $1.86^{+0.05}_{-0.04}$ & $0.10^{+0.05}_{-0.10}$ & $0.31^{+0.03}_{-0.05}$ & - & 0.015 \\
        \hline
        & $f_{\mathrm{diff}}(z\rightarrow0)$ (Fiducial) & \multicolumn{5}{c}{$0.655$} \\
        & $f_{\mathrm{diff}}(z\rightarrow0)$ (NoBH) & \multicolumn{5}{c}{$0.604$} \\
        \hline\hline

        \multirow{12}{*}{DPL-Exp}
        & $f_{\mathrm{IGM}}$:& $\kappa$ & $\tau$ & $\zeta$ & $\gamma$ & $\chi^2$/DOF\\
        & Fiducial (FoF) & $8.89^{+2.21}_{-8.80} \times 10^{-7}$ & $18.94^{+5.11}_{-1.50}$ & $4.84^{+0.12}_{-0.34}$ & $0.67^{+4.34}_{-3.84}$ & 0.100 \\
        & NoBH (FoF) & $8.31^{+7.20}_{-2.05} \times 10^{-7}$ & $18.22^{+3.47}_{-1.16}$ & $4.92^{+0.08}_{-0.25}$ & $0.66^{+4.09}_{-3.83}$ & 0.083 \\
        & $R_{\mathrm{cut}} = R_{200}$ (Fiducial) & $8.99^{+4.27}_{-5.12} \times 10^{-7}$ & $25.32^{+1.38}_{-2.72}$ & $4.45^{+0.10}_{-0.19}$ & $0.87^{+3.68}_{-3.40}$ & 0.015 \\
        & $R_{\mathrm{cut}} = R_{200}$ (NoBH) & $1.51^{+3.20}_{-6.22} \times 10^{-6}$ & $20.60^{+1.32}_{-1.66}$ & $4.55^{+0.18}_{-0.36}$ & $0.68^{+3.69}_{-3.39}$ & 0.040 \\
        & $R_{\mathrm{cut}} = 2R_{200}$ (Fiducial) & $3.18^{+1.88}_{-2.42}\times10^{-6}$ & $23.29^{+2.43}_{-12.80}$ & $4.04^{+0.42}_{-2.32}$ & $1.03^{+3.43}_{-0.68}$ & 0.030 \\
        & $R_{\mathrm{cut}} = 2R_{200}$ (NoBH) & $3.20^{+9.71}_{-3.19} \times 10^{-4}$ & $14.55^{+7.47}_{-8.19}$ & $3.01^{+1.54}_{-1.74}$ & $1.00^{+3.54}_{-0.27}$ & 0.052 \\
        & $R_{\mathrm{cut}} = 3R_{200}$ (Fiducial) & $0.18^{+0.11}_{-0.12}$ & $8.18^{+3.74}_{-1.38}$ & $0.67^{+0.38}_{-0.19}$ & $2.03^{+0.99}_{-1.56}$ & 0.127 \\
        & $R_{\mathrm{cut}} = 3R_{200}$ (NoBH) & $0.16^{+0.12}_{-0.12}$ & $7.28^{+3.79}_{-1.36}$ & $0.74^{+0.47}_{-0.21}$ & $1.89^{+0.968}_{-1.37}$ & 0.123 \\
        \cline{2-7}
        &$f_{\mathrm{diff}}$ (Fiducial) & $1.97^{+0.00}_{-1.97}$ &$1.73^{+44.10}_{-0.10}\times10^{-8}$  & $0.03^{+4.97}_{-0.01}$& $9.82^{+4.82}_{-9.80}$&0.016 \\
        &$f_{\mathrm{diff}}$ (NoBH) & $1.92^{+0.01}_{-1.32}$ &$1.12^{+15.91}_{-1.03}\times10^{-8}$  & $0.03^{+4.97}_{-0.01}$& $9.30^{+4.41}_{-9.80}$&0.016 \\
        \hline
        &$f_{\mathrm{diff}}(z\rightarrow0)$ (Fiducial) & \multicolumn{5}{c}{$0.706$} \\
        &$f_{\mathrm{diff}}(z\rightarrow0)$ (NoBH) & \multicolumn{5}{c}{$0.684$} \\
        \hline\hline
    \end{tabular}
    \label{tab:fitting_parameters}
\end{table*}

\begin{table*}[htbp]
\centering
\caption{mNFW fitting parameters for Dark Matter and Gas Density profiles across halo mass bins.}
\renewcommand{\arraystretch}{1.3}
\begin{tabular}{cc|ccccc}
\hline
\multicolumn{2}{c|}{} & \multicolumn{5}{c}{Mass Range ($M_\odot$)} \\
Profile & Parameter
& $10^{10.5}$--$10^{11.5}$ & $10^{11.5}$--$10^{12.5}$ & $10^{12.5}$--$10^{13.5}$ & $10^{13.5}$--$10^{14.5}$ & $10^{14.5}$--$10^{15.5}$ \\
\cline{3-7}
\multirow{4}{*}{\shortstack{Dark\\Matter}}
& $y_0$    & $4.89^{+3.48}_{-3.30}$ & $5.18^{+3.28}_{-3.45}$ & $5.04^{+3.34}_{-3.31}$ & $4.18^{+3.76}_{-2.96}$ & $4.40^{+3.64}_{-3.02}$ \\
& $\alpha$ & $-1.06^{+3.04}_{-0.46}$ & $-0.77^{+2.81}_{-0.48}$ & $-0.29^{+2.71}_{-0.74}$ & $-0.31^{+1.06}_{-0.50}$ & $0.01^{+1.50}_{-0.66}$ \\
& $C$    & $9.99^{+6.87}_{-6.80}$ & $9.71^{+7.00}_{-6.78}$ & $9.91^{+6.86}_{-6.76}$ & $11.40^{+5.93}_{-6.88}$ & $11.38^{+5.99}_{-6.91}$ \\
& $\beta$  & $0.06^{+0.68}_{-1.34}$ & $-0.01^{+0.67}_{-1.28}$ & $0.34^{+0.82}_{-1.35}$ & $0.33^{+1.19}_{-1.47}$ & $-0.48^{+1.10}_{-1.10}$ \\
\hline
\multirow{5}{*}{Gas} 
& $y_0$    & $4.83^{+3.50}_{-3.39}$ & $5.41^{+3.15}_{-3.51}$ & $5.40^{+3.20}_{-3.40}$ & $4.52^{+3.59}_{-2.91}$ & $4.58^{+3.57}_{-3.07}$ \\
& $\alpha$ & $-1.29^{+3.69}_{-1.62}$ & $-0.85^{+0.31}_{-0.27}$ & $-0.33^{+0.48}_{-0.31}$ & $0.52^{+1.08}_{-0.64}$ & $1.05^{+1.63}_{-0.95}$ \\
& $C$    & $10.11^{+6.75}_{-6.67}$ & $9.35^{+7.23}_{-6.64}$ & $9.29^{+7.22}_{-6.68}$ & $11.33^{+5.95}_{-6.63}$ & $11.56^{+5.78}_{-6.49}$ \\
& $\beta$  & $0.85^{+0.78}_{-1.48}$ & $-0.99^{+0.77}_{-0.69}$ & $-0.77^{+0.88}_{-0.82}$ & $0.35^{+1.14}_{-1.45}$ & $-0.66^{+1.32}_{-0.91}$ \\
& $f_{\mathrm{gas}}$ & $0.034^{+0.019}_{-0.012}$ & $0.035^{+0.019}_{-0.012}$ & $0.082^{+0.049}_{-0.030}$ & $0.142^{+0.081}_{-0.052}$ & $0.155^{+0.091}_{-0.055}$ \\

\hline
\end{tabular}
\label{tab:mnfw_vertical_compact}
\end{table*}

\begin{table}[htbp]
\centering
\caption{Comparison between theoretical $\fhalo$ and simulated $f_{\mathrm{CGM}}$}
\begin{tabular}{cccc}
\hline
Redshift $z$ & $\fhalo^{\mathrm{Press}}$ & $\fhalo^{\mathrm{Tinker}}$ & $f_{\mathrm{CGM}}^{\mathrm{sim}}$ \\
\hline
0.000  & 0.1899 & 0.1407 & 0.1662 \\
0.106  & 0.1758 & 0.1318 & 0.1683 \\
0.287  & 0.1527 & 0.1176 & 0.1619 \\
0.508  & 0.1268 & 0.0991 & 0.1493 \\
0.754  & 0.1021 & 0.0805 & 0.1378 \\
1.041  & 0.0790 & 0.0624 & 0.1230 \\
\hline
\end{tabular}
\label{tab:fhalo_vs_fCGM}
\end{table}
    
\section{Redshift Evolution of Baryons in Different Cosmological Components}
\label{app:baryon_evolution}
In this appendix, we present a comprehensive view of the redshift evolution of baryonic mass fractions across different cosmological components and classification schemes.

\begin{figure*}[htbp]
    \centering
    \includegraphics[width=1\textwidth]{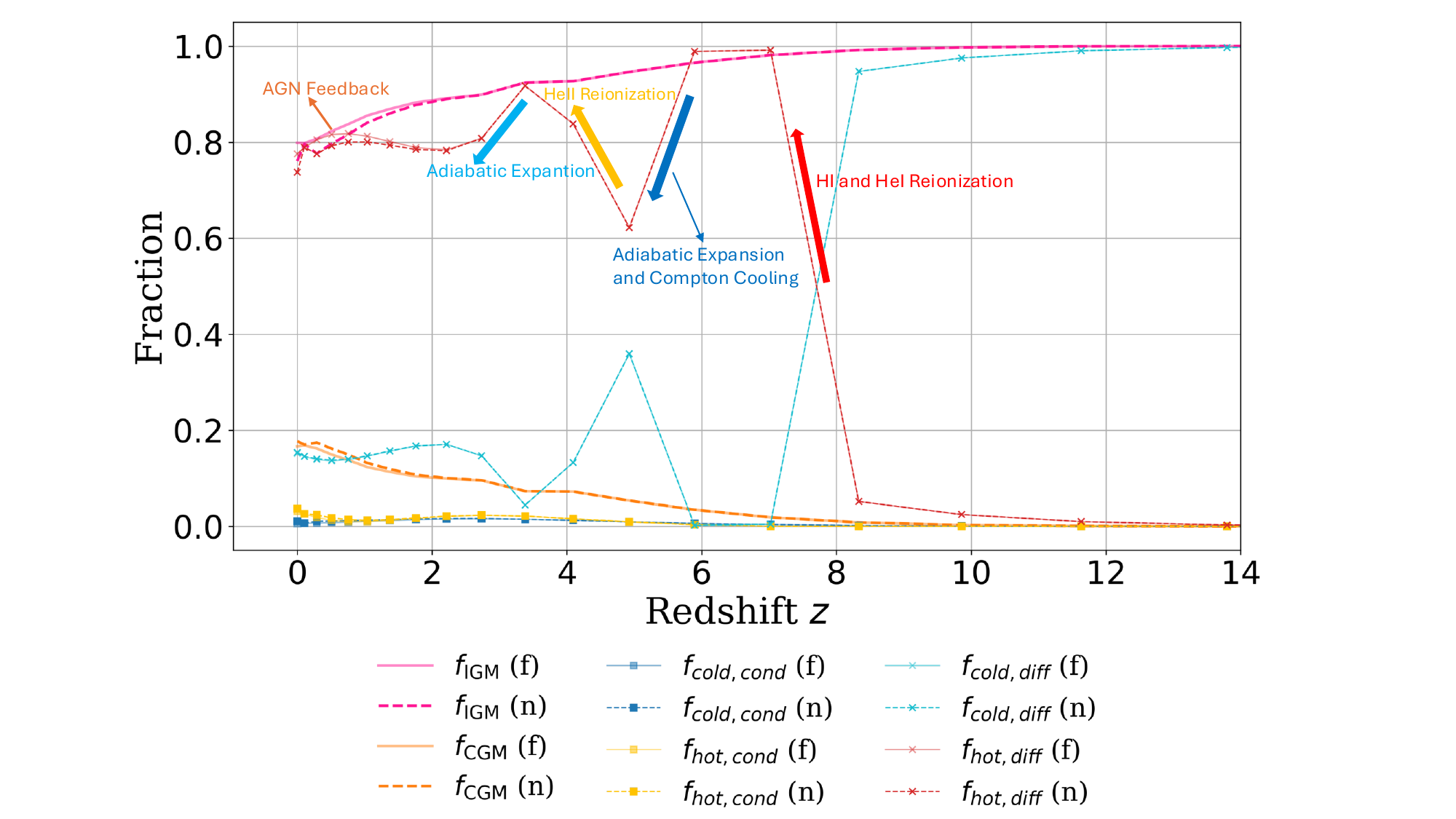} 
    \caption{Redshift evolution of baryonic mass fractions under two classification schemes using a CGM cutoff of $R_{\mathrm{cut}} = R_{200}$. Solid and dashed lines represent results from the fiducial and NoBH models, respectively. For the structure-based classification, the IGM and CGM are shown as pink and orange lines (without markers). For the phase-based classification, square markers denote condensed phases and \texttt{x} symbols denote diffuse phases. Cold and hot components are colored using cool and warm tones, respectively: deep blue for $f_{\mathrm{cold,\,cond}}$, yellow for $f_{\mathrm{hot,\,cond}}$, cyan-green for $f_{\mathrm{cold,\,IGM}}$, and bright red for $f_{\mathrm{hot,\,IGM}}$.}
    \label{fig:f_gas_evolution}
\end{figure*}

\begin{figure*}[htbp]
    \centering
    \includegraphics[width=0.8\textwidth]{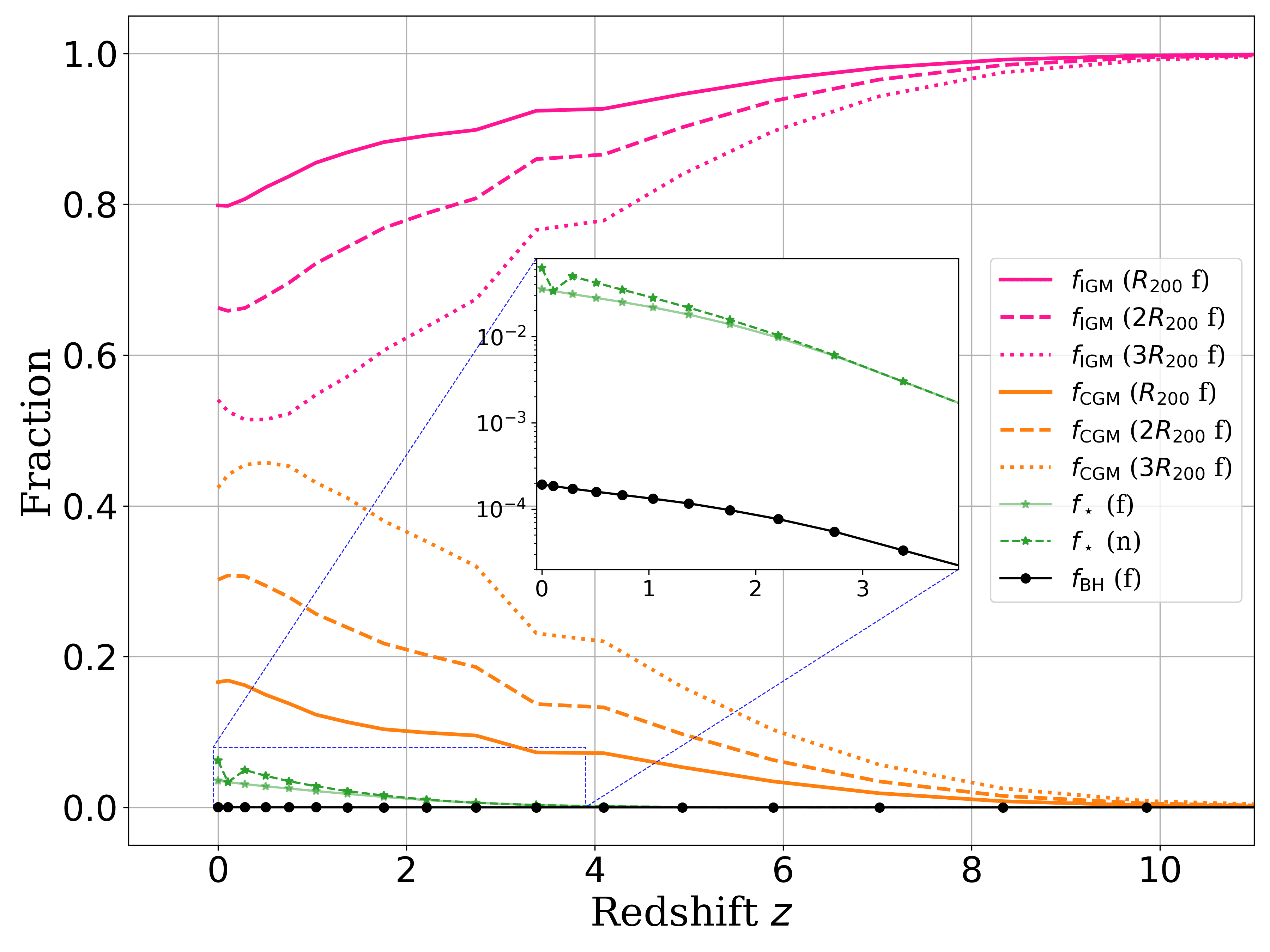} 
    \caption{Redshift evolution of baryon mass fractions in various cosmological components. The curves for $f_{\mathrm{CGM}}$ (orange) and $f_{\mathrm{IGM}}$ (pink) are plotted for different $R_{\mathrm{cut}}$ values: solid ($R_{200}$), dashed ($2R_{200}$), and dotted ($3R_{200}$). The evolution of $f_{\star}$ (green, five-pointed star markers) and $f_{\mathrm{BH}}$ (black, solid circles) is also shown for both fiducial and NoBH models. The inset panel zooms in on the redshift range $z < 4$ to highlight the detailed evolution of the stellar and black hole mass fractions.}
    \label{fig:f_structure_evolution}
\end{figure*}

\textit{IGM--CGM evolution:} Figure~\ref{fig:f_gas_evolution} shows the evolution of gas baryon fractions under both phase-based and structure-based classification. With a CGM cutoff of $R_{\mathrm{cut}} = R_{200}$, we find that $f_{\mathrm{IGM}}$ decreases from nearly unity at high redshift to $\lesssim 0.8$ at $z=0$, while $f_{\mathrm{CGM}}$ increases from zero to $\lesssim 0.2$. This reflects the gradual incorporation of intergalactic gas into halos. The impact of AGN feedback on baryon redistribution appears mainly below $z \lesssim 2$ and is relatively modest in our CROCODILE simulations.

\textit{Early phase evolution ($z > 8$):} At early times, $f_{\mathrm{cold, diff}}$ dominates due to the absence of collapsed halo structures and the overall low temperature of the gas. As redshift decreases below $z \sim 13$, $f_{\mathrm{cold, diff}}$ begins to decline slowly, while $f_{\mathrm{hot, diff}}$ shows a gradual increase. This transition reflects mild heating caused by adiabatic compression in regions of growing density perturbations, although the overall composition of the baryon phase remains largely unchanged during this epoch.

\textit{Reionization ($z \sim 8 \rightarrow 6$):} Reionization starts at $z \sim 8$, causing a rapid and nearly complete conversion of cold diffuse gas ($f_{\mathrm{cold,diff}}$) into hot diffuse gas ($f_{\mathrm{hot,diff}}$) by $z \sim 7$. After this point, both components stabilize, indicating the end of reionization and the establishment of a fully ionized diffuse medium.

\textit{Thermal evolution ($z \approx 6$ to 2):} After hydrogen reionization completes around $z \sim 6$, the diffuse IGM cools through adiabatic expansion and Compton scattering off the CMB, leading to a decline in $f_{\mathrm{hot,diff}}$ and a corresponding rise in $f_{\mathrm{cold,diff}}$.
From $z \sim 5$ to $z \sim 3.2$, HeII reionization driven by quasars gradually reheats the IGM (whose effect is included via the uniform UV background radiation), causing $f_{\mathrm{hot,diff}}$ to increase again.
Below $z \sim 3.2$, after the completion of HeII reionization, the IGM enters another cooling phase due to adiabatic expansion.
As a result, $f_{\mathrm{hot,diff}}$ begins to decline while $f_{\mathrm{cold,diff}}$ correspondingly rises.
We also note that this thermal transition is sensitive to the adopted temperature threshold: lowering the cutoff from $10000$ K to $8000$ K significantly suppresses the apparent oscillatory feature.

\textit{AGN feedback at low redshift ($z < 2$):} In this phase, AGN feedback becomes increasingly important and begins to heat and expel gas from condensed structures back into the diffuse phase. This effect is reflected by comparing the fiducial and NoBH simulations:
in the fiducial run, both $f_{\mathrm{hot,cond}}$ and $f_{\mathrm{cold,cond}}$ show lower values, while $f_{\mathrm{hot,diff}}$ increases correspondingly.
Thus, the apparent transition is not primarily a result of redshift evolution, but rather a manifestation of AGN-driven redistribution of baryons.

\textit{Structure-based view:} Figure~\ref{fig:f_structure_evolution} highlights how the evolution of $f_{\mathrm{CGM}}$ and $f_{\mathrm{IGM}}$ depends on $R_{\mathrm{cut}}$. For $R_{\mathrm{cut}} = 3R_{200}$, both fractions tend toward $\sim 0.5$ at $z = 0$. This emphasizes that the observationally inferred $\langle f_{\mathrm{diff,\,obs}} \rangle$ from FRBs is not a snapshot of the instantaneous IGM fraction at redshift $z$, but rather a LoS integrated average over all contributing structures at $z' < z$. At low redshift, many FRBs intersect no halos at all, resulting in a dominance of the IGM contribution to DM. The inset panel further shows that AGN feedback suppresses stellar growth in the fiducial model, while black hole growth remains limited throughout cosmic time.

\section{Fitting Parameters for the Models}
\label{app:fit_params}

Table~\ref{tab:fitting_parameters} summarizes the best-fit parameters for the C-Exp, CPL-Exp, and DPL-Exp models.

\section{Origin of FRBs}
\label{APP:FRB-Origin}
The formation mechanism of FRBs remains an active field of research, with magnetars being the leading candidate. This theory is supported by observations of the magnetar SGR 1935+2154, which is associated with a similar burst, FRB 200428 \citep{Bochenek2020}. However, other formation channels may contribute to the overall FRB population, albeit to a lesser extent.

One alternative formation scenario involves binary neutron star (BNS) mergers \citep{ZhangB2020ApJ...890L..24Z, Wang2020}. While BNS mergers have been proposed as a potential FRB production mechanism, their event rate is too low to explain the observed FRB population. \citet{Lipunov2014} estimated a BNS merger rate of approximately one per galaxy per 1000 years, far below the observed FRB event rate. Additionally, \citet{Zhang2020ApJ...893...44Z} suggested that only $\sim 6\%$ of repeating FRBs could originate from magnetars formed through BNS mergers. These results reinforce the idea that BNS mergers contribute only a minor fraction of all FRBs.

Although BNS mergers are unlikely to be a dominant FRB formation channel, they may generate offset FRB sources due to the high kick velocities imparted to neutron star binaries during supernova explosions. These kicks displace the progenitor system from the galactic center and lead to FRB sources located at significant distances from their host galaxies' centers, introducing a unique observational signature.

Investigating these offset FRBs provides insights into the kinematic evolution of FRB progenitors and the diversity of their host environments. Although BNS mergers may contribute to some offset FRBs, other mechanisms could also produce such spatial distributions. Their spatial distribution, while not the dominant formation pathway, offers a complementary probe of FRB progenitor mechanisms.
\clearpage 

\label{lastpage}
\end{CJK}
\end{document}